\documentclass[twocolumn,secnumarabic,amssymb, nobibnotes, aps, prl,
superscriptaddress, nobalancelastpage, floatfix]{revtex4}

 \usepackage{graphicx}
\usepackage{amsmath} \usepackage{braket} \usepackage{epsfig}
\usepackage{upgreek}
\usepackage{textcomp}
\usepackage[free-standing-units]{siunitx}
\usepackage{xcolor}
\usepackage{placeins}
\usepackage{multirow}
\usepackage{bm}

\newcolumntype{N}{>{\centering\arraybackslash}m{1.3in}}
\newcolumntype{M}{>{\centering\arraybackslash}m{1.0in}}
\newcolumntype{G}{>{\centering\arraybackslash}m{0.5in}}
\newcolumntype{R}{>{\raggedleft\arraybackslash}m{0.4in}}

\newcommand{\tot}{\ensuremath{\sigma_{tot}}}

\newcommand{\rxn}{\ensuremath{\sigma_{rxn}}}
\newcommand{\el}{\ensuremath{\frac{\lowercase{d}\sigma}{\lowercase{d}\Omega}}}

\newcommand{\oSix}{\ensuremath{^{16}}O}

\newcommand{\oEight}{\ensuremath{^{18}}O}
\newcommand{\oSixEight}{\ensuremath{^{16,18}}O}

\newcommand{\caForty}{\ensuremath{^{40}}C\lowercase{a}}
\newcommand{\caEight}{\ensuremath{^{48}}C\lowercase{a}}
\newcommand{\caAughtEight}{\ensuremath{^{40,48}}C\lowercase{a}}

\newcommand{\niEight}{\ensuremath{^{58}}N\lowercase{i}}

\newcommand{\niFour}{\ensuremath{^{64}}N\lowercase{i}}
\newcommand{\niEightFour}{\ensuremath{^{58,64}}N\lowercase{i}}

\newcommand{\rhThree}{\ensuremath{^{103}}R\lowercase{h}}

\newcommand{\snTwelve}{\ensuremath{^{112}}S\lowercase{n}}

\newcommand{\snFour}{\ensuremath{^{124}}S\lowercase{n}}
\newcommand{\snTwelveFour}{\ensuremath{^{112,124}}S\lowercase{n}}

\newcommand{\pbEight}{\ensuremath{^{208}}P\lowercase{b}}

\newcommand{\sOne}{s\ensuremath{_{1/2}}}
\newcommand{\pThree}{p\ensuremath{_{3/2}}}
\newcommand{\pOne}{p\ensuremath{_{1/2}}}
\newcommand{\dFive}{d\ensuremath{_{5/2}}}
\newcommand{\dThree}{d\ensuremath{_{3/2}}}
\newcommand{\fSeven}{f\ensuremath{_{7/2}}}
\newcommand{\fFive}{f\ensuremath{_{5/2}}}
\newcommand{\gNine}{g\ensuremath{_{9/2}}}

\newcommand{\hEleven}{h\ensuremath{_{11/2}}}

\begin{document}

\begin{abstract}
    The neutron total cross sections \tot\ of $^{16,18}$O,
    $^{58,64}$Ni, $^{103}$Rh, and $^{112,124}$Sn have been measured at the Los Alamos
    Neutron Science Center (LANSCE) from low to intermediate energies (3 $\leq E_{lab}
    \leq$ 450 MeV) by
    leveraging waveform-digitizer technology. The \tot\ relative differences between
    isotopes are presented, revealing additional information about
    the isovector components needed for an accurate optical-model (OM)
    description away from stability. Digitizer-enabled \tot-measurement
    techniques are discussed and a series of uncertainty-quantified dispersive optical model (DOM)
    analyses using these new data is presented, validating the use of the DOM for modeling light
    systems (\oSixEight) and systems with open neutron shells (\niEightFour\ and \snTwelveFour).
    The valence-nucleon spectroscopic factors extracted for each isotope reaffirm the usefulness of
    high-energy proton reaction cross sections for characterizing depletion from the mean-field
    expectation.
    \end{abstract}

\title{Isotopically resolved neutron total cross sections at
intermediate energies}

\author{C.~D.~Pruitt}  \email[Corresponding author: ]{pruitt9@llnl.gov}
\altaffiliation{Present Address: \textit{Lawrence Livermore National Laboratory, Livermore, CA
94550}}
\affiliation{Department of Chemistry, Washington University, St. Louis, MO 63130}
\author{R.~J.~Charity}
\affiliation{Department of Chemistry, Washington University, St. Louis, MO 63130}
\author{L.~G.~Sobotka}
\affiliation{Department of Chemistry, Washington University, St. Louis, MO 63130}
\affiliation{Department of Physics, Washington University, St. Louis, MO 63130}
\author{J.~M.~Elson}
\affiliation{Department of Chemistry, Washington University, St. Louis, MO 63130}
\author{D. E.~M.~Hoff}  
\altaffiliation{Present Address: \textit{Department of Physics, University of
        Massachusetts-Lowell, Lowell, MA USA 01854}}
\affiliation{Department of Chemistry, Washington University, St. Louis, MO 63130}
\author{K.~W.~Brown}
\affiliation{Department of Chemistry, Washington University, St. Louis, MO 63130}
\affiliation{National Superconducting Cyclotron Laboratory, Departments of Physics and
Astronomy, Michigan State University, East Lansing, MI 48824, USA}
\author{M.C.~Atkinson}
\altaffiliation{Present Address: \textit{TRIUMF, Vancouver, BC V6T 2A3, Canada}}
\affiliation{Department of Physics, Washington University, St. Louis, MO 63130}
\author{W.H.~Dickhoff}
\affiliation{Department of Physics, Washington University, St. Louis, MO 63130}

\author{H. Y. Lee}
\author{M. Devlin}
\author{N. Fotiades}
\author{S. Mosby}
\affiliation{Los Alamos National Laboratory, Los Alamos, NM 87545, USA}
\maketitle

\section{Introduction}
Neutron scattering is a direct, Coulomb-insensitive tool for probing the nuclear
environment. The simplest neutron-nucleus interaction quantity is 
the neutron total cross section, \tot, which provides information about
nuclear size and the ratio of elastic-to-inelastic components of nucleon 
scattering. Additionally, \tot\ data are thought to be tightly correlated with
a variety of structural nuclear properties of great interest
including the neutron skin of neutron-rich nuclei
\cite{Mahzoon2017} and thus the density dependence of the symmetry energy $L$,
an essential equation-of-state input for neutron-star
structure calculations \cite{Fattoyev2012, Vinas2014, Brown2000}.

\begin{figure}
    \includegraphics[width=\linewidth]{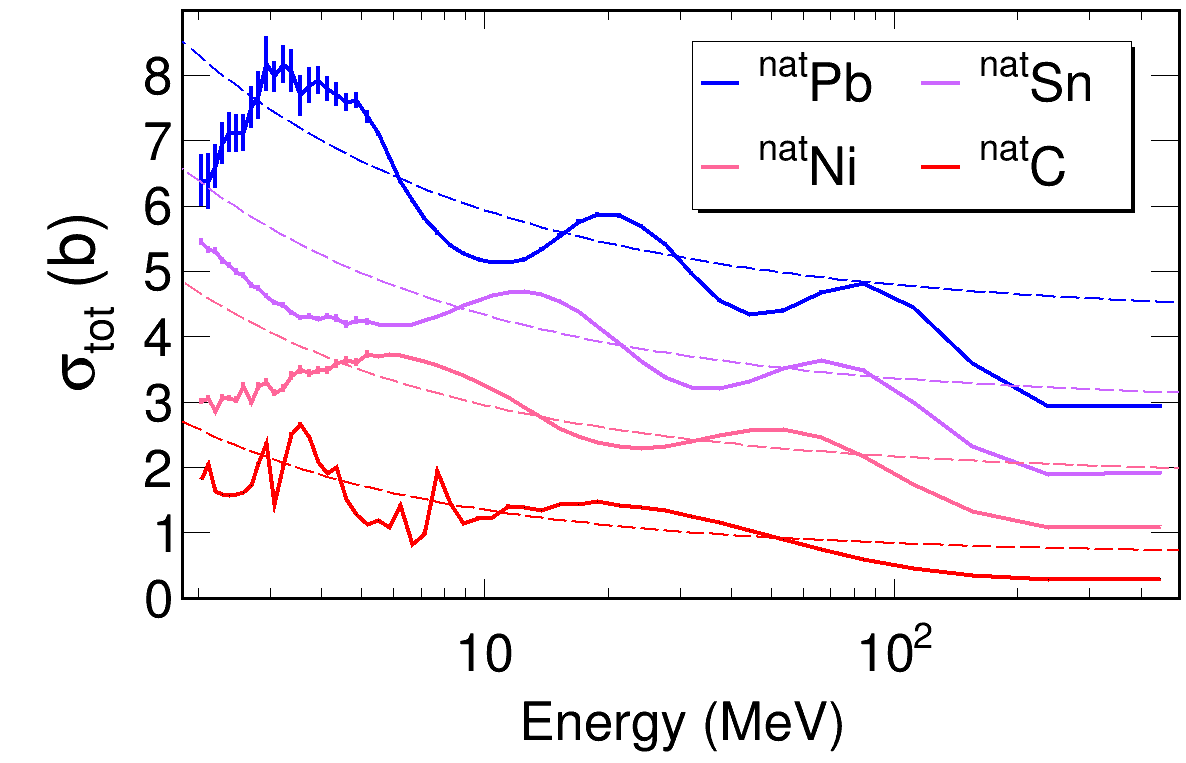}
    \caption{
        Experimental \tot\ data are shown from 2-500
        MeV for nuclides from $A$=12 to $A$=208
        \protect\cite{Finlay1993, Schwartz1974, Poenitz1983, Abfalterer2000, Abfalterer2001}.
        Predictions for \tot\ given by the ``strongly absorbing sphere'' (SAS)
        model [Eq. (\ref{SASAbsolute})], are shown as thin dashed lines for each nucleus.
        Regular oscillations about the SAS model are visible
        as is the trend for the oscillation
        maxima and minima to shift to \textit{higher} energies as $A$ is increased.
    }
    \label{SASphereVsExperiment}
\end{figure}

In the crude ``strongly-absorbing-sphere'' (SAS) approximation, where a target
nucleus absorbs incident neutrons passing within a nuclear radius,
\tot\ depends solely on the target nucleus size and the energy of the incident neutron:
\begin{equation} \label{SASAbsolute}
    \sigma_{tot}(E) = 2\pi(R + \lambdabar)^{2},
\end{equation}
where $R=r_{0}A^{\frac{1}{3}}$ and $\lambdabar$ is the reduced wavelength
of the incident neutron with energy $E$ in the center of mass \cite{Fernbach1949, Satchler1980}. 
While on \textit{average}, experimental \tot\ data comport with this na\"ive
model, the most prominent feature of experimental \tot\ data is the oscillatory
behavior centered about the average of Eq. (\ref{SASAbsolute}), visible in Fig.
\ref{SASphereVsExperiment}. Peterson \cite{Peterson1962} interpreted these oscillations as the 
result of a phase shift between neutron partial waves passing \textit{around} the 
nucleus (thus undergoing no phase shift) and waves passing
\textit{through} the nuclear potential, where they are refracted and exhibit a 
retardation of phase (an illustration is available in \cite{Satchler1980}).
This explanation was termed the 
``nuclear Ramsauer effect'' by Carpenter and Wilson \cite{Carpenter1959} based on 
the analogous effect seen in electron scattering on noble gases.

Following Angeli and Csikai \cite{Angeli1970}, this explanation can be
incorporated by imbuing the strongly-absorbing-sphere relations
with a sinusoidal term:
\begin{equation} \label{OscillatoryModel}
    \tot = 2\pi (R+\lambdabar)^{2}(1 - \rho \cos(\delta))
\end{equation}
where $\rho = e^{-\operatorname{Im}(\Delta)}$ and $\delta =
\operatorname{Re}(\Delta)$, $\Delta$ being the phase difference between a
partial wave traveling
around and traveling through the nucleus. The large amplitude of the
oscillations suggests that elastic scattering accounts for a
significant fraction of the total cross section, in turn implying a 
larger mean free path for neutrons through the nucleus 
than might otherwise be expected in the absence of Pauli blocking
\cite{Mohr1955, Feshbach1958}.
If we approximate the nucleus with a
real spherical potential of radius $R$ and depth $U$, the total phase shift $\delta$ is:
\begin{equation} \label{phaseShift}
    \delta =
    \frac{\overline{C}\left(\left[{\frac{E+U}{E}}\right]^{\frac{1}{2}}-1\right)}{\lambdabar}
\end{equation}
where $\overline{C} = \frac{4}{3}R$ is the average chord length through the
sphere \cite{Angeli1970}. Rearranging Eq. (\ref{phaseShift}) in terms of $A$ and $E$ and
discarding leading constants yields:
\begin{equation}
    \delta \propto A^{\frac{1}{3}}\times\left(\sqrt{E+U}-\sqrt{E}\right)
\end{equation}
This form reveals an important relation: as $A$ is increased, to maintain constant 
phase $\delta$, $E$ must also increase \cite{Satchler1980, Peterson1962}. 
This is contrary to a typical resonance condition where an integer number of wavelengths
are fit inside a potential; in that case, to maintain constant phase as $A$ is increased,
$E$ must be decreased. Thus these \tot\ oscillations have been referred to as
``anti-resonances'' or ``echoes'' \cite{Satchler1980, McVoy1967}.
Other authors \cite{Ahmad1973} have
exposed weaknesses in Angeli and Csikai's interpretation of
Eq. (\ref{OscillatoryModel}) and have provided a more general semi-empirical
equation for \tot. However, Eq. (\ref{OscillatoryModel}) is a valuable starting
point for connecting \tot\ with the depth and shape of the nuclear
potential as experienced by neutrons.

By including additional surface, spin-orbit, and other terms, OMs have been 
used to successfully reproduce the general features of all manner of single-nucleon scattering 
data across the chart of nuclides up to several hundred MeV \cite{Perey1976,
CH89, KoningDelaroche}. However, despite the excellent agreement with experiment, OMs
involve the interaction of many partial waves with many sometimes-opaque terms
in the potential, complicating intuitive understanding of the underlying
physics at play. In particular, the isovector components of optical potentials
are quite difficult to constrain as they depend on both proton and neutron 
scattering data, one or both of which are often unavailable. For example,
when Dietrich et al. conducted an analysis of neutron total cross section
differences between W isotopes, including standard isovector terms in their
optical potential \textit{worsened} the reproduction of experimental
relative differences, an illustration of how poorly these isovector components are known \cite{Dietrich2003}.

With these considerations in mind, our present goal is twofold: first, to
provide new isotopically resolved \tot\ data useful for identifying the 
dependence of optical 
potential terms on nuclear asymmetry; and second, conduct a DOM analysis of these new
\tot\ data along with a large corpus of scattering
and bound-state data to extract veiled structural quantities (e.g. neutron skin
thicknesses and spectroscopic factors, or SFs) for several cornerstone, closed-proton-shell nuclei.
Key findings of this DOM analysis are presented in the companion Letter \cite{Pruitt2020PRL}.

\section{Experimental Considerations}
By scattering secondary radioactive beams off of hydrogen targets in inverse
kinematics, proton-scattering experiments are possible even on highly unstable
nuclides. Because neutrons themselves must be generated as a
secondary radioactive beam, neutron-scattering experiments are restricted to
normal kinematics and \tot\ measurements are possible only for relatively stable
nuclides that can be formed into a target. At present, \tot\ measurements above
the resonance region on nuclides with short half-lives (shorter than the timescale of
days) are technically infeasible for this reason, though a handful have been carried out on
samples with half-lives in the tens to thousands of years \cite{Poenitz1983,
Phillips1980, Foster1971}.

Traditionally, \tot\ measurements have relied on analog-electronics techniques for recording
events, techniques that suffer from a large per-event deadtime of
up to several $\micro\second$. For a typical analog intermediate-energy \tot\ measurement
with dozens or hundreds of energy bins, achieving statistical uncertainty at the
level of 1\% requires a thick sample to attenuate a sizable fraction of the
incident neutron flux. If cross sections are in the 1-10 barn range, this means
sample masses of tens of grams \cite{Finlay1993, Abfalterer2001}.
Producing an isotopically enriched sample of this size is often
prohibitively expensive. As a result, there is a dearth of \tot\ data on
isotopically resolved targets from 1-300 MeV, even for
closed-shell isotopes of special importance like $^{3,4}$He, \oEight, \niFour,
\snTwelveFour, and $^{204,206}$Pb (see Fig. 1.3 in \cite{PruittPhDThesis}).

Recent developments in waveform digitizer technology have made it
possible to reduce the per-event deadtime by an order of magnitude or more,
enabling a corresponding reduction in the necessary sample size. In 2008, we
embarked on a campaign of \tot\ measurements on isotopically enriched samples
using these new technical capabilities,
starting with $^{40,48}$Ca from $15 \leq E_{lab} \leq 300$ MeV \cite{Shane2010}.
The data from that measurement were incorporated into several
DOM analyses \cite{Mueller2011, Mahzoon2014,
MahzoonPhDThesis} that yielded proton and neutron SFs, charge
radii, and initial estimates of the neutron skins \cite{Mahzoon2017}
for these nuclei.
Here we significantly expand on that effort by providing \tot\ results for
the important closed-shell nuclides
$^{16,18}$O, $^{58,64}$Ni, and $^{112,124}$Sn. We also present a measurement
on a very thin sample of the naturally monoisotopic $^{103}$Rh to demonstrate that
\tot\ experiments over a broad energy range using only a few grams of material are feasible.

\section{Experimental Details}
\begin{figure}
    \includegraphics[width=0.45\textwidth]{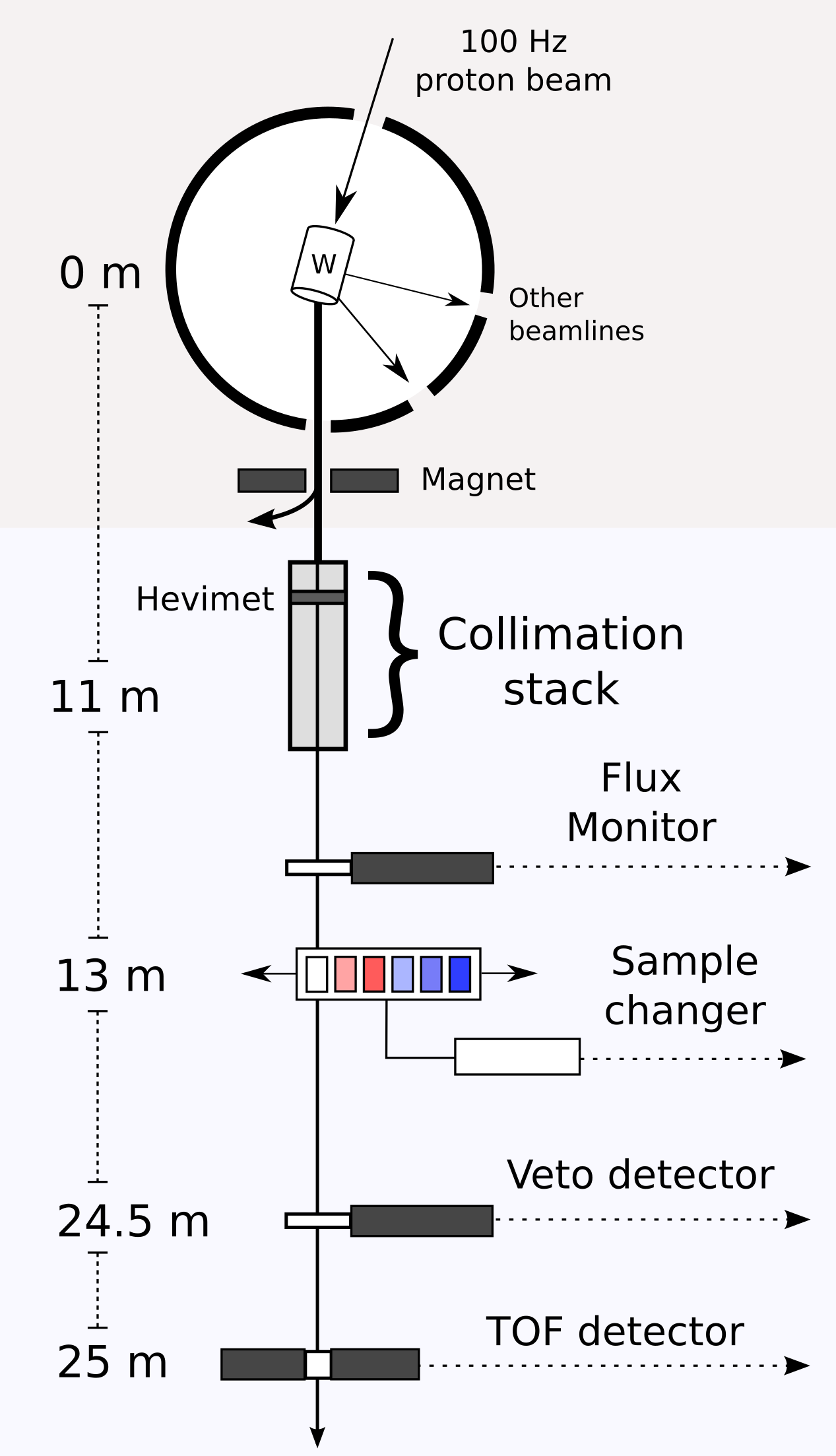}
    \caption{Experimental configuration at WNR facility.
        Samples are cycled into and out of the beam
        using a linear actuator with a period of 150 seconds. Times-of-flight (TOFs) are
    determined by the TOF detector and used to calculate neutron energies.}
    \label{ExperimentalApparatus}
\end{figure}

All \tot\ measurements were carried out at the 15R
beamline at the Weapons Neutron Research (WNR) facility of the Los Alamos
Neutron Science Center (LANSCE) during the 2016 and
2017 run cycles. Our experiment was modeled on previous
\tot\ measurements at WNR \cite{Finlay1993,Abfalterer2001,Shane2010}.
At WNR,
broad-spectrum neutrons up
to $\approx$700 MeV are generated by impinging proton pulses onto a water-cooled, 7.5
cm-long tungsten target (Fig. \ref{ExperimentalApparatus}). Before the beam
enters the experimental area, a
permanent magnet deflects all charged particles generated by the proton pulses, 
allowing only neutrons and $\gamma$ rays to reach the experimental area. At the
entrance to the experimental area, the beam was collimated to 0.200 inches using steel
donuts with a total thickness of 24 inches. In addition, the $\gamma$-ray content of the beam
was suppressed using a plug of Hevimet (90\% W, 6\% Ni, 4\% Cu by weight)
at the upstream entrance of the collimation stack.
After collimation, the beam passed successively through a flux 
monitor, the sample of interest, a veto detector, and finally the 
time-of-flight (TOF) detector approximately 25 meters from the neutron source.
All detectors consisted of BC-400 fast scintillating plastic mated with 
photomultiplier tubes (PMTs) and encased in either a plastic or
an aluminum housing. The flux monitor and veto detector each had
scintillator thicknesses of 0.25 inch and the TOF detector had a
scintillator thickness of 1 inch. Signals from all detectors and
the target changer were relayed to a 500-MHz CAEN DT-5730 waveform digitizer
running custom software. To improve time resolution, the TOF detector used two
PMTs (one left, one right) mated to the same plastic scintillator and the PMTs' signals were 
summed before digitization.

\begin{figure}
    \includegraphics[width=\linewidth]{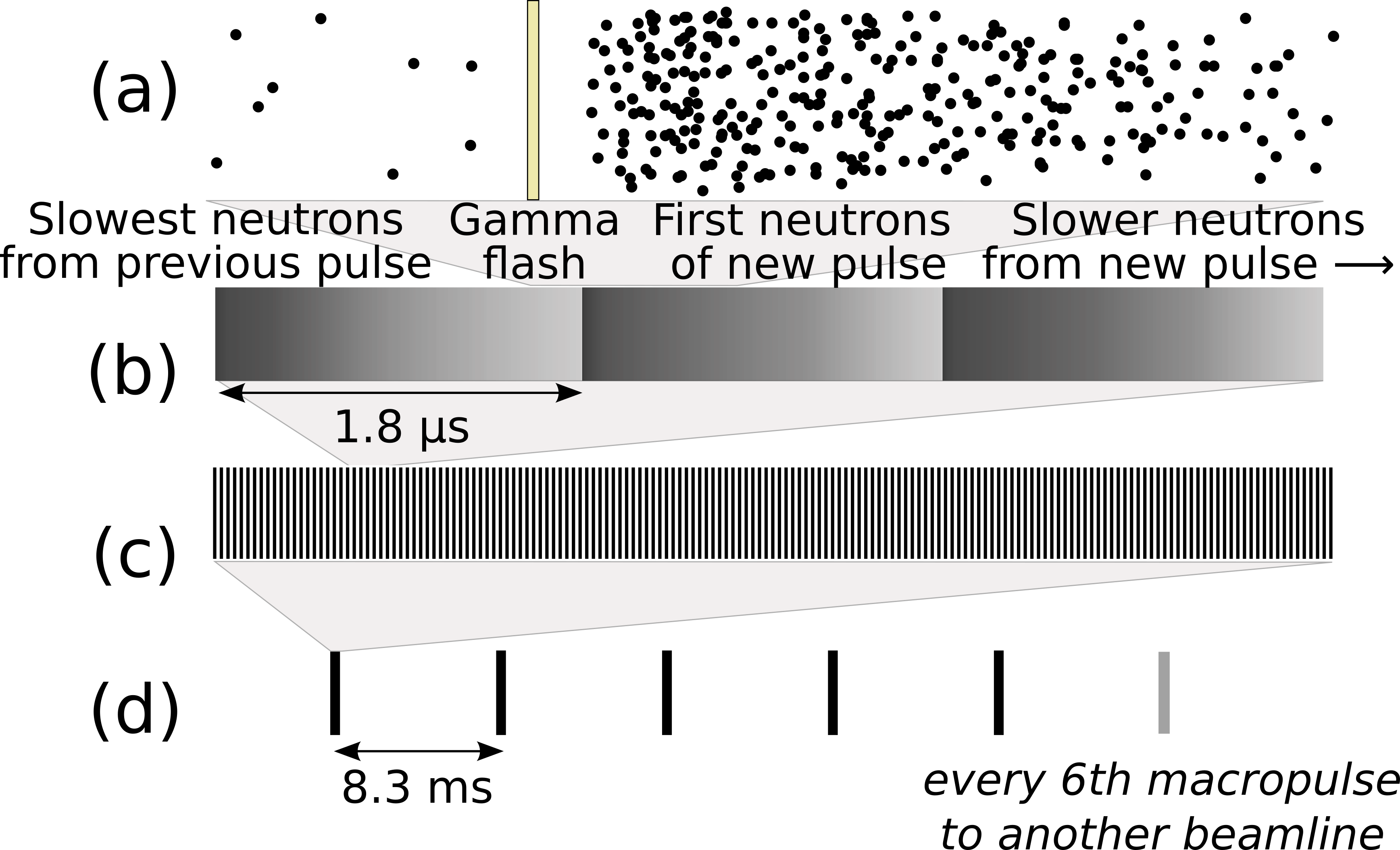}
    \caption{Neutron-beam structure at WNR facility.
        ``Macropulses'' of protons (d) are delivered to
        WNR's tungsten Target 4, where they generate neutrons by spallation.
        Each macropulse consists of
        $\approx$350 proton ``micropulses'' (c). Neutrons
        from each micropulse (b) disperse in
        time as they travel along the flight path so that $\gamma$ rays and high-energy 
    neutrons catch up to low-energy ones from the previous pulse (a).}
    \label{BeamStructure}
\end{figure}

The particular neutron beam structure at WNR dictates the energy range
achievable for \tot\ measurements (Fig. \ref{BeamStructure}).
Proton pulse trains, called ``macropulses'', are delivered to the tungsten target at 120 Hz.
Each macropulse consists of $\approx$350 individual proton pulses, called
``micropulses'', spaced 1.8 
\micro\second\ apart. Each micropulse consists of a single proton packet
that generates $\gamma$ rays and neutrons within a tight
temporal-spatial range. As neutrons from this micropulse travel along the beam path, 
high-energy neutrons separate in time from lower-energy neutrons so that neutron
energy can be determined by standard TOF techniques (see \cite{Moore1980} for details).
Because the $\gamma$ rays and high-energy neutrons from later micropulses can
overtake slower neutrons from an earlier micropulse, the distance of the TOF
detector from the neutron source determines both the minimum neutron energy that can be 
unambiguously resolved and the maximum instantaneous neutron flux, critical to correcting
for per-event deadtime.

A programmable sample changer with six positions
was used to cycle each sample into the beam at a regular interval of 150 seconds 
per sample. Once per macropulse, an analog signal from the sample changer
was recorded to indicate its current position.
The flux monitor was used to correct for variations in beam flux between 
macropulses. The veto detector suppressed events from charged-particle production 
in the samples and in air along the flight path.

Custom digitizer software was used to run the 
digitizer in two complementary modes, referred to as ``DPP mode'' and ``waveform 
mode''. In DPP mode, triggers were initiated by the digitizer's onboard
peak-sensing firmware. For each trigger, several quantities were recorded: the trigger 
timestamp, two charge integrals over the detected peak with different
integration ranges (32 ns for the short integral, 100 ns for the long integral),
and a 96-ns portion of the raw digitized waveform, referred to as a ``wavelet''.
DPP mode was used for the vast majority of the 
experiment and accounts for $\approx$99\% of the total data volume. In waveform mode, 
the digitizer performs no peak-sensing and was externally triggered. Upon 
triggering, the trigger timestamp and a very long wavelet (60 $\micro\second$) 
were recorded. While waveform mode data accounts for only $\approx$1\% of the total data, 
the instantaneous data rate is much higher than in DPP 
mode because hundreds of $\micro\second$ of consecutive waveform samples are 
stored. Roughly once every three seconds, the digitizer was switched to 
waveform mode for one macropulse, then switched back to DPP mode as quickly as
possible (10-40 ms, depending on run configuration).  

Except for the O and Rh samples, all samples were prepared as right
cylinders 8.25 mm in diameter and ranging from 10-27 mm in length (see
Table \ref{SampleCharacteristics} for sample characteristics). 
For each element studied, a natural-abundance sample
was also prepared as were two natural C
samples and a natural Pb sample, useful for benchmarking against
literature data. The samples
were inserted into styrofoam sleeves and seated in the cradles of the sample
changer. This design minimizes the amount of non-target mass proximate to the
neutron beam path. Our samples were generally
much smaller than those used in previous measurements;
for example, the Ni and Sn samples used in \cite{Abfalterer2001,
Finlay1993} had areal densities of 1.515 and 0.5475
mol/cm$^{2}$, respectively, 12.7 and 6.5 times larger than for our
Ni and Sn samples. 

\begin{table}[tb]
    \centering
    \caption{Physical characteristics of samples used for neutron \tot\ measurements.
    The relevant ``sample thickness'' for cross section calculations is the areal density of nuclei
        $\rho_{A}$, equal to the volumetric number density times
        the length of the sample. For liquid
        samples H$_{2}^{\text{nat}}$O, D$_{2}^{\text{nat}}$O, and H$_{2}^{18}$O,
        the length and diameter given are for the interior of the vessels
        used to hold the samples and the masses listed are calculated based on 
        literature values for the density of each sample at 25 C. Isotopic natural abundances (NA) and
        the abundances in our enriched samples (SA) are provided for reference.}

    \begin{tabular}{c c c c c c c}
        \small Isotope & Length & Diam. & Mass & $\rho_{A}$ & NA & SA\\
        \small & (mm) & (mm) & (g) & (mol/cm$^{2}$) & (\%) & (\%)\\
        \hline
        $^{\text{nat}}$C& 13.66(2)& 8.260(5)& 1.2363& 0.1921(1)& -& -\\
        $^{\text{nat}}$C& 27.29(2)& 8.260(5)& 2.4680& 0.3835(2)& -& -\\
        \\
        H$_{2}$O& 20.00(1)& 8.92(1)& 1.2461& 0.1107(3)& -& - \\
        D$_{2}$O& 20.00(1)& 8.92(1)& 1.3852& 0.1107(3)& 0.02& 99.9\\
        H$_{2}^{18}$O& 20.00(1)& 8.92(1)& 1.3844& 0.1107(3)& 0.20& 99.9\\
        \\
        $^{58}$Ni& 7.97(3)& 8.18(2)& 3.6438& 0.1197(3)& 68.1& 99.6 \\
        $^{\text{nat}}$Ni& 8.00(3)& 8.20(2 & 3.6898& 0.1192(3)& - & -\\
        $^{64}$Ni& 7.96(2)& 8.20(4)& 3.9942& 0.1192(6)& 0.93& 92.2\\
        \\
        $^{103}$Rh& 2.03(1)& 10.20(2)& 2.8359& 0.02426(4)& 100& 99.9\\
        \\
        $^{112}$Sn& 13.65(3)& 8.245(5)& 4.9720& 0.08332(5)& 0.97& 99.9\\
        $^{\text{nat}}$Sn& 13.68(3)& 8.245(5)& 5.3263& 0.08414(5)& - & -\\
        $^{124}$Sn& 13.73(3)& 8.245(5)& 5.5492& 0.08399(5)& 5.79& 99.9\\
        \\
        $^{\text{nat}}$Pb& 10.07(2)& 8.27(1)& 6.130& 0.05508(6)& -& -\\
        \hline
    \end{tabular}
    \centering
        \label{SampleCharacteristics}
\end{table}

The O isotopes were prepared as water samples to increase the areal density
of atoms and for ease of handling. Each water sample was contained by a
cylindrical brass vessel with thin brass endcaps (0.002 inches), and an
empty brass vessel served as the blank. \oSixEight\
cross sections were calculated by
subtracting the well-known H cross section from the raw H$_{2}$O results.
We used H \tot\  data sets from Clement et al. \cite{Clement1972} and Abfalterer
et al. \cite{Abfalterer2001}, which together cover the range $0.5 \leq E_n \leq 500$ MeV
and are in excellent agreement where their energy ranges overlap. In light of
the additional uncertainty inherent to this subtractive \tot\ determination,
we prepared a deuterated water sample,
from which the literature \tot\ for D$_{2}$ could be subtracted, to serve as an additional
cross-check. Due to
the poor machining properties of Rh, the \rhThree\ sample
was prepared by purchasing and stacking a series of thin discs rather than by
manufacturing a fused cylinder. These discs were held in place
by a cylindrical plastic case with open ends.

\section{Experimental Analysis}
The quantity of interest, \tot, is related to the flux
loss through a sample by:
\begin{equation}
    I_{t} = I_{0}e^{-{\ell\rho_{A}\sigma_{tot}}}
\end{equation}
or, equivalently,
\begin{equation}
    \tot = -\frac{1}{\ell\rho_{A}}\ln\left(\frac{I_{t}}{I_{0}}\right)
\end{equation}
where $I_{0}$ is the neutron flux entering the sample, $I_{t}$ is the neutron
flux transmitted through the sample without interaction, $\rho_{A}$ is the number
density of nuclei in the sample, and $\ell$ is the sample length. For thin
or low-density samples, flux attenuation through the sample will be small
(e.g., 13\% for our Ni samples at 100 MeV) and a large number
of counts will be required to determine the cross section to high
precision.

\begin{figure}
    \includegraphics[width=\linewidth]{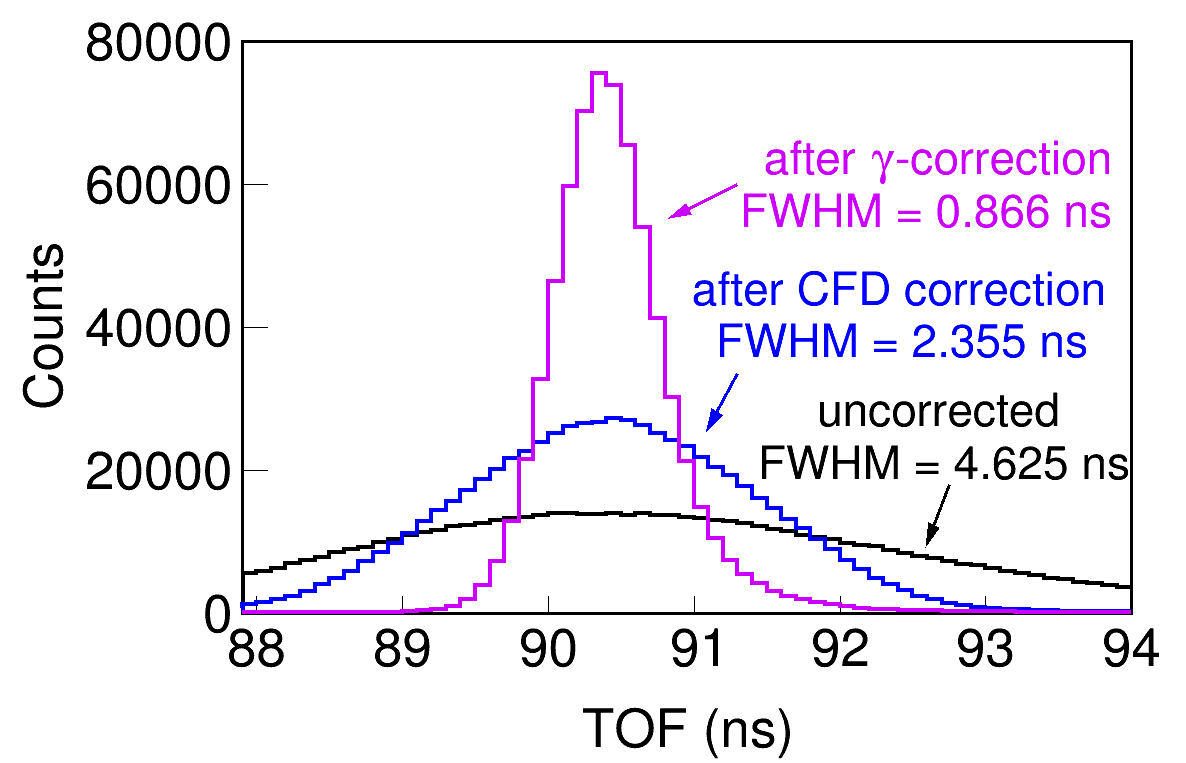}
    \caption{The effects of timing corrections on the $\gamma$-ray
        peak of a typical run are shown. The uncorrected spectrum is shown in black,
        the spectrum after correction with our software CFD is shown in blue,
        and the spectrum after correction with both our software CFD and
        $\gamma$-averaging is shown in magenta. For this run, the final $\gamma$-ray peak 
        FWHM after both corrections is 0.866 ns, comparable to the precision we
        achieved in our Ca study \cite{Shane2010}, which also employed $\gamma$-averaging.}
    \label{TimingCorrectionStudy}
\end{figure}

Two post-processing steps were used to improve TOF-detector timing resolution (see Fig.
\ref{TimingCorrectionStudy}). First, the waveform for each TOF-detector event
was passed through a software constant-fraction discriminator (CFD) logic, improving 
precision by a factor of two. Second, a $\gamma$-ray-averaging procedure (cf. \cite{Shane2010}) was
used to improve the precision of each micropulse start time. The final corrected TOF resolution (taken
as the FWHM of the $\gamma$-ray peak in the TOF spectra) ranged from
0.60-0.90 ns over the series of \tot\ measurements.
This is comparable to the resolution from 
our digitizer-mediated \tot\ measurement on Ca isotopes in 2008 \cite{Shane2010}.
For context, for a 100-MeV neutron and a TOF detector distance of 25 meters, a TOF 
uncertainty of 0.80 ns translates to an energy resolution of $\approx$900 keV.
For neutrons below $\approx$20 MeV, the TOF time resolution worsens because the traversal time 
through the 1-inch thickness of the TOF detector becomes non-negligible.
However, because the TOF of these neutrons is already very long (several hundred ns or
longer) the relative energy resolution ($\frac{\Delta E}{E}$) is
superior at low energies. As an example from one of our runs, a 5 MeV neutron with
a 0.82 ns detector-traversal time and an inherent TOF resolution of 0.80 ns
has an energy uncertainty of 13 keV. These energy uncertainties
have been propagated through subsequent analysis into our \tot\ results below.

Calculating the neutron energy requires knowledge of the flight-path
distance to high precision. We determined this distance by calculating 
putative \tot\ data for $^{\text{nat}}$C from 3-15 MeV from our measurement and 
comparing the resonance peaks in this region with high-precision literature data
sets. From this study, the mean TOF distance was determined as 2709 $\pm$1
\centi\meter\ for the Ni and Rh run configuration and 2554
$\pm$1 \centi\meter\ for the Sn and O run configuration.

Before cross sections could be tabulated, the per-event deadtime had to be
modeled and corrected for. Because events are not processed
instantaneously, there is a brief period
after each trigger during which the digitizer is busy processing that trigger.
Any newly arriving events in this period will be ignored,
privileging events arriving earlier and thus distorting
TOF spectra and resulting cross sections. This busy period is referred to as the
``analytic'' or ``per-event'' deadtime and can be corrected for according to standard 
techniques
\cite{Moore1980}. An additional complication is the possibility of flux
variation between micropulses. If there is no variation, the fraction of time
that the digitizer is dead for a given time bin $i$ can be calculated \cite{Moore1980}:
\begin{equation}
    F_{i} = \sum^{N-1}_{j=0} R_{(i-j)\text{ mod N}}\times P_{j}
    \label{DeadtimeEquation}
\end{equation}
where $N$ is the number of time bins in the micropulse, $R_{x}$ is the rate of
detected events per micropulse in bin $x$, and $P_{j}$ is the probability that the
digitizer is still busy from a trigger $j$ bins ago.
If the variation in beam flux is significant, a more advanced formula can be
used; however, an examination of our flux-per-micropulse data showed
very little flux variance across macropulses, except during the first 10\%
of the micropulses within each macropulse. In the final analysis we discarded these first
10\% and used the simpler Eq. (\ref{DeadtimeEquation}) to calculate the dead time fraction.

\begin{figure}
    \includegraphics[width=\linewidth]{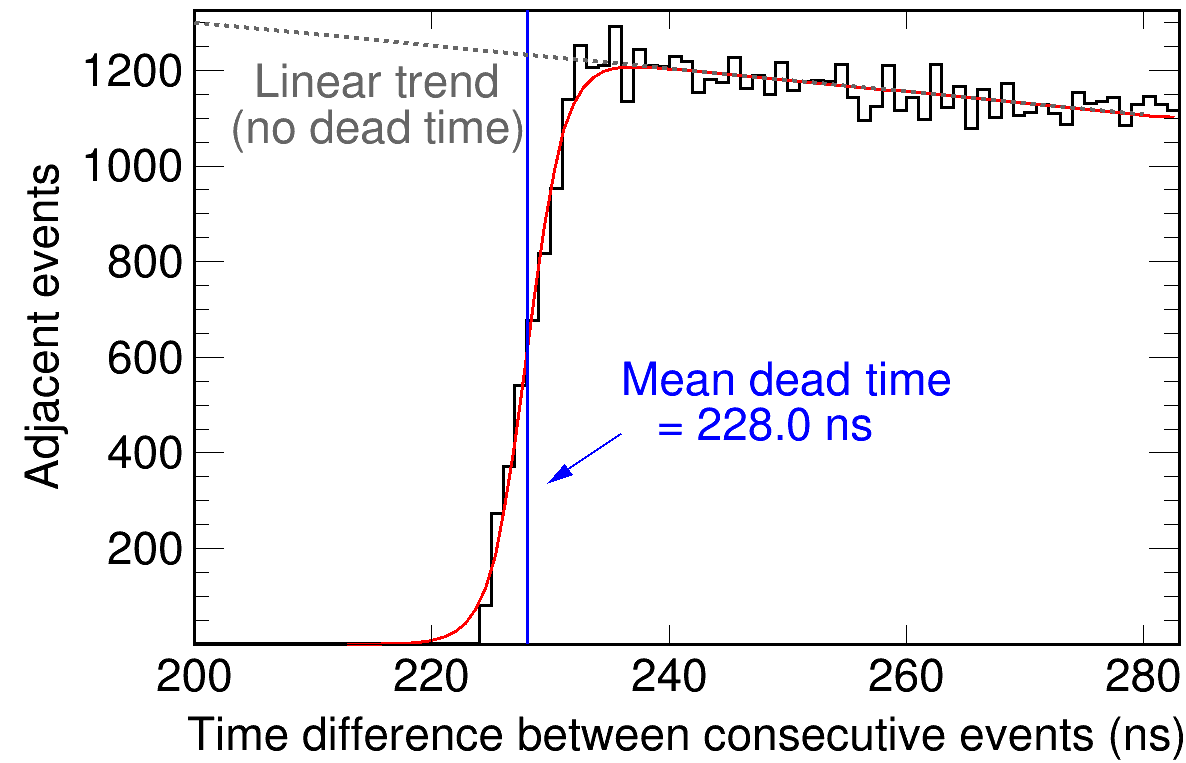}
    \caption{The time difference between adjacent TOF-detector
    events for a single run is plotted (black histogram). Below a certain
minimum time difference (the ``deadtime''), no events are recorded. A logistic
fit (red line) models the detector's deadtime response and is used to generate a
deadtime correction. The underlying linearly decreasing count rate (gray dashed
line) is incorporated into the logistic model. From the fit, a mean deadtime of
228.1 ns was extracted for the Sn and O run configurations (a similar
procedure was used to recover a deadtime of 159.7 ns for the Ni and Rh
run configurations).}
    \label{TimeDifferenceBetweenEvents}
\end{figure}

To model the experimentally observed probability-dead, $P_{j}$,
we fitted a logistic function to the observed spectrum for time
differences between consecutive events (Fig.
\ref{TimeDifferenceBetweenEvents}). For a given bin $i$, the fraction of time that the 
digitizer is dead, $F_{i}$, is a discrete convolution of the
\textit{measured} TOF spectrum with $P_{j}$. Note that except for the first and
last micropulses in a macropulse, all micropulses are consecutive, so deadtime effects can
``wrap around'' from the end of one micropulse to the next. For these wrap-around
contributions (that is, $j>i$), the (mod $N$) term ensures that the bin referred
to by $i-j$ is non-negative.

Because trigger processing is done in firmware onboard the digitizer,
the per-event deadtimes affecting our
measurement were reduced to between 150-230 \nano\second.
After we calculated the average probability-dead for each time bin,
the total number of events \textit{detected} in that bin, $N_{d}[i]$, could be
corrected to recover the \textit{true} number of events that would have been
detected in the absence of a per-event deadtime:
\begin{equation}
    N_{t}[i] = -\ln\left[1-\frac{\frac{N_{d}[i]}{M}}{(1-F_{i})}\right]\times M
\end{equation}
where $M$ is the total number of micropulse periods. At large TOFs (low energies) 
the correction is as low as a few percent,
but at small TOFs (high energies), the digitizer is often still dead
from the $\gamma$-ray flash and high-energy neutrons. In this regime
the correction can be quite large ($\approx$20\% for our Ni/Rh runs,
and $\approx$40\% for our Sn/O runs). Still, the corrections needed for our measurement
are far smaller than the typical analytic deadtime corrections required
with the deadtime mitigation scheme of previous analog measurements \cite{Finlay1993,
Abfalterer2001}.

In addition to analytic deadtime, there is an additional deadtime effect associated with 
digitizer readout to the data acquisition computer (DAQ). During data
collection, each pair of digitizer channels shares a common buffer for storing events.
After several seconds of acquisition, the digitizer begins readout at which time the
acquisition is paused and buffer contents are read out to the DAQ. However,
because each buffer is independently read out to the DAQ, it is possible that buffers
could be emptied and readied for new acquisition at slightly different times
(10-40 ms apart), and a mismatch could develop between the number of macropulses
seen on different channels. Such run-time interactions between the firmware and USB
traffic of the DAQ were difficult to characterize, but we estimate that they might cause a 
systematic error of a few tenths of one percent in the number of macropulses seen
by different channels, depending on the user-defined 
threshold and the buffer size. This effect could contribute to the discrepancy at the
highest energies ($>$100 MeV) between our results and past analog-enabled
measurements.

During analysis, it was noted that occasionally (1 in 400 macropulses), one or two 
adjacent macropulses would have an abnormally small number of events. The frequency
of these ``data dropouts'' was similar to the rate of
switching between DPP and waveform modes; we suspect it is related to edge
case behavior right before or after a mode switch. To mitigate this issue,
we threw out any macropulse that had less than 50\% of the average event rate in either the
flux monitor or TOF detector channel.

\begin{figure}
    \includegraphics[width=\linewidth]{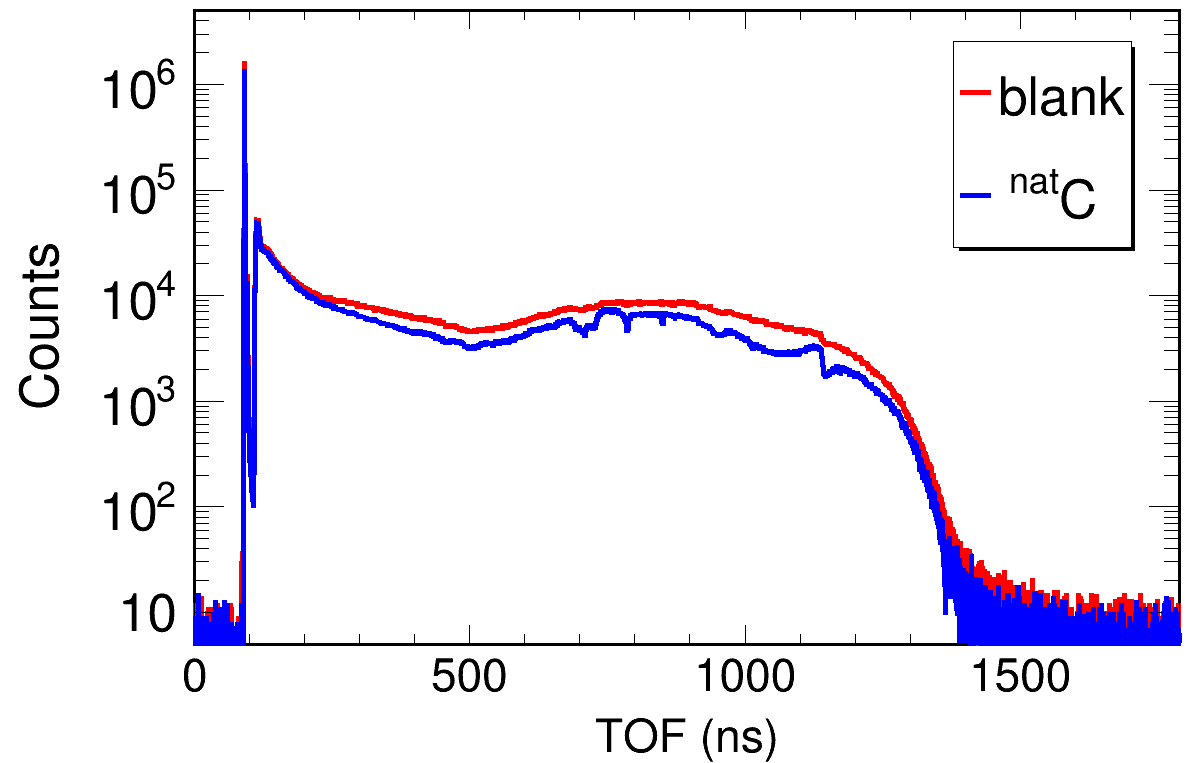}
    \caption{TOF spectra after the analytic deadtime correction and
        the veto and integrated charge gating for the blank sample (in
        red) and the $^{\text{nat}}$C sample (in blue), from the Ni/Rh experiment.
        The $\gamma$-ray peak is visible as a sharp spike at 90 ns, followed by
        the highest-energy neutrons at 130 ns.}
    \label{ExampleTOFSpectrum}
\end{figure}

After applying these corrections, the veto and integrated charge gates were applied to 
all events and surviving events were populated into TOF spectra (Fig.
\ref{ExampleTOFSpectrum}). Next, room background was subtracted (responsible for 0.1\% to 
1\% of event rate, depending on energy) and spectra were mapped to the energy domain.

From these energy spectra, the raw cross sections were calculated, bin-wise, as follows:
\begin{equation}\label{RawTCSEquation}
    \tot = -\frac{1}{\ell\rho_{A}}
    \ln \left(\frac{I_{0}}{I_{s}}\times\frac{M_{s}}{M_{0}}\right)
\end{equation}
where $I_{0}$/$I_{s}$ is the ratio of counts in the energy spectra between 
the blank and sample, $M_{s}$/$M_{0}$ is the ratio of counts in the
monitor detector between the sample and blank (for flux normalization).

Finally, two isotope-dependent corrections were applied to the raw cross
sections. First, because the blank sample contains air and not vacuum,
the cross section of air must be added to each sample's cross section.
Second, the cross section for $^{64}$Ni was corrected for the isotopic enrichment of our
sample (92.2\%) using our measured $^{\text{nat}}$Ni cross section. All other isotopes were 
sufficiently pure such that the impurity correction was negligible.

\begin{figure}[tb]
    \includegraphics[width=\linewidth]{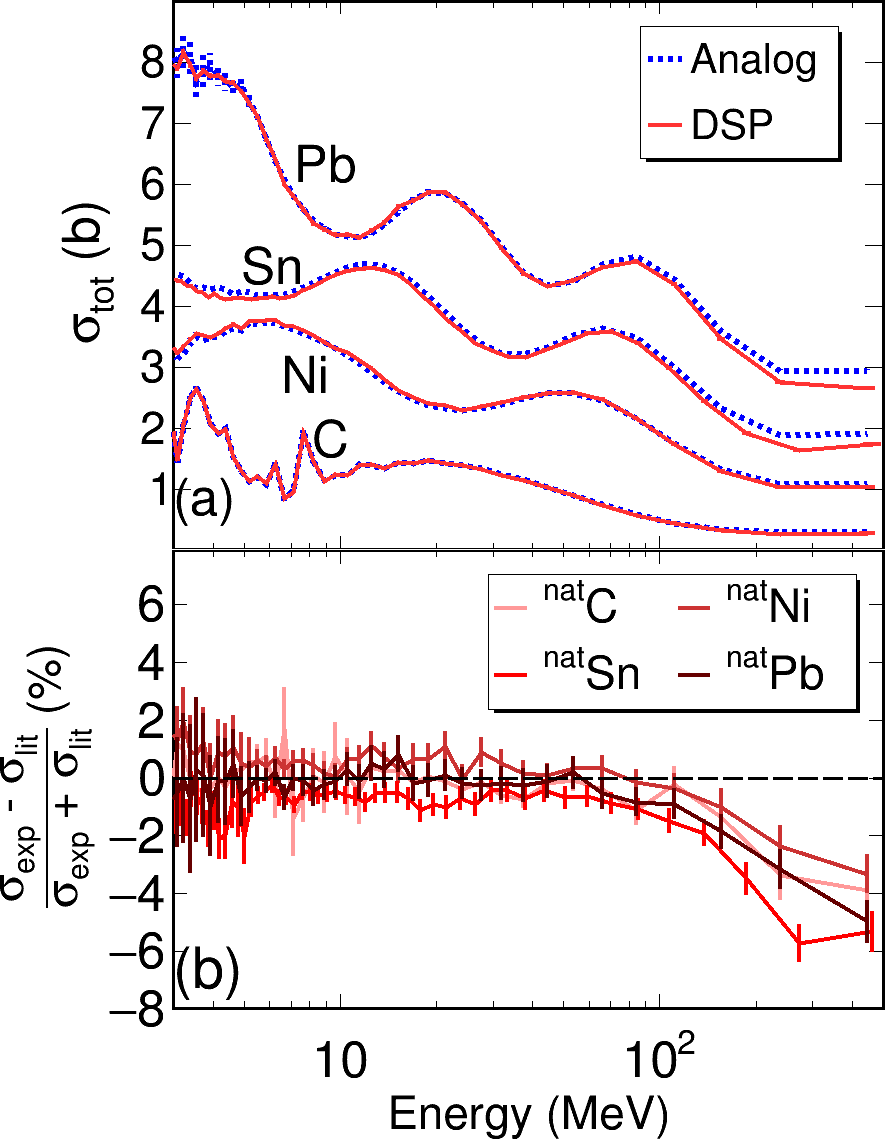}
    \caption{(a) A comparison of literature data (taken with analog
    techniques) and our results (signals processed with a digitizer, or ``DSP'')
    for natural C, Ni, Sn, and Pb. The absolute cross sections are shown from
    3-500 MeV. (b) Relative differences between the literature data and
    our data are shown in percent. From 3-100 MeV, our data are fully consistent with the
    literature but above 100 MeV, a difference arises, peaking at
    $\approx$5\% at 300 MeV.}
    \label{LiteratureBenchmarking}
\end{figure}

To validate our analysis, we first benchmarked our \tot\ measurements of natural samples
($^{\text{nat}}$C, $^{\text{nat}}$Ni, $^{\text{nat}}$Sn, and
$^{\text{nat}}$Pb) against the high-precision data sets on natural samples from
\cite{Finlay1993} and \cite{Abfalterer2001} (Fig.
\ref{LiteratureBenchmarking}). Our natural sample results
are in excellent agreement with 
these previous results from 3-100 MeV and show slight deviation above 100 MeV (a
relative difference of up to 5\% at 300 MeV), suggesting a small systematic
error at high energies in one or both approaches when the instantaneous neutron
flux is highest. As an additional diagnostic, we compared 
\tot\ results from our long and short natural carbon targets and
found excellent agreement, within 1\% throughout the measured energy domain.

\begin{figure}[tb]
    \centering
    \includegraphics[width=\linewidth]{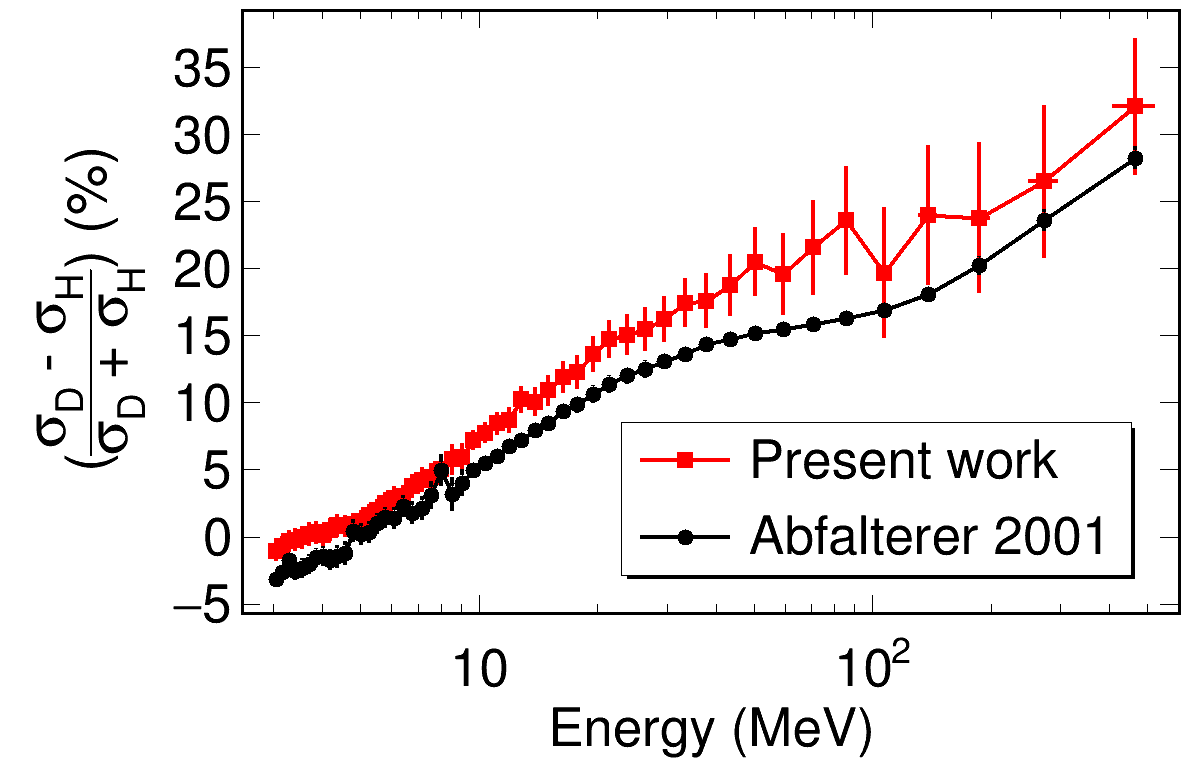}
    \caption[\tot\ relative difference between deuterium and hydrogen from our measurement]
    {The \tot\ relative difference between deuterium and hydrogen,
        as calculated by subtraction of our O \tot\ results from
        D$_{2}$O and H$_{2}$O. Data from our measurement are shown as red
        squares; the data of Abfalterer et al. \cite{Abfalterer1998}, which were generated using
        CH$_{2}$, C$_{8}$H$_{18}$, and D$_{2}$O targets, are shown as black circles.}
    \label{DtoH}
\end{figure}

Extracting the \oSixEight\ \tot\ required subtraction of the
well-measured \tot\ for H. To better characterize
the additional systematic uncertainty
associated with this subtractive analysis, we subtracted our measured
values for \oSix\ neutron \tot\ from our raw D$_{2}$O and H$_{2}$O data and 
calculated the D-to-H relative difference. A
comparison of our D-to-H relative difference with that of
\cite{Abfalterer1998} is shown in Fig. \ref{DtoH}.
Our results differ systematically from the previous (analog) measurement by 2-3\% throughout the
energy range, comparable to the 2\% systematic difference between our final
\oSix\ neutron \tot\ results and those of \cite{Abfalterer2001}. The size and
uniformity of these systematic differences is consistent with
a combination of slight ($\approx$1\%) normalization errors
in some or all of the H, D, O, and C neutron \tot\ results
from our measurement or in the literature data.

\section{Experimental Results}
\begin{figure*}[tb]
    \centering
    \includegraphics[width=\textwidth]{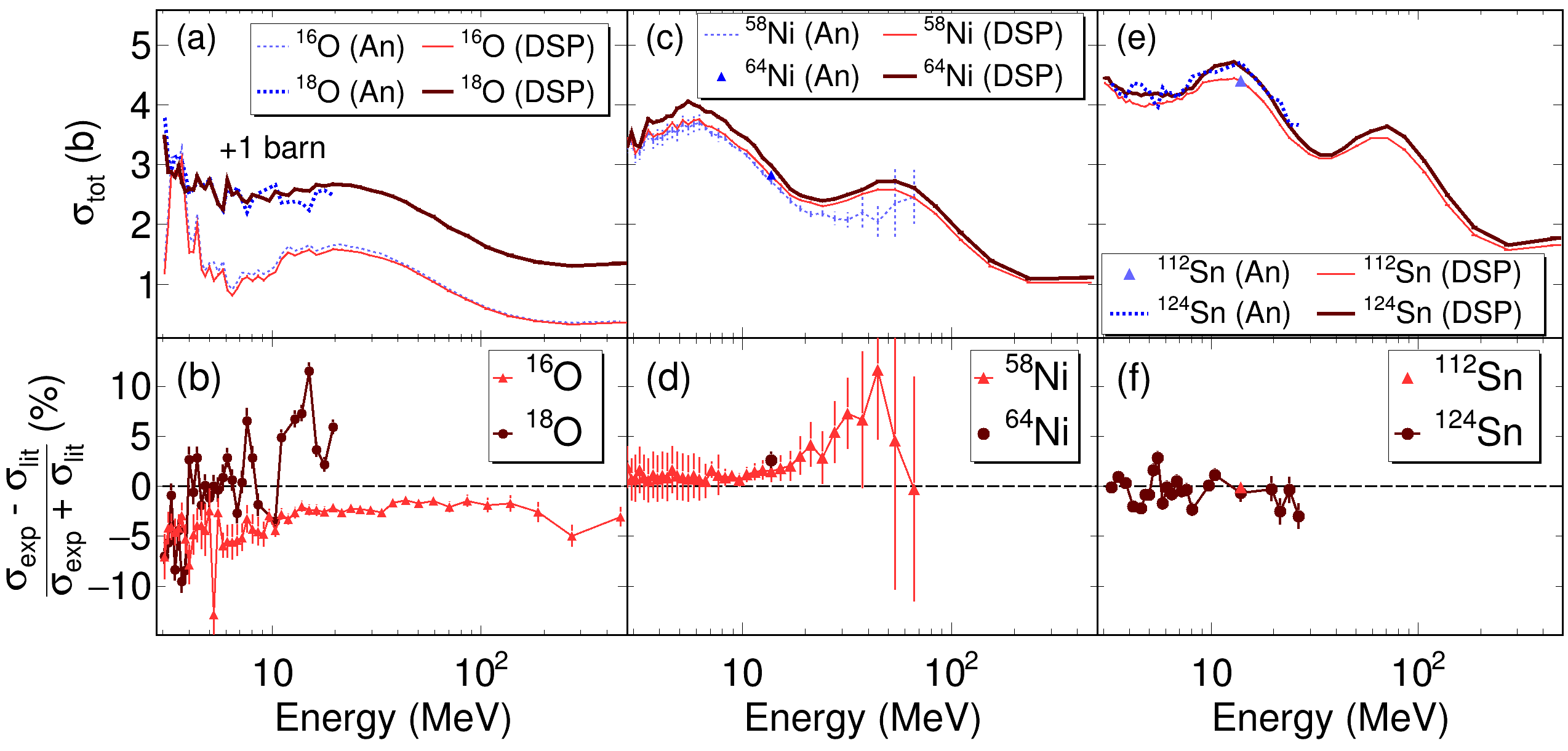}
    \caption[Neutron \tot\ for \oSixEight, \niEightFour, and \snTwelveFour: our results and literature data]
    {Neutron \tot\ for \oSixEight, \niEightFour, and \snTwelveFour: our results 
        and literature data.  In the upper three panels, our digitizer-measured
        isotopic results are shown in red and
        corresponding analog-measured literature data \cite{Finlay1993, 
        Perey1972, Vaughn1965, Salisbury1965, Perey1993, Dukarevich1967,
        Harper1982, Timokhov1989, Rapaport1980} are shown in blue.
        The data for \oEight\ have been
        shifted up by 1 barn for visibility.
        The lower three panels show residuals between our data and the
        literature data shown in the upper panels.
    }
    \label{SixPanel}
\end{figure*}

Our absolute \tot\ results for O, Ni, and Sn isotopic targets are shown in Fig.
\ref{SixPanel}. Results for Rh are shown in Fig. \ref{TwoPanelRh}.
Literature isotopic \tot\ measurements
(where they exist) are shown alongside our results for comparison.
Residuals between our data and any existing literature data are also shown.
In each figure, the literature data sets have been rebinned to match the bin
structure of our data to facilitate comparison. In regions with a low density of
states where individual resonances are visible (e.g., $^{\text{nat}}$C
below 10 MeV), this rebinning washes out the fine structure of the
cross sections.

Except for the already well-measured \oSix, our new data significantly
extend knowledge of the neutron \tot\ for each sample. In the cases of \oEight,
\niEight, \rhThree, and \snFour, almost no previous data were available
above 20 MeV. Our new data are in good agreement with the previous
measurements where available. In the cases of the rare isotopes \niFour\ and \snTwelve,
data were available at only one energy, 14.1 MeV, from a study from more than 50
years ago \cite{Dukarevich1967} and our measurement is in excellent agreement, within 2-3\%.

\begin{figure*}[tb]
    \centering
    \includegraphics[width=\textwidth]{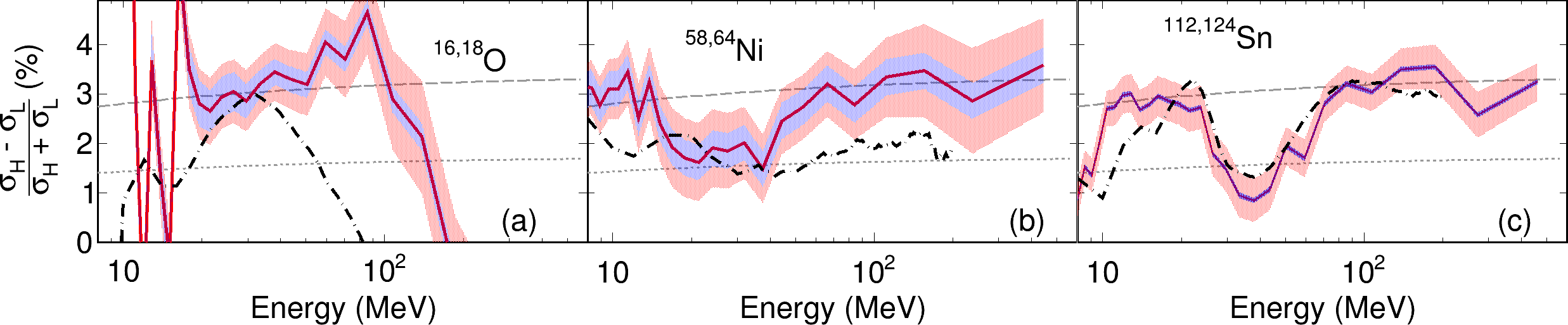}
    \caption[\oSixEight, \niEightFour, \snTwelveFour\ neutron \tot\ relative difference]
    {
        \oSixEight, \niEightFour, \snTwelveFour\ neutron \tot\ relative differences
        from our measurement. In each panel, the colored bands indicate
        regions of 1$\sigma$-uncertainty due to target thickness imprecision (blue) and from both target
        thickness and statistics (red). The gray dashed lines show the 
        prediction for the \tot\ relative difference per the strongly absorbing 
        sphere (SAS) model of Eq. (\ref{SASAbsolute}), which assumes a simple 
        $A^{\frac{1}{3}}$ size scaling for the nuclear radius. The gray dotted lines show the SAS model prediction but with an
        $A^{\frac{1}{6}}$ size scaling. The black dash-dotted lines shows the \tot\ relative
        differences from the median parameter values of the O, Ni, and Sn DOM analyses performed in
        this work (detailed in the following section). 
    }
    \label{ThreePanelRelDiff}
\end{figure*}

\begin{figure}[tb]
    \centering
    \includegraphics[width=0.5\textwidth]{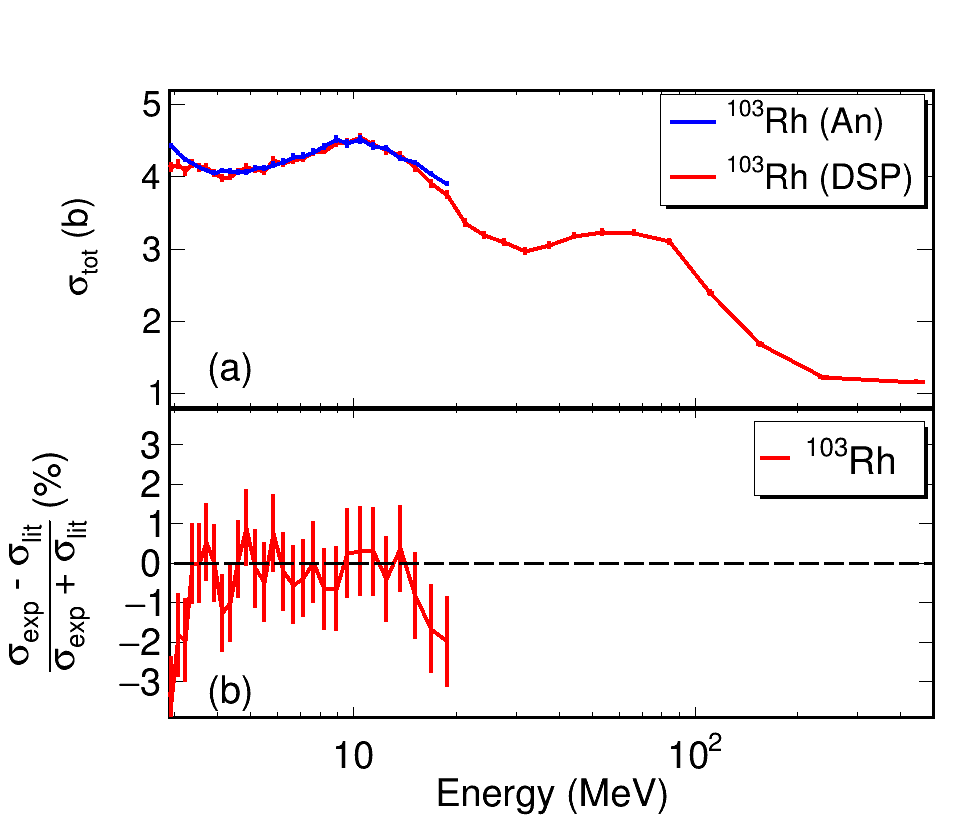}
    \caption[Neutron \tot\ for \rhThree: our results and literature data]
    {
        Neutron \tot\ for \rhThree: our results and literature data.
        In panel (a), our digitizer-measured results are shown in red and corresponding
        analog-measured literature data \cite{Poenitz1983} are shown in blue. Panel (b) shows the residuals
        between our data and the literature data, where it exists.
    }
    \label{TwoPanelRh}
\end{figure}

Our results for relative differences between isotopic pairs \oSixEight,
\niEightFour, and \snTwelveFour\ are shown in Fig. \ref{ThreePanelRelDiff}. For
\oSixEight\ [Fig. \ref{ThreePanelRelDiff}(a)], the purely isoscalar
SAS model [Eq. (\ref{SASAbsolute})] grossly reproduces the relative
difference below 100 MeV, but fails completely above 100 MeV. Near 200
MeV, the \oEight\ \tot\ crosses over that of \oSix\ resulting in a negative
relative difference, in keeping with the Ramsauer-logic expectation of Eq.
(\ref{OscillatoryModel}) that \tot\ oscillation minima shift to higher
energies as $A$ is increased. In the relative difference subfigures for 
\niEightFour\ and \snTwelveFour\ [Fig. \ref{ThreePanelRelDiff}(b) and \ref{ThreePanelRelDiff}(c)],
the average \tot\ values are below the
SAS model trend ($r \propto A^{\frac{1}{3}}$), shown by the dashed lines. 
The well-known $r \propto A^{\frac{1}{6}}$ trend in Sn isotope-shift data 
\cite{Anselment1986} is also shown for reference and
underpredicts the relative differences. In the DOM analyses presented below, we fit only
absolute \tot\ data and did not directly fit these relative 
differences. Still, the relative differences between our individual DOM fits for
\niEightFour\ and \snTwelveFour\ (black dashed-dotted lines) show overall agreement
with the experimental relative differences, especially for the Sn relative difference. For the
\oSixEight\ relative difference, there is an obvious phase mismatch between the oscillations of DOM calculation
and the experimental data. This mismatch is symptomatic of a slight DOM overestimation of the \oSix\
radius (0.02 fm), which nudges the DOM-calculated \oSix\ \tot\ rightward so that the \oEight\
crossover occurs at too low an energy. As was noted by Dietrich et al. in
their study of \tot\ relative differences in W isotopes, a \textit{simultaneous} OM
analysis along the entire isotopic chain, as in \cite{Mueller2011},
may be required to realize the full isovector-constraining power latent in the relative differences.


\section{DOM Analysis}
The DOM is a phenomenological Green's-function
framework enabling a simultaneous and self-consistent analysis of nuclear
structure and reaction data. An essential feature of the DOM is the enforcement of a dispersion
relation between the complex components of the self-energy across the entire
energy domain, allowing structural data from below the Fermi energy
(e.g., charge densities, bound levels) to help constrain the potential above,
and data from above the Fermi energy (e.g., elastic, reaction, and total
cross sections) to help constrain the potential
below. Using our new \tot\ data for \oSixEight, \niEightFour, and \snTwelveFour, we
performed a simultaneous fit on each isotopic pair and also
revisited \caAughtEight\ and \pbEight. Compared to previous DOM
analyses \cite{Mueller2011, Atkinson2018, Mahzoon2014, Mahzoon2017},
we employ an updated version of the DOM that has been generalized for use with
any combination of near-spherical even-even nuclei. Partial occupation of neutron open shells,
as for the neutron \dFive\ valence shell in \oEight, is accommodated using the level's energy $E$ and
the pairing parameter $\Delta$:
\begin{equation}
    \begin{split}
        \Delta(N,Z) \equiv \frac{1}{4}[B(N-2,Z)-3B(N-1,Z) \\
        +3B(N,Z)-B(N+1,Z)],
    \end{split}
\end{equation}
where $B(N,Z)$ is the binding energy of the nucleus with $N$ neutrons and $Z$ protons.
Occupation for the level is split into upper ($n_{+}$) and lower ($n_{-}$) components:
\begin{equation}
    n_{\pm} = \frac{1}{2}(1\pm\frac{\chi}{\s}),
\end{equation}
where $\chi \equiv E-\epsilon_{F}$, $s \equiv (\chi^{2}+\Delta^{2})^{\frac{1}{2}}$.
Only the lower (occupied) component is included in calculations of bound-state quantities
(e.g., total particle number, binding energy).

In the appendices, we provide the functional forms
used to define the potential (Appendix A), optimized parameter values with
uncertainties (Appendix B), and figures showing the quality
of the DOM reproduction to each experimental data set (Appendix C).
The other major methodological difference is the use of Markov-Chain Monte Carlo
(MCMC) for parameter optimization, discussed below. 

For additional details on the underlying DOM formalism, see \cite{Mahaux1991,
Dickhoff2018}. To calculate cross sections from the self-energy,
the standard R-matrix approach was used \cite{LaneThomas}. Except where indicated, experimental data used for
fitting are the same as in \cite{PruittPhDThesis}. To situate the reader, we describe the
corpus of experimental data and DOM results for \oSixEight\ in full detail. The
experimental data used and fit quality for \caAughtEight, \niEightFour,
\snTwelveFour, and \pbEight\ are similar in quantity and quality and only key differences are noted. For
systematics of neutron skins and binding energies, see companion Letter \cite{Pruitt2020PRL}.

\subsection{\oSix\ experimental data used in DOM analysis} \label{o16Results}
For protons, twenty-eight differential elastic cross
sections data sets and twenty analyzing power data sets from 10-200 MeV were
incorporated. Only three proton reaction cross section data sets, ranging from 20-65 MeV,
were available. As an added constraint, we used systematic trends
from the comprehensive proton \rxn\ review of Carlson \cite{Carlson1996} to generate proton
\rxn\ pseudo-data from 70-200 MeV, which were included in the fit.
These pseudo-data are shown as gray open symbols in the proton \rxn\ figures in
Appendix C. For neutrons, ten differential elastic cross section
data sets from 10 MeV to 95 MeV, a single
neutron reaction cross section data point at 14 MeV, and our newly measured \tot\ results
for $^{16}$O were included. In all, over sixty experimental nucleon scattering
data sets were used to constrain the \oSix\ parameters.

In addition to nucleon scattering data, several sectors of bound-state data were
included in the fit. Neutron (proton) 0\pOne\ and 0\dFive\
single-particle level energies were
assigned according to the nucleon separation energies of $^{16}$O and
$^{17}$O isotopes ($^{16}$O, $^{17}$F isotopes) \cite{AME2016}.
Charge density distributions were taken from the compilation of \cite{DeVries1987}.
Since the time of that compilation,
new experiments (particularly muonic-atom measurements) have improved the precision
of many root-mean-square (rms) charge radii by roughly an order of magnitude \cite{Angeli2013}.
To account for these improved data, we rescaled the distributions from
\cite{DeVries1987} to recover the updated
rms charge radii while still conserving particle number. We also fitted directly
to the updated rms charge radii of \cite{Angeli2013}.
Because the DOM self-energy does not necessarily conserve particle number, we
included the ``experimental'' proton and neutron numbers of eight as part of the
fit. Lastly, the total binding energy of \oSix\ from \cite{AME2016} was
included as a constraint. 

\subsection{\oEight\ experimental data used in DOM analysis}
Extensive proton elastic scattering data for \oEight\ was
available from the EXFOR database. Twenty-eight proton elastic differential cross
sections were included ranging from 10-200 MeV.
Unfortunately, no proton reaction cross section data were available at all in
the relevant range of 10-200 MeV. As with \oSix, we generated proton reaction cross section pseudo-data from
systematic trends in \cite{Carlson1996} from 70-200 MeV. On the neutron side, two
differential elastic cross section data sets were included, at 14 and 24 MeV,
but no analyzing powers were available. One datum for the neutron reaction cross
section, at 14.1 MeV, was incorporated as well.
Our \tot\ results for \oEight\ were the
sole neutron total cross section data used in the fit. The energies of the
proton and neutron 0\pOne\ and 0\dFive\ single-particle
levels were assigned according to the same procedure used for \oSix.

Unlike \oSix, for \oEight, no charge density distribution was available from
\cite{DeVries1987}. To approximate it, we rescaled the charge density
distribution used for \oSix\ to give the \oEight\ rms charge radius of
\cite{Angeli2013} while preserving eight units of charge.
As with \oSix, we also fitted to the
experimental rms charge radius directly, to the particle numbers $N$ and $Z$, and the
total binding energy.

\subsection{MCMC analysis}

Several aspects of the DOM potential make optimization challenging.
Even with the reduced number of potential parameters used in this work (42 for \pbEight\ and 43 for
all other pairwise fits) compared to past DOM studies (for example, 60 or more in \cite{Mahzoon2017}),
we found that classical gradient-descent methods were inappropriate for reliably searching the
parameter space. A recent study \cite{King2019} systematically compared Bayesian optical model
optimization techniques to frequentist ones, the type almost universally used in previous analyses,
and found that traditional algorithms may be overconfident in their parameter estimation.
To avoid these problems, we used the affine-invariant MCMC library,
\texttt{emcee} \cite{Foreman-Mackey2013}, for optimization and uncertainty characterization.
For an in-depth introduction to applied MCMC, see \cite{Sharma2017}. 

In the ensemble-sampling approach, several hundred
``walkers'' are first randomly initialized in parameter space for each isotopic system to be 
fitted. At each subsequent step $t$ during the
random walk, each walker's position is updated from $\vec{x}_{t} \rightarrow
\vec{x}_{t+1}$ either by accepting a new position $\vec{x}'$ with probability:
\begin{equation}\label{AcceptanceEquation}
    p(\vec{x}\rightarrow \vec{x}') = \min(1,\frac{U(\vec{x}'|D)}{U(\vec{x}|D)}),
\end{equation}
or by remaining in the same position $\vec{x}$ with probability
$1-p(\vec{x}\rightarrow \vec{x}')$. New positions are proposed according
to the stretch-move proposal distribution of \cite{Goodman2010} (for our stretch
move scaling, we used $\alpha = 1.3$ instead of the default $\alpha = 2.0$, which improved the
typical acceptance fraction from around 5\% to 15\%). In Eq.
(\ref{AcceptanceEquation}), the utility of a parameter vector conditional on the
experimental data $U(\vec{x}|D)$ was defined according to Bayes rule (omitting the
evidence term):
\begin{equation}
    U(\vec{x}|D) \propto L(D|\vec{x}) \times P(\vec{x}),
\end{equation}
where $D$ is the full set of constraining experimental data. The parameter prior distribution
$P(\vec{x})$ was specified as uniform over a
physically reasonable range for each parameter. For example, the
diffusenesses of all Woods-Saxon potential geometry terms were restricted
to 0.4-1.0 fm. Other more sophisticated choices for the prior distribution (e.g., broad truncated
Gaussians) were tested and had little impact on the resulting posterior distributions.
The likelihood function was defined as a least-squares function over all data sectors $d$:
\begin{equation} \label{LikelihoodFunction}
    L(D|\vec{x}) = \sum_{d} \frac{1}{N_{d}} \sum_{i=1}^{N_{d}}
    \left(\frac{y^{calc}_{d,i}-y^{exp}_{d,i}}{\sigma^{calc}_{d,i}+\sigma^{exp}_{d,i}}\right)^{2},
\end{equation}
where
\begin{itemize}
    \item $N_{d}$ is the number of experimental data points in a data sector $d$,
    \item $y^{calc, exp}_{d,i}$ are the calculated and experimental values,
        respectively, for the $i^{th}$ datum of sector $d$,
    \item $\sigma^{calc, exp}_{d,i}$ are the assigned model and experimental errors,
        respectively, for the $i^{th}$ datum of sector $d$.
\end{itemize}
Appendix A shows the parameter definitions and prior distributions used in the present
analysis.

Due to the choice of functional form and finite model basis size,
DOM predictions for nuclear observables suffer from inherent model error.
For example, many previous OM analyses tend to easily reproduce low-angle
experimental \el\ data taken at
lower scattering energies but are increasingly discrepant with the data
at high energies and at backward angles, where the predicted cross sections may
differ from experimental results by an order of magnitude or more. This discrepancy indicates a
deficiency in the potential form of the OM; ignoring it can
lead to drastic underestimation of variances of extracted
quantities. In this investigation, we found that the inclusion of reasonable model
discrepancy terms in our utility function improved the visual fit to experimental data while
broadening parameter uncertainties, in keeping with the methodological
findings of \cite{Brynjarsdottir2014}. Table \ref{ModelDiscrepancyTable} shows the model error terms we used 
for each data sector. We assigned model error for each data set
according to how well preliminary fits could reproduce differing regions of each
data sector, the flexibility of the functional forms, and intuition
from the successes and failures of past OM analyses.
In principle, the form of these model error terms could also be treated as random
variables to be sampled over during MCMC, but due to computational limitations and the
already-challenging size of the DOM parameter space, we elected to fix the model error terms.
After $N$ samples have been taken from the posterior distribution,
a subset can be used to estimate the true parameter distributions,
and physics results calculated for each sample. Ensuring that this subset is
representative of the true posterior is discussed in the next section.
\begin{table}[tb]
    \centering
    \caption{Model error terms for each data sector used in the
        MCMC utility function. For terms with units of \%, the
        model error was calculated as a percentage of the experimental data point magnitude.
        For \el\, the model error increased
        linearly with respect to the scattering angle in the
        center-of-mass frame with units of \% per degree. 
        $\epsilon_{nlj}$ are the single-particle energies for valence nucleons as calculated
        from separation energies in \cite{AME2016}.
        $r_{rms}$ is the root-mean-square charge radius and
        $\rho_{q}$ is the charge density distribution.}
    \renewcommand{\arraystretch}{1.2}
    \begin{tabular}{ c c c c c c c c c}
        \el\ & $A$ & \tot\ & \rxn\ & $\epsilon_{nlj}$ & $BE/A$ & $N,Z$ & $r_{rms}$ & $\rho_{q}$ \\ \relax
        (\%/$\degree$) & (-) & (\%) & (\%) & (MeV) & (\%) & (-) & (fm) & (\%) \\
        \hline
        0.25 & 0.10 & 0.25 & 0.25 & 0.10 & 5.0 & 0.10 & 0.005 & 1.0 \\
    \end{tabular}
        \label{ModelDiscrepancyTable}
\end{table}

Following \cite{Foreman-Mackey2013} we attempted an
autocorrelation analysis to test for convergence
and estimate the number of independent samples we had collected for each nucleus.
Because of computational limitations on the number of walkers and steps used to approximate
the posteriors, posterior estimation involves a finite MCMC sampling error.
The integrated autocorrelation time for a physics feature $f$, denoted $\tau_{f}$, represents
the number of steps required for a walker to produce a new, decorrelated posterior sample for the
feature that is independent of the previous independent sample. In an ideal MCMC analysis, $\tau_{f}$ could be
accurately computed for each physics quantity and the MCMC sampling error could be robustly estimated.
In practice, we found this to be computationally infeasible
for the DOM parameter space. For example, in preliminary analysis of \oEight, we were
able to perform $N=31000$ steps for each of 336 walkers (more than 100,000
CPU-hours in total). Over this domain, we calculated the
integrated autocorrelation time for each potential parameter $p$, denoted $\tau_{p}$, 
to be roughly 2800 steps. Assuming a $N > 100\tau_{p}$ rule-of-thumb condition for convergence of the
$\tau_{p}$ estimate near its true value, the decorrelation time appears to be extremely
long. In other words, from $\tau_{p}$ alone, we could not exclude the
possibility that the parameters had not yet fully ``settled'' in the region of their
optimal values and begun independent sampling of the parameter posteriors. We note that the true
$\tau_{f}$ \textit{could} be considerably smaller than $\tau_{p}$ due to the highly correlated
nature of DOM parameter space.

To proceed, we applied several commonsense tests to
judge whether our parameter and extracted-quantity estimates were accurate.
First, we sampled as long as possible and used as many parallel walkers as
possible, given our computational resources. From time to time during sampling,
we analyzed the mean walker positions and the mean walker position likelihood
as a function of sampling step. Encouragingly, for all nuclei walkers quickly
converged on a common region (within 1000 samples) and their mean parameter
values stabilized soon afterward (within 10000 samples), suggesting that walkers
were sampling a reasonably optimal subspace. At this point, we considered the chain
tentatively converged. As an additional test, we re-started sampling from a different
(uniformly random) initial position for each nucleus and found that a similar
optimal subspace was reached, again within roughly 1000 samples, indicating that
our results are independent of the initial walker positions. Finally, for a
``converged'' chain, we calculated extracted physics quantities (e.g., neutron
skins, scattering cross sections) for all walkers at several intervals
to confirm that their mean values were stable. Again using \oSix\ and \oEight\ as
an example, we found their mean neutron skin values varied by less
than 0.001 and 0.01 fm, respectively, over several thousand sampling steps late
in sampling. Out of caution (and given our expectation of very large autocorrelation times)
we used only the terminal sample for each walker chain to produce
the results presented here and in the companion Letter \cite{Pruitt2020PRL}.
In the end, we expect that additional sampling could slightly
reduce the estimated variance of each extracted quantity but have a negligible
effect on the mean values. For all quantities derived from MCMC analysis, the estimated
16\textsuperscript{th}, 50\textsuperscript{th}, and 84\textsuperscript{th} posterior percentile
values are denoted as $50^{84}_{16}$. The range between the 16\textsuperscript{th} and
84\textsuperscript{th} percentiles corresponds to a 1$\sigma$-uncertainty range if the posteriors
are assumed to be Gaussian. The median values and ranges for each parameter for each isotope
system are listed in Appendix B.


\subsection{Fit results on $^{16,18}$O}
Figure \ref{DOM_o16} in Appendix C shows
the DOM fit of \oSix\ and experimental data. The experimental proton \rxn, neutron \el, \tot, 
and \rxn\, charge density distribution, binding energy per nucleon,
and \pOne\ and \dFive\ single-particle energy data are all well-reproduced
suggesting that the DOM is effective for modeling nuclei as light as $A$=16.
Almost all experimental proton \el\ data are accurately reproduced by the DOM
calculations with the exception of an overprediction of cross sections
at backward angles and high energies, a regime known to be challenging from past
OM analyses. In addition, the median DOM-generated rms charge radius, 2.72 fm, slightly exceeds
the experimental value of 2.70 fm. Taken together with the \oSixEight\ relative difference results in
panel (a) of Fig. \ref{ThreePanelRelDiff}, these overestimations indicate that the traditional OM assumption of radial
proportionality with $A^{1/3}$ must be tweaked for a better description of \oSix.

To reproduce the \oSix\ proton \rxn\
pseudo-data generated from \cite{Carlson1996}, a larger volume imaginary term
was required above 100 MeV, which in turn reduced the spectroscopic strength for
the valence $\pi$ and $\nu$ \pOne\ nucleons by roughly 0.05.
We also note the importance of the charge density distribution for determining the 
magnitude of the imaginary strength below the Fermi energy. For example, in test
fits where the charge density was not included as a constraint, most of the
negative imaginary strength was concentrated in the surface term
between $-30 < E < \epsilon_{F}$ MeV, and the tail of the charge density was
overpredicted. With the charge density included as a constraint,
the imaginary surface magnitude shrank by a factor of two and the volume term
grew to compensate, pushing nucleon density deeper in energy space and
increasing the binding energy closer to the experimental value.

While all data sectors contributed at least some information not fully
captured by any other sector, the proton \rxn, neutron \tot, and charge density
provided the most stringent constraints on the self-energy. The analyzing powers
were the most difficult sector of experimental data to
reproduce, with moderate deviations visible from 10-15 MeV for both protons and
neutrons and above 100 MeV for protons [Figs. \ref{DOM_o16}(b) and \ref{DOM_o16}(d)].
Some of the difficulty with the analyzing powers is attributable to our neglecting of
an imaginary spin-orbit term in the DOM potential used in this work, a choice
made due to the unreasonable unbounded growth of the imaginary spin-orbit term
as $\ell$ grows in the traditional $\ell\cdot\sigma$ definition used in
\cite{KoningDelaroche}. In a future analysis we intend to quantitatively
investigate the importance of the imaginary spin-orbit term and to compare
different options for its functional form.

Figure \ref{DOM_o18} in Appendix C shows the \oEight\
experimental data and the DOM fit.
The paucity of \oEight\ experimental data presented a challenge for our analysis.
To constrain the negative-energy domain
of the potential, the only unambiguous experimental data were the neutron
and proton separation energies and the overall binding energy.
As with \oSix, broad agreement with experimental data was achieved for
experimental proton and neutron \el\ data, the neutron \tot, rms charge radius,
binding energy per nucleon, and \pOne\ and \dFive\ single-particle energy data.
The artificially scaled charge density and proton \rxn\ data were also
easily reproduced. Due to the deterioration of systematic trends from
\cite{Carlson1996} below 70 MeV, we did not generate proton \rxn\ pseudo-data
for lower energies, so the positive-energy surface term of the potential
was largely unconstrained in this important area.

\begin{table*}[t]
    \centering
    \renewcommand{\arraystretch}{1.2}
    \caption{Spectroscopic factors for valence proton ($\pi$) and neutron ($\nu$) levels,
    extracted from our DOM analysis. The 16\textsuperscript{th}, 
    50\textsuperscript{th}, and 84\textsuperscript{th} percentile values of the MCMC-generated
    posterior distributions are reported as $50^{84}_{16}$.}
    \begin{tabular}{c c c c c c c c c c c c}
        \multicolumn{3}{c}{Isotope} & \oSix & \oEight & \caForty & \caEight & \niEight & \niFour & \snTwelve & \snFour & \pbEight \\
        \hline
        \multirow{2}{*}{\large$\pi$} & & Level & 0\pOne & 0\pOne & 0\dThree & 0\dThree & 0\fSeven &
        0\fSeven & 0\gNine & 0\gNine & 2\sOne \\
        & & SF & $0.64^{0.70}_{0.58}$ & $0.59^{0.66}_{0.53}$ & $0.63^{0.70}_{0.55}$ &
        $0.62^{0.70}_{0.55}$ & $0.59^{0.65}_{0.55}$ & $0.57^{0.63}_{0.52}$ &
        $0.55^{0.61}_{0.52}$ & $0.56^{0.62}_{0.52}$ & $0.64^{0.70}_{0.58}$ \\
        \hline
        \multirow{2}{*}{\large$\nu$} & & Level & 0\pOne & 0\dFive & 0\dThree & 0\fSeven & 1\pThree &
        1\pThree & 1\dFive & 0\hEleven & 1\fFive \\
        & & SF & $0.63^{0.71}_{0.57}$ & $0.83^{0.79}_{0.87}$ & $0.62^{0.70}_{0.55}$ &
        $0.72^{0.77}_{0.65}$ & $0.72^{0.76}_{0.69}$ & $0.68^{0.75}_{0.64}$ &
        $0.65^{0.70}_{0.60}$ & $0.64^{0.70}_{0.59}$ & $0.67^{0.73}_{0.60}$ \\
    \end{tabular}
        \label{SpectroscopicFactors}
\end{table*}

In symmetric \oSix, the proton and neutron potentials were identical except for
the Coulomb interaction, so the neutron \tot\ data provided information about
both the proton and neutron imaginary strength at positive energies. For
\oEight, this expectation of symmetric potentials was inapplicable, making proton
\rxn\ data essential for fixing the positive-energy imaginary
strength for protons. In principle, \oEight\ proton and neutron
differential elastic scattering cross sections about 100 MeV could jointly
yield some information about the asymmetry-dependence of the imaginary strength for
\oEight, but no neutron elastic scattering data were available above 24 MeV. For
a better characterization of this nucleus, even a single proton \rxn\ datum
between 10 and 50 MeV would be valuable.

\subsection{Fit results for \\ \caAughtEight, \niEightFour, \snTwelveFour, and \pbEight}
Figures \ref{DOM_ca40}-\ref{DOM_pb208} in Appendix C show
\caAughtEight, \niFour, \snTwelveFour, and \pbEight\ experimental data and the DOM fits. 
The availability of single-nucleon scattering data for
\caAughtEight, \niEightFour, \snTwelveFour, and \pbEight\ followed the same trends as that
for \oSixEight: 
plentiful proton differential elastic scattering data, moderate coverage for
neutron differential elastic cross sections and proton reaction cross sections on abundant isotopes
(\caForty, \niEight, and \pbEight), with little-to-no coverage for neutron scattering
or proton reaction cross section data on rare isotopes (\caEight, \niFour, \snTwelve, \snFour).
For \snTwelve\ and \snFour, however, even proton elastic scattering data sets were sparse and no 
data above 50 MeV were available, making our newly collected neutron \tot\ data especially valuable
in constraining the potential. For \caForty\ and \pbEight, experimental proton reaction cross 
section data were available up to 200 MeV; for the other isotopes, proton reaction cross section 
pseudo-data (discussed in the \oSixEight\ subsections) were used as a constraint.
As for \oEight, no charge-density parameterization was available for \snTwelve\ in 
\cite{DeVries1987}, so we rescaled the available \snFour\ distribution to reproduce the \snTwelve\
charge radius.

Generally, all sectors of experimental data were well-reproduced; exceptions include
the high-angle (above 120$\degree$) proton elastic scattering data for \caForty\ and
\pbEight, where data sets were available up to 200 MeV, and the single-particle energies for 
neutron open shells in \snTwelveFour\ (see Figs. \ref{DOM_sn112} and \ref{DOM_sn124}), where 
several levels are partially filled and clustered near the Fermi surface. Achieving more accurate single-particle 
energies while preserving particle number accuracy may require a more sophisticated treatment of pairing.
Our new neutron \tot\ data were well-reproduced across the board, typically within 2\% of the experimental value,
by the DOM fits, suggesting that our Lane-like parameterization of the potential's asymmetry dependence [Eqs.
(\ref{HFDepthAsymmDependence}-\ref{ImSurAsymmDependence})] is a promising starting point for extrapolation away from stability.
We note that because \pbEight\ was fit on its own without an isotopic partner, initial fits showed
that the asymmetry-dependence of the HF radius term was too poorly constrained to yield reliable neutron skin
results; in the final treatment, this term was disabled for \pbEight.

\subsection{Discussion}
Table \ref{SpectroscopicFactors} shows DOM-calculated SFs for 
valence proton and neutron levels for all nine systems. Significant depletion from the mean-field expectation
appears even in the light systems \oSixEight. In the present study, the extracted proton SFs
show only a very weak dependence on neutron-richness within each isotopic pair,
in keeping with the weak dependence extracted in $(e,e'p)$ and transfer reaction studies
and at odds with knockout-reaction analyses that recover a strong asymmetry-dependence
\cite{Tostevin2014, Dickhoff2018}.
The recent DOM analyses of \cite{Atkinson2018, Atkinson2019} identified proton reaction cross sections above roughly 100 MeV as 
important for their successful reproduction of \caAughtEight\ $(e,e'p)$ cross sections without arbitrary SF rescaling. Compared to the
present work, these analyses found a much larger reduction of valence proton
SFs in \caEight\ with respect to \caForty, indicative of an SF asymmetry dependence somewhere between the weak
dependence deduced from transfer reactions and the very strong dependence from knockout reactions. 

To understand the differences between these analyses, we conducted several diagnostic runs with artificially
scaled Carlson pseudo-data in \caEight. These diagnostic runs confirmed that fitting to appropriate
high-energy proton reaction cross sections leads to larger \caEight\ proton imaginary strength both far above and far below the
Fermi energy, an effect already seen in previous DOM work. However, the growth we observed in the imaginary potential was
more modest compared to previous treatments, potentially explaining the weaker asymmetry-dependent
SF reduction. We also note that in the present
work, the high-energy neutron total cross sections and proton reaction cross sections appeared to have
little impact on other extracted quantities such as neutron skins, as had been previously hypothesized
for the neutron skin of \caEight\ \cite{Mahzoon2017}. We conclude that the different
methodological choices, especially the focus of this work on simultaneous fitting of isotope pairs,
is responsible for the differences in these asymmetry-dependent quantities. To further clarify the
situation, the potentials of the present work should be used to generate $(e,e'p)$ cross
sections that can be compared to the previous findings of \cite{Atkinson2019}.

Surprisingly, despite the extensive proton and neutron elastic scattering data for
\oSix, \caForty, and \pbEight, the extracted spectroscopic factor distributions and parameter
uncertainties for these isotopes are just as wide as for those systems with barely any available 
elastic scattering data, such as \niFour.
We tentatively conclude that the elastic scattering data we used are very weak constraints on the 
all-important imaginary terms of the optical potential, at least for the stable, spherical systems 
discussed here. Unfortunately, this suggests that elastic scattering measurements in inverse
kinematics on radioactive beams are of diminishing utility for extrapolating optical potentials
away from $\beta$-stability. A program of proton reaction cross section
and neutron total cross section
measurements on radioactive targets could be useful for understanding the
potential's near-Fermi-level asymmetry dependence but is experimentally daunting.
Instead, a two-pronged approach may be required. On the experimental side, proton 
reaction and neutron total cross section measurements on stable isotopic chains can help identify
which asymmetry-dependence forms are justifiable for increasingly asymmetric systems.
On the theoretical side, sensitivity studies are needed
to clarify how bound-state data on highly asymmetric systems connect to scattering cross sections.

Lastly, a few systematics in optical potential parameter values are worth mention.
For most of the parameters, there was minimal variation with nuclear size or asymmetry, 
suggesting that a global DOM treatment using the functional forms we have
selected is achievable. The radial term for the real central potential
($\mathbf{r_{1}}$) and for the positive-energy imaginary volume and surface ({$\mathbf{r_{4}^{+},
r_{5}^{+}}$) are nearly constant among \caAughtEight, \niEightFour, \snTwelveFour, and \pbEight, but the values for
\oSixEight\ show moderate deviations, another indication that the geometric form of the
potential is insufficient for light systems. As a consequence of the limited negative-energy data
available for fitting, the negative energy geometric terms ($\mathbf{r_{4}^{-}, r_{5}^{-}, a_{4}^{-},
a_{5}^{-}}$) show large variation. The nonlocalities for the negative imaginary components
are systematically larger than those for the positive imaginary components. This suggests that while
traditional OMs have been able to successfully reproduce positive-energy scattering data
with strictly local potentials, description of hole properties requires 
true nonlocal character in the negative-energy potential. In practice, we found it impossible to 
simultaneously reproduce charge density distributions, binding energies, and scattering data 
unless the central potential and at least the volume imaginary terms were equipped
with a nonlocality. In the end, for simplicity and generality, each element of the
potential (except Coulomb) was treated nonlocally, but it is unclear which particular data
are most important for constraining these several nonlocalities. As one moves further from stability
to systems with even less (or no) scattering data available, the risk of
overfitting will loom until this issue is resolved.

In preliminary fits, the imaginary volume magnitude ($\mathbf{A_{4}^{-}}$)
component of the potential was shown to be strongly sensitive to the inclusion of the binding 
energy as a constraint during fitting. We expect the asymmetry-dependence of this term,
($\mathbf{A_{vol,asym}^{-}}$), to impact DOM-based predictions
of the Ca, Ni, and Sn neutron driplines (as in \cite{Mueller2011}), though in this work, this dependence was very poorly
constrained due to the absence of experimental asymmetry-dependent data probing the most deeply
bound nucleons. Because they encode information about how protons and neutrons
share energy throughout the nucleus, experimental neutron-skin thicknesses could
provide this kind of valuable information. For the Ca, Ni, Sn, and Pb fits,
the median positive-energy surface imaginary magnitude
($\mathbf{A_{sur,asymm}^{+}}$) is positive, indicating enhancement in proton surface imaginary
strength with increasing neutron richness, and a corresponding decrease for neutron surface
imaginary strength. Of course, the nuclei under study in the present work are stable; the trend
for nuclei with large asymmetries, relevant for the r-process neutron-capture rate, is unknown.

\section{Conclusion}
By adopting a digitizer-driven
approach, we measured \tot\ on the important closed-shell nuclides
$^{16,18}$O, $^{58,64}$Ni, and $^{112,124}$Sn across more than two orders of
magnitude in energy (3-450 MeV). Except at the highest energies, our results
on natural targets are in good agreement with previous analog-mediated measurements
that required 10-20 times more target material. 

Using these new data and a suite of scattering and bound-state literature data
on \oSixEight, \niEightFour, and \snTwelveFour,
we extracted DOM potentials capable of reproducing a diverse range of scattering
and structural data for both neutrons and protons, validating the use of the
DOM away from doubly closed shells from $A$=16 to $A$=208, though with indications
that the traditional $A^{1/3}$ radial dependence may require modification for light systems.
These analyses further indicate that simultaneous fits of isotopically resolved neutron \tot,
proton \rxn, and charge-density distribution data on isotopic partners provide a more
stringent constraint on the asymmetry-dependence of both real and imaginary components.

\section{Acknowledgements}
This work is supported by the U.S. Department of Energy, Office of Science, 
Office of Nuclear Physics under award numbers DE-FG02-87ER-40316,
by the U.S. National Science Foundation under grants PHY-1613362 and PHY-1912643, 
and by the National Nuclear Security Administration of the U.S. Department of
Energy at Los Alamos National Laboratory under Contract No.  89233218CNA000001.
C.D.P. acknowledges support from the
U.S. Department of Energy SCGSR Program (2014 and 2016 solicitations) and the
National Nuclear Security Administration through the Center for Excellence in Nuclear
Training and University Based Research (CENTAUR) under grant number DE-NA0003841.
Computations were performed in part using the facilities of the Washington University
Center for High Performance Computing, which were partially provided through NIH
grant S10 OD018091, and in part under the auspices of the U.S. Department of Energy
by Lawrence Livermore National Laboratory under Contract DE-AC52-07NA27344.

\bibliography{references}

\begin{thebibliography}{59}
\expandafter\ifx\csname natexlab\endcsname\relax\def\natexlab#1{#1}\fi
\expandafter\ifx\csname bibnamefont\endcsname\relax
  \def\bibnamefont#1{#1}\fi
\expandafter\ifx\csname bibfnamefont\endcsname\relax
  \def\bibfnamefont#1{#1}\fi
\expandafter\ifx\csname citenamefont\endcsname\relax
  \def\citenamefont#1{#1}\fi
\expandafter\ifx\csname url\endcsname\relax
  \def\url#1{\texttt{#1}}\fi
\expandafter\ifx\csname urlprefix\endcsname\relax\def\urlprefix{URL }\fi
\providecommand{\bibinfo}[2]{#2}
\providecommand{\eprint}[2][]{\url{#2}}

\bibitem[{\citenamefont{Mahzoon et~al.}(2017)\citenamefont{Mahzoon, Atkinson,
  Charity, and Dickhoff}}]{Mahzoon2017}
\bibinfo{author}{\bibfnamefont{M.~H.} \bibnamefont{Mahzoon}},
  \bibinfo{author}{\bibfnamefont{M.~C.} \bibnamefont{Atkinson}},
  \bibinfo{author}{\bibfnamefont{R.~J.} \bibnamefont{Charity}},
  \bibnamefont{and} \bibinfo{author}{\bibfnamefont{W.~H.}
  \bibnamefont{Dickhoff}}, \bibinfo{journal}{Phys. Rev. Lett.}
  \textbf{\bibinfo{volume}{119}}, \bibinfo{pages}{222503}
  (\bibinfo{year}{2017}),
  \urlprefix\url{https://link.aps.org/doi/10.1103/PhysRevLett.119.222503}.

\bibitem[{\citenamefont{Fattoyev and Piekarewicz}(2012)}]{Fattoyev2012}
\bibinfo{author}{\bibfnamefont{F.~J.} \bibnamefont{Fattoyev}} \bibnamefont{and}
  \bibinfo{author}{\bibfnamefont{J.}~\bibnamefont{Piekarewicz}},
  \bibinfo{journal}{Phys. Rev. C} \textbf{\bibinfo{volume}{86}},
  \bibinfo{pages}{015802} (\bibinfo{year}{2012}),
  \urlprefix\url{https://link.aps.org/doi/10.1103/PhysRevC.86.015802}.

\bibitem[{\citenamefont{Vi{\~{n}}as et~al.}(2014)\citenamefont{Vi{\~{n}}as,
  Centelles, Roca-Maza, and Warda}}]{Vinas2014}
\bibinfo{author}{\bibfnamefont{X.}~\bibnamefont{Vi{\~{n}}as}},
  \bibinfo{author}{\bibfnamefont{M.}~\bibnamefont{Centelles}},
  \bibinfo{author}{\bibfnamefont{X.}~\bibnamefont{Roca-Maza}},
  \bibnamefont{and} \bibinfo{author}{\bibfnamefont{M.}~\bibnamefont{Warda}},
  \bibinfo{journal}{Eur. J. Phys. A} \textbf{\bibinfo{volume}{50}},
  \bibinfo{pages}{27} (\bibinfo{year}{2014}),
  \urlprefix\url{https://doi.org/10.1140/epja/i2014-14027-8}.

\bibitem[{\citenamefont{Brown}(2000)}]{Brown2000}
\bibinfo{author}{\bibfnamefont{B.~A.} \bibnamefont{Brown}},
  \bibinfo{journal}{Phys. Rev. Lett.} \textbf{\bibinfo{volume}{85}},
  \bibinfo{pages}{5296} (\bibinfo{year}{2000}),
  \urlprefix\url{https://link.aps.org/doi/10.1103/PhysRevLett.85.5296}.

\bibitem[{\citenamefont{Fernbach et~al.}(1949)\citenamefont{Fernbach, Serber,
  and Taylor}}]{Fernbach1949}
\bibinfo{author}{\bibfnamefont{S.}~\bibnamefont{Fernbach}},
  \bibinfo{author}{\bibfnamefont{R.}~\bibnamefont{Serber}}, \bibnamefont{and}
  \bibinfo{author}{\bibfnamefont{T.~B.} \bibnamefont{Taylor}},
  \bibinfo{journal}{Phys. Rev.} \textbf{\bibinfo{volume}{75}},
  \bibinfo{pages}{1352} (\bibinfo{year}{1949}),
  \urlprefix\url{https://link.aps.org/doi/10.1103/PhysRev.75.1352}.

\bibitem[{\citenamefont{Satchler}(1980)}]{Satchler1980}
\bibinfo{author}{\bibfnamefont{G.~R.} \bibnamefont{Satchler}},
  \emph{\bibinfo{title}{Introduction to Nuclear Reactions}}
  (\bibinfo{publisher}{John Wiley And Sons}, \bibinfo{year}{1980}).

\bibitem[{\citenamefont{Peterson}(1962)}]{Peterson1962}
\bibinfo{author}{\bibfnamefont{J.~M.} \bibnamefont{Peterson}},
  \bibinfo{journal}{Phys. Rev.} \textbf{\bibinfo{volume}{125}},
  \bibinfo{pages}{955} (\bibinfo{year}{1962}),
  \urlprefix\url{https://link.aps.org/doi/10.1103/PhysRev.125.955}.

\bibitem[{\citenamefont{Finlay et~al.}(1993)\citenamefont{Finlay, Abfalterer,
  Fink, Montei, Adami, Lisowski, Morgan, and Haight}}]{Finlay1993}
\bibinfo{author}{\bibfnamefont{R.~W.} \bibnamefont{Finlay}},
  \bibinfo{author}{\bibfnamefont{W.~P.} \bibnamefont{Abfalterer}},
  \bibinfo{author}{\bibfnamefont{G.}~\bibnamefont{Fink}},
  \bibinfo{author}{\bibfnamefont{E.}~\bibnamefont{Montei}},
  \bibinfo{author}{\bibfnamefont{T.}~\bibnamefont{Adami}},
  \bibinfo{author}{\bibfnamefont{P.~W.} \bibnamefont{Lisowski}},
  \bibinfo{author}{\bibfnamefont{G.~L.} \bibnamefont{Morgan}},
  \bibnamefont{and} \bibinfo{author}{\bibfnamefont{R.~C.}
  \bibnamefont{Haight}}, \bibinfo{journal}{Phys. Rev. C}
  \textbf{\bibinfo{volume}{47}}, \bibinfo{pages}{237} (\bibinfo{year}{1993}),
  \urlprefix\url{http://dx.doi.org/10.1103/PhysRevC.47.237}.

\bibitem[{\citenamefont{Schwartz et~al.}(1974)\citenamefont{Schwartz, Schrack,
  and Heaton~II}}]{Schwartz1974}
\bibinfo{author}{\bibfnamefont{R.~B.} \bibnamefont{Schwartz}},
  \bibinfo{author}{\bibfnamefont{R.~A.} \bibnamefont{Schrack}},
  \bibnamefont{and} \bibinfo{author}{\bibfnamefont{H.~T.}
  \bibnamefont{Heaton~II}}, \bibinfo{type}{Tech. Rep.} \bibinfo{number}{138},
  \bibinfo{institution}{National Bureau of Standards} (\bibinfo{year}{1974}).

\bibitem[{\citenamefont{Poenitz and Whalen}(1983)}]{Poenitz1983}
\bibinfo{author}{\bibfnamefont{W.~P.} \bibnamefont{Poenitz}} \bibnamefont{and}
  \bibinfo{author}{\bibfnamefont{J.~F.} \bibnamefont{Whalen}},
  \bibinfo{type}{Tech. Rep.} \bibinfo{number}{80},
  \bibinfo{institution}{Argonne National Laboratory} (\bibinfo{year}{1983}).

\bibitem[{\citenamefont{Abfalterer et~al.}(2000)\citenamefont{Abfalterer,
  Finlay, and Grimes}}]{Abfalterer2000}
\bibinfo{author}{\bibfnamefont{W.~P.} \bibnamefont{Abfalterer}},
  \bibinfo{author}{\bibfnamefont{R.~W.} \bibnamefont{Finlay}},
  \bibnamefont{and} \bibinfo{author}{\bibfnamefont{S.~M.}
  \bibnamefont{Grimes}}, \bibinfo{journal}{Phys. Rev. C}
  \textbf{\bibinfo{volume}{62}}, \bibinfo{pages}{064312}
  (\bibinfo{year}{2000}),
  \urlprefix\url{https://link.aps.org/doi/10.1103/PhysRevC.62.064312}.

\bibitem[{\citenamefont{Abfalterer et~al.}(2001)\citenamefont{Abfalterer,
  Bateman, Dietrich, Finlay, Haight, and Morgan}}]{Abfalterer2001}
\bibinfo{author}{\bibfnamefont{W.~P.} \bibnamefont{Abfalterer}},
  \bibinfo{author}{\bibfnamefont{F.~B.} \bibnamefont{Bateman}},
  \bibinfo{author}{\bibfnamefont{F.~S.} \bibnamefont{Dietrich}},
  \bibinfo{author}{\bibfnamefont{R.~W.} \bibnamefont{Finlay}},
  \bibinfo{author}{\bibfnamefont{R.~C.} \bibnamefont{Haight}},
  \bibnamefont{and} \bibinfo{author}{\bibfnamefont{G.~L.}
  \bibnamefont{Morgan}}, \bibinfo{journal}{Phys. Rev. C}
  \textbf{\bibinfo{volume}{63}}, \bibinfo{pages}{044608}
  (\bibinfo{year}{2001}),
  \urlprefix\url{http://dx.doi.org/10.1103/PhysRevC.63.044608}.

\bibitem[{\citenamefont{Carpenter and Wilson}(1959)}]{Carpenter1959}
\bibinfo{author}{\bibfnamefont{S.~G.} \bibnamefont{Carpenter}}
  \bibnamefont{and} \bibinfo{author}{\bibfnamefont{R.}~\bibnamefont{Wilson}},
  \bibinfo{journal}{Phys. Rev.} \textbf{\bibinfo{volume}{114}},
  \bibinfo{pages}{510} (\bibinfo{year}{1959}),
  \urlprefix\url{http://journals.aps.org/pr/pdf/10.1103/PhysRev.114.510}.

\bibitem[{\citenamefont{Angeli and Csikai}(1970)}]{Angeli1970}
\bibinfo{author}{\bibfnamefont{I.}~\bibnamefont{Angeli}} \bibnamefont{and}
  \bibinfo{author}{\bibfnamefont{J.}~\bibnamefont{Csikai}},
  \bibinfo{journal}{Nucl. Phys. A} \textbf{\bibinfo{volume}{158}},
  \bibinfo{pages}{389} (\bibinfo{year}{1970}),
  \urlprefix\url{http://www.sciencedirect.com/science/article/pii/0375947470901909}.

\bibitem[{\citenamefont{Mohr}(1955)}]{Mohr1955}
\bibinfo{author}{\bibfnamefont{C.~B.~O.} \bibnamefont{Mohr}},
  \bibinfo{journal}{Proc. Phys. Soc. A} \textbf{\bibinfo{volume}{68}},
  \bibinfo{pages}{340} (\bibinfo{year}{1955}),
  \urlprefix\url{http://stacks.iop.org/0370-1298/68/i=4/a=410}.

\bibitem[{\citenamefont{Feshbach}(1958)}]{Feshbach1958}
\bibinfo{author}{\bibfnamefont{H.}~\bibnamefont{Feshbach}},
  \bibinfo{journal}{Ann. Rev. Nucl. Part. Sci.} \textbf{\bibinfo{volume}{8}},
  \bibinfo{pages}{49} (\bibinfo{year}{1958}),
  \urlprefix\url{https://doi.org/10.1146/annurev.ns.08.120158.000405}.

\bibitem[{\citenamefont{McVoy}(1967)}]{McVoy1967}
\bibinfo{author}{\bibfnamefont{K.~W.} \bibnamefont{McVoy}},
  \bibinfo{journal}{Ann. Sci.} \textbf{\bibinfo{volume}{43}},
  \bibinfo{pages}{91} (\bibinfo{year}{1967}),
  \urlprefix\url{http://www.sciencedirect.com/science/article/pii/000349166790293X}.

\bibitem[{\citenamefont{Ahmad et~al.}(1973)\citenamefont{Ahmad, Bano, and
  Saharia}}]{Ahmad1973}
\bibinfo{author}{\bibfnamefont{I.}~\bibnamefont{Ahmad}},
  \bibinfo{author}{\bibfnamefont{N.}~\bibnamefont{Bano}}, \bibnamefont{and}
  \bibinfo{author}{\bibfnamefont{A.~N.} \bibnamefont{Saharia}},
  \bibinfo{journal}{Pramana - J. Phys.} \textbf{\bibinfo{volume}{1}},
  \bibinfo{pages}{188} (\bibinfo{year}{1973}),
  \urlprefix\url{https://link.springer.com/article/10.1007/BF02847190}.

\bibitem[{\citenamefont{Perey and Perey}(1976)}]{Perey1976}
\bibinfo{author}{\bibfnamefont{C.~M.} \bibnamefont{Perey}} \bibnamefont{and}
  \bibinfo{author}{\bibfnamefont{F.~G.} \bibnamefont{Perey}},
  \bibinfo{journal}{Atom. Data Nucl. Data Tables} \textbf{\bibinfo{volume}{17}}
  (\bibinfo{year}{1976}).

\bibitem[{\citenamefont{Varner et~al.}(1991)\citenamefont{Varner, Thompson,
  McAbee, Ludwig, and Clegg}}]{CH89}
\bibinfo{author}{\bibfnamefont{R.~L.} \bibnamefont{Varner}},
  \bibinfo{author}{\bibfnamefont{W.~J.} \bibnamefont{Thompson}},
  \bibinfo{author}{\bibfnamefont{T.~L.} \bibnamefont{McAbee}},
  \bibinfo{author}{\bibfnamefont{E.~J.} \bibnamefont{Ludwig}},
  \bibnamefont{and} \bibinfo{author}{\bibfnamefont{T.~B.} \bibnamefont{Clegg}},
  \bibinfo{journal}{Phys. Rep.} \textbf{\bibinfo{volume}{201}},
  \bibinfo{pages}{57} (\bibinfo{year}{1991}),
  \urlprefix\url{http://www.sciencedirect.com/science/article/pii/037015739190039O}.

\bibitem[{\citenamefont{Koning and Delaroche}(2003)}]{KoningDelaroche}
\bibinfo{author}{\bibfnamefont{A.~J.} \bibnamefont{Koning}} \bibnamefont{and}
  \bibinfo{author}{\bibfnamefont{J.~P.} \bibnamefont{Delaroche}},
  \bibinfo{journal}{Nucl. Phys. A} \textbf{\bibinfo{volume}{713}},
  \bibinfo{pages}{231 } (\bibinfo{year}{2003}),
  \urlprefix\url{http://www.sciencedirect.com/science/article/pii/S0375947402013210}.

\bibitem[{\citenamefont{Dietrich et~al.}(2003)\citenamefont{Dietrich, Anderson,
  Bauer, Grimes, Finlay, Abfalterer, Bateman, Haight, Morgan, Bauge
  et~al.}}]{Dietrich2003}
\bibinfo{author}{\bibfnamefont{F.~S.} \bibnamefont{Dietrich}},
  \bibinfo{author}{\bibfnamefont{J.~D.} \bibnamefont{Anderson}},
  \bibinfo{author}{\bibfnamefont{R.~W.} \bibnamefont{Bauer}},
  \bibinfo{author}{\bibfnamefont{S.~M.} \bibnamefont{Grimes}},
  \bibinfo{author}{\bibfnamefont{R.~W.} \bibnamefont{Finlay}},
  \bibinfo{author}{\bibfnamefont{W.~P.} \bibnamefont{Abfalterer}},
  \bibinfo{author}{\bibfnamefont{F.~B.} \bibnamefont{Bateman}},
  \bibinfo{author}{\bibfnamefont{R.~C.} \bibnamefont{Haight}},
  \bibinfo{author}{\bibfnamefont{G.~L.} \bibnamefont{Morgan}},
  \bibinfo{author}{\bibfnamefont{E.}~\bibnamefont{Bauge}},
  \bibnamefont{et~al.}, \bibinfo{journal}{Phys. Rev. C}
  \textbf{\bibinfo{volume}{67}}, \bibinfo{pages}{044606}
  (\bibinfo{year}{2003}),
  \urlprefix\url{https://link.aps.org/doi/10.1103/PhysRevC.67.044606}.

\bibitem[{\citenamefont{Pruitt et~al.}(2020)\citenamefont{Pruitt, Charity,
  Sobotka, Atkinson, and Dickhoff}}]{Pruitt2020PRL}
\bibinfo{author}{\bibfnamefont{C.~D.} \bibnamefont{Pruitt}},
  \bibinfo{author}{\bibfnamefont{R.~J.} \bibnamefont{Charity}},
  \bibinfo{author}{\bibfnamefont{L.~G.} \bibnamefont{Sobotka}},
  \bibinfo{author}{\bibfnamefont{M.~C.} \bibnamefont{Atkinson}},
  \bibnamefont{and} \bibinfo{author}{\bibfnamefont{W.~H.}
  \bibnamefont{Dickhoff}}, \bibinfo{journal}{Phys. Rev. Lett.}
  \textbf{\bibinfo{volume}{125}}, \bibinfo{pages}{102501}
  (\bibinfo{year}{2020}).

\bibitem[{\citenamefont{Phillips et~al.}(1980)\citenamefont{Phillips, Berman,
  and Seagrave}}]{Phillips1980}
\bibinfo{author}{\bibfnamefont{T.~W.} \bibnamefont{Phillips}},
  \bibinfo{author}{\bibfnamefont{B.~L.} \bibnamefont{Berman}},
  \bibnamefont{and} \bibinfo{author}{\bibfnamefont{J.~D.}
  \bibnamefont{Seagrave}}, \bibinfo{journal}{Phys. Rev. C}
  \textbf{\bibinfo{volume}{22}}, \bibinfo{pages}{384} (\bibinfo{year}{1980}),
  \urlprefix\url{https://link.aps.org/doi/10.1103/PhysRevC.22.384}.

\bibitem[{\citenamefont{Foster and Glasgow}(1971)}]{Foster1971}
\bibinfo{author}{\bibfnamefont{D.~G.} \bibnamefont{Foster}} \bibnamefont{and}
  \bibinfo{author}{\bibfnamefont{D.~W.} \bibnamefont{Glasgow}},
  \bibinfo{journal}{Phys. Rev. C} \textbf{\bibinfo{volume}{3}},
  \bibinfo{pages}{576} (\bibinfo{year}{1971}),
  \urlprefix\url{https://link.aps.org/doi/10.1103/PhysRevC.3.576}.

\bibitem[{\citenamefont{Pruitt}(2019)}]{PruittPhDThesis}
\bibinfo{author}{\bibfnamefont{C.~D.} \bibnamefont{Pruitt}}, Ph.D. thesis,
  \bibinfo{school}{Washington University in St Louis} (\bibinfo{year}{2019}).

\bibitem[{\citenamefont{Shane et~al.}(2010)\citenamefont{Shane, Charity, Elson,
  Sobotka, Devlin, Fotiades, and O`Donnell}}]{Shane2010}
\bibinfo{author}{\bibfnamefont{R.}~\bibnamefont{Shane}},
  \bibinfo{author}{\bibfnamefont{R.~J.} \bibnamefont{Charity}},
  \bibinfo{author}{\bibfnamefont{J.~M.} \bibnamefont{Elson}},
  \bibinfo{author}{\bibfnamefont{L.~G.} \bibnamefont{Sobotka}},
  \bibinfo{author}{\bibfnamefont{M.}~\bibnamefont{Devlin}},
  \bibinfo{author}{\bibfnamefont{N.}~\bibnamefont{Fotiades}}, \bibnamefont{and}
  \bibinfo{author}{\bibfnamefont{J.~M.} \bibnamefont{O`Donnell}},
  \bibinfo{journal}{Nucl. Instrum. Meth.} \textbf{\bibinfo{volume}{614}},
  \bibinfo{pages}{468} (\bibinfo{year}{2010}),
  \urlprefix\url{http://dx.doi.org/10.1016/j.nima.2010.01.005}.

\bibitem[{\citenamefont{Mueller et~al.}(2011)\citenamefont{Mueller, Charity,
  Shane, Sobotka, Waldecker, Dickhoff, Crowell, Esterline, Fallin, Howell
  et~al.}}]{Mueller2011}
\bibinfo{author}{\bibfnamefont{J.~M.} \bibnamefont{Mueller}},
  \bibinfo{author}{\bibfnamefont{R.~J.} \bibnamefont{Charity}},
  \bibinfo{author}{\bibfnamefont{R.}~\bibnamefont{Shane}},
  \bibinfo{author}{\bibfnamefont{L.~G.} \bibnamefont{Sobotka}},
  \bibinfo{author}{\bibfnamefont{S.~J.} \bibnamefont{Waldecker}},
  \bibinfo{author}{\bibfnamefont{W.~H.} \bibnamefont{Dickhoff}},
  \bibinfo{author}{\bibfnamefont{A.~S.} \bibnamefont{Crowell}},
  \bibinfo{author}{\bibfnamefont{J.~H.} \bibnamefont{Esterline}},
  \bibinfo{author}{\bibfnamefont{B.}~\bibnamefont{Fallin}},
  \bibinfo{author}{\bibfnamefont{C.~R.} \bibnamefont{Howell}},
  \bibnamefont{et~al.}, \bibinfo{journal}{Phys. Rev. C}
  \textbf{\bibinfo{volume}{83}}, \bibinfo{pages}{064605}
  (\bibinfo{year}{2011}),
  \urlprefix\url{https://link.aps.org/doi/10.1103/PhysRevC.83.064605}.

\bibitem[{\citenamefont{Mahzoon et~al.}(2014)\citenamefont{Mahzoon, Charity,
  Dickhoff, Dussan, and Waldecker}}]{Mahzoon2014}
\bibinfo{author}{\bibfnamefont{M.~H.} \bibnamefont{Mahzoon}},
  \bibinfo{author}{\bibfnamefont{R.~J.} \bibnamefont{Charity}},
  \bibinfo{author}{\bibfnamefont{W.~H.} \bibnamefont{Dickhoff}},
  \bibinfo{author}{\bibfnamefont{H.}~\bibnamefont{Dussan}}, \bibnamefont{and}
  \bibinfo{author}{\bibfnamefont{S.~J.} \bibnamefont{Waldecker}},
  \bibinfo{journal}{Phys. Rev. Lett.} \textbf{\bibinfo{volume}{112}},
  \bibinfo{pages}{162503} (\bibinfo{year}{2014}),
  \urlprefix\url{https://link.aps.org/doi/10.1103/PhysRevLett.112.162503}.

\bibitem[{\citenamefont{Mahzoon}(2015)}]{MahzoonPhDThesis}
\bibinfo{author}{\bibfnamefont{M.}~\bibnamefont{Mahzoon}}, Ph.D. thesis,
  \bibinfo{school}{Washington University in St Louis} (\bibinfo{year}{2015}),
  \urlprefix\url{http://libproxy.wustl.edu/login?url=https://search.proquest.com/docview/1749780826?accountid=15159}.

\bibitem[{\citenamefont{Moore}(1980)}]{Moore1980}
\bibinfo{author}{\bibfnamefont{M.~S.} \bibnamefont{Moore}},
  \bibinfo{journal}{Nucl. Instrum. Meth.} \textbf{\bibinfo{volume}{169}},
  \bibinfo{pages}{245 } (\bibinfo{year}{1980}),
  \urlprefix\url{http://www.sciencedirect.com/science/article/pii/0029554X80901299}.

\bibitem[{\citenamefont{Clement et~al.}(1972)\citenamefont{Clement, Stoler,
  Goulding, and Fairchild}}]{Clement1972}
\bibinfo{author}{\bibfnamefont{J.~M.} \bibnamefont{Clement}},
  \bibinfo{author}{\bibfnamefont{P.}~\bibnamefont{Stoler}},
  \bibinfo{author}{\bibfnamefont{C.~A.} \bibnamefont{Goulding}},
  \bibnamefont{and} \bibinfo{author}{\bibfnamefont{R.~W.}
  \bibnamefont{Fairchild}}, \bibinfo{journal}{Nucl. Phys. A}
  \textbf{\bibinfo{volume}{183}}, \bibinfo{pages}{51} (\bibinfo{year}{1972}),
  \urlprefix\url{http://dx.doi.org/10.1016/0375-9474(72)90930-X}.

\bibitem[{\citenamefont{Abfalterer et~al.}(1998)\citenamefont{Abfalterer,
  Bateman, Dietrich, Elster, Finlay, Gl{\"o}ckle, Golak, Haight, H{\"u}ber,
  Morgan et~al.}}]{Abfalterer1998}
\bibinfo{author}{\bibfnamefont{W.~P.} \bibnamefont{Abfalterer}},
  \bibinfo{author}{\bibfnamefont{F.~B.} \bibnamefont{Bateman}},
  \bibinfo{author}{\bibfnamefont{F.~S.} \bibnamefont{Dietrich}},
  \bibinfo{author}{\bibfnamefont{C.}~\bibnamefont{Elster}},
  \bibinfo{author}{\bibfnamefont{R.~W.} \bibnamefont{Finlay}},
  \bibinfo{author}{\bibfnamefont{W.}~\bibnamefont{Gl{\"o}ckle}},
  \bibinfo{author}{\bibfnamefont{J.}~\bibnamefont{Golak}},
  \bibinfo{author}{\bibfnamefont{R.~C.} \bibnamefont{Haight}},
  \bibinfo{author}{\bibfnamefont{D.}~\bibnamefont{H{\"u}ber}},
  \bibinfo{author}{\bibfnamefont{G.~L.} \bibnamefont{Morgan}},
  \bibnamefont{et~al.}, \bibinfo{journal}{Phys. Rev. Lett.}
  \textbf{\bibinfo{volume}{81}} (\bibinfo{year}{1998}).

\bibitem[{\citenamefont{Dukarevich et~al.}(1967)\citenamefont{Dukarevich,
  Dyumin, and Kaminker}}]{Dukarevich1967}
\bibinfo{author}{\bibfnamefont{Y.~V.} \bibnamefont{Dukarevich}},
  \bibinfo{author}{\bibfnamefont{A.~N.} \bibnamefont{Dyumin}},
  \bibnamefont{and} \bibinfo{author}{\bibfnamefont{D.~M.}
  \bibnamefont{Kaminker}}, \bibinfo{journal}{Nucl. Phys. A}
  \textbf{\bibinfo{volume}{92}}, \bibinfo{pages}{433} (\bibinfo{year}{1967}),
  \urlprefix\url{http://dx.doi.org/10.1016/0375-9474(67)90228-X}.

\bibitem[{\citenamefont{Anselment et~al.}(1986)\citenamefont{Anselment, Bekk,
  Hanser, Hoeffgen, Meisel, Goring, Rebel, and Schatz}}]{Anselment1986}
\bibinfo{author}{\bibfnamefont{M.}~\bibnamefont{Anselment}},
  \bibinfo{author}{\bibfnamefont{K.}~\bibnamefont{Bekk}},
  \bibinfo{author}{\bibfnamefont{A.}~\bibnamefont{Hanser}},
  \bibinfo{author}{\bibfnamefont{H.}~\bibnamefont{Hoeffgen}},
  \bibinfo{author}{\bibfnamefont{G.}~\bibnamefont{Meisel}},
  \bibinfo{author}{\bibfnamefont{S.}~\bibnamefont{Goring}},
  \bibinfo{author}{\bibfnamefont{H.}~\bibnamefont{Rebel}}, \bibnamefont{and}
  \bibinfo{author}{\bibfnamefont{G.}~\bibnamefont{Schatz}},
  \bibinfo{journal}{Phys. Rev. C} \textbf{\bibinfo{volume}{34}},
  \bibinfo{pages}{1052} (\bibinfo{year}{1986}).

\bibitem[{\citenamefont{Perey et~al.}(1972)\citenamefont{Perey, Love, and
  Kinney}}]{Perey1972}
\bibinfo{author}{\bibfnamefont{F.~G.} \bibnamefont{Perey}},
  \bibinfo{author}{\bibfnamefont{T.~A.} \bibnamefont{Love}}, \bibnamefont{and}
  \bibinfo{author}{\bibfnamefont{W.~E.} \bibnamefont{Kinney}},
  \bibinfo{type}{Tech. Rep.} \bibinfo{number}{4823}, \bibinfo{institution}{Oak
  Ridge National Lab} (\bibinfo{year}{1972}).

\bibitem[{\citenamefont{Vaughn et~al.}(1965)\citenamefont{Vaughn, Grench,
  Imhof, Rowland, and Walt}}]{Vaughn1965}
\bibinfo{author}{\bibfnamefont{F.~J.} \bibnamefont{Vaughn}},
  \bibinfo{author}{\bibfnamefont{H.~A.} \bibnamefont{Grench}},
  \bibinfo{author}{\bibfnamefont{W.~L.} \bibnamefont{Imhof}},
  \bibinfo{author}{\bibfnamefont{J.~H.} \bibnamefont{Rowland}},
  \bibnamefont{and} \bibinfo{author}{\bibfnamefont{M.}~\bibnamefont{Walt}},
  \bibinfo{journal}{Nucl. Phys.} \textbf{\bibinfo{volume}{64}},
  \bibinfo{pages}{336} (\bibinfo{year}{1965}),
  \urlprefix\url{http://dx.doi.org/10.1016/0029-5582(65)90361-5}.

\bibitem[{\citenamefont{Salisbury et~al.}(1965)\citenamefont{Salisbury, Fossan,
  and Vaughn}}]{Salisbury1965}
\bibinfo{author}{\bibfnamefont{S.~R.} \bibnamefont{Salisbury}},
  \bibinfo{author}{\bibfnamefont{D.~B.} \bibnamefont{Fossan}},
  \bibnamefont{and} \bibinfo{author}{\bibfnamefont{F.~J.}
  \bibnamefont{Vaughn}}, \bibinfo{journal}{Nucl. Phys.}
  \textbf{\bibinfo{volume}{64}}, \bibinfo{pages}{343} (\bibinfo{year}{1965}),
  \urlprefix\url{http://dx.doi.org/10.1016/0029-5582(65)90362-7}.

\bibitem[{\citenamefont{Perey et~al.}(1993)\citenamefont{Perey, Perey, Harvey,
  Hill, Larson, Macklin, and Larson}}]{Perey1993}
\bibinfo{author}{\bibfnamefont{C.~M.} \bibnamefont{Perey}},
  \bibinfo{author}{\bibfnamefont{F.~G.} \bibnamefont{Perey}},
  \bibinfo{author}{\bibfnamefont{J.~A.} \bibnamefont{Harvey}},
  \bibinfo{author}{\bibfnamefont{N.~W.} \bibnamefont{Hill}},
  \bibinfo{author}{\bibfnamefont{N.~M.} \bibnamefont{Larson}},
  \bibinfo{author}{\bibfnamefont{R.~L.} \bibnamefont{Macklin}},
  \bibnamefont{and} \bibinfo{author}{\bibfnamefont{D.~C.}
  \bibnamefont{Larson}}, \bibinfo{journal}{Phys. Rev. C}
  \textbf{\bibinfo{volume}{47}}, \bibinfo{pages}{1143} (\bibinfo{year}{1993}),
  \urlprefix\url{http://dx.doi.org/10.1103/PhysRevC.47.1143}.

\bibitem[{\citenamefont{Harper et~al.}(1982)\citenamefont{Harper, Godfrey, and
  Weil}}]{Harper1982}
\bibinfo{author}{\bibfnamefont{R.~W.} \bibnamefont{Harper}},
  \bibinfo{author}{\bibfnamefont{T.~W.} \bibnamefont{Godfrey}},
  \bibnamefont{and} \bibinfo{author}{\bibfnamefont{J.~L.} \bibnamefont{Weil}},
  \bibinfo{journal}{Phys. Rev. C} \textbf{\bibinfo{volume}{26}},
  \bibinfo{pages}{1432} (\bibinfo{year}{1982}),
  \urlprefix\url{http://dx.doi.org/10.1103/PhysRevC.26.1432}.

\bibitem[{\citenamefont{Timokhov et~al.}(1989)\citenamefont{Timokhov, Bokhovko,
  Isakov, Kazakov, Kononov, Manturov, Poletaev, and Pronyaev}}]{Timokhov1989}
\bibinfo{author}{\bibfnamefont{V.~M.} \bibnamefont{Timokhov}},
  \bibinfo{author}{\bibfnamefont{M.~V.} \bibnamefont{Bokhovko}},
  \bibinfo{author}{\bibfnamefont{A.~G.} \bibnamefont{Isakov}},
  \bibinfo{author}{\bibfnamefont{L.~E.} \bibnamefont{Kazakov}},
  \bibinfo{author}{\bibfnamefont{V.~N.} \bibnamefont{Kononov}},
  \bibinfo{author}{\bibfnamefont{G.~N.} \bibnamefont{Manturov}},
  \bibinfo{author}{\bibfnamefont{E.~D.} \bibnamefont{Poletaev}},
  \bibnamefont{and} \bibinfo{author}{\bibfnamefont{V.~G.}
  \bibnamefont{Pronyaev}}, \bibinfo{journal}{Yad. Fiz.}
  \textbf{\bibinfo{volume}{50}}, \bibinfo{pages}{609} (\bibinfo{year}{1989}).

\bibitem[{\citenamefont{Rapaport et~al.}(1980)\citenamefont{Rapaport, Mirzaa,
  Hadizadeh, Bainum, and Finlay}}]{Rapaport1980}
\bibinfo{author}{\bibfnamefont{J.}~\bibnamefont{Rapaport}},
  \bibinfo{author}{\bibfnamefont{M.}~\bibnamefont{Mirzaa}},
  \bibinfo{author}{\bibfnamefont{M.}~\bibnamefont{Hadizadeh}},
  \bibinfo{author}{\bibfnamefont{D.~E.} \bibnamefont{Bainum}},
  \bibnamefont{and} \bibinfo{author}{\bibfnamefont{R.~W.}
  \bibnamefont{Finlay}}, \bibinfo{journal}{Nucl. Phys. A}
  \textbf{\bibinfo{volume}{341}}, \bibinfo{pages}{56} (\bibinfo{year}{1980}),
  \urlprefix\url{http://dx.doi.org/10.1016/0375-9474(80)90361-9}.

\bibitem[{\citenamefont{Atkinson et~al.}(2018)\citenamefont{Atkinson, Blok,
  Lapik\'as, Charity, and Dickhoff}}]{Atkinson2018}
\bibinfo{author}{\bibfnamefont{M.~C.} \bibnamefont{Atkinson}},
  \bibinfo{author}{\bibfnamefont{H.~P.} \bibnamefont{Blok}},
  \bibinfo{author}{\bibfnamefont{L.}~\bibnamefont{Lapik\'as}},
  \bibinfo{author}{\bibfnamefont{R.~J.} \bibnamefont{Charity}},
  \bibnamefont{and} \bibinfo{author}{\bibfnamefont{W.~H.}
  \bibnamefont{Dickhoff}}, \bibinfo{journal}{Phys. Rev. C}
  \textbf{\bibinfo{volume}{98}}, \bibinfo{pages}{044627}
  (\bibinfo{year}{2018}),
  \urlprefix\url{https://link.aps.org/doi/10.1103/PhysRevC.98.044627}.

\bibitem[{\citenamefont{Mahaux and Sartor}(1991)}]{Mahaux1991}
\bibinfo{author}{\bibfnamefont{C.}~\bibnamefont{Mahaux}} \bibnamefont{and}
  \bibinfo{author}{\bibfnamefont{R.}~\bibnamefont{Sartor}},
  \bibinfo{journal}{Adv. Nucl. Phys.} \textbf{\bibinfo{volume}{20}},
  \bibinfo{pages}{1} (\bibinfo{year}{1991}).

\bibitem[{\citenamefont{Dickhoff and Charity}(2018)}]{Dickhoff2018}
\bibinfo{author}{\bibfnamefont{W.~H.} \bibnamefont{Dickhoff}} \bibnamefont{and}
  \bibinfo{author}{\bibfnamefont{R.~J.} \bibnamefont{Charity}},
  \bibinfo{journal}{Prog. Part. Nucl. Phys.}  (\bibinfo{year}{2018}).

\bibitem[{\citenamefont{Lane and Thomas}(1958)}]{LaneThomas}
\bibinfo{author}{\bibfnamefont{A.~M.} \bibnamefont{Lane}} \bibnamefont{and}
  \bibinfo{author}{\bibfnamefont{R.~G.} \bibnamefont{Thomas}},
  \bibinfo{journal}{Rev. Mod. Phys.} \textbf{\bibinfo{volume}{30}},
  \bibinfo{pages}{257} (\bibinfo{year}{1958}),
  \urlprefix\url{https://link.aps.org/doi/10.1103/RevModPhys.30.257}.

\bibitem[{\citenamefont{Carlson}(1996)}]{Carlson1996}
\bibinfo{author}{\bibfnamefont{R.~F.} \bibnamefont{Carlson}},
  \bibinfo{journal}{Atom. Data Nucl. Data Tables}
  \textbf{\bibinfo{volume}{63}}, \bibinfo{pages}{93 } (\bibinfo{year}{1996}),
  \urlprefix\url{http://www.sciencedirect.com/science/article/pii/S0092640X96900108}.

\bibitem[{\citenamefont{Wang et~al.}(2017)\citenamefont{Wang, Audi, Kondev,
  Huang, Naimi, and Xi}}]{AME2016}
\bibinfo{author}{\bibfnamefont{M.}~\bibnamefont{Wang}},
  \bibinfo{author}{\bibfnamefont{G.}~\bibnamefont{Audi}},
  \bibinfo{author}{\bibfnamefont{F.~G.} \bibnamefont{Kondev}},
  \bibinfo{author}{\bibfnamefont{W.}~\bibnamefont{Huang}},
  \bibinfo{author}{\bibfnamefont{S.}~\bibnamefont{Naimi}}, \bibnamefont{and}
  \bibinfo{author}{\bibfnamefont{X.}~\bibnamefont{Xi}}, \bibinfo{journal}{Chin.
  Phys. C} \textbf{\bibinfo{volume}{41}}, \bibinfo{pages}{030003}
  (\bibinfo{year}{2017}).

\bibitem[{\citenamefont{Vries et~al.}(1987)\citenamefont{Vries, Jager, and
  Vries}}]{DeVries1987}
\bibinfo{author}{\bibfnamefont{H.~D.} \bibnamefont{Vries}},
  \bibinfo{author}{\bibfnamefont{C.~W.~D.} \bibnamefont{Jager}},
  \bibnamefont{and} \bibinfo{author}{\bibfnamefont{C.~D.} \bibnamefont{Vries}},
  \bibinfo{journal}{Atom. Data Nucl. Data Tables}
  \textbf{\bibinfo{volume}{36}}, \bibinfo{pages}{495} (\bibinfo{year}{1987}),
  \urlprefix\url{https://www.sciencedirect.com/science/article/pii/0092640X87900131}.

\bibitem[{\citenamefont{Angeli and Marinova}(2013)}]{Angeli2013}
\bibinfo{author}{\bibfnamefont{I.}~\bibnamefont{Angeli}} \bibnamefont{and}
  \bibinfo{author}{\bibfnamefont{K.~P.} \bibnamefont{Marinova}},
  \bibinfo{journal}{Atom. Data Nucl. Data Tables}
  \textbf{\bibinfo{volume}{99}}, \bibinfo{pages}{69 } (\bibinfo{year}{2013}),
  \urlprefix\url{http://www.sciencedirect.com/science/article/pii/S0092640X12000265}.

\bibitem[{\citenamefont{King et~al.}(2019)\citenamefont{King, Lovell,
  Neufcourt, and Nunes}}]{King2019}
\bibinfo{author}{\bibfnamefont{G.~B.} \bibnamefont{King}},
  \bibinfo{author}{\bibfnamefont{A.~E.} \bibnamefont{Lovell}},
  \bibinfo{author}{\bibfnamefont{L.}~\bibnamefont{Neufcourt}},
  \bibnamefont{and} \bibinfo{author}{\bibfnamefont{F.~M.} \bibnamefont{Nunes}},
  \bibinfo{journal}{Phys. Rev. Lett.} \textbf{\bibinfo{volume}{122}},
  \bibinfo{pages}{232502} (\bibinfo{year}{2019}).

\bibitem[{\citenamefont{Foreman-Mackey
  et~al.}(2013)\citenamefont{Foreman-Mackey, Hogg, Lang, and
  Goodman}}]{Foreman-Mackey2013}
\bibinfo{author}{\bibfnamefont{D.}~\bibnamefont{Foreman-Mackey}},
  \bibinfo{author}{\bibfnamefont{D.~W.} \bibnamefont{Hogg}},
  \bibinfo{author}{\bibfnamefont{D.}~\bibnamefont{Lang}}, \bibnamefont{and}
  \bibinfo{author}{\bibfnamefont{J.}~\bibnamefont{Goodman}},
  \bibinfo{journal}{Publ. Astron. Soc. Pac.} \textbf{\bibinfo{volume}{125}},
  \bibinfo{pages}{306–312} (\bibinfo{year}{2013}),
  \urlprefix\url{http://dx.doi.org/10.1086/670067}.

\bibitem[{\citenamefont{Sharma}(2017)}]{Sharma2017}
\bibinfo{author}{\bibfnamefont{S.}~\bibnamefont{Sharma}},
  \bibinfo{journal}{Ann. Rev. Astron. Astrophys.}
  \textbf{\bibinfo{volume}{55}}, \bibinfo{pages}{213} (\bibinfo{year}{2017}).

\bibitem[{\citenamefont{Goodman and Weare}(2010)}]{Goodman2010}
\bibinfo{author}{\bibfnamefont{J.}~\bibnamefont{Goodman}} \bibnamefont{and}
  \bibinfo{author}{\bibfnamefont{J.}~\bibnamefont{Weare}},
  \bibinfo{journal}{Commun. Appl. Math. Comput. Sci.}
  \textbf{\bibinfo{volume}{5}}, \bibinfo{pages}{65} (\bibinfo{year}{2010}),
  \urlprefix\url{https://doi.org/10.2140/camcos.2010.5.65}.

\bibitem[{\citenamefont{Brynjarsdóttir and
  O'Hagan}(2014)}]{Brynjarsdottir2014}
\bibinfo{author}{\bibfnamefont{J.}~\bibnamefont{Brynjarsdóttir}}
  \bibnamefont{and} \bibinfo{author}{\bibfnamefont{A.}~\bibnamefont{O'Hagan}},
  \bibinfo{journal}{Inverse Problems} \textbf{\bibinfo{volume}{30}}
  (\bibinfo{year}{2014}).

\bibitem[{\citenamefont{Tostevin and Gade}(2014)}]{Tostevin2014}
\bibinfo{author}{\bibfnamefont{J.~A.} \bibnamefont{Tostevin}} \bibnamefont{and}
  \bibinfo{author}{\bibfnamefont{A.}~\bibnamefont{Gade}},
  \bibinfo{journal}{Phys. Rev. C} \textbf{\bibinfo{volume}{90}},
  \bibinfo{pages}{057602} (\bibinfo{year}{2014}),
  \urlprefix\url{https://link.aps.org/doi/10.1103/PhysRevC.90.057602}.

\bibitem[{\citenamefont{Atkinson and Dickhoff}(2019)}]{Atkinson2019}
\bibinfo{author}{\bibfnamefont{M.~C.} \bibnamefont{Atkinson}} \bibnamefont{and}
  \bibinfo{author}{\bibfnamefont{W.~H.} \bibnamefont{Dickhoff}},
  \bibinfo{journal}{Phys. Lett. B} \textbf{\bibinfo{volume}{798}},
  \bibinfo{pages}{135027} (\bibinfo{year}{2019}),
  \urlprefix\url{http://www.sciencedirect.com/science/article/pii/S037026931930749X}.

\bibitem[{\citenamefont{Perey and Buck}(1962)}]{Perey1962}
\bibinfo{author}{\bibfnamefont{F.}~\bibnamefont{Perey}} \bibnamefont{and}
  \bibinfo{author}{\bibfnamefont{B.}~\bibnamefont{Buck}},
  \bibinfo{journal}{Nucl. Phys.} \textbf{\bibinfo{volume}{32}},
  \bibinfo{pages}{353 } (\bibinfo{year}{1962}),
  \urlprefix\url{http://www.sciencedirect.com/science/article/pii/0029558262903450}.

\bibitem[{\citenamefont{Charity et~al.}(2007)\citenamefont{Charity, Mueller,
  Sobotka, and Dickhoff}}]{Charity2006}
\bibinfo{author}{\bibfnamefont{R.~J.} \bibnamefont{Charity}},
  \bibinfo{author}{\bibfnamefont{J.~M.} \bibnamefont{Mueller}},
  \bibinfo{author}{\bibfnamefont{L.~G.} \bibnamefont{Sobotka}},
  \bibnamefont{and} \bibinfo{author}{\bibfnamefont{W.~H.}
  \bibnamefont{Dickhoff}}, \bibinfo{journal}{Phys. Rev. C}
  \textbf{\bibinfo{volume}{76}}, \bibinfo{pages}{044314}
  (\bibinfo{year}{2007}),
  \urlprefix\url{https://link.aps.org/doi/10.1103/PhysRevC.76.044314}.

\end{thebibliography}


\begin{thebibliography}{32} \expandafter\ifx\csname
        natexlab\endcsname\relax\def\natexlab#1{#1}\fi \expandafter\ifx\csname
        bibnamefont\endcsname\relax \def\bibnamefont#1{#1}\fi
        \expandafter\ifx\csname bibfnamefont\endcsname\relax
        \def\bibfnamefont#1{#1}\fi \expandafter\ifx\csname
        citenamefont\endcsname\relax \def\citenamefont#1{#1}\fi
        \expandafter\ifx\csname url\endcsname\relax \def\url#1{\texttt{#1}}\fi
        \expandafter\ifx\csname urlprefix\endcsname\relax\def\urlprefix{URL
        }\fi \providecommand{\bibinfo}[2]{#2}
        \providecommand{\eprint}[2][]{\url{#2}}
\end{thebibliography}

\clearpage
\appendix \label{DOMFunctionalForms}
\section{Appendix A: Definition of DOM Potential}
\subsection{Functional Forms}
Before giving the full parameterization, we identify a few standard functional
forms. Radial dependences are defined by a Woods-Saxon shape or a derivative:
\begin{equation} \label{WoodsSaxon}
    \begin{split}
        f_{vol}(r; r_{0}, a) & = \frac{-1}{1+e^{(r-R)/a}},\\
        \\
        f_{sur}(r; r_{0}, a) & = \frac{1}{r}\frac{d}{dr}f_{vol}(r; r_{0}, a).
    \end{split}
\end{equation}
$R$ is the nuclear radius, calculated as $R = r_{0}A^{\frac{1}{3}}$.
The sign of the potential is such that the Woods-Saxon form
provides an attractive interaction. For nonlocalities, we use a Gaussian
nonlocality first proposed by \cite{Perey1962}:
\begin{equation}
    N(r, r';\beta) = \frac{1}{\pi^{\frac{3}{2}}\beta^{3}}
    e^{-(r-r')^{2}/{\beta^{2}}},
\end{equation}
where $\beta$ sets the Gaussian width.
The energy-dependences of the imaginary components is based on the functional form of \cite{Charity2006}:
\begin{equation} \label{omega}
    \omega_{n}(E; A, B, C) = \Theta(X)A\frac{X^{n}}{X^{n}+B^{n}}
\end{equation}
where
\begin{equation*}
    X = |E-\epsilon_{F}|-C\\
\end{equation*}
and $\Theta(X)$ is the Heaviside step function. 

For symmetric nuclei, the same potential was used for protons and neutrons,
excepting Coulomb. For asymmetric nuclei, we introduced five asymmetry-dependent terms.
For all energy dependences, the energy domain was $\epsilon_{F}$-300
 \mega\electronvolt\ to $\epsilon_{F}$+200 \mega\electronvolt.

The irreducible self-energy (optical potential) used in this work is defined
\begin{equation} \label{SelfEnergyBreakdown}
        \Sigma^{*}(\alpha,\beta;E) =  \Sigma_{s}^{*}(\alpha,\beta)
        + \Sigma_{im}^{*}(\alpha,\beta;E) + \Sigma_{d}^{*}(\alpha,\beta;E).
\end{equation}
The energy-independent real part $\Sigma_{s}(\alpha,\beta)$ and 
energy-dependent imaginary part $\Sigma_{im}^{*}(\alpha,\beta)$
parameterizations are given in the following two subsections. The
dispersive correction term $\Sigma_{d}^{*}*\alpha,\beta;E)$ is completely
determined by an integral over the imaginary part [Eq. (3) of
\cite{Mahzoon2014}]. All free parameters that are fit via MCMC sampling are
typeset in \textbf{bold}.

\subsection{Real Part}
The energy-independent real part of the self-energy consists of a nonlocal Hartree-Fock and
a spin-orbit component (plus a local Coulomb term if the nucleon in question is a proton):
\begin{equation}
    \Sigma_{s}(r,r') =
    \Sigma_{HF}(r,r')+V_{so}(r,r')+V_{C}(r)\delta(r-r').
\end{equation}
The Coulomb potential is calculated using the same experimentally derived charge
density distributions (see \cite{DeVries1987}) used in fitting.
The Hartree-Fock component $V_{HF}$ has two subcomponents:
\begin{equation} \label{HFWBEquation}
    \Sigma_{HF}(r,r') = V_{vol}(r,r') + V_{wb}(r),
\end{equation}
where the nonlocal Hartree-Fock volume term $V_{vol}(r,r')$, is defined as
a Woods-Saxon form coupled to a Gaussian nonlocality:
\begin{equation} \label{RealVolume}
    V_{vol}(r,r') = -\mathbf{V_{1}}{\times}f_{vol}(r; \mathbf{r_{1}}, \mathbf{a_{1}})
    {\times}N(r,r';\boldsymbol{\beta_{1}}).
\end{equation}
The local Hartree-Fock wine-bottle
term $V_{wb}$, named for resemblance to the dimple at the bottom of a wine
bottle, is defined as a Gaussian centered at the nuclear origin,
\begin{equation}
    V_{wb}(r) = \mathbf{V_{2}}{\times}e^{r^{2}/\boldsymbol{\sigma_{2}}^{2}}.
\end{equation}

The real spin-orbit component $V_{so}$ is defined using a derivative-Woods-Saxon
shape in keeping with the expectation that the spin-orbit coupling is strongest near the
nuclear surface:
\begin{equation} \label{RealSOEquation}
    \begin{split}
        V_{so}(r,r') = \left(\frac{\hbar}{m_{\pi}c}\right)^{2}
        \mathbf{V_{3}}\times\frac{1}{r}f_{sur}(r;\mathbf{r_{3}}, \mathbf{a_{3}})\\
        {\times}N(r,r';\boldsymbol{\beta_{3}}) {\times}(\ell\cdot\sigma).
    \end{split}
\end{equation}
The leading constant $\left(\frac{\hbar}{m_{\pi}c}\right)^{2}$ is taken to be 2.0 fm$^{2}$
\cite{MahzoonPhDThesis}. In total, there are ten free parameters for the
symmetric real part of the potential.

\subsection{Imaginary Part}
The imaginary part of the potential is comprised of independent surface and volume terms
both above and below the Fermi surface:
\begin{equation}
    \Sigma_{im}^{*}(r,r',E) =
    \Sigma_{vol}^{\pm}(r,r',E)
    + \Sigma_{sur}^{\pm}(r,r',E),
\end{equation}
where the volume and surface components are defined:
\begin{equation}
    \begin{split}
        \Sigma_{vol}^{\pm}(r,r',E)
        = W_{vol}^{\pm}(E) & {\times}f_{vol}(r; \mathbf{r_{4}^{\pm}},
        \mathbf{a_{4}^{\pm}})\\
        & {\times}N(r,r'; \boldsymbol{\beta_{4}^{\pm}}),\\
        \Sigma_{sur}^{\pm}(r,r',E)
        = 4\mathbf{a_{5}}W_{sur}^{\pm}(E) & {\times}f_{sur}(r;
        \mathbf{r_{5}^{\pm}}, \mathbf{a_{5}^{\pm}})\\
        & {\times} N(r,r';\boldsymbol{\beta_{5}^{\pm}}).
    \end{split}
\end{equation}
The terms labeled with $+$ determine the potential above $\epsilon_{F}$, and the terms labeled
with $-$ determine the potential below $\epsilon_{F}$.
The energy dependence of the imaginary volume terms read:
\begin{equation} \label{ImagVolume}
    W_{vol}^{\pm}(E) = \mathbf{A_{4}^{\pm}}\left[\frac{(E_{\Delta})^{4}}
    {(E_{\Delta})^{4}+(\mathbf{B_{4}^{\pm}})^{4}} + W_{NM}^{\pm}(E)\right],
\end{equation}
where $E_{\Delta} = |E-\epsilon_{F}|$ and
\begin{equation}
    \begin{split}
        W_{NM}^{+}(E) & = {\boldsymbol{\alpha_{4}}}\left[\sqrt{E} +
            \frac{(\epsilon_{F}+\mathbf{E^{+}_{4}})^{\frac{3}{2}}}{2E}
        -\frac{3}{2}\sqrt{\epsilon_{F}+\mathbf{E^{+}_{4}}}\right],\\
        W_{NM}^{-}(E) & = \frac{(\epsilon_{F}-E-\mathbf{E_{4}^{-}})^{2}}
        {(\epsilon_{F}-E-\mathbf{E_{4}^{-}})^{2}+(\mathbf{E_{4}^{-}})^{2}}.
        \end{split}
    \end{equation}
The terms $W_{NM}^{\pm}$ are asymmetric above and below the Fermi surface and are modeled after
nuclear-matter calculations. They account for the decreasing phase space at negative energies
and the increasing phase space at positive energies.
The energy-dependence of the imaginary surface terms read:
\begin{equation} \label{ImagSurface}
    W_{sur}^{\pm}(E) = \omega_{4}(E, \mathbf{A_{5}^{\pm}}, \mathbf{B_{5}^{\pm}}, 0)
    - \omega_{2}(E, \mathbf{A_{5}^{\pm}}, \mathbf{B_{5}^{'\pm}}, \mathbf{C_{5}^{\pm}})
\end{equation}
In total, there are
thirteen free parameters for the symmetric imaginary volume terms of the
potential and fourteen free parameters for the symmetric imaginary surface terms
of the potential. Thus for symmetric nuclei, thirty-seven real and imaginary
parameters were used.

\subsection{Parameterization of Asymmetry Dependence}
For asymmetric nuclei,
the parametric forms must be modified to account for the different potential experienced by 
protons and neutrons. For the real central potential, the depth $\mathbf{V_{1}}$ and radius
$\mathbf{r_{1}}$ from Eq. (\ref{RealVolume}) were allowed to vary linearly with asymmetry:
\begin{equation} \label{HFDepthAsymmDependence}
    \mathbf{V_{1}} \Rightarrow \begin{cases}
        \mathbf{V_{1}} + \mathbf{V_{asym}}\times\frac{N-Z}{A} & \text{for protons}\\
        \mathbf{V_{1}} - \mathbf{V_{asym}}\times\frac{N-Z}{A} & \text{for neutrons},
    \end{cases}
\end{equation}
\begin{equation} \label{HFRadAsymmDependence}
    \mathbf{r_{1}} \Rightarrow \begin{cases}
        \mathbf{r_{1}} + \mathbf{r_{asym}}\times\frac{N-Z}{A} & \text{for protons}\\
        \mathbf{r_{1}} - \mathbf{r_{asym}}\times\frac{N-Z}{A} & \text{for neutrons}.
    \end{cases}
\end{equation}
The magnitude of the energy-dependence for the imaginary surface and volume potentials,
$\mathbf{A_{4}^{\pm}}$ and $\mathbf{A_{5}^{\pm}}$ from Eqs. (\ref{ImagVolume}) and
(\ref{ImagSurface}), were also allowed to vary with linearly with asymmetry:
\begin{equation} \label{ImVolAsymmDependence}
    \mathbf{A_{4}^{\pm}} \Rightarrow \begin{cases}
        \mathbf{A_{4}^{\pm}} + \mathbf{A_{vol, asym}^{\pm}}\times\frac{N-Z}{A} & \text{for protons}\\
        \mathbf{A_{4}^{\pm}} - \mathbf{A_{vol, asym}^{\pm}}\times\frac{N-Z}{A} & \text{for neutrons},
    \end{cases}
\end{equation}
\begin{equation} \label{ImSurAsymmDependence}
    \mathbf{A_{5}^{\pm}} \Rightarrow \begin{cases}
        \mathbf{A_{5}^{\pm}} + \mathbf{A_{sur, asym}^{\pm}}\times\frac{N-Z}{A} & \text{for protons}\\
        \mathbf{A_{5}^{\pm}} - \mathbf{A_{sur, asym}^{\pm}}\times\frac{N-Z}{A} & \text{for neutrons}.
    \end{cases}
\end{equation}
There should be no confusion between $\mathbf{A_{4,5}^{\pm}}$, $A$ (the total number
of nucleons), and the analyzing power. 
With these six additional asymmetry-dependent terms,
the total number of free parameters used for
fitting asymmetric nuclei in the present work totals forty-three.

\clearpage
\onecolumngrid
\appendix \label{DOMParameterValues}
\section{Appendix B: Parameter Values for DOM Potential}
Parameter labels correspond to those in the equations of Appendix A. 
For each parameter, the prior distribution was defined to be uniform with minimum and maximum values
listed in columns 2 and 3 of each table. For each nucleus, the
16\textsuperscript{th}, 50\textsuperscript{th}, and
84\textsuperscript{th} percentile values for each estimated parameter
distribution are listed. The format is $\text{50}^{\text{84}}_{\text{16}}$.
For \pbEight, the asymmetry-dependent HF radius term ($\mathbf{r_{asym}}$) was disabled during fitting.
\begin{table}[htbp]
    \caption{Real central potential parameters}
    \bgroup
\def\arraystretch{1.5}%
\begin{tabular}{ c c c c c c c c c c} 
\textbf{Par.} & \textbf{Min} & \textbf{Max} &                \textbf{Units} & \textbf{Eq.}& \textbf{$\mathbf{^{16,18}}$O}& \textbf{$\mathbf{^{40,48}}$Ca}& \textbf{$\mathbf{^{58,64}}$Ni}& \textbf{$\mathbf{^{112,124}}$Sn}& \textbf{$\mathbf{^{208}}$Pb}\\
 \hline 
$\mathbf{V_{1}}$ & 50 & 150 & MeV & 19 & $112.0^{124.8}_{100.1}$ & $101.6^{111.3}_{92.3}$ & $103.4^{115.8}_{92.5}$ & $108.7^{119.0}_{98.2}$ & $102.6^{120.4}_{91.0}$\\ 
$\mathbf{V_{asym}}$ & -100 & 200 & MeV & 27 & $-10.66^{34.39}_{-49.61}$ & $40.58^{53.81}_{28.47}$ & $-17.32^{8.71}_{-43.29}$ & $24.59^{43.08}_{4.09}$ & $30.36^{42.05}_{20.18}$\\ 
$\mathbf{r_{1}}$ & 0.6 & 1.6 & fm & 19 & $0.99^{1.03}_{0.95}$ & $1.10^{1.13}_{1.07}$ & $1.09^{1.12}_{1.06}$ & $1.11^{1.14}_{1.09}$ & $1.12^{1.16}_{1.09}$\\ 
$\mathbf{r_{asym}}$ & -1.0 & 1.0 & fm & 28 & $0.10^{0.30}_{-0.11}$ & $-0.01^{0.05}_{-0.10}$ & $0.34^{0.45}_{0.21}$ & $-0.04^{0.05}_{-0.13}$ & -\\ 
$\mathbf{a_{1}}$ & 0.4 & 1.0 & fm & 19 & $0.51^{0.56}_{0.46}$ & $0.58^{0.63}_{0.54}$ & $0.60^{0.64}_{0.56}$ & $0.48^{0.58}_{0.42}$ & $0.68^{0.75}_{0.60}$\\ 
$\mathbf{\beta_{1}}$ & 0.5 & 1.5 & fm & 19 & $1.05^{1.13}_{0.96}$ & $1.14^{1.20}_{1.06}$ & $1.10^{1.19}_{1.02}$ & $1.17^{1.23}_{1.12}$ & $1.14^{1.23}_{1.06}$\\ 
$\mathbf{V_{2}}$ & 0 & 50 & MeV & 20 & $27.76^{43.62}_{10.72}$ & $26.00^{42.79}_{7.53}$ & $24.68^{40.64}_{7.01}$ & $29.51^{44.77}_{10.54}$ & $25.50^{42.30}_{8.48}$\\ 
$\mathbf{\sigma_{2}}$ & 0 & 3 & fm & 20 & $0.11^{0.20}_{0.04}$ & $0.16^{0.25}_{0.05}$ & $0.17^{0.26}_{0.05}$ & $0.26^{0.33}_{0.21}$ & $0.17^{0.27}_{0.07}$\\ 
\\ 
\end{tabular}
\egroup

    \caption{Imaginary central potential parameters}
    \bgroup
\def\arraystretch{1.5}%
\begin{tabular}{ c c c c c c c c c c} 
\textbf{Par.} & \textbf{Min} & \textbf{Max} &                \textbf{Units} & \textbf{Eq.}& \textbf{$\mathbf{^{16,18}}$O}& \textbf{$\mathbf{^{40,48}}$Ca}& \textbf{$\mathbf{^{58,64}}$Ni}& \textbf{$\mathbf{^{112,124}}$Sn}& \textbf{$\mathbf{^{208}}$Pb}\\
 \hline 
$\mathbf{A_{4}^{+}}$ & 0 & 60 & MeV & 24 & $34.24^{49.21}_{23.22}$ & $23.14^{34.87}_{16.87}$ & $25.69^{41.18}_{15.61}$ & $25.60^{36.94}_{20.18}$ & $26.46^{35.55}_{19.27}$\\ 
$\mathbf{B_{4}^{+}}$ & 0 & 200 & MeV & 24 & $71.20^{86.82}_{56.97}$ & $74.90^{95.95}_{56.62}$ & $77.01^{97.72}_{51.18}$ & $53.22^{66.99}_{43.60}$ & $65.46^{77.13}_{52.45}$\\ 
$\mathbf{r_{4}^{+}}$ & 0.6 & 1.6 & fm & 23 & $0.92^{1.16}_{0.72}$ & $1.19^{1.31}_{1.03}$ & $1.34^{1.44}_{1.20}$ & $1.23^{1.32}_{1.14}$ & $1.28^{1.33}_{1.22}$\\ 
$\mathbf{a_{4}^{+}}$ & 0.4 & 1.0 & fm & 23 & $0.82^{0.94}_{0.64}$ & $0.78^{0.93}_{0.60}$ & $0.65^{0.83}_{0.50}$ & $0.78^{0.93}_{0.62}$ & $0.68^{0.84}_{0.54}$\\ 
$\mathbf{\beta_{4}^{+}}$ & 0.5 & 1.5 & fm & 23 & $0.62^{0.75}_{0.53}$ & $0.59^{0.67}_{0.53}$ & $0.73^{0.82}_{0.64}$ & $0.68^{0.73}_{0.62}$ & $0.60^{0.67}_{0.54}$\\ 
$\mathbf{A_{4}^{-}}$ & 0 & 60 & MeV & 24 & $10.05^{24.98}_{3.83}$ & $34.26^{51.39}_{15.06}$ & $28.15^{37.91}_{18.70}$ & $30.56^{42.31}_{19.87}$ & $38.00^{51.09}_{26.52}$\\ 
$\mathbf{B_{4}^{-}}$ & 0 & 200 & MeV & 24 & $130.3^{177.1}_{70.0}$ & $110.5^{153.9}_{63.2}$ & $79.0^{125.2}_{38.7}$ & $72.7^{117.2}_{34.3}$ & $105.8^{159.3}_{53.9}$\\ 
$\mathbf{r_{4}^{-}}$ & 0.6 & 1.6 & fm & 23 & $1.09^{1.36}_{0.80}$ & $0.96^{1.14}_{0.77}$ & $1.00^{1.15}_{0.83}$ & $0.91^{1.07}_{0.79}$ & $1.12^{1.23}_{0.99}$\\ 
$\mathbf{a_{4}^{-}}$ & 0.4 & 1.0 & fm & 23 & $0.72^{0.90}_{0.52}$ & $0.61^{0.81}_{0.45}$ & $0.70^{0.87}_{0.51}$ & $0.80^{0.94}_{0.64}$ & $0.56^{0.76}_{0.44}$\\ 
$\mathbf{\beta_{4}^{-}}$ & 0.5 & 1.5 & fm & 23 & $1.02^{1.35}_{0.76}$ & $0.98^{1.31}_{0.74}$ & $1.00^{1.29}_{0.80}$ & $1.03^{1.31}_{0.83}$ & $1.15^{1.39}_{0.88}$\\ 
$\mathbf{\alpha_{4}}$ & 0 & 0.5 & - & 25 & $0.16^{0.29}_{0.06}$ & $0.20^{0.28}_{0.11}$ & $0.13^{0.29}_{0.03}$ & $0.18^{0.26}_{0.11}$ & $0.20^{0.30}_{0.11}$\\ 
$\mathbf{E_{4}^{+}}$ & 50 & 200 & MeV & 25 & $109.5^{160.5}_{71.7}$ & $109.9^{157.7}_{74.2}$ & $105.6^{160.6}_{61.4}$ & $90.0^{125.8}_{65.7}$ & $132.2^{178.0}_{85.1}$\\ 
$\mathbf{E_{4}^{-}}$ & 50 & 200 & MeV & 25 & $104.7^{143.9}_{77.4}$ & $101.3^{131.6}_{74.1}$ & $114.2^{144.5}_{85.2}$ & $127.9^{170.2}_{97.0}$ & $135.1^{171.3}_{97.3}$\\ 
$\mathbf{A_{vol,asym}^{+}}$ & -100 & 200 & MeV & 29 & $37.72^{76.02}_{9.43}$ & $11.39^{28.73}_{-0.39}$ & $8.47^{30.29}_{-13.51}$ & $7.53^{18.04}_{-4.93}$ & $17.44^{29.66}_{6.43}$\\ 
$\mathbf{A_{vol,asym}^{-}}$ & -100 & 200 & MeV & 29 & $131.1^{180.2}_{37.7}$ & $7.9^{117.8}_{-63.5}$ & $-10.39^{66.67}_{-61.63}$ & $-8.86^{70.86}_{-59.46}$ & $-9.27^{50.04}_{-66.04}$\\ 
\\ 
\end{tabular}
\egroup
\end{table}
\clearpage
\begin{table}[htbp!]
    \caption{Imaginary surface potential parameters}
    \bgroup
\def\arraystretch{1.5}%
\begin{tabular}{ c c c c c c c c c c} 
\textbf{Par.} & \textbf{Min} & \textbf{Max} &                \textbf{Units} & \textbf{Eq.}& \textbf{$\mathbf{^{16,18}}$O}& \textbf{$\mathbf{^{40,48}}$Ca}& \textbf{$\mathbf{^{58,64}}$Ni}& \textbf{$\mathbf{^{112,124}}$Sn}& \textbf{$\mathbf{^{208}}$Pb}\\
 \hline 
$\mathbf{A_{5}^{+}}$ & 0 & 50 & MeV & 26 & $24.18^{34.54}_{17.24}$ & $23.57^{31.91}_{16.31}$ & $25.22^{34.17}_{17.14}$ & $31.63^{41.32}_{22.73}$ & $32.98^{41.90}_{22.15}$\\ 
$\mathbf{B_{5}^{+}}$ & 0 & 50 & MeV & 26 & $21.96^{24.14}_{19.70}$ & $21.73^{24.57}_{18.83}$ & $18.34^{20.60}_{15.74}$ & $18.89^{21.90}_{16.54}$ & $18.42^{20.93}_{15.79}$\\ 
$\mathbf{B_{5}^{'+}}$ & 0 & 50 & MeV & 26 & $28.73^{37.59}_{20.87}$ & $41.40^{47.79}_{31.75}$ & $31.92^{40.20}_{23.80}$ & $29.08^{38.47}_{21.91}$ & $41.14^{47.18}_{31.04}$\\ 
$\mathbf{C_{5}^{+}}$ & 0 & 10 & MeV & 26 & $4.78^{8.31}_{1.81}$ & $5.76^{8.65}_{1.79}$ & $6.68^{8.83}_{3.29}$ & $3.01^{6.85}_{0.87}$ & $6.36^{8.62}_{2.75}$\\ 
$\mathbf{r_{5}^{+}}$ & 0.6 & 1.6 & fm & 23 & $1.38^{1.48}_{1.21}$ & $1.21^{1.31}_{1.07}$ & $1.22^{1.30}_{1.10}$ & $1.22^{1.29}_{1.08}$ & $1.22^{1.26}_{1.16}$\\ 
$\mathbf{a_{5}^{+}}$ & 0.4 & 1.0 & fm & 23 & $0.59^{0.79}_{0.49}$ & $0.73^{0.86}_{0.63}$ & $0.66^{0.80}_{0.56}$ & $0.67^{0.80}_{0.57}$ & $0.61^{0.76}_{0.51}$\\ 
$\mathbf{\beta_{5}^{+}}$ & 0.5 & 1.5 & fm & 23 & $1.04^{1.36}_{0.72}$ & $1.13^{1.36}_{0.84}$ & $0.99^{1.27}_{0.72}$ & $0.96^{1.25}_{0.72}$ & $0.87^{1.07}_{0.67}$\\ 
$\mathbf{A_{5}^{-}}$ & 0 & 50 & MeV & 26 & $23.02^{34.32}_{13.93}$ & $38.61^{47.59}_{23.44}$ & $24.85^{36.31}_{12.25}$ & $26.04^{34.54}_{17.03}$ & $35.08^{45.70}_{24.41}$\\ 
$\mathbf{B_{5}^{-}}$ & 0 & 50 & MeV & 26 & $11.78^{14.83}_{9.09}$ & $13.49^{18.26}_{9.98}$ & $9.07^{11.06}_{7.23}$ & $9.16^{11.28}_{7.51}$ & $15.77^{21.66}_{11.11}$\\ 
$\mathbf{B_{5}^{'-}}$ & 0 & 50 & MeV & 26 & $33.48^{44.79}_{21.44}$ & $36.32^{46.11}_{22.61}$ & $32.15^{43.99}_{20.96}$ & $28.47^{39.31}_{18.61}$ & $34.49^{43.95}_{23.61}$\\ 
$\mathbf{C_{5}^{-}}$ & 0 & 10 & MeV & 26 & $6.47^{9.02}_{3.35}$ & $6.24^{8.57}_{1.88}$ & $5.84^{8.71}_{2.54}$ & $5.51^{8.68}_{1.70}$ & $7.07^{9.24}_{4.03}$\\ 
$\mathbf{r_{5}^{-}}$ & 0.6 & 1.6 & fm & 23 & $0.76^{0.91}_{0.64}$ & $0.82^{0.93}_{0.67}$ & $0.78^{0.97}_{0.63}$ & $1.10^{1.14}_{1.02}$ & $1.01^{1.08}_{0.88}$\\ 
$\mathbf{a_{5}^{-}}$ & 0.4 & 1.0 & fm & 23 & $0.47^{0.57}_{0.42}$ & $0.51^{0.62}_{0.43}$ & $0.62^{0.74}_{0.48}$ & $0.53^{0.68}_{0.44}$ & $0.64^{0.85}_{0.50}$\\ 
$\mathbf{\beta_{5}^{-}}$ & 0.5 & 1.5 & fm & 23 & $1.17^{1.40}_{0.92}$ & $1.24^{1.39}_{0.98}$ & $1.12^{1.31}_{0.90}$ & $1.12^{1.34}_{0.91}$ & $0.91^{1.17}_{0.71}$\\ 
$\mathbf{A_{sur,asym}^{+}}$ & -100 & 200 & MeV & 30 & $-22.10^{15.74}_{-64.99}$ & $20.11^{45.79}_{2.93}$ & $9.42^{40.11}_{-25.44}$ & $54.32^{80.91}_{29.32}$ & $27.45^{55.00}_{6.87}$\\ 
$\mathbf{A_{sur,asym}^{-}}$ & -100 & 200 & MeV & 30 & $48.2^{142.3}_{-52.4}$ & $-7.68^{31.56}_{-47.07}$ & $12.92^{54.11}_{-28.07}$ & $11.35^{37.52}_{-16.09}$ & $-4.79^{24.43}_{-32.12}$\\ 
\\ 
\end{tabular}
\egroup
    \vspace{0.3em}
    \caption{Spin-orbit parameters}
    \bgroup
\def\arraystretch{1.5}%
\begin{tabular}{ c c c c c c c c c c} 
\textbf{Par.} & \textbf{Min} & \textbf{Max} &                \textbf{Units} & \textbf{Eq.}& \textbf{$\mathbf{^{16,18}}$O}& \textbf{$\mathbf{^{40,48}}$Ca}& \textbf{$\mathbf{^{58,64}}$Ni}& \textbf{$\mathbf{^{112,124}}$Sn}& \textbf{$\mathbf{^{208}}$Pb}\\
 \hline 
$\mathbf{V_{3}}$ & 0 & 20 & MeV & 21 & $10.44^{12.64}_{8.57}$ & $12.07^{13.93}_{10.36}$ & $13.48^{16.00}_{11.28}$ & $9.99^{12.49}_{8.00}$ & $13.05^{16.62}_{10.03}$\\ 
$\mathbf{r_{3}}$ & 0.6 & 1.6 & fm & 21 & $0.89^{1.00}_{0.79}$ & $0.93^{1.02}_{0.81}$ & $1.05^{1.14}_{0.90}$ & $1.05^{1.14}_{0.97}$ & $1.14^{1.20}_{1.05}$\\ 
$\mathbf{a_{3}}$ & 0.4 & 1.0 & fm & 21 & $0.60^{0.72}_{0.49}$ & $0.68^{0.79}_{0.57}$ & $0.68^{0.85}_{0.55}$ & $0.60^{0.77}_{0.46}$ & $0.77^{0.90}_{0.61}$\\ 
$\mathbf{\beta_{3}}$ & 0.5 & 1.5 & fm & 21 & $0.59^{0.80}_{0.53}$ & $0.63^{0.75}_{0.54}$ & $0.74^{1.00}_{0.58}$ & $0.83^{1.08}_{0.59}$ & $0.77^{1.05}_{0.60}$\\ 
\\ 
\end{tabular}
\egroup
\end{table}
\appendix \label{DOMFitResults}
\section{Appendix C: DOM Fit Comparison to Experimental Data}
Figures \ref{DOM_o16}-\ref{DOM_pb208} show the data sectors used to constrain the DOM potential.
Experimental scattering cross sections are shown as points with associated experimental error bars
in panels (a) through (f) of each figure. Experimental bound-state data are shown as bands in panels 
(g) through (j). DOM calculations for each data sector are plotted as 1$\sigma$ and 2$\sigma$ 
uncertainty bands. References for each data set are provided in Appendix B of \cite{PruittPhDThesis}.

Panels (a) and (c) show proton \el\ and analyzing powers from 10-200 MeV.
Panels (b) and (d) show neutron \el\ and analyzing powers from 10-200 MeV.
For visibility, data sets at different energies are offset vertically and
colored according to the scattering energy. Panels (e) show proton \rxn\ data. Experimental data
are plotted as black points and pseudo-data generated from \cite{Carlson1996} are plotted as gray 
open circles. Panels (f) show the neutron \tot\ and \rxn. The charge distributions of panels (g) are 
derived from the compilation of \cite{DeVries1987} (see comments in \textit{DOM Analysis} section), and 
are displayed with an arbitrary 1\% uncertainty band in black.
In panels (h), single-particle energies $\epsilon_{nlj}$ are shown as
horizontal lines. In the ``calc'' column, DOM-calculated single-particle energies are plotted; the
height of each rectangle spans the 1$\sigma$ calculated uncertainty for that level. Panels (i)
show DOM-calculated charge radii; the experimental charge radius is displayed
using dark gray and light gray bands representing 1$\sigma$ and 2$\sigma$ uncertainties,
respectively. Panels (j) show the DOM-calculated binding energy per nucleon; the experimental value is shown with a thin gray band.

\begin{figure*}[!htb]
    \centering
    \begin{minipage}{0.4\linewidth}
        \centering
        \includegraphics[width=\linewidth]{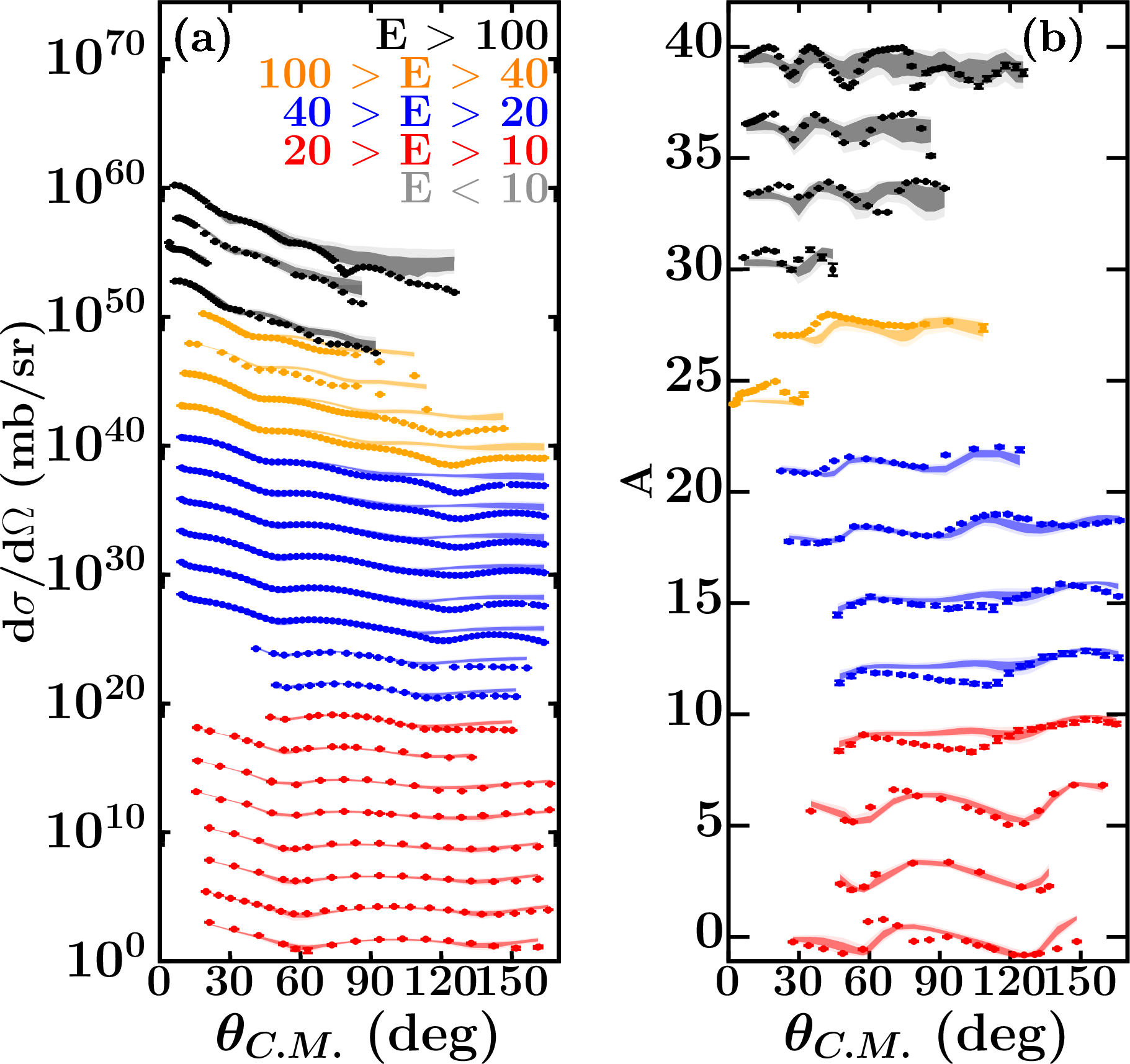}
        \label{DOM_o16_proton_elastic}
    \end{minipage}\hspace{6pt}
    \begin{minipage}{0.4\linewidth}
        \centering
        \includegraphics[width=\linewidth]{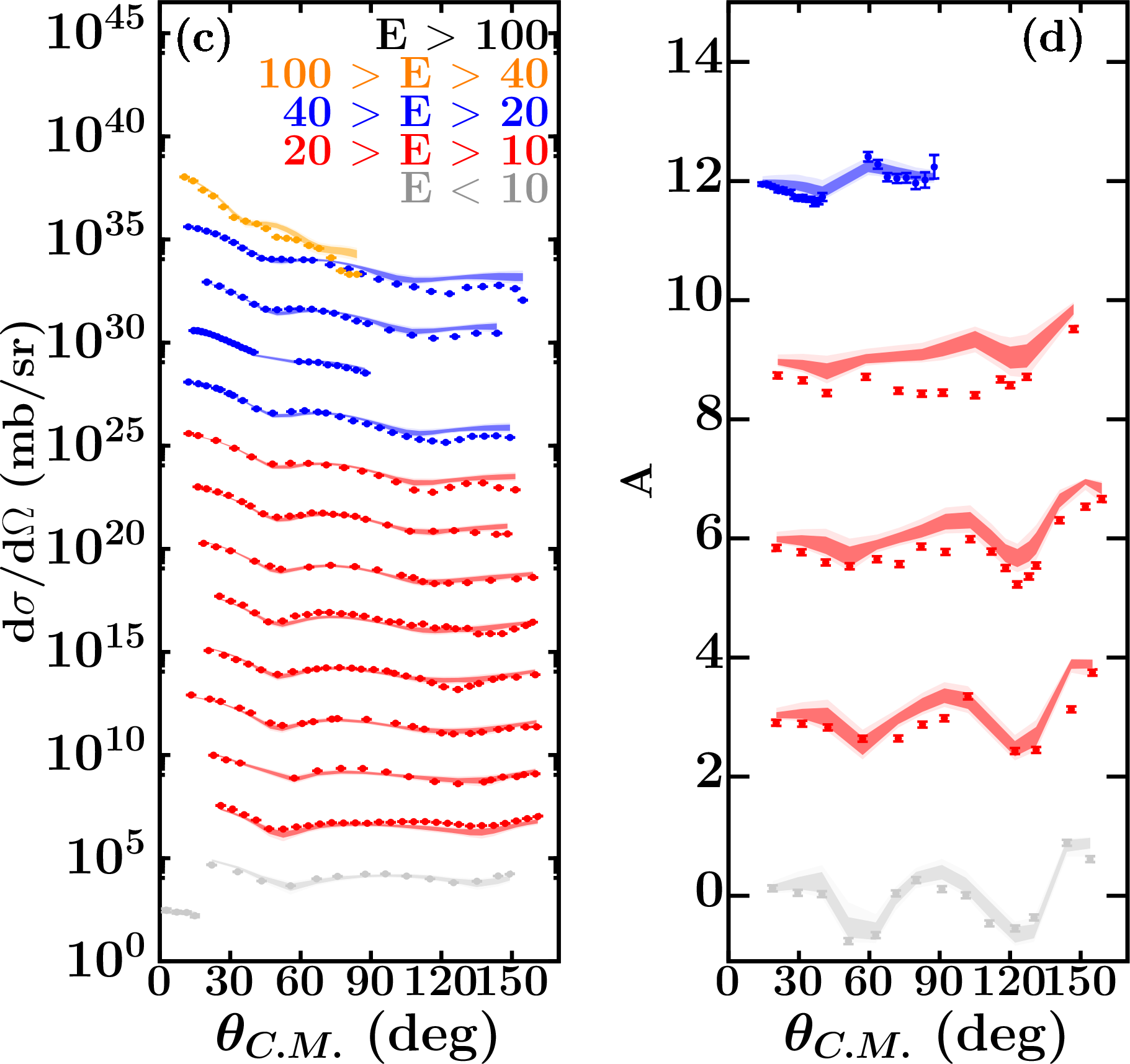}
        \label{DOM_o16_neutron_elastic}
    \end{minipage}
    \centering
    \begin{minipage}{0.4\linewidth}
        \centering
        \includegraphics[width=\linewidth]{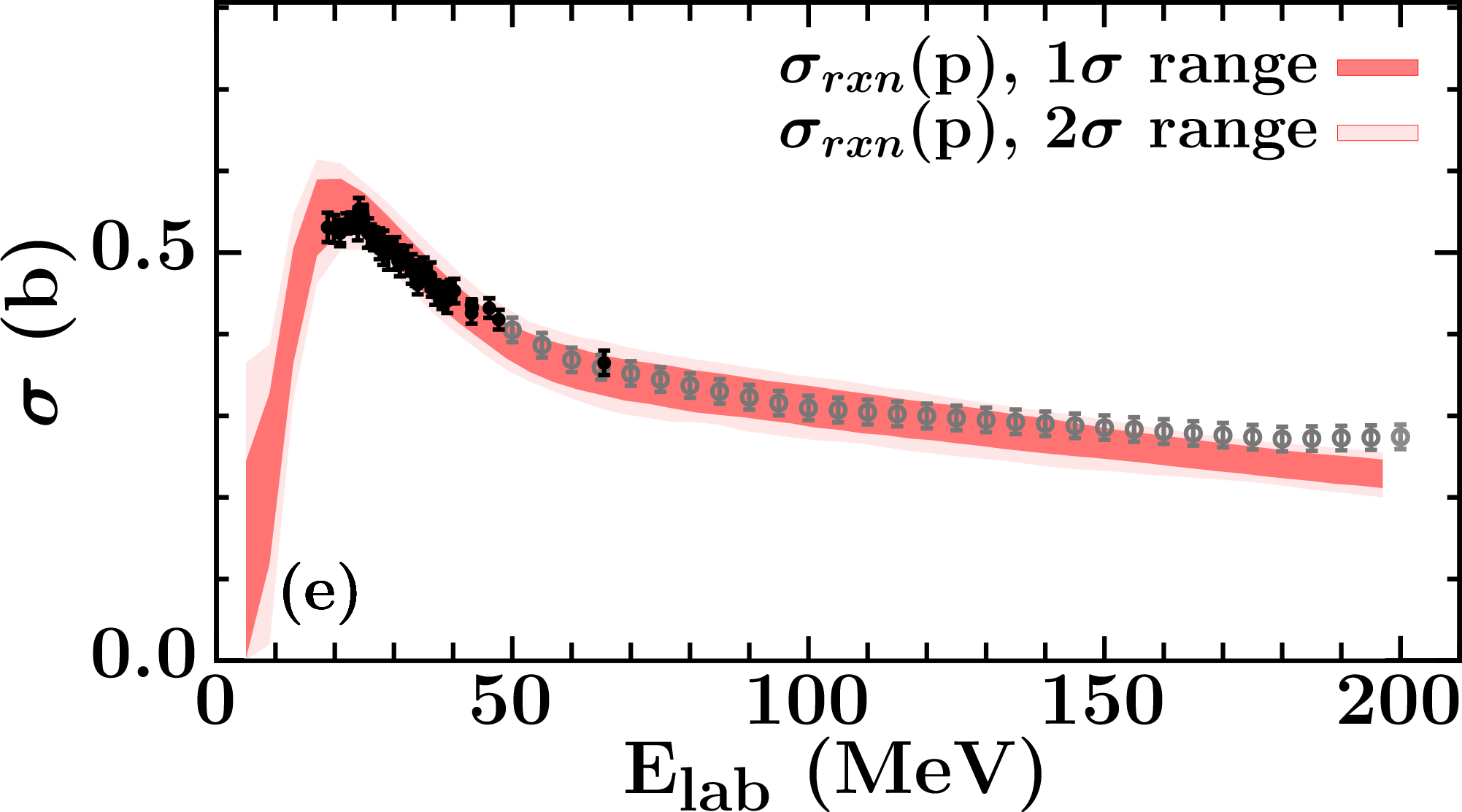}
        \label{DOM_o16_proton_inelastic}
    \end{minipage}\hspace{6pt}
    \begin{minipage}{0.4\linewidth}
        \centering
        \includegraphics[width=\linewidth]{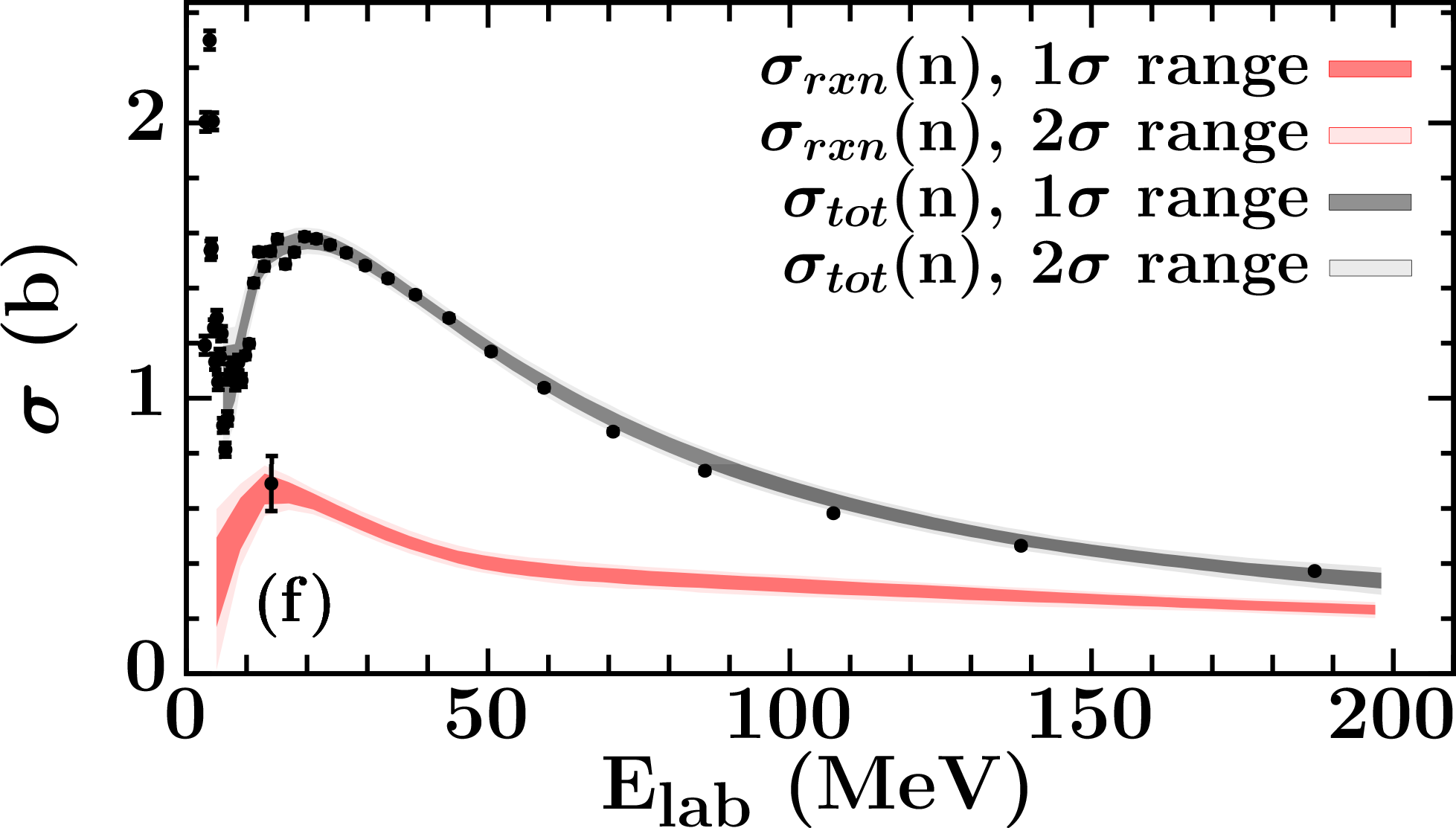}
        \label{DOM_o16_neutron_inelastic}
    \end{minipage}
    \centering
    \begin{minipage}{0.4\linewidth}
        \centering
        \includegraphics[width=\linewidth]{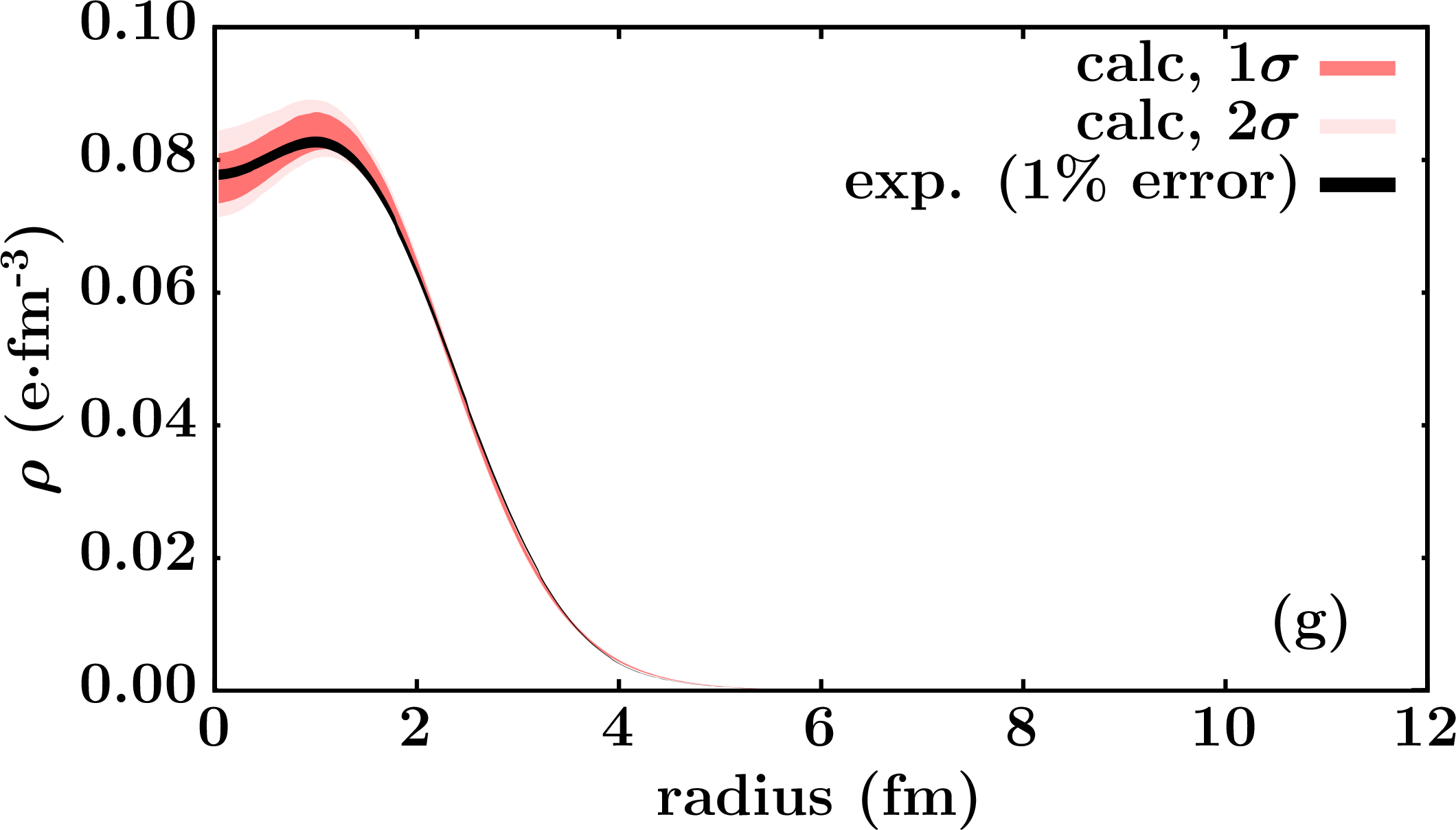}
        \label{DOM_o16_chargeDensity}
    \end{minipage}\hspace{6pt}
    \begin{minipage}{0.4\linewidth}
        \centering
        \includegraphics[width=\linewidth]{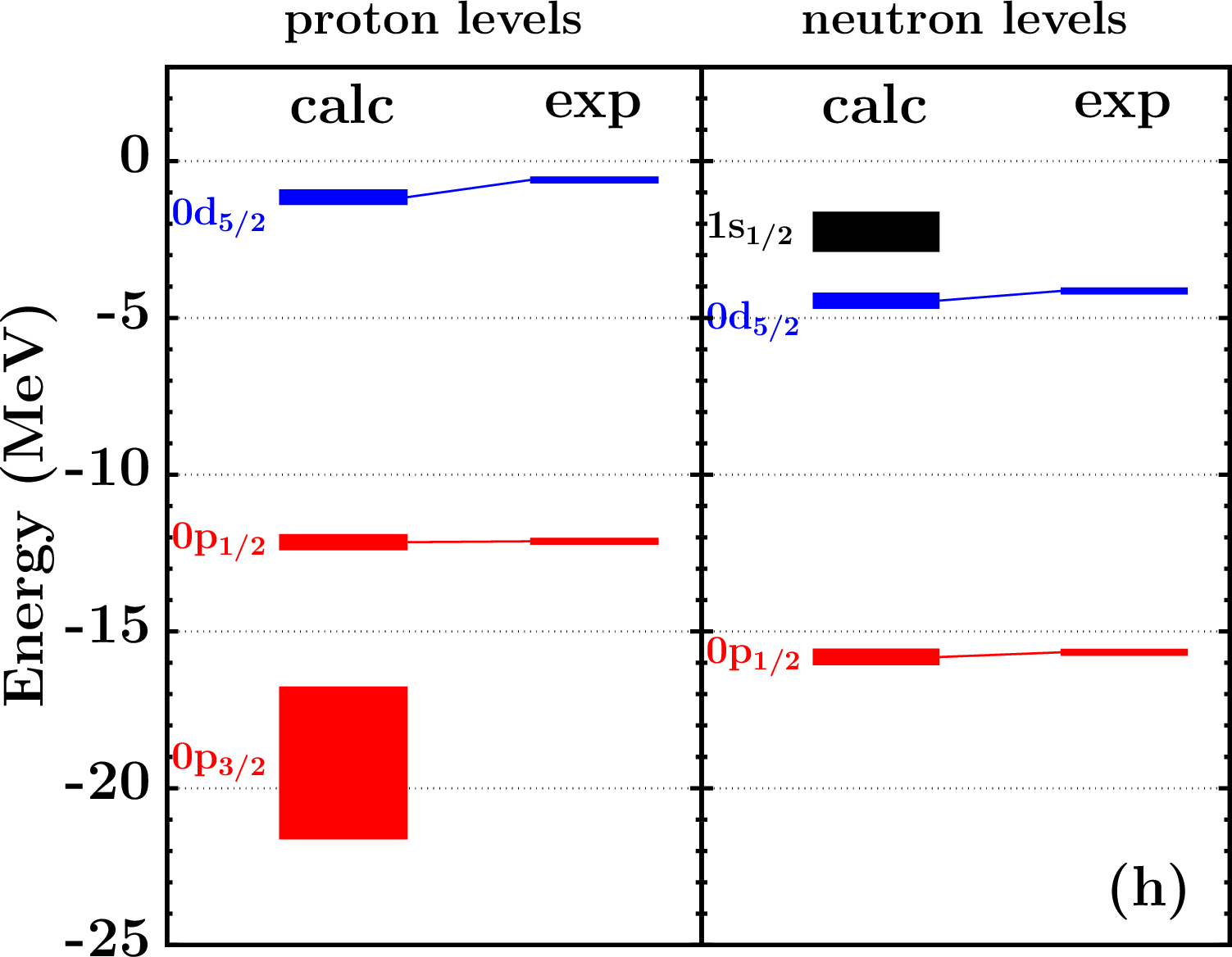}
        \label{DOM_o16_SPLevels}
    \end{minipage}
    \begin{minipage}{0.4\linewidth}
        \centering
        \includegraphics[width=\linewidth]{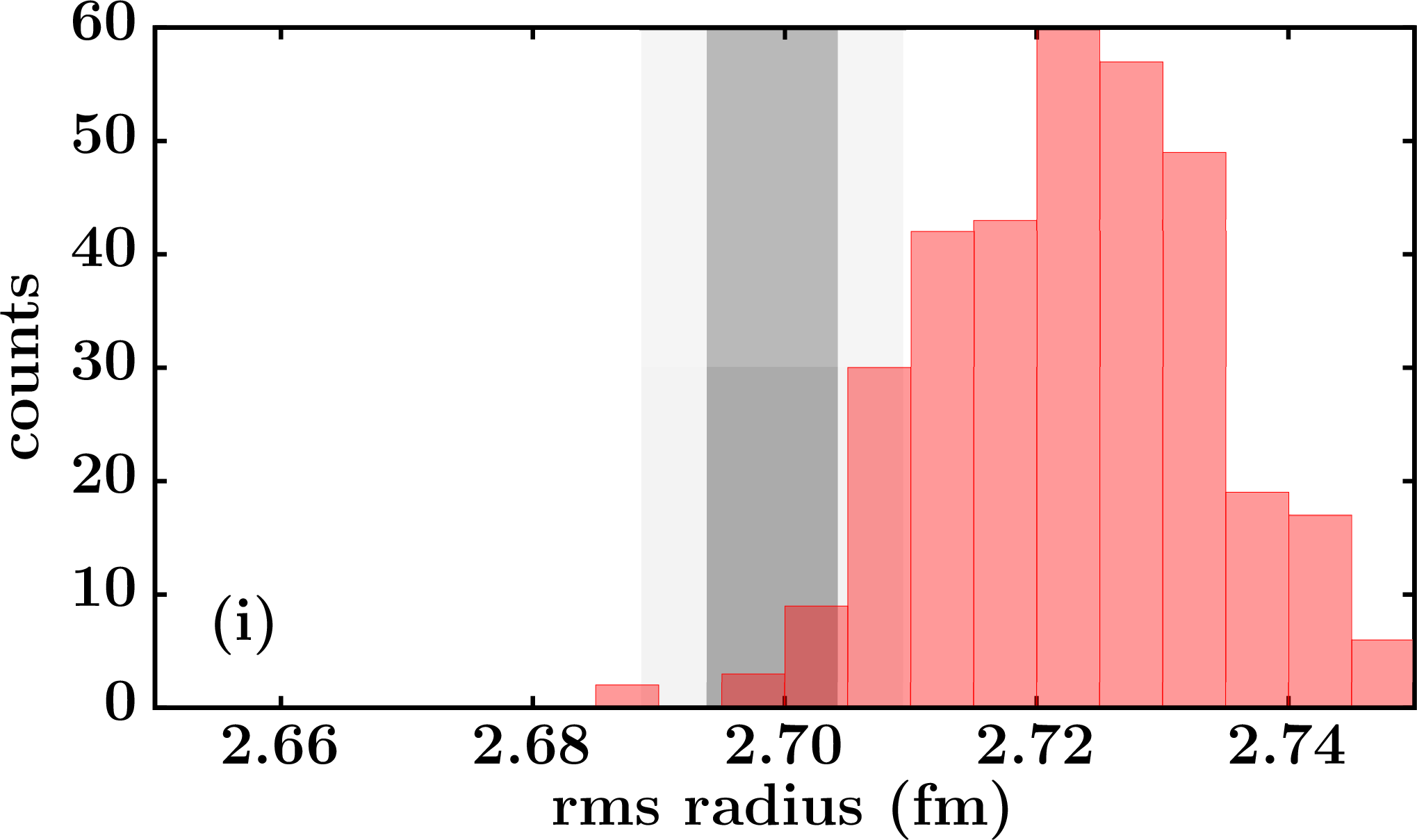}
        \label{DOM_o16_RMSRadius}
    \end{minipage}\hspace{6pt}
    \begin{minipage}{0.4\linewidth}
        \centering
        \includegraphics[width=\linewidth]{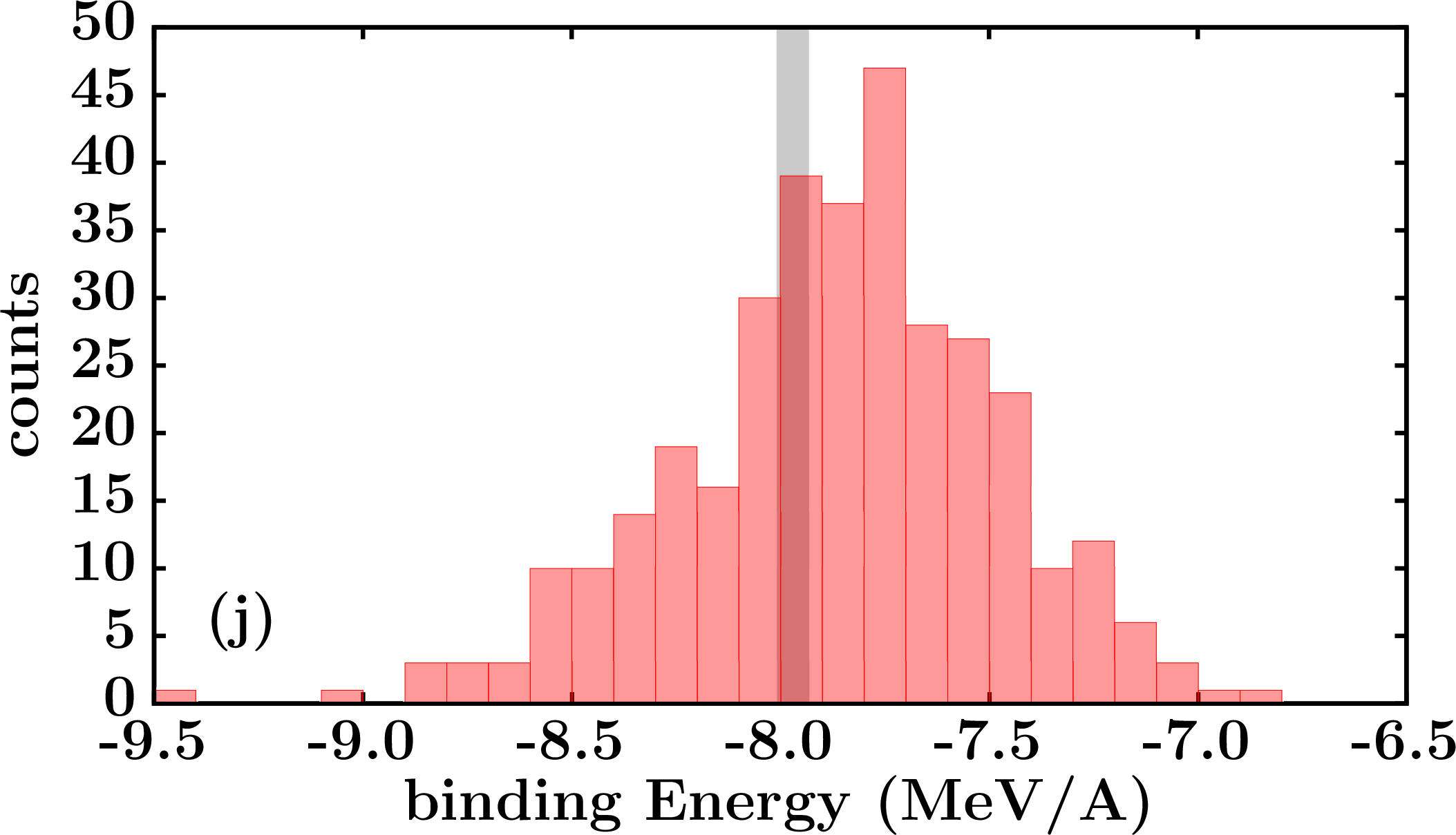}
        \label{DOM_o16_BE}
    \end{minipage}
    \caption{\oSix: constraining experimental data and DOM fit. See introduction of
    Appendix C for description.}
    \label{DOM_o16}
\end{figure*}

\begin{figure*}[!htb]
    \centering
    \begin{minipage}{0.4\linewidth}
        \centering
        \includegraphics[width=\linewidth]{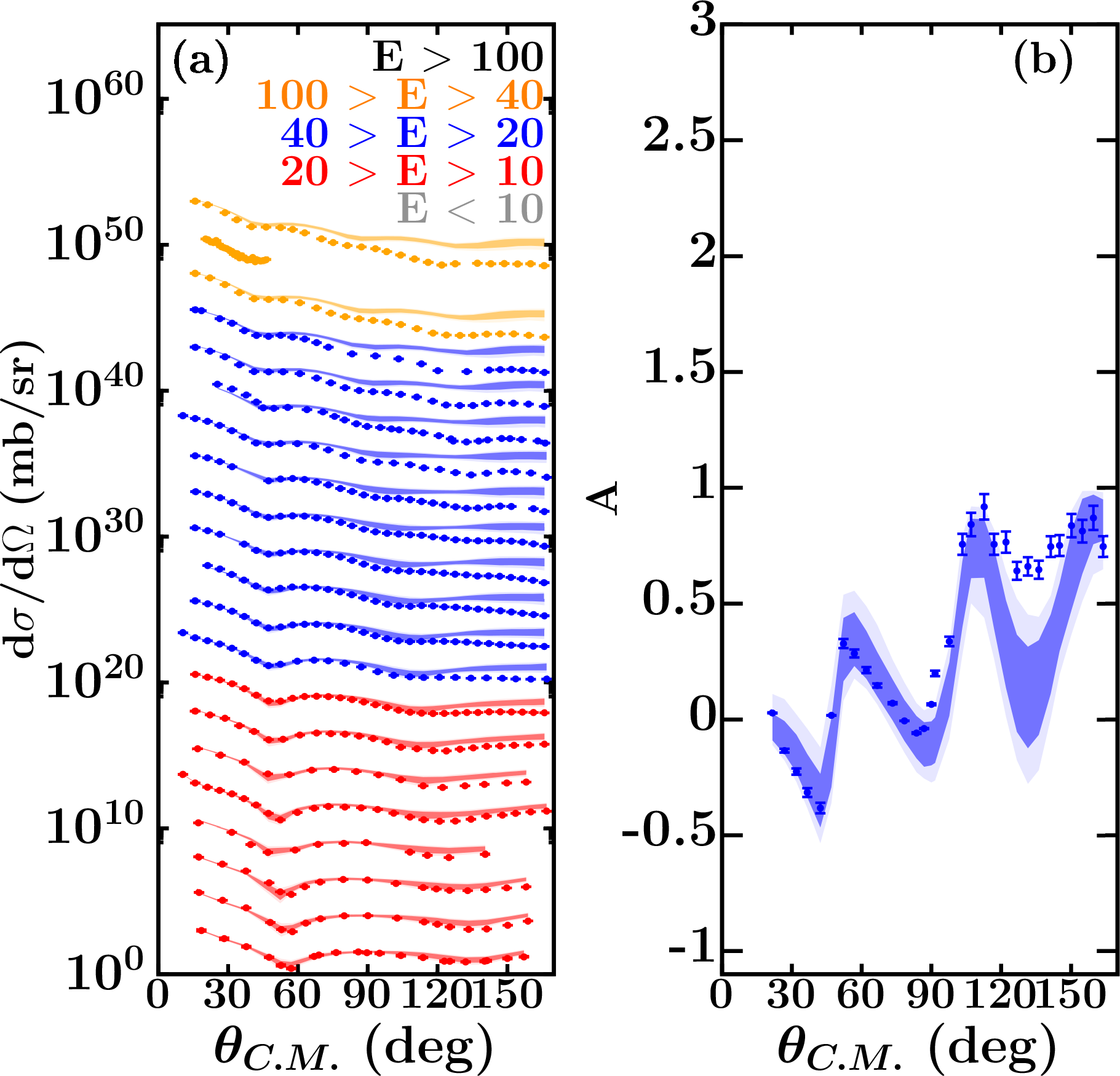}
        \label{DOM_o18_proton_elastic}
    \end{minipage}\hspace{6pt}
    \begin{minipage}{0.4\linewidth}
        \centering
        \vspace{-10pt}
        \begin{minipage}[c]{0.5\linewidth}
            \centering
                \includegraphics[width=\linewidth]{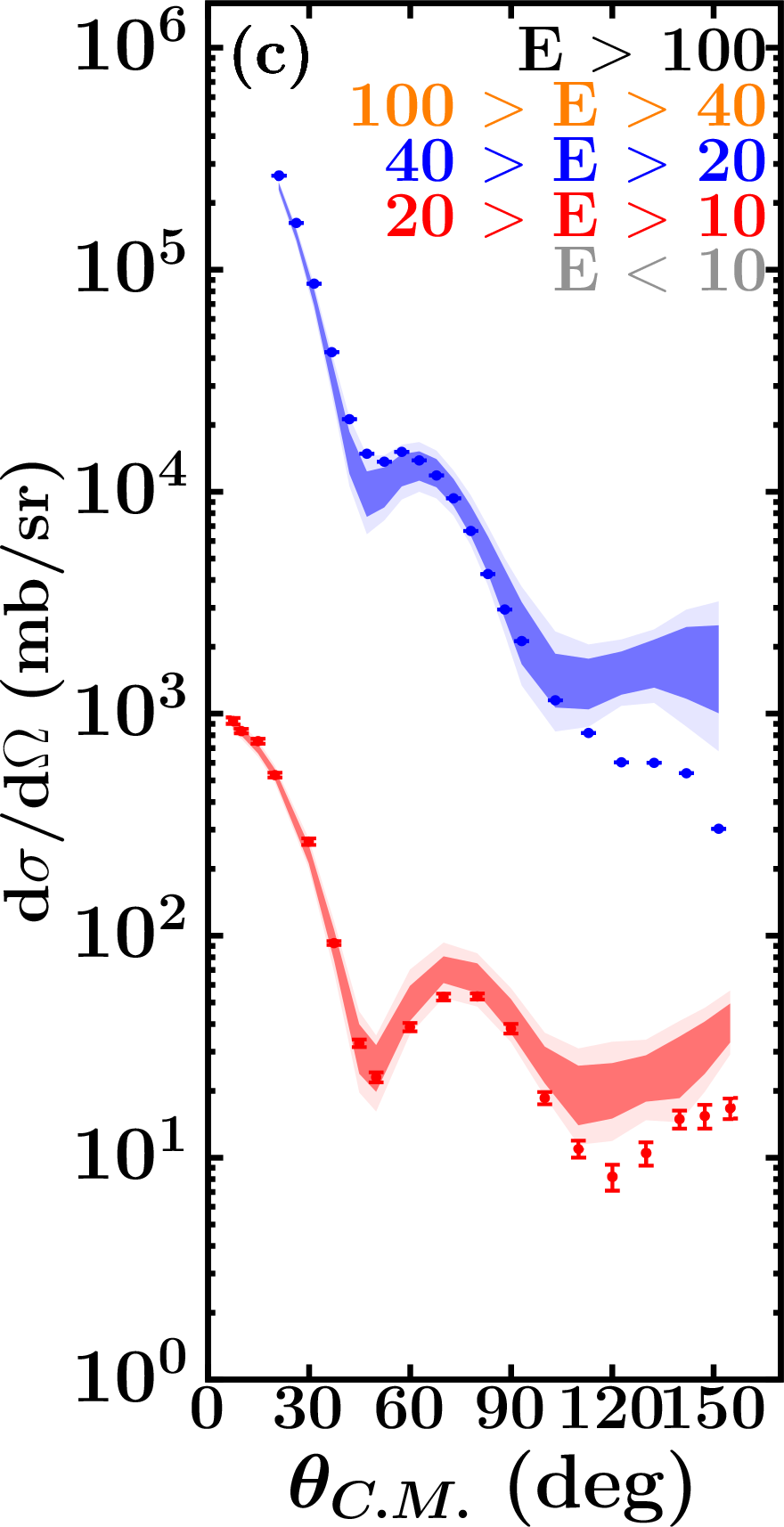}
        \end{minipage}
        \begin{minipage}[c]{0.45\linewidth}
            \centering
            No \oEight\ neutron \\
            analyzing powers\\
            were available
        \end{minipage}
        \label{DOM_o18_neutron_elastic}
    \end{minipage}
    \centering
    \begin{minipage}{0.4\linewidth}
        \centering
        \includegraphics[width=\linewidth]{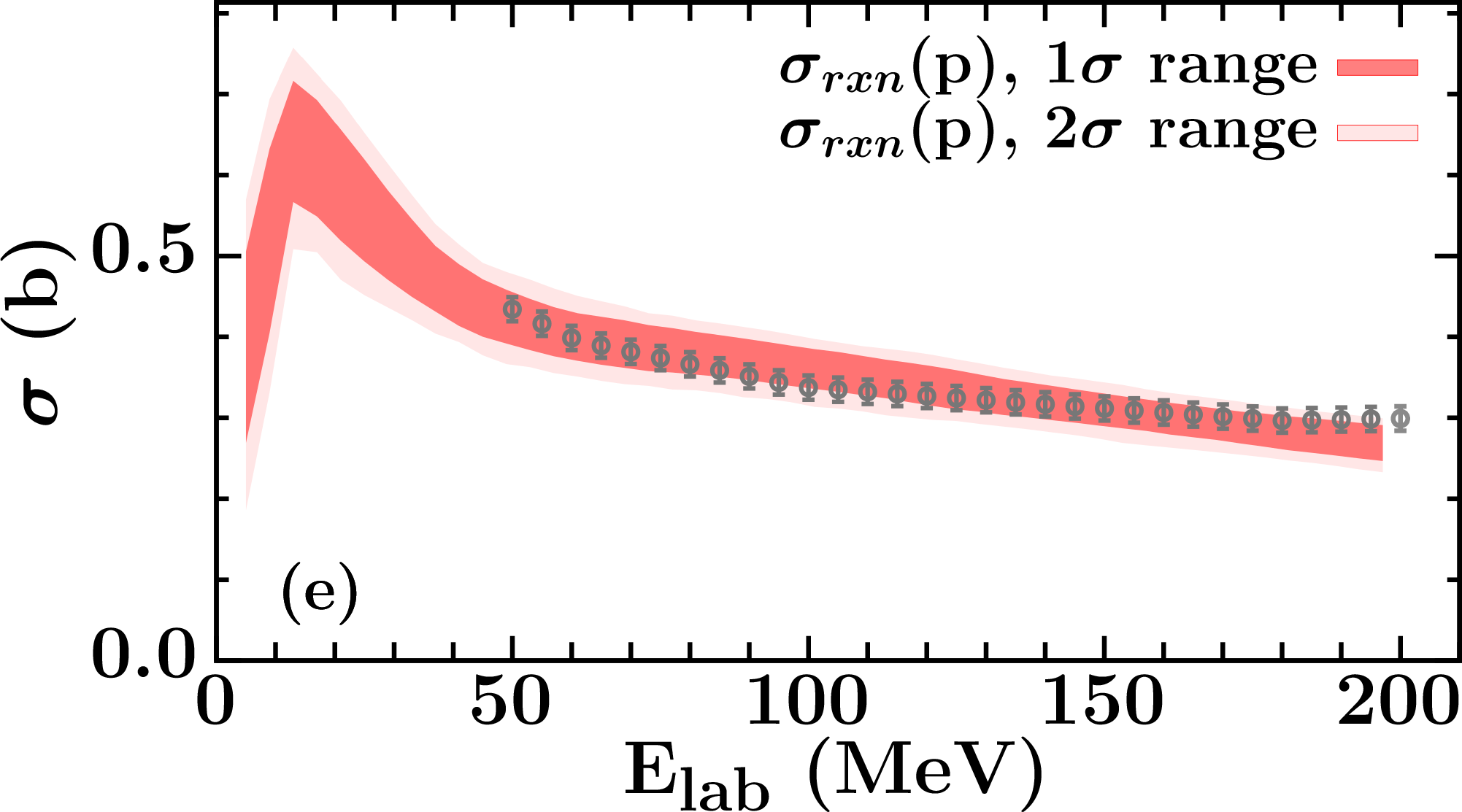}
        \label{DOM_o18_proton_inelastic}
    \end{minipage}\hspace{6pt}
    \begin{minipage}{0.4\linewidth}
        \centering
        \includegraphics[width=\linewidth]{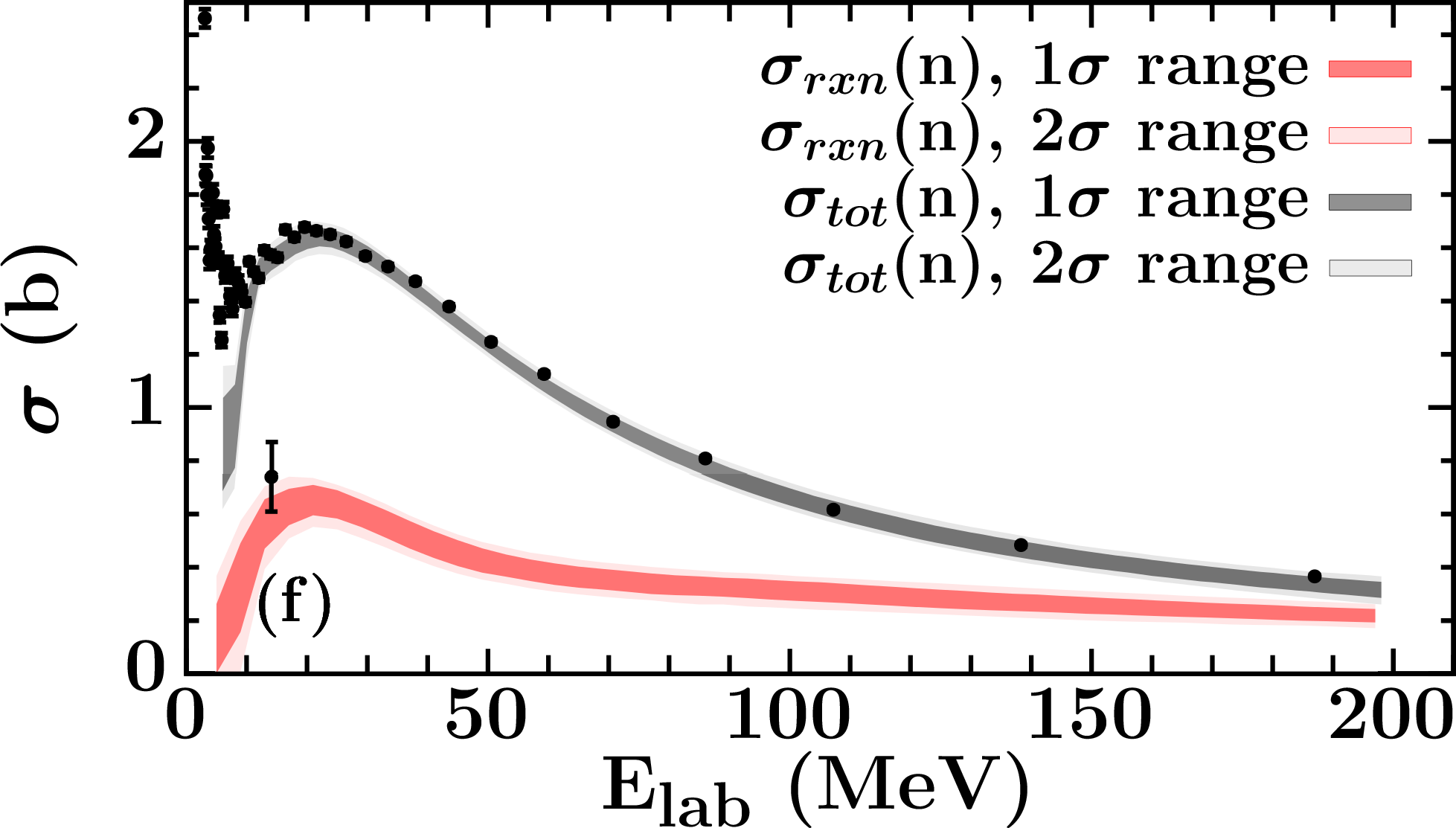}
        \label{DOM_o18_neutron_inelastic}
    \end{minipage}
    \centering
    \begin{minipage}{0.4\linewidth}
        \centering
        \includegraphics[width=\linewidth]{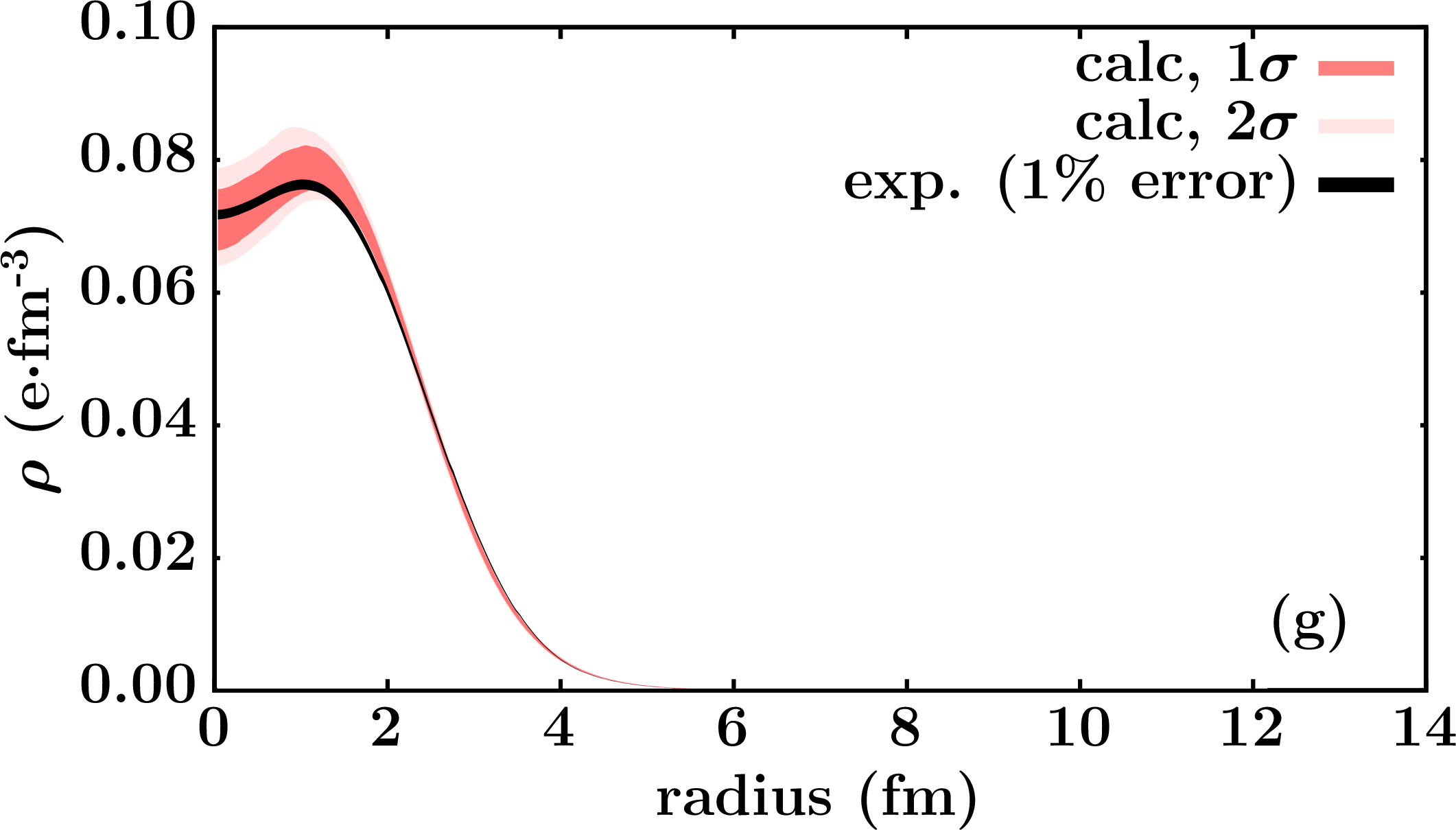}
        \label{DOM_o18_chargeDensity}
    \end{minipage}\hspace{6pt}
    \begin{minipage}{0.4\linewidth}
        \centering
        \includegraphics[width=\linewidth]{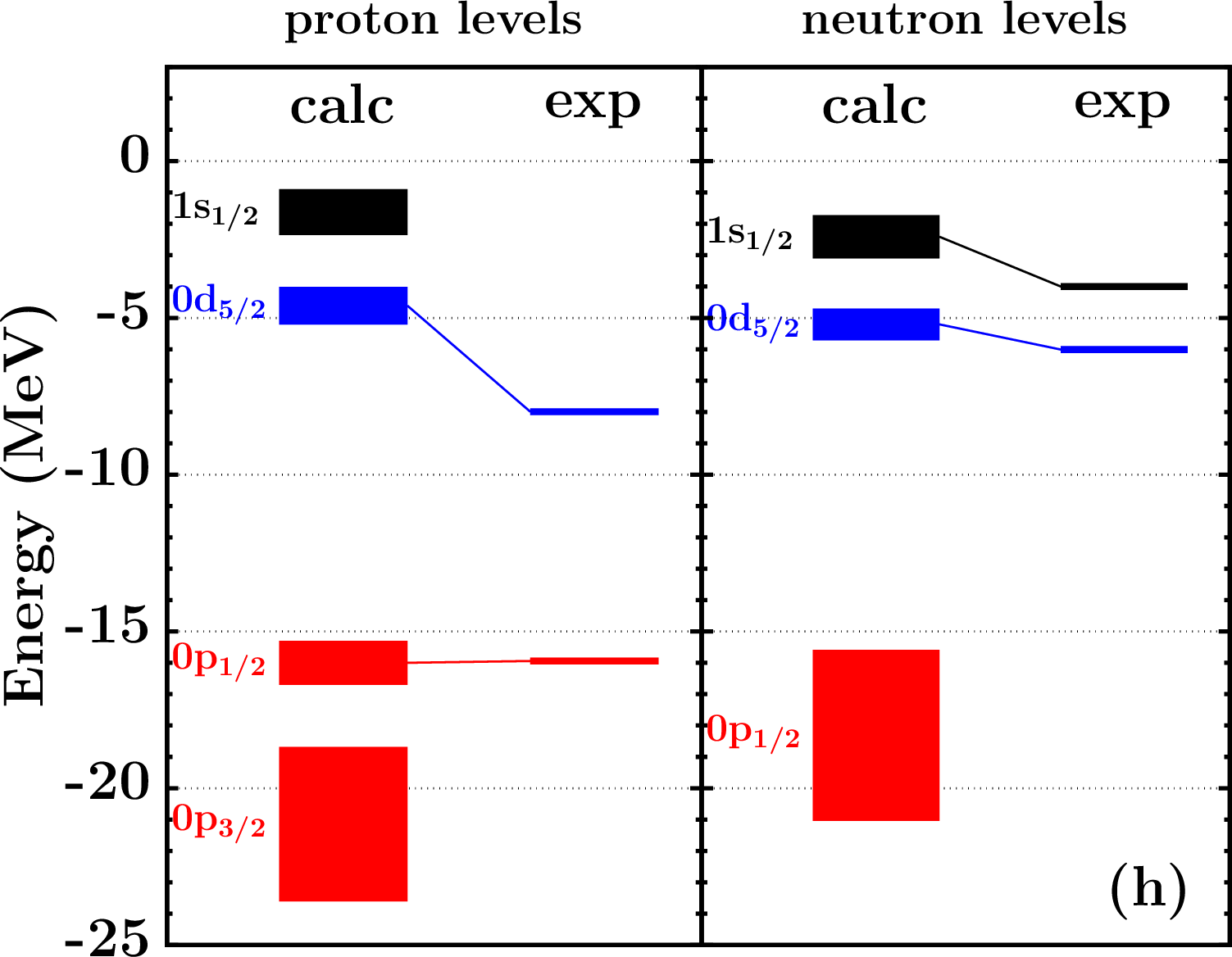}
        \label{DOM_o18_SPLevels}
    \end{minipage}
    \begin{minipage}{0.4\linewidth}
        \centering
        \includegraphics[width=\linewidth]{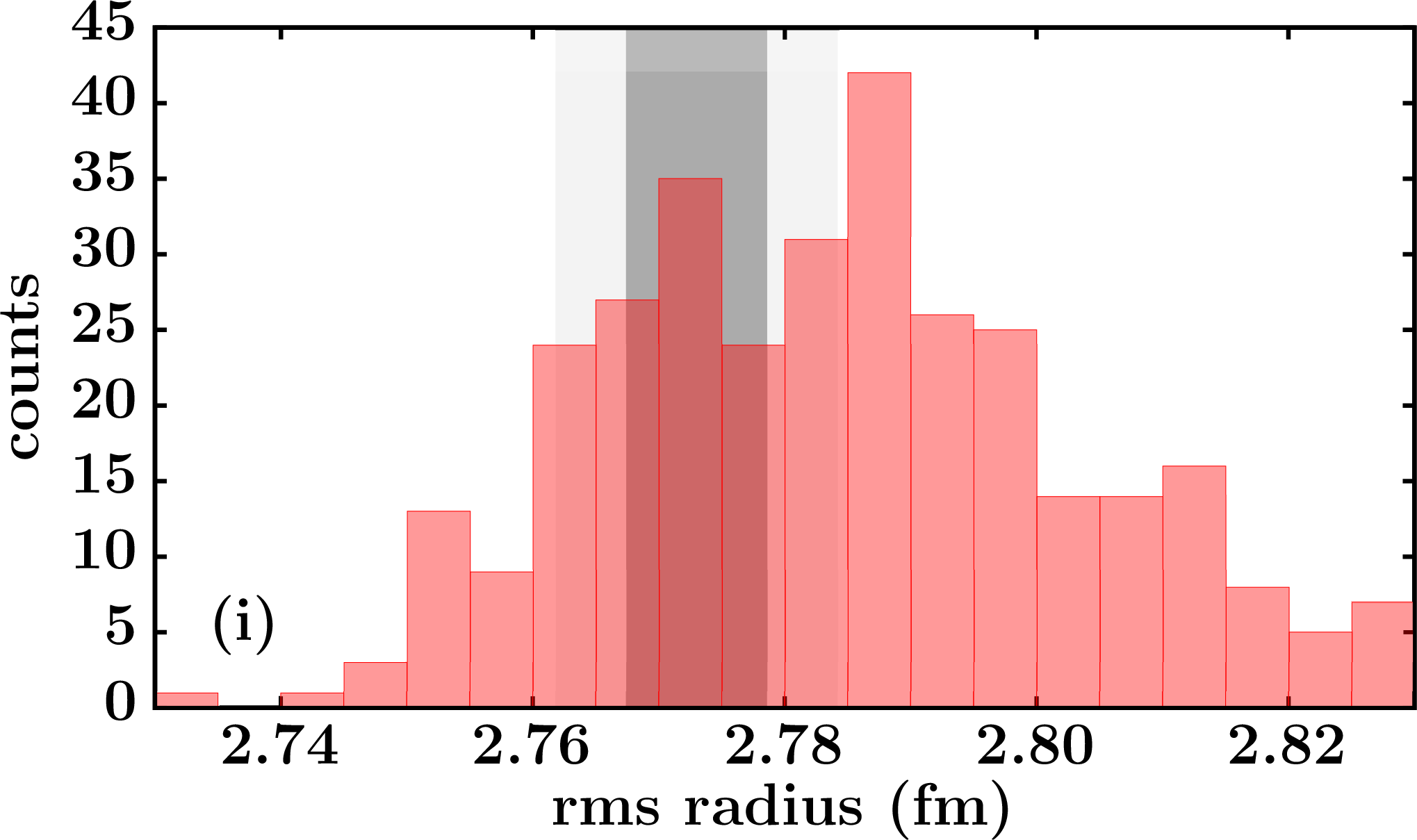}
        \label{DOM_o18_RMSRadius}
    \end{minipage}\hspace{6pt}
    \begin{minipage}{0.4\linewidth}
        \centering
        \includegraphics[width=\linewidth]{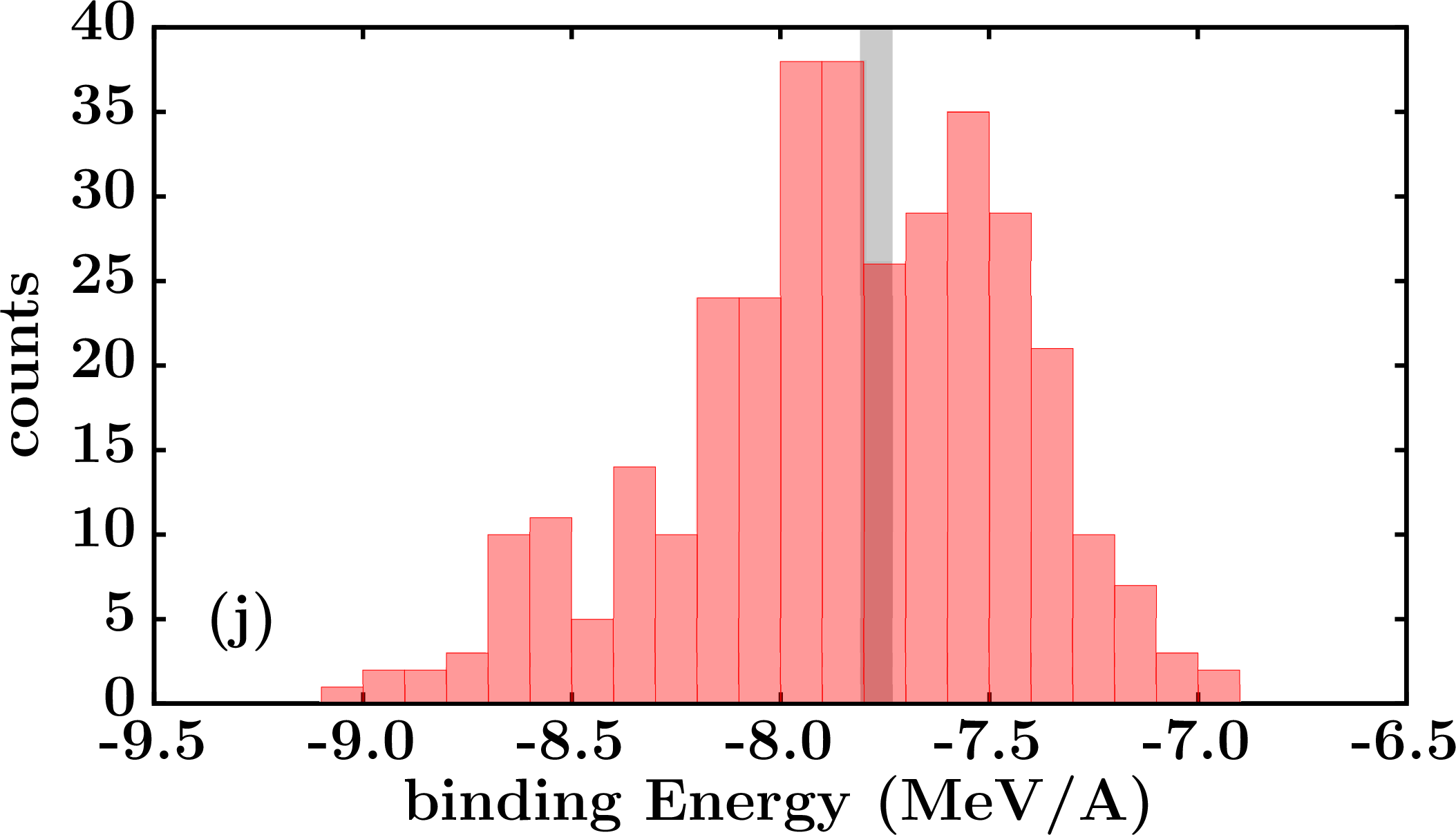}
        \label{DOM_o18_BE}
    \end{minipage}
    \caption{\oEight: constraining experimental data and DOM fit. See introduction of
    Appendix C for description.}
    \label{DOM_o18}
\end{figure*}

\begin{figure*}[!htb]
    \centering
    \begin{minipage}{0.4\linewidth}
        \centering
        \includegraphics[width=\linewidth]{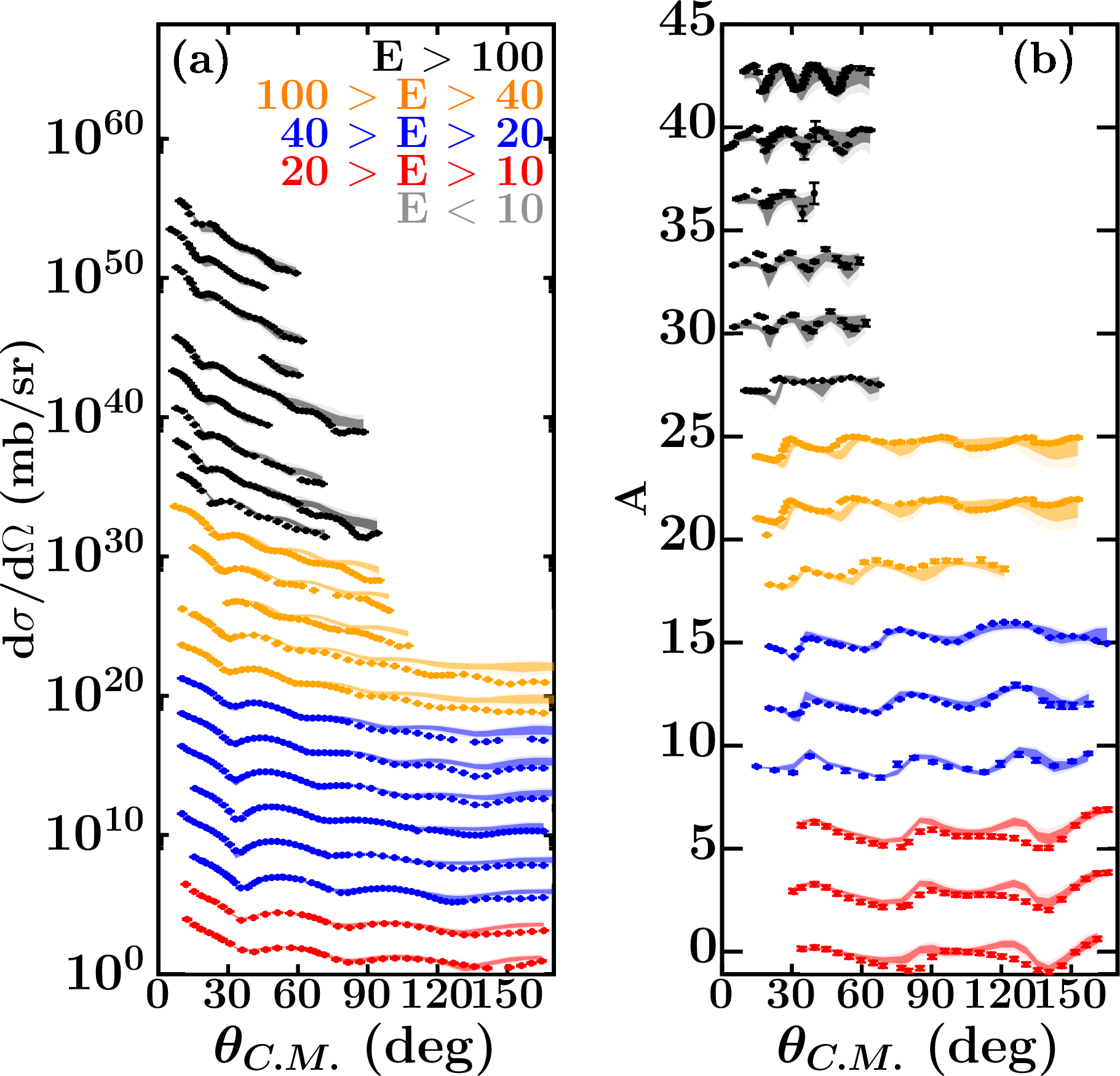}
        \label{DOM_ca40_proton_elastic}
    \end{minipage}\hspace{6pt}
    \begin{minipage}{0.4\linewidth}
        \centering
        \includegraphics[width=\linewidth]{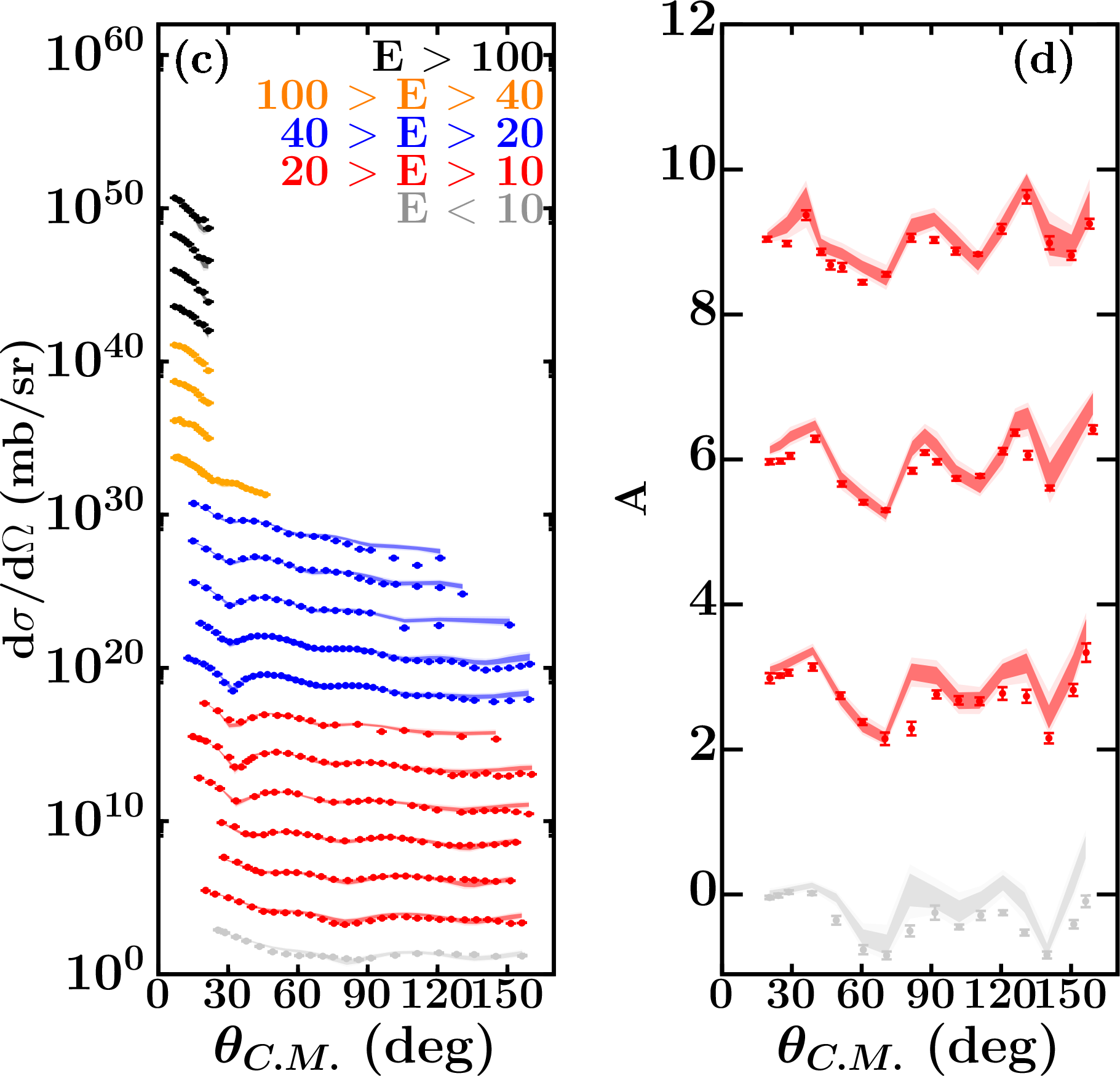}
        \label{DOM_ca40_neutron_elastic}
    \end{minipage}
    \centering
    \begin{minipage}{0.4\linewidth}
        \centering
        \includegraphics[width=\linewidth]{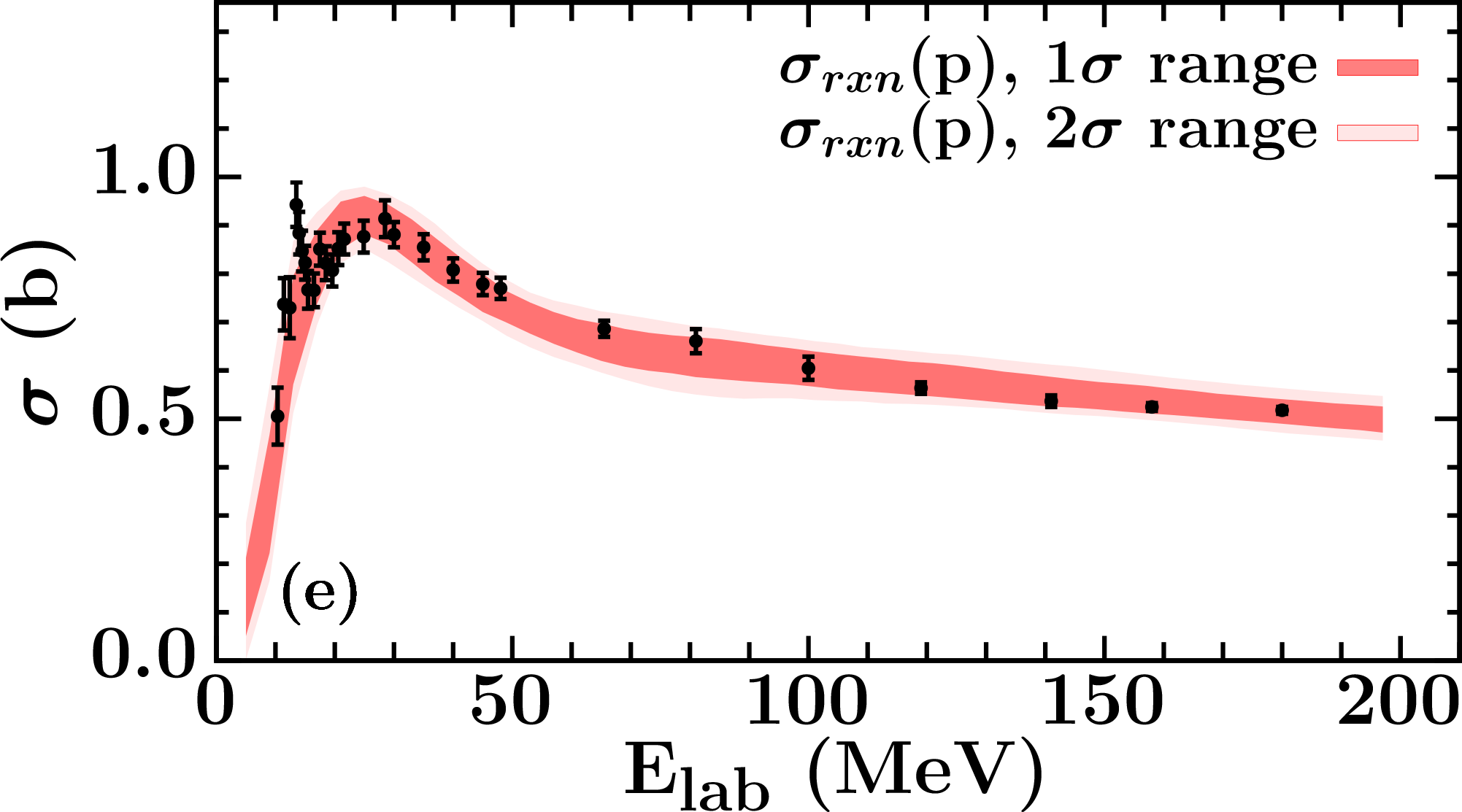}
        \label{DOM_ca40_proton_inelastic}
    \end{minipage}\hspace{6pt}
    \begin{minipage}{0.4\linewidth}
        \centering
        \includegraphics[width=\linewidth]{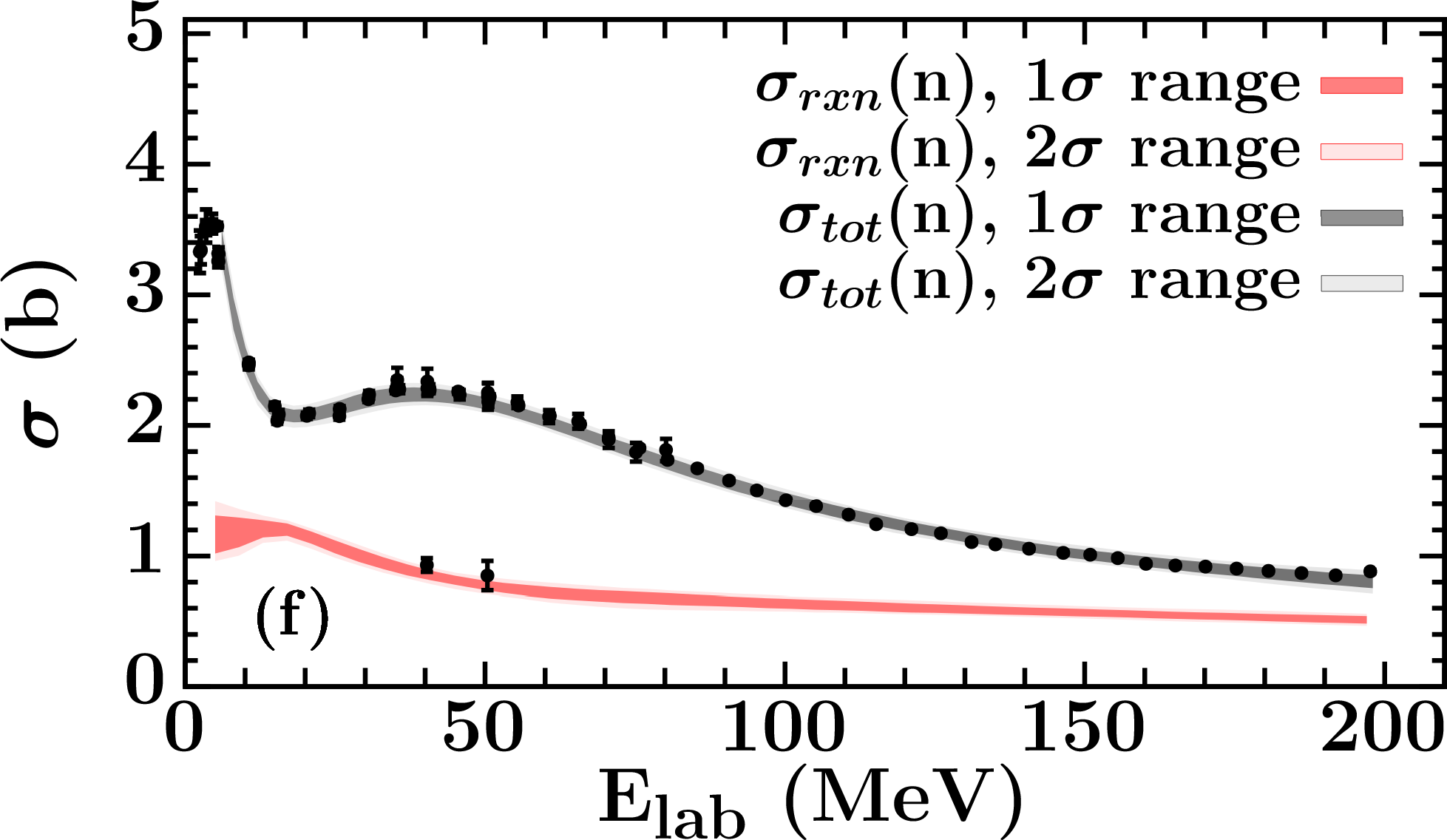}
        \label{DOM_ca40_neutron_inelastic}
    \end{minipage}
    \centering
    \begin{minipage}{0.4\linewidth}
        \centering
        \includegraphics[width=\linewidth]{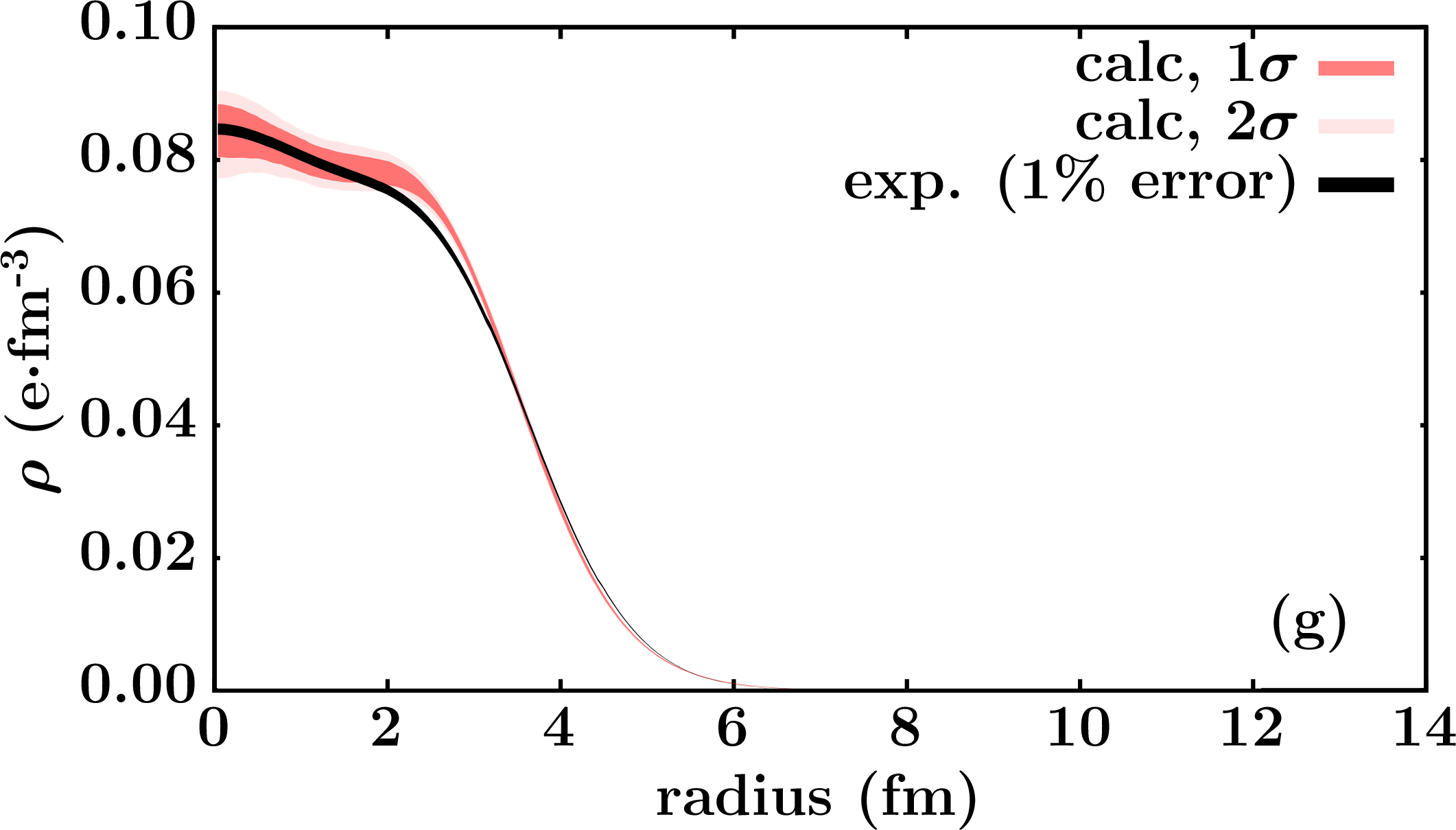}
        \label{DOM_ca40_chargeDensity}
    \end{minipage}\hspace{6pt}
    \begin{minipage}{0.4\linewidth}
        \centering
        \includegraphics[width=\linewidth]{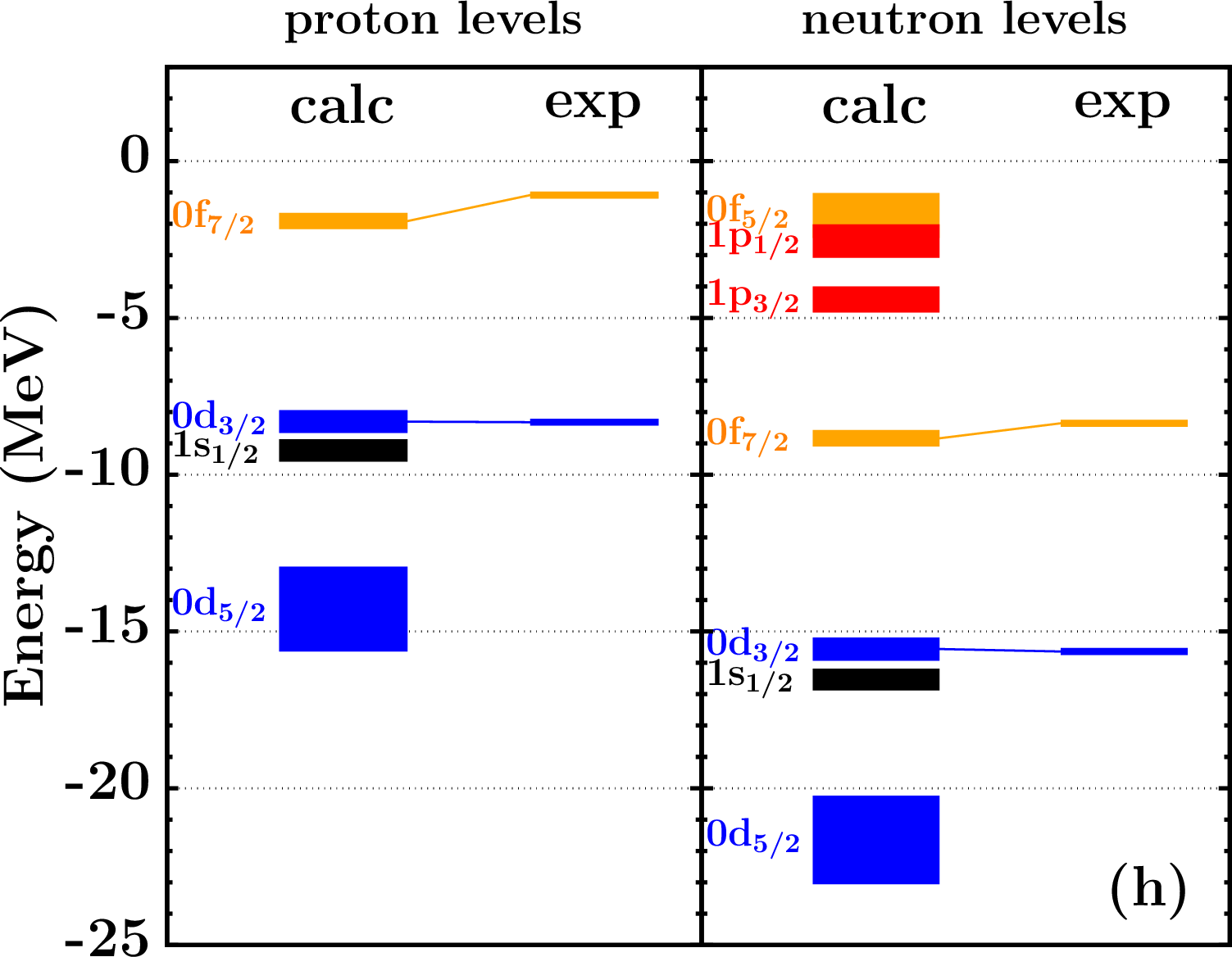}
        \label{DOM_ca40_SPLevels}
    \end{minipage}
    \begin{minipage}{0.4\linewidth}
        \centering
        \includegraphics[width=\linewidth]{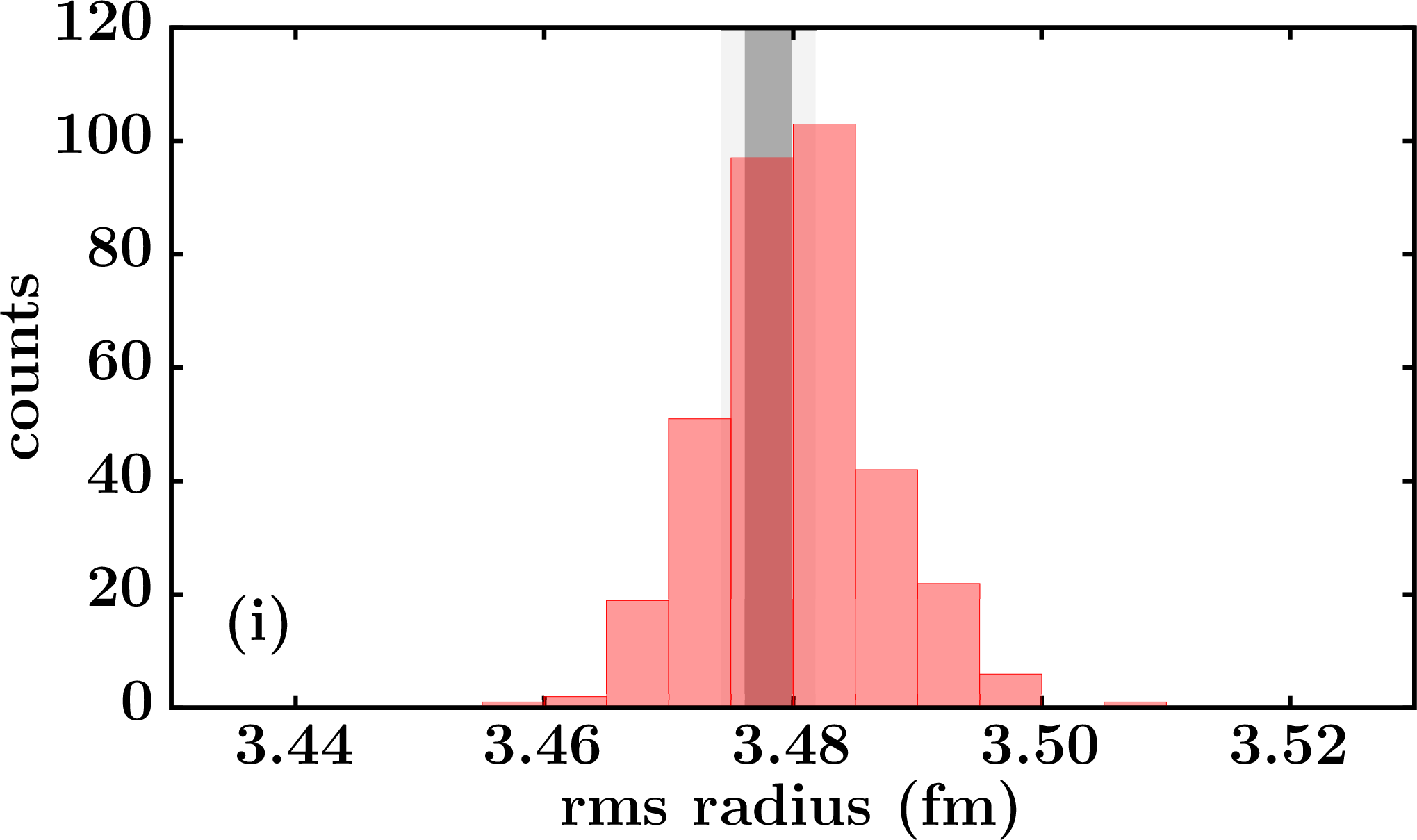}
        \label{DOM_ca40_RMSRadius}
    \end{minipage}\hspace{6pt}
    \begin{minipage}{0.4\linewidth}
        \centering
        \includegraphics[width=\linewidth]{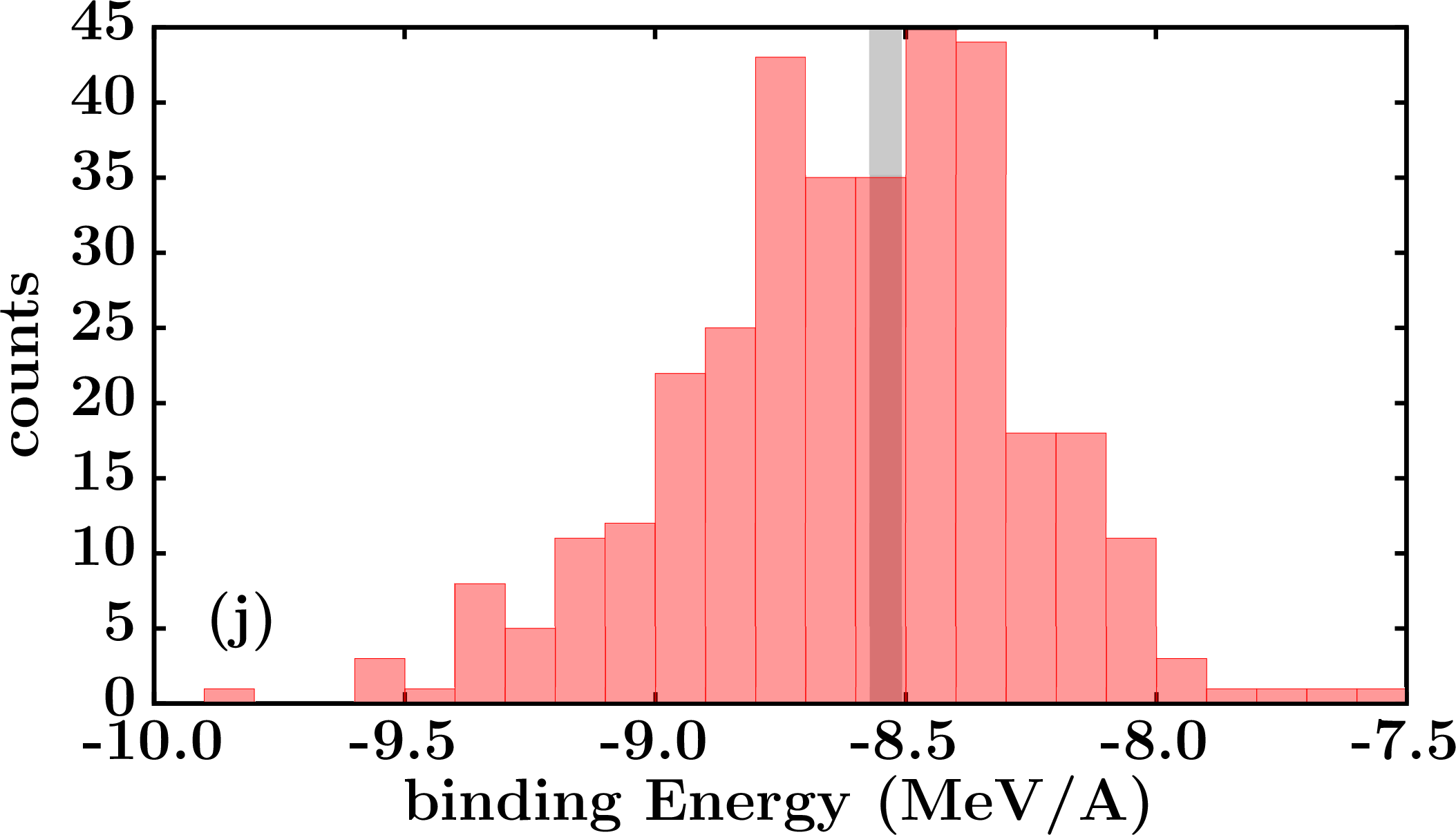}
        \label{DOM_ca40_BE}
    \end{minipage}
    \caption{\caForty: constraining experimental data and DOM fit. See introduction of
    Appendix C for description.}
    \label{DOM_ca40}
\end{figure*}

\begin{figure*}[!htb]
    \centering
    \begin{minipage}{0.4\linewidth}
        \centering
        \includegraphics[width=\linewidth]{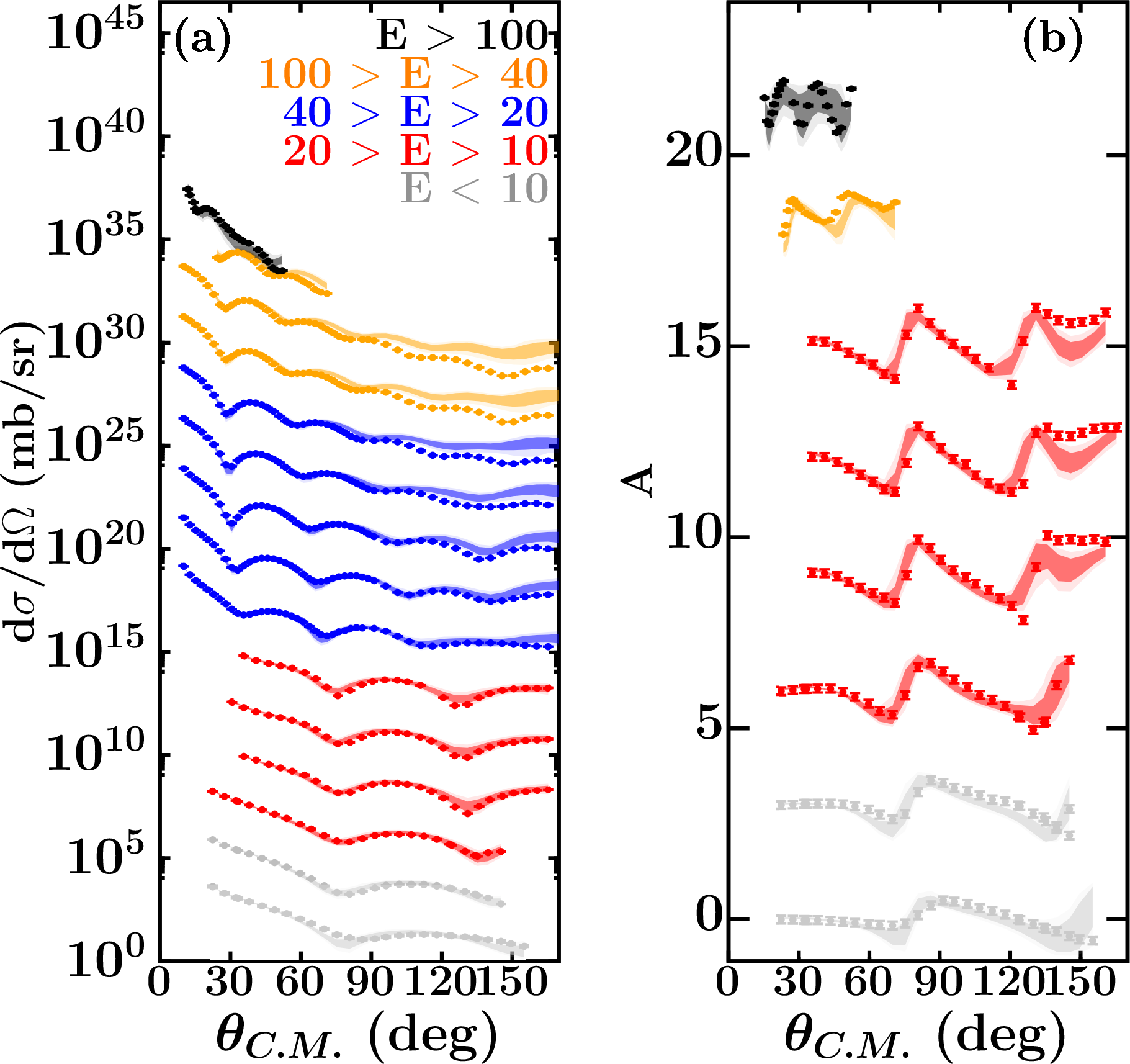}
        \label{DOM_ca48_proton_elastic}
    \end{minipage}\hspace{6pt}
    \begin{minipage}{0.4\linewidth}
        \centering
        \vspace{-10pt}
        \begin{minipage}[c]{0.5\linewidth}
            \centering
                \includegraphics[width=\linewidth]{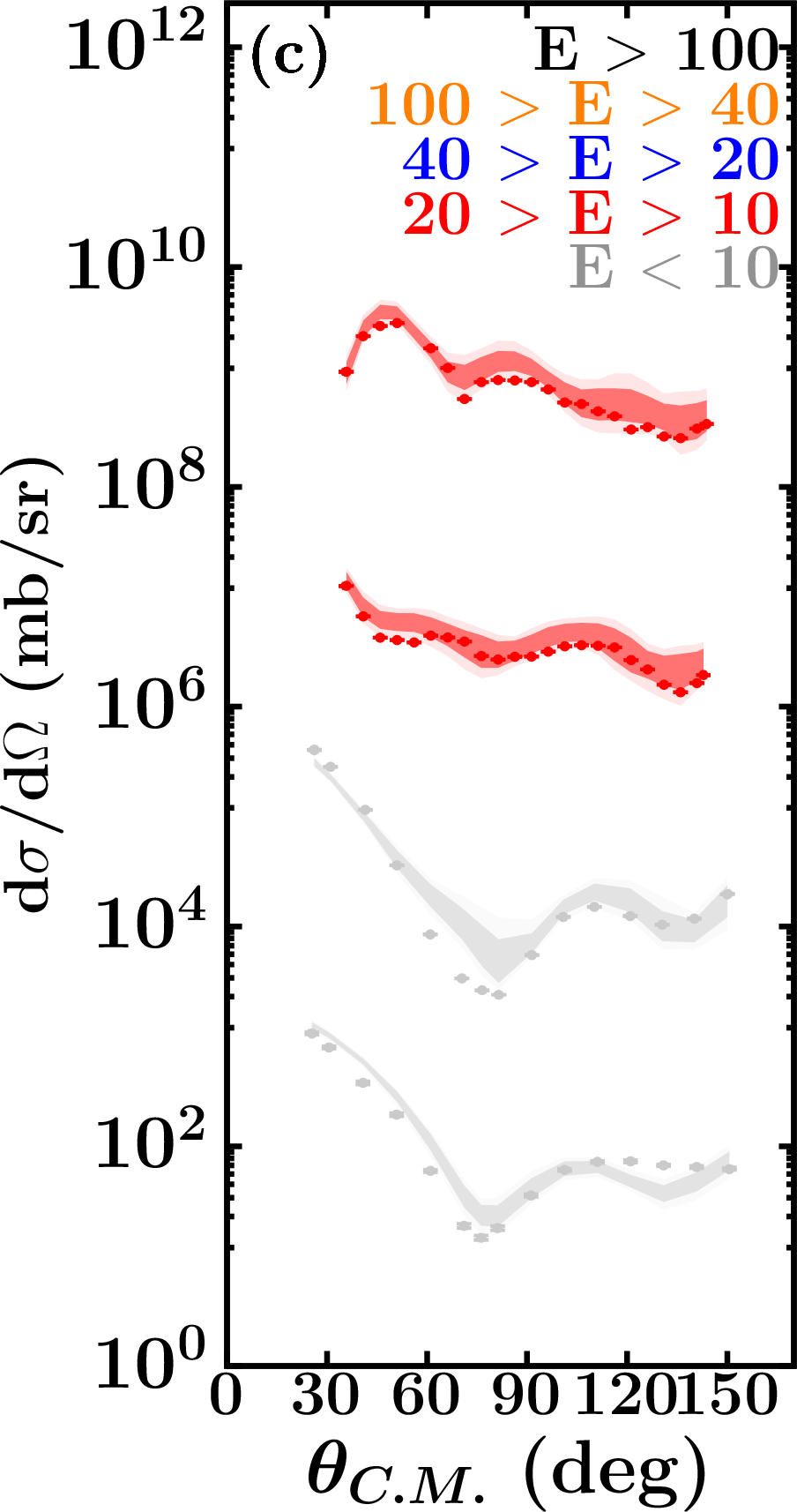}
        \end{minipage}
        \begin{minipage}[c]{0.45\linewidth}
            \centering
            No \caEight\ neutron \\
            analyzing powers \\
            were available
        \end{minipage}
        \label{DOM_ca48_neutron_elastic}
    \end{minipage}
    \centering
    \begin{minipage}{0.4\linewidth}
        \centering
        \includegraphics[width=\linewidth]{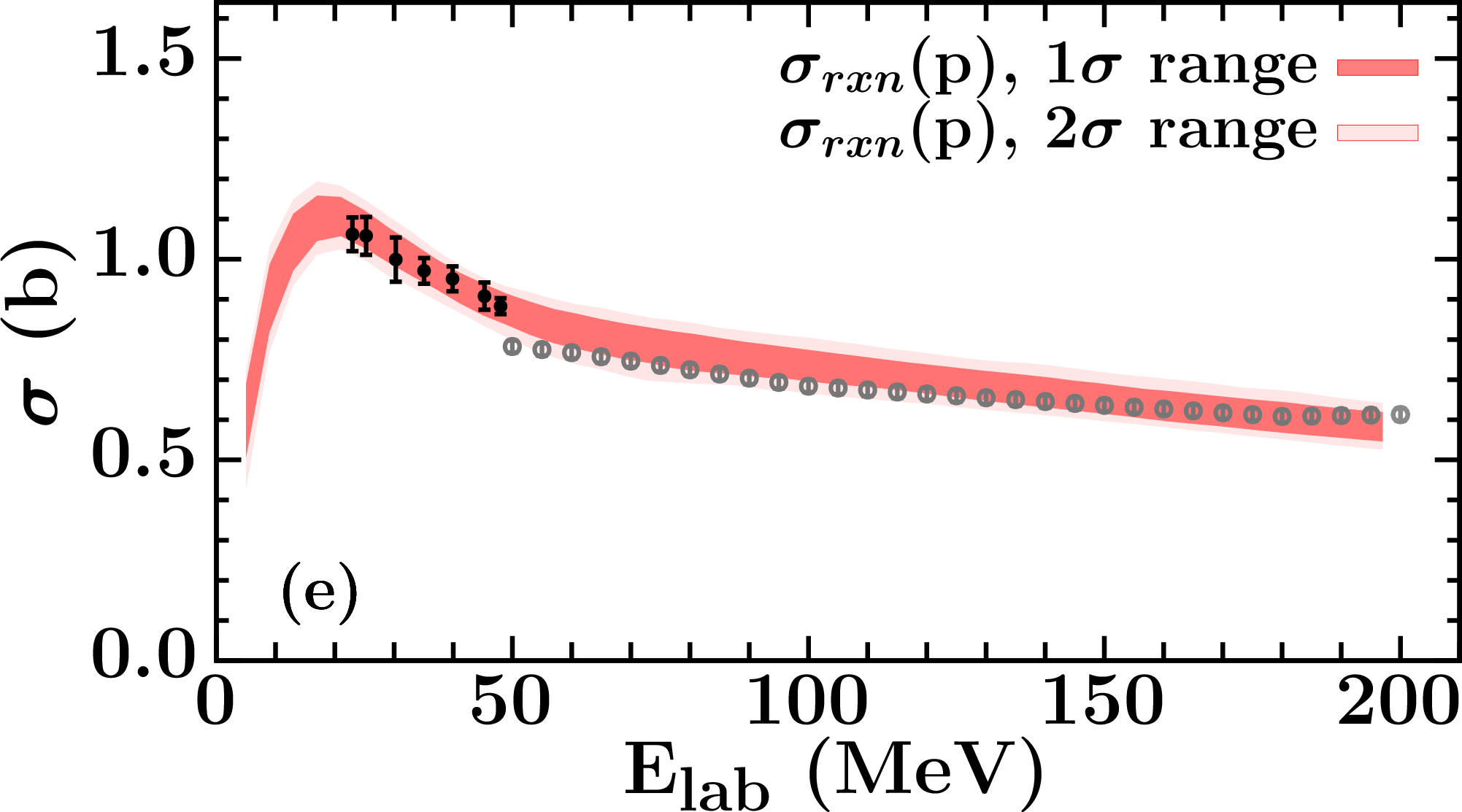}
        \label{DOM_ca48_proton_inelastic}
    \end{minipage}\hspace{6pt}
    \begin{minipage}{0.4\linewidth}
        \centering
        \includegraphics[width=\linewidth]{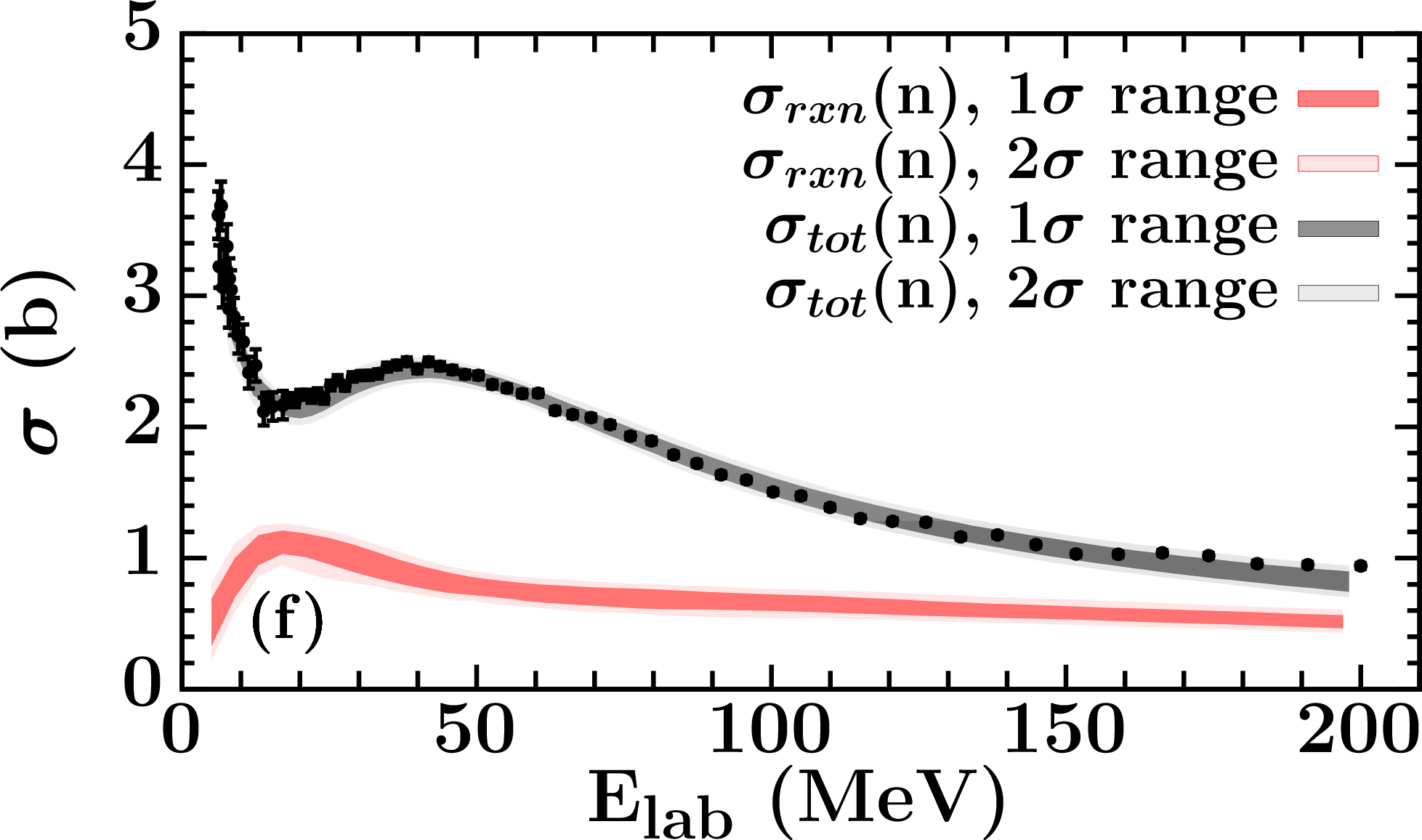}
        \label{DOM_ca48_neutron_inelastic}
    \end{minipage}
    \centering
    \begin{minipage}{0.4\linewidth}
        \centering
        \includegraphics[width=\linewidth]{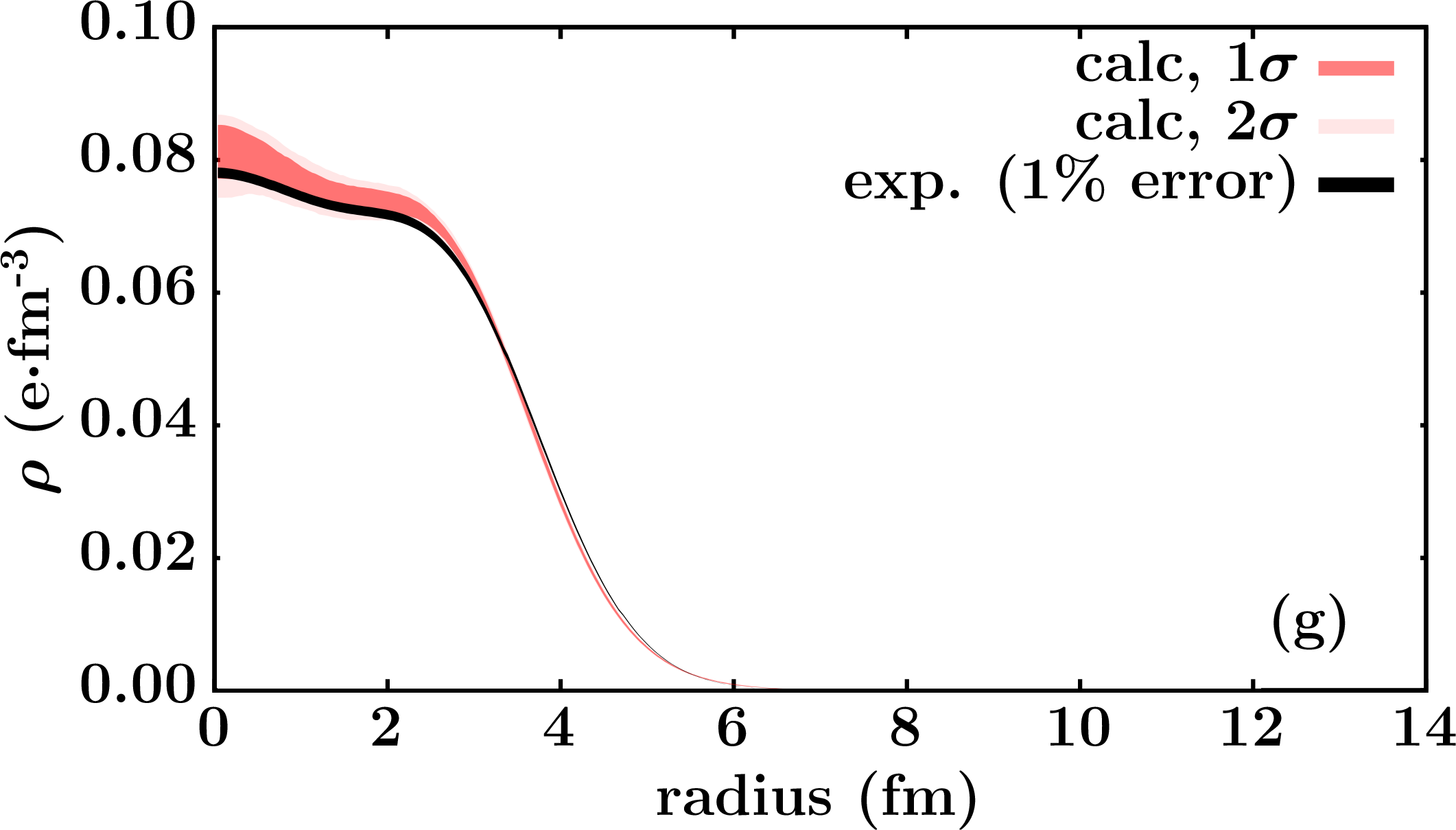}
        \label{DOM_ca48_chargeDensity}
    \end{minipage}\hspace{6pt}
    \begin{minipage}{0.4\linewidth}
        \centering
        \includegraphics[width=\linewidth]{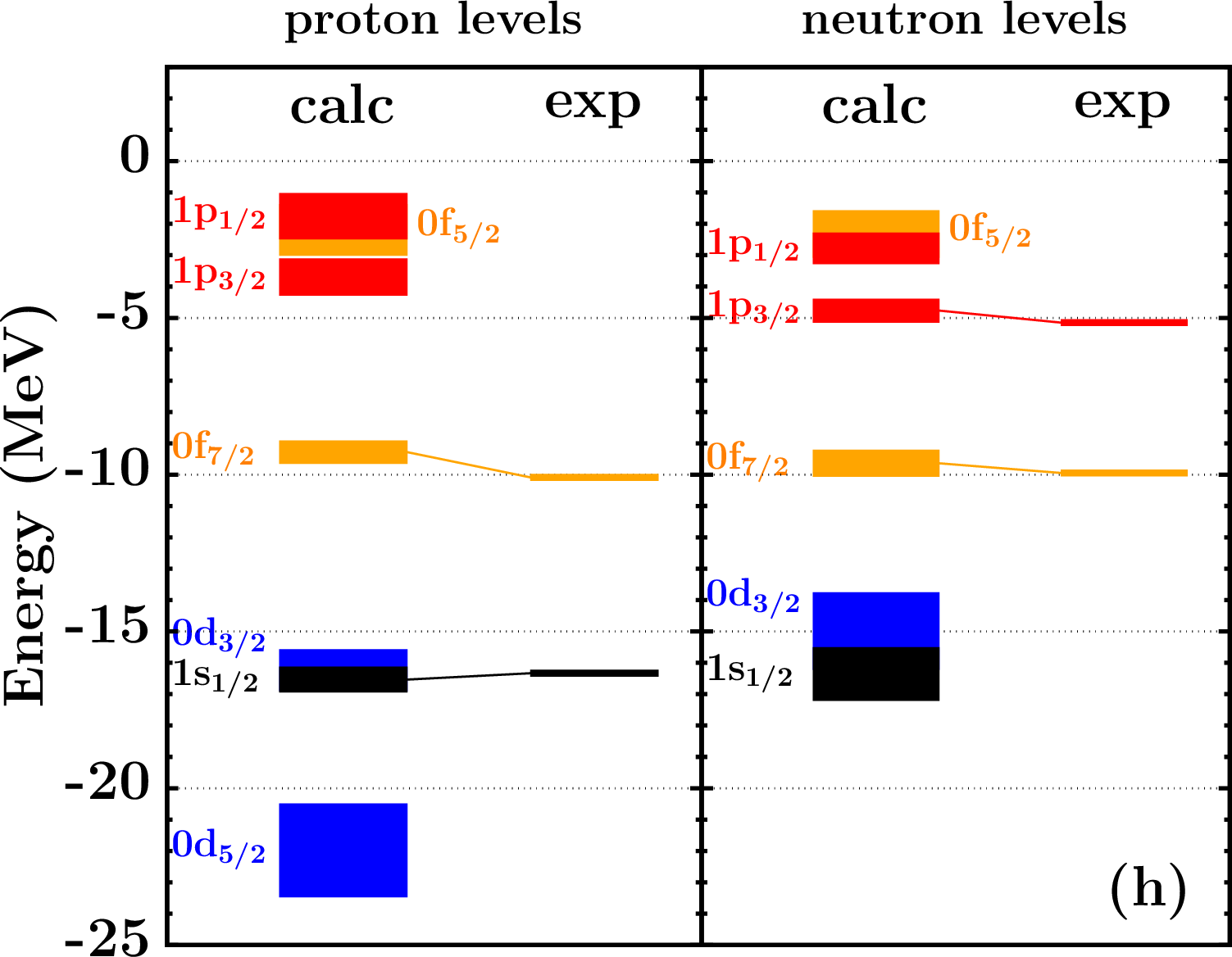}
        \label{DOM_ca48_SPLevels}
    \end{minipage}
    \begin{minipage}{0.4\linewidth}
        \centering
        \includegraphics[width=\linewidth]{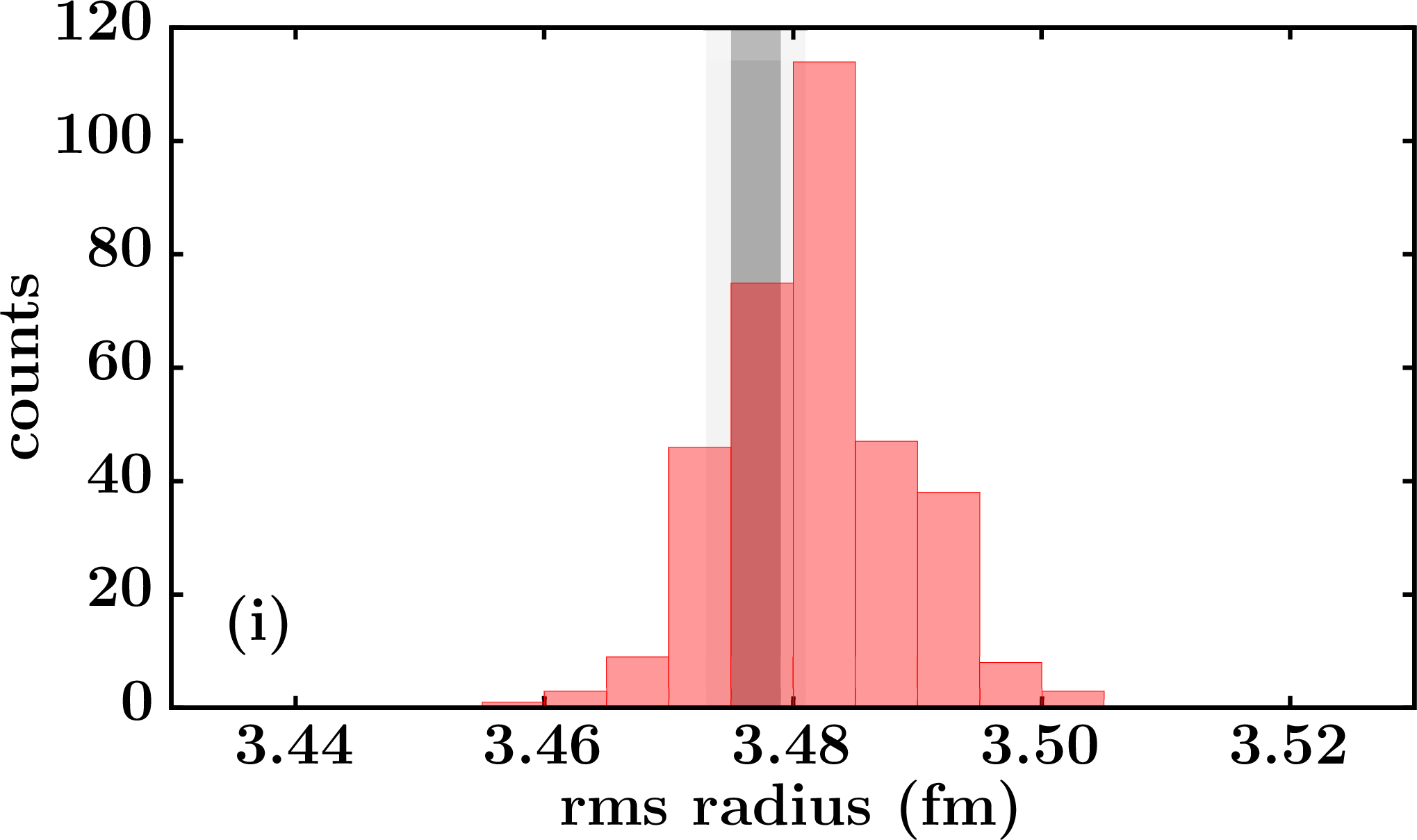}
        \label{DOM_ca48_RMSRadius}
    \end{minipage}\hspace{6pt}
    \begin{minipage}{0.4\linewidth}
        \centering
        \includegraphics[width=\linewidth]{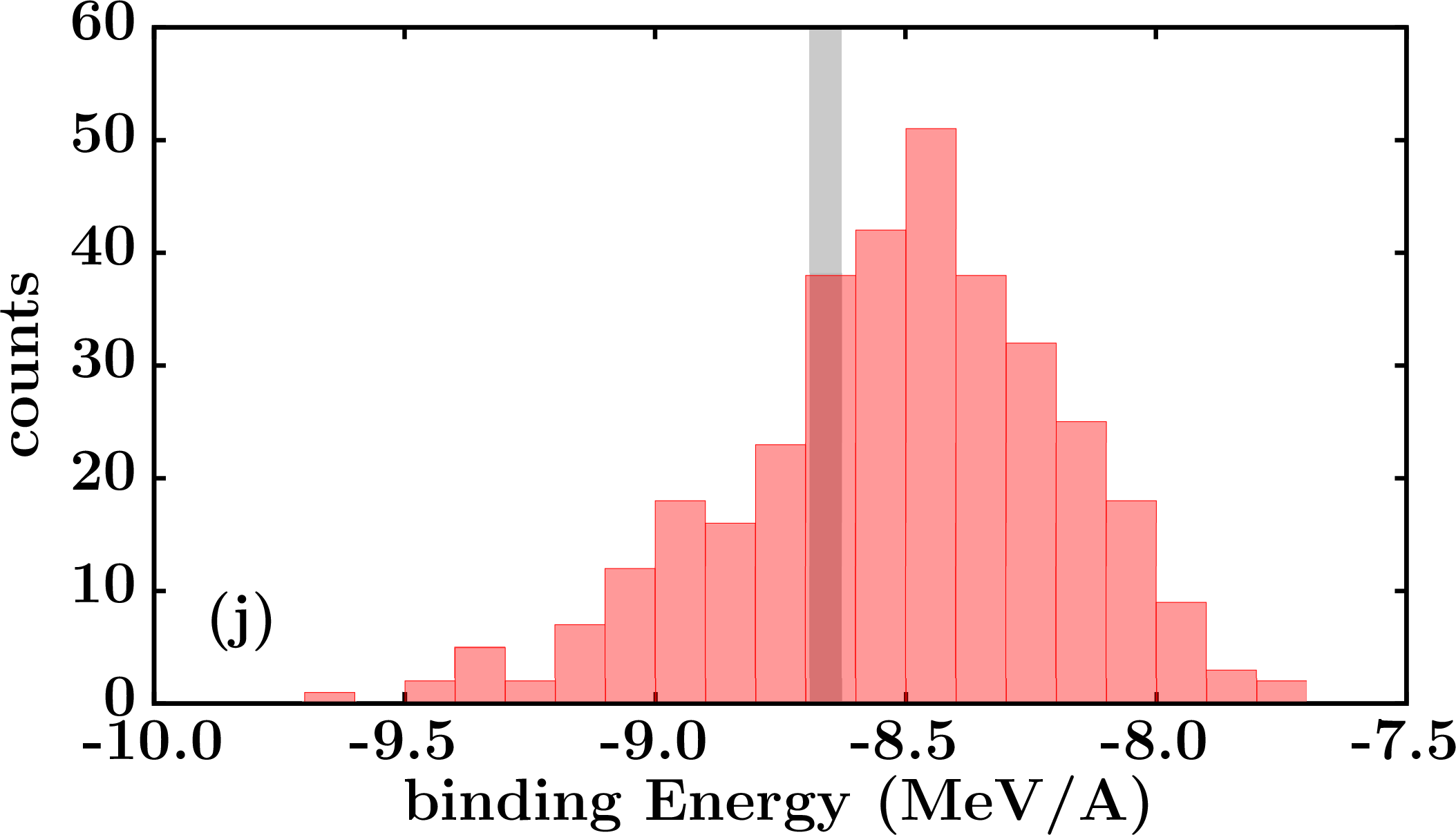}
        \label{DOM_ca48_BE}
    \end{minipage}
    \caption{\caEight: constraining experimental data and DOM fit. See introduction of
    Appendix C for description.}
    \label{DOM_ca48}
\end{figure*}

\begin{figure*}[!htb]
    \centering
    \begin{minipage}{0.4\linewidth}
        \centering
        \includegraphics[width=\linewidth]{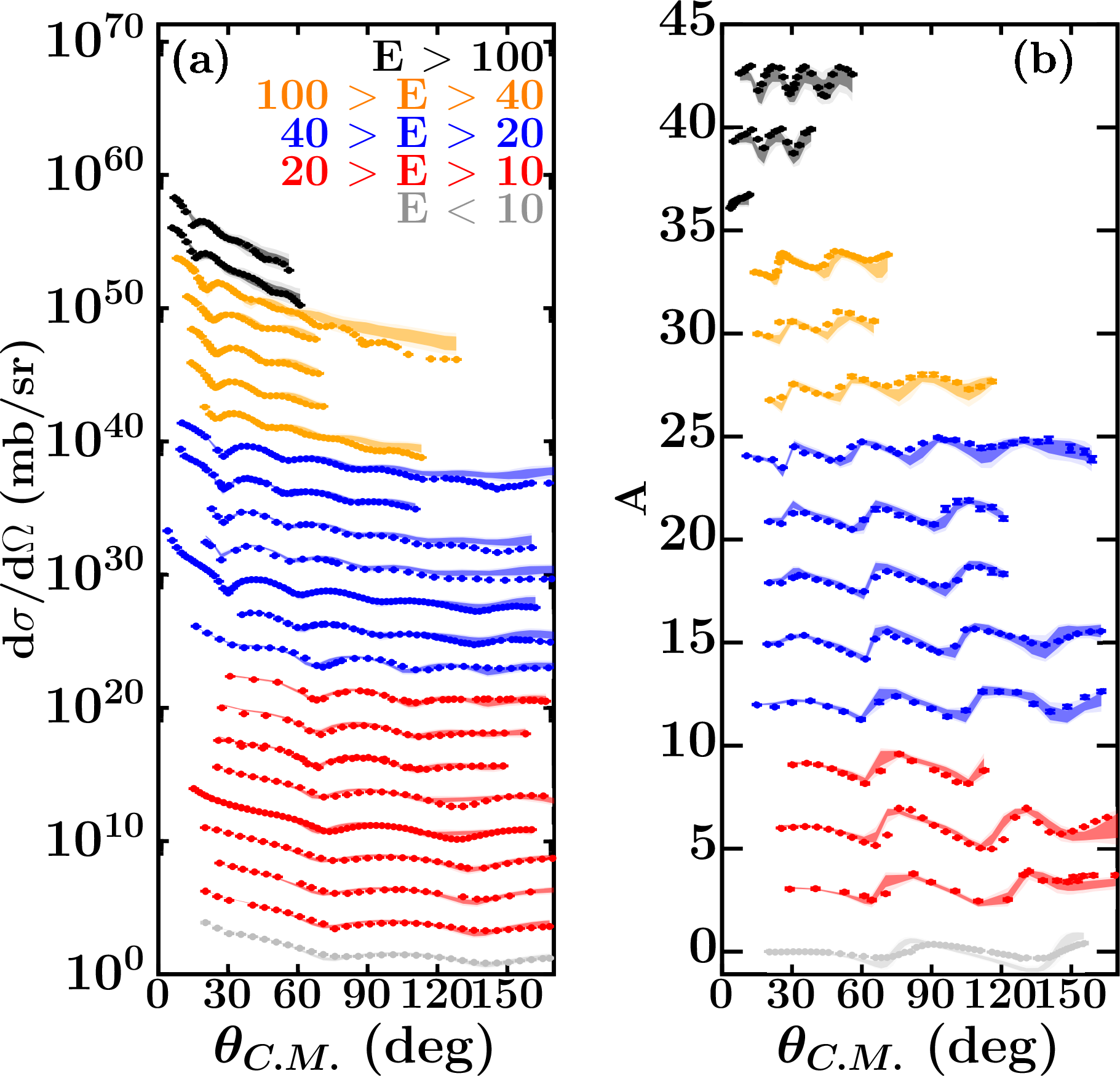}
        \label{DOM_ni58_proton_elastic}
    \end{minipage}\hspace{6pt}
    \begin{minipage}{0.4\linewidth}
        \centering
        \includegraphics[width=\linewidth]{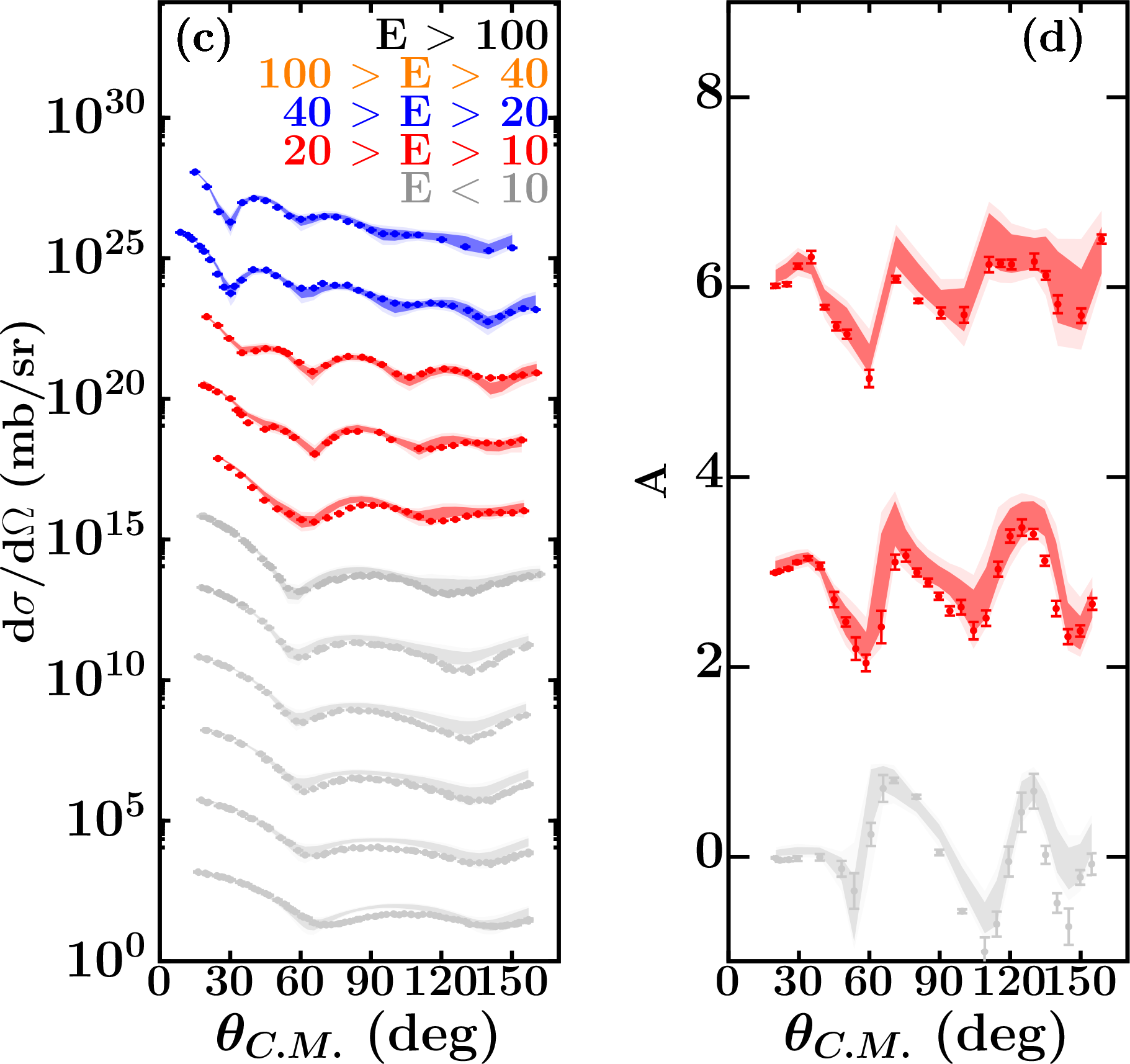}
        \label{DOM_ni58_neutron_elastic}
    \end{minipage}
    \centering
    \begin{minipage}{0.4\linewidth}
        \centering
        \includegraphics[width=\linewidth]{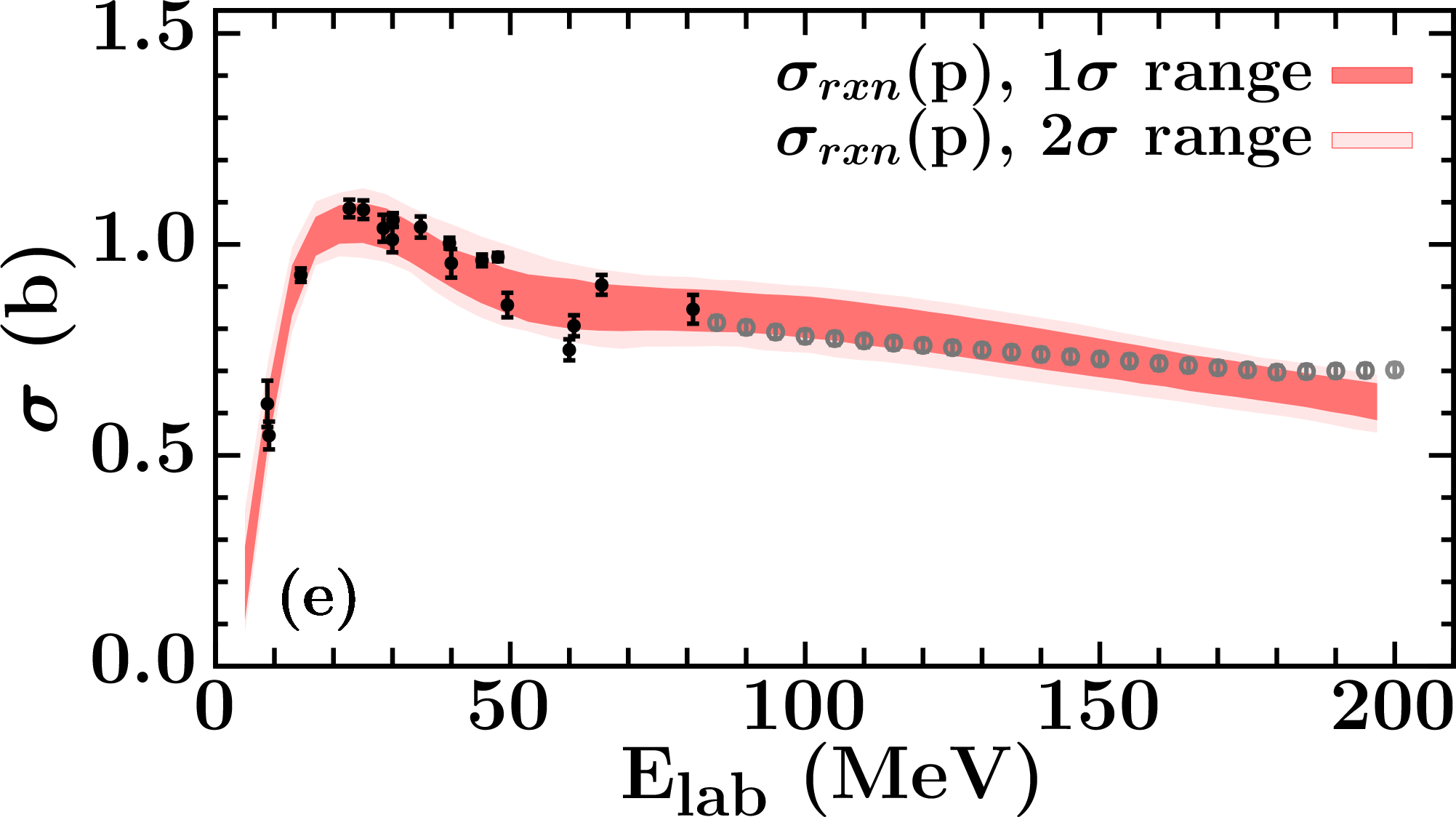}
        \label{DOM_ni58_proton_inelastic}
    \end{minipage}\hspace{6pt}
    \begin{minipage}{0.4\linewidth}
        \centering
        \includegraphics[width=\linewidth]{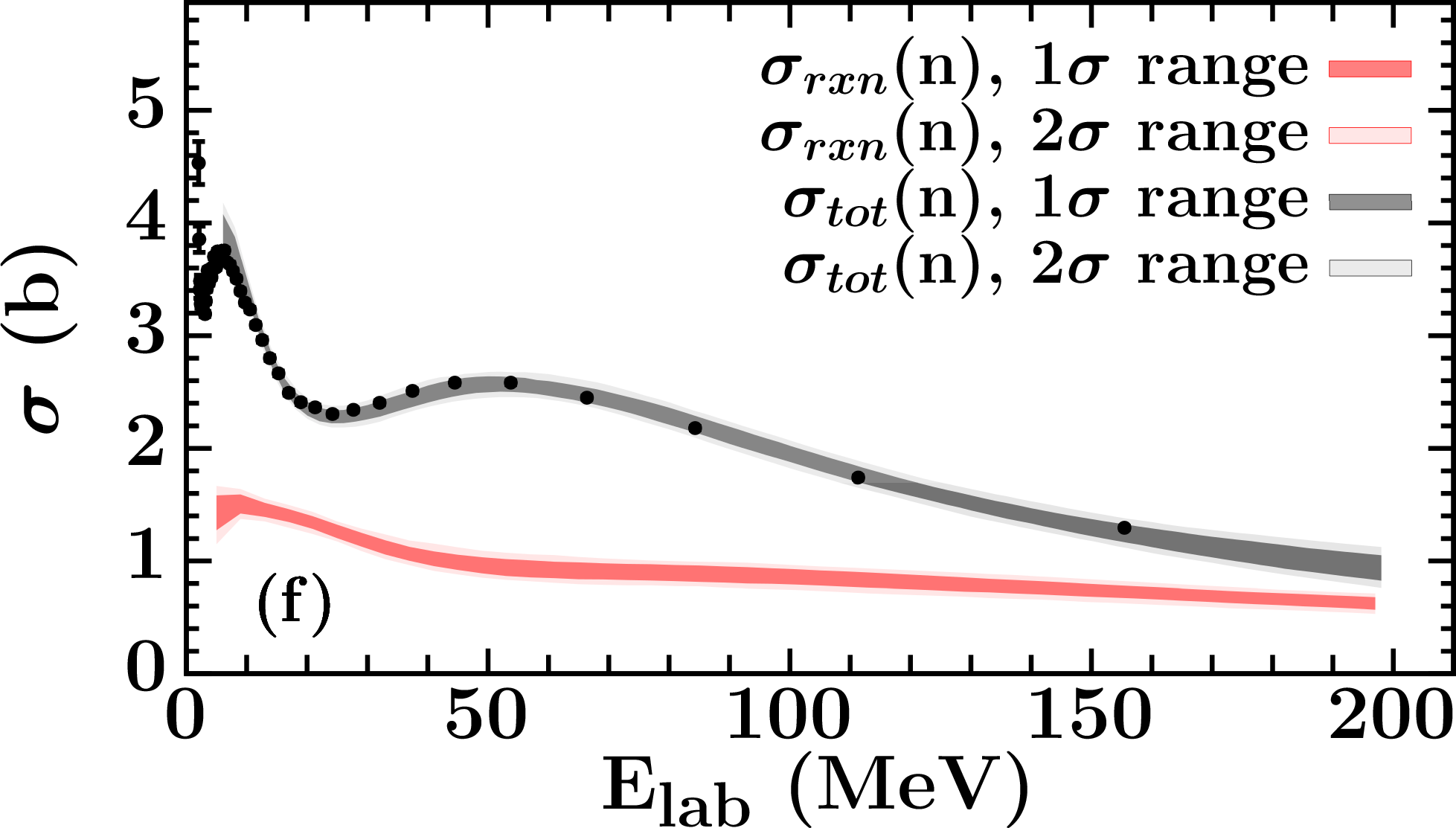}
        \label{DOM_ni58_neutron_inelastic}
    \end{minipage}
    \centering
    \begin{minipage}{0.4\linewidth}
        \centering
        \includegraphics[width=\linewidth]{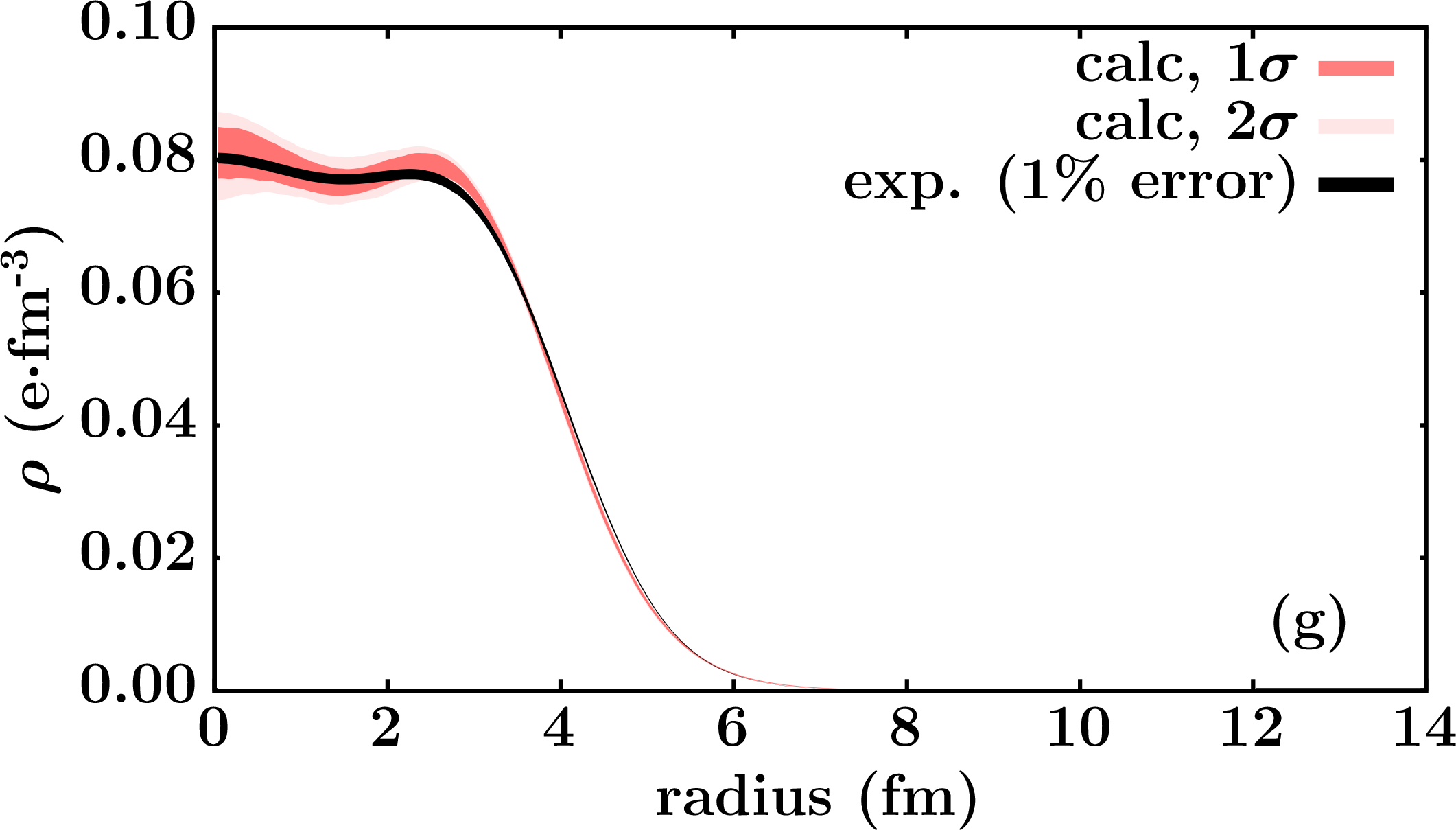}
        \label{DOM_ni58_chargeDensity}
    \end{minipage}\hspace{6pt}
    \begin{minipage}{0.4\linewidth}
        \centering
        \includegraphics[width=\linewidth]{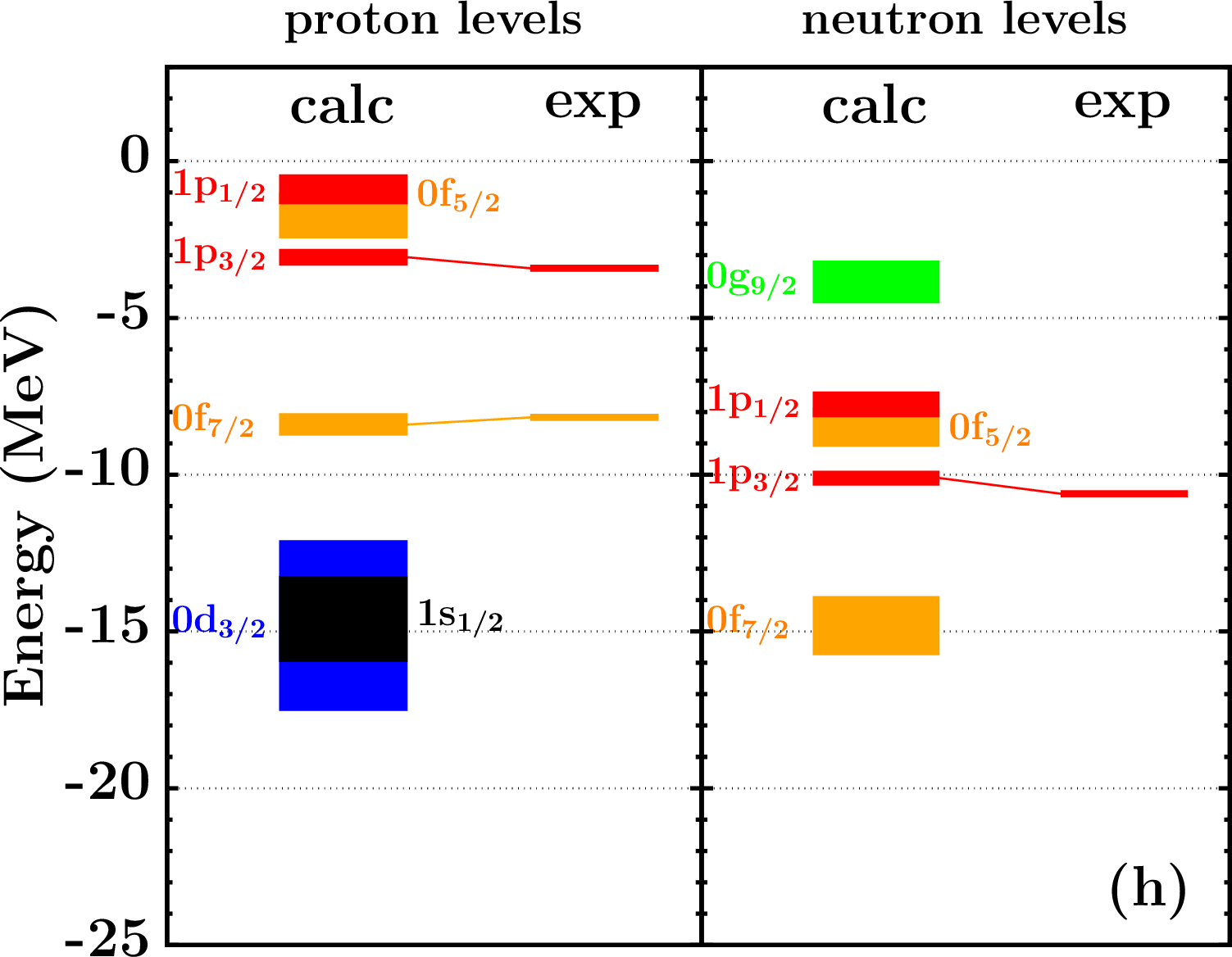}
        \label{DOM_ni58_SPLevels}
    \end{minipage}
    \begin{minipage}{0.4\linewidth}
        \centering
        \includegraphics[width=\linewidth]{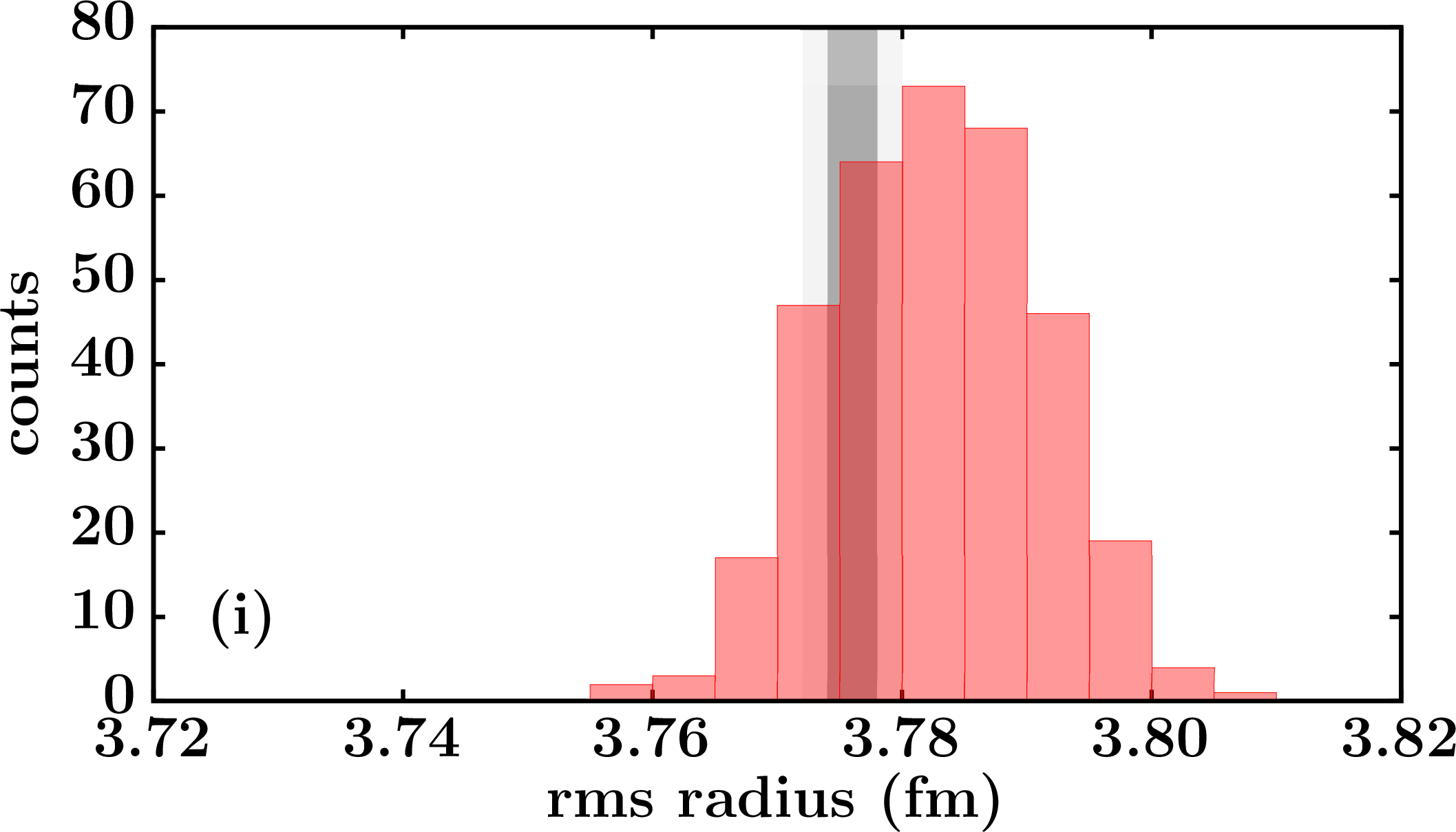}
        \label{DOM_ni58_RMSRadius}
    \end{minipage}\hspace{6pt}
    \begin{minipage}{0.4\linewidth}
        \centering
        \includegraphics[width=\linewidth]{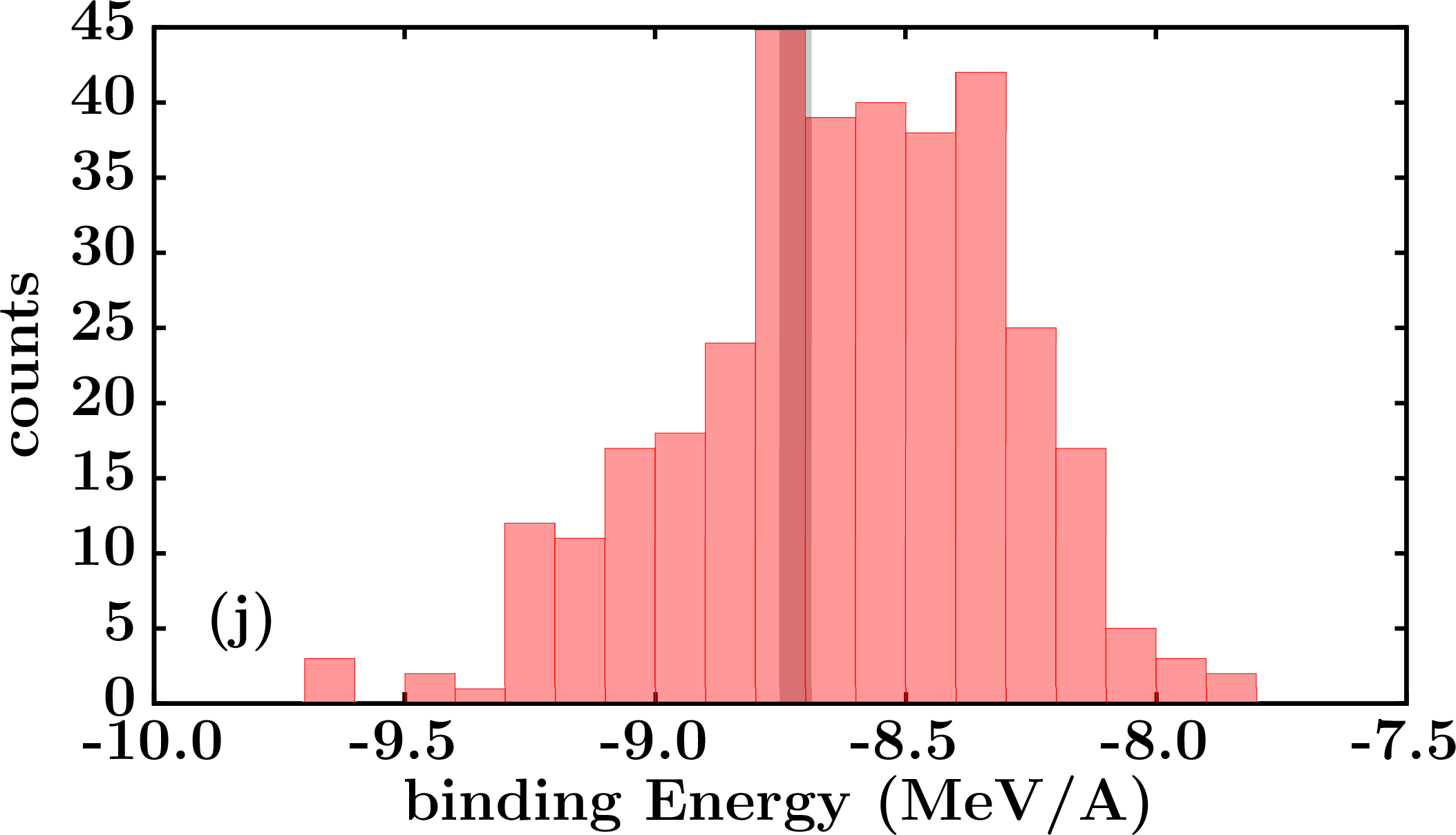}
        \label{DOM_ni58_BE}
    \end{minipage}
    \caption{\niEight: constraining experimental data and DOM fit. See introduction of
    Appendix C for description.}
    \label{DOM_ni58}
\end{figure*}

\begin{figure*}[!htb]
    \centering
    \begin{minipage}{0.4\linewidth}
        \centering
        \includegraphics[width=\linewidth]{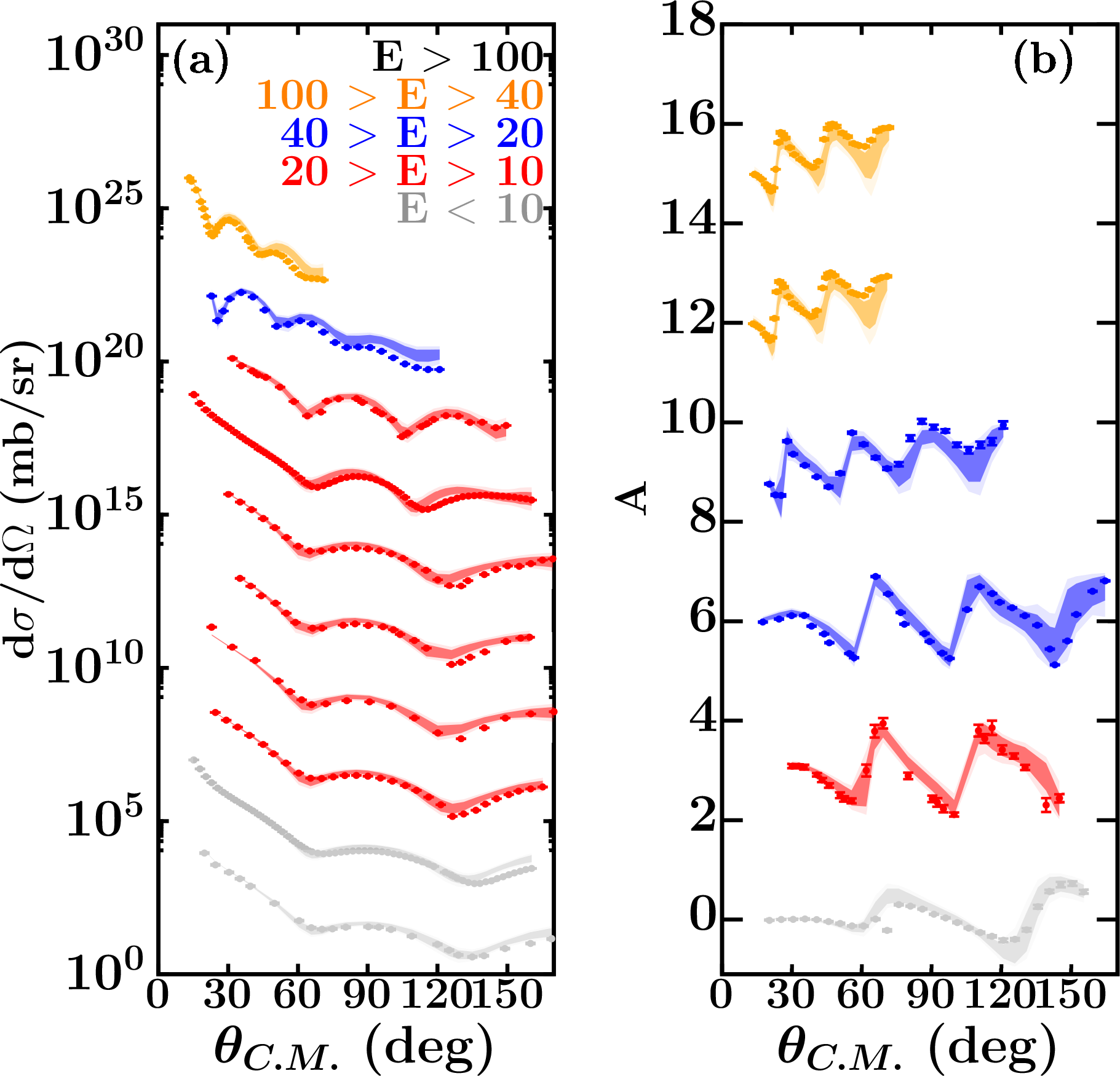}
        \label{DOM_ni64_proton_elastic}
    \end{minipage}\hspace{6pt}
    \begin{minipage}{0.4\linewidth}
        \centering
        \vspace{-10pt}
        \begin{minipage}[c]{0.5\linewidth}
            \centering
                \includegraphics[width=\linewidth]{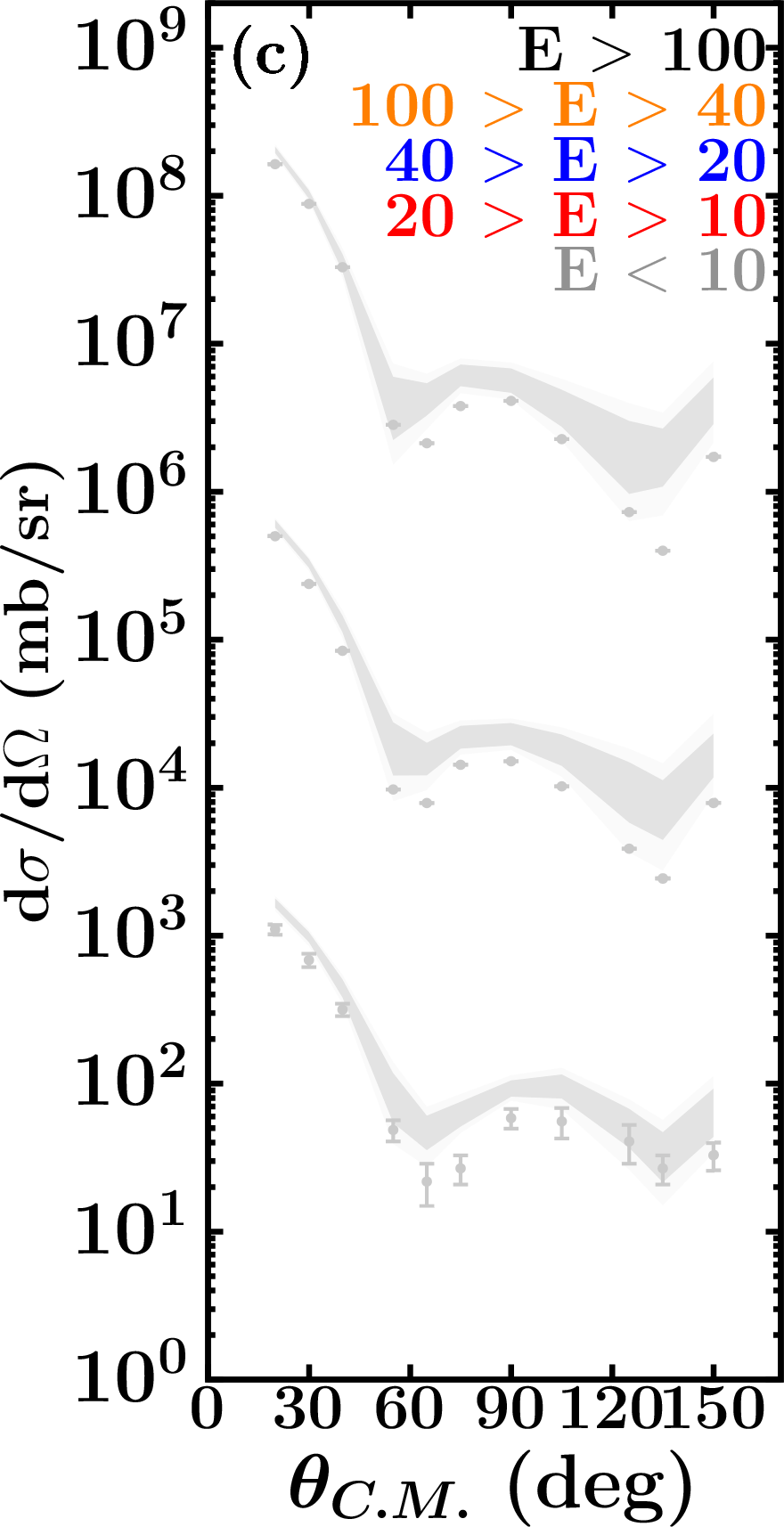}
        \end{minipage}
        \begin{minipage}[c]{0.45\linewidth}
            \centering
            No \niFour\ neutron \\
            analyzing powers \\
            were available
        \end{minipage}
        \label{DOM_ni64_neutron_elastic}
    \end{minipage}
    \centering
    \begin{minipage}{0.4\linewidth}
        \centering
        \includegraphics[width=\linewidth]{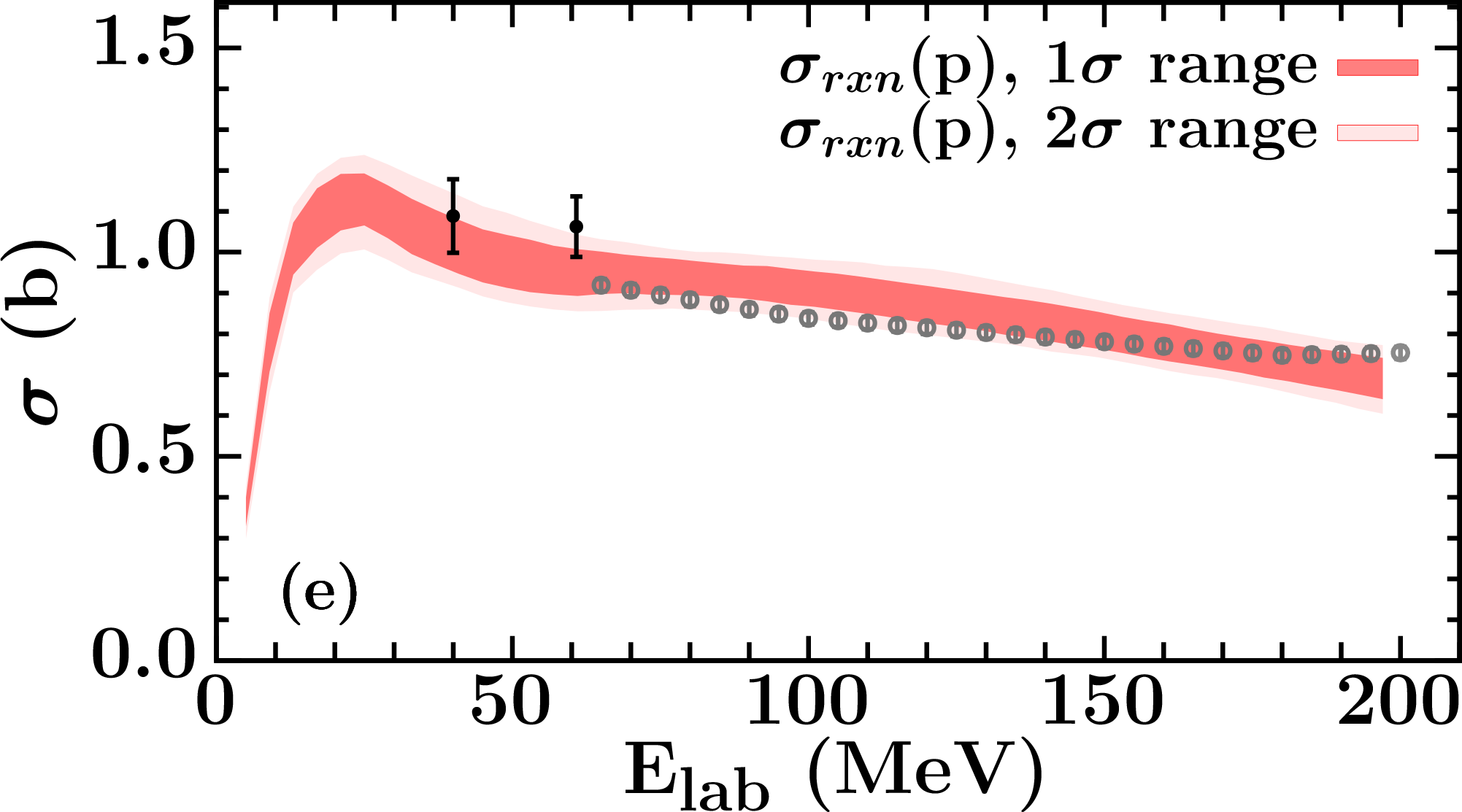}
        \label{DOM_ni64_proton_inelastic}
    \end{minipage}\hspace{6pt}
    \begin{minipage}{0.4\linewidth}
        \centering
        \includegraphics[width=\linewidth]{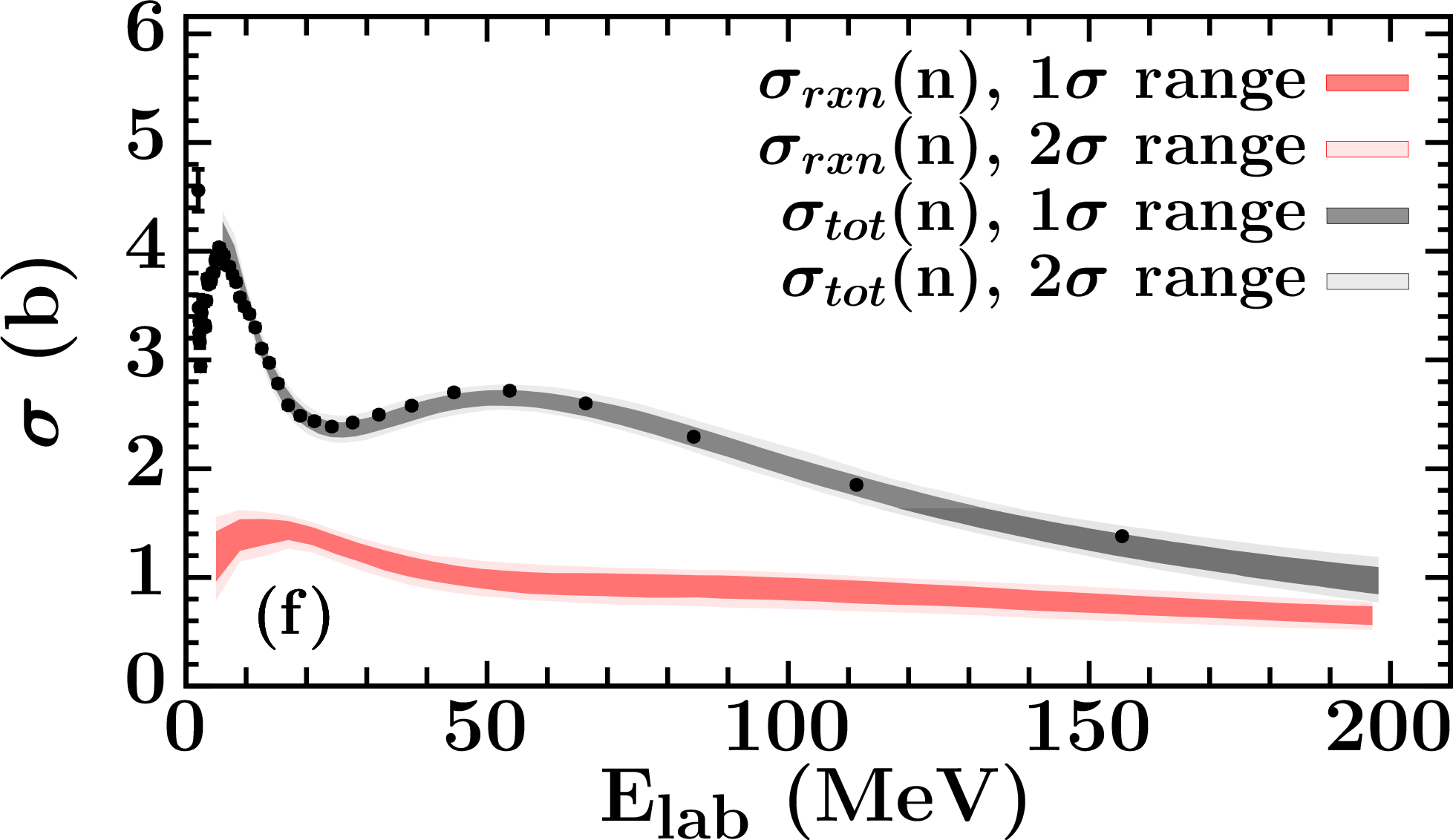}
        \label{DOM_ni64_neutron_inelastic}
    \end{minipage}
    \centering
    \begin{minipage}{0.4\linewidth}
        \centering
        \includegraphics[width=\linewidth]{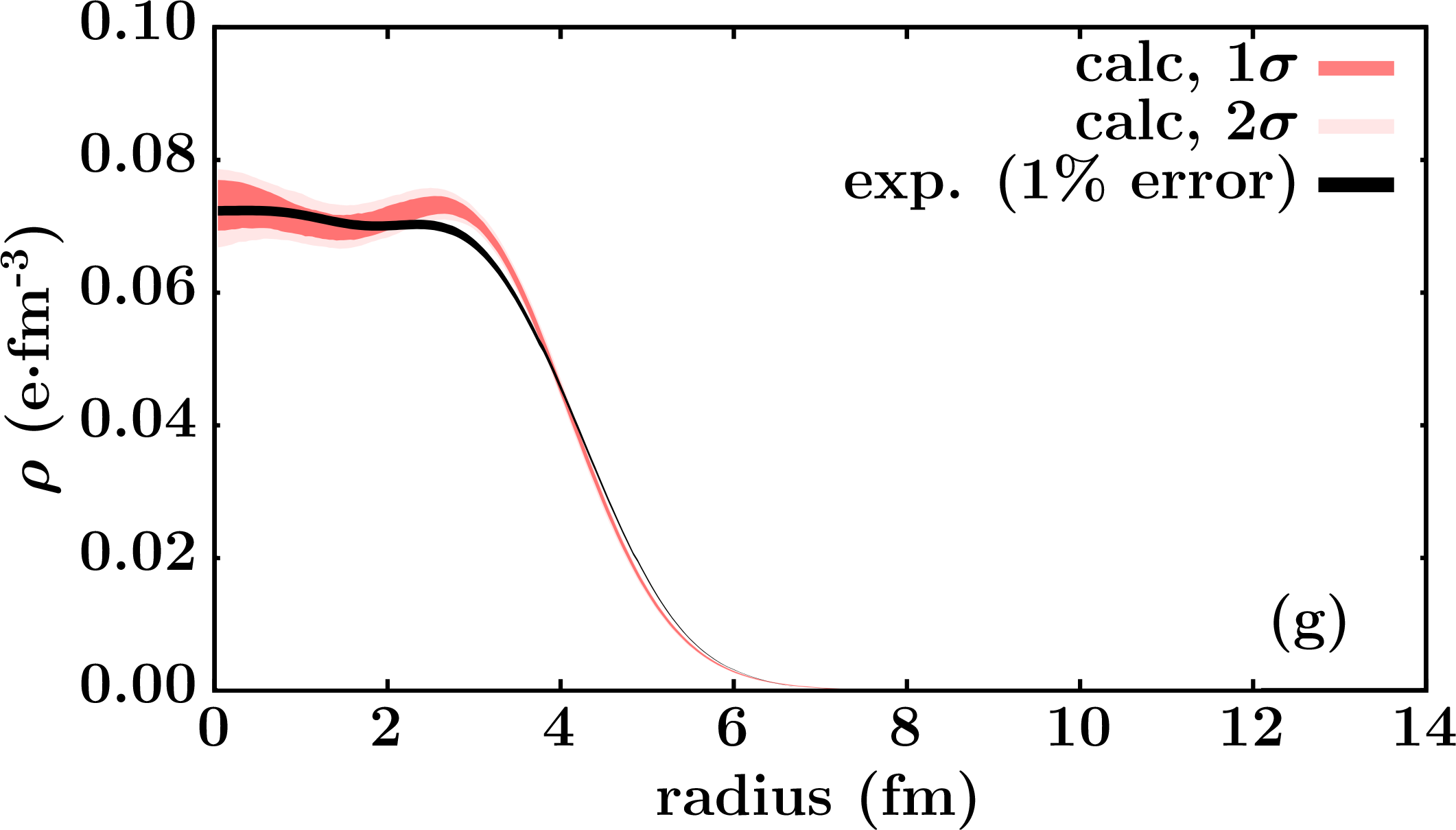}
        \label{DOM_ni64_chargeDensity}
    \end{minipage}\hspace{6pt}
    \begin{minipage}{0.4\linewidth}
        \centering
        \includegraphics[width=\linewidth]{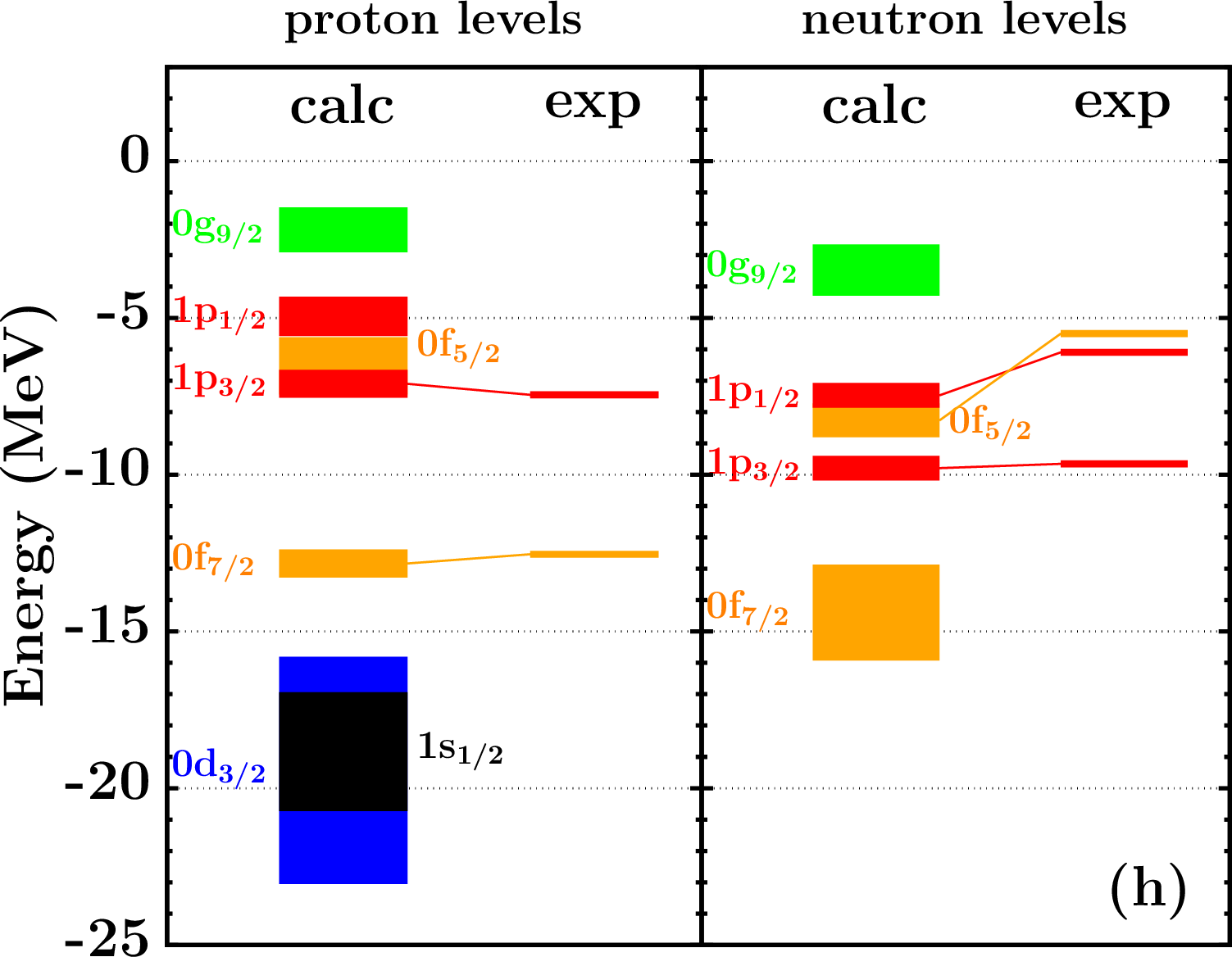}
        \label{DOM_ni64_SPLevels}
    \end{minipage}
    \begin{minipage}{0.4\linewidth}
        \centering
        \includegraphics[width=\linewidth]{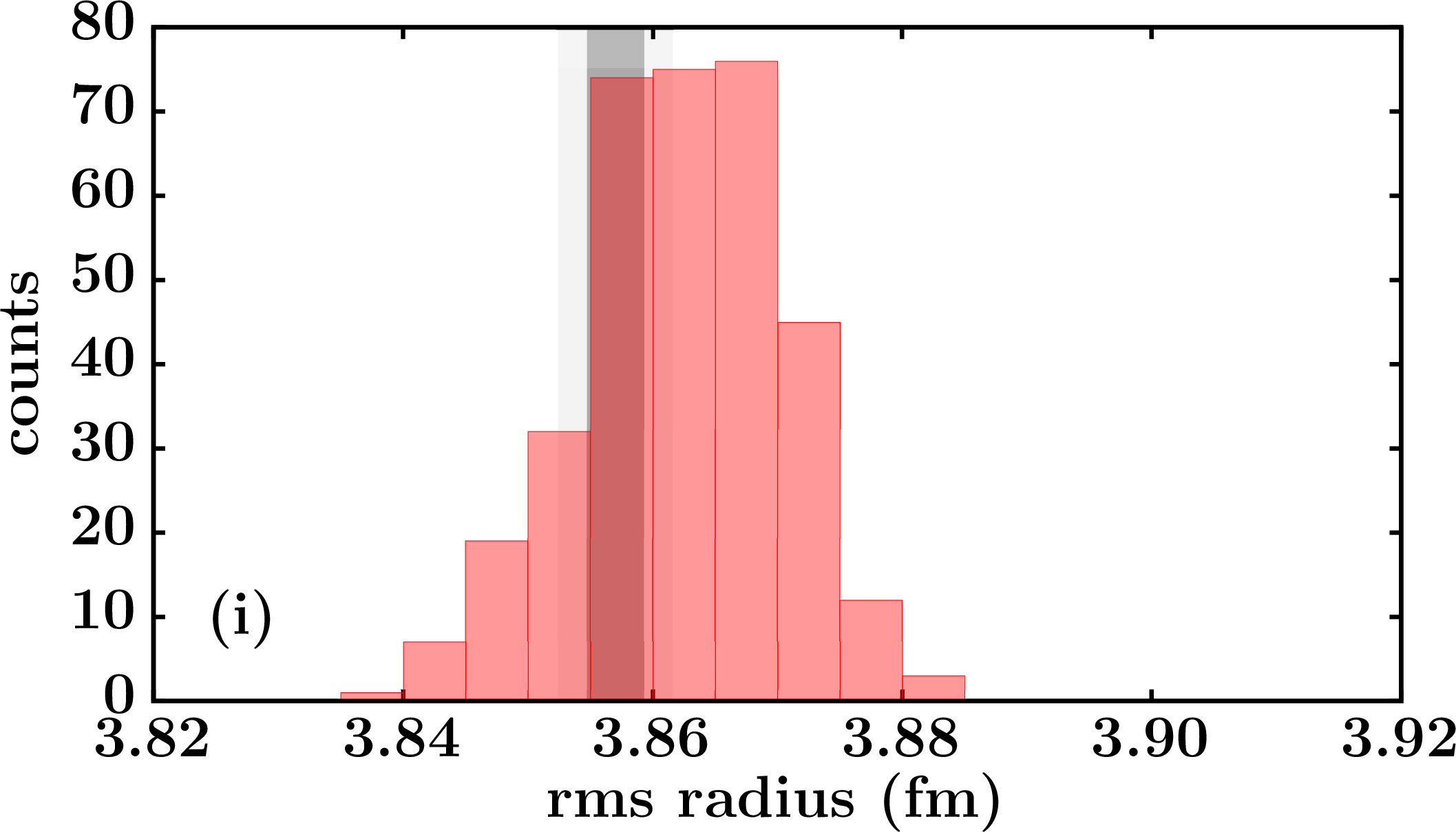}
        \label{DOM_ni64_RMSRadius}
    \end{minipage}\hspace{6pt}
    \begin{minipage}{0.4\linewidth}
        \centering
        \includegraphics[width=\linewidth]{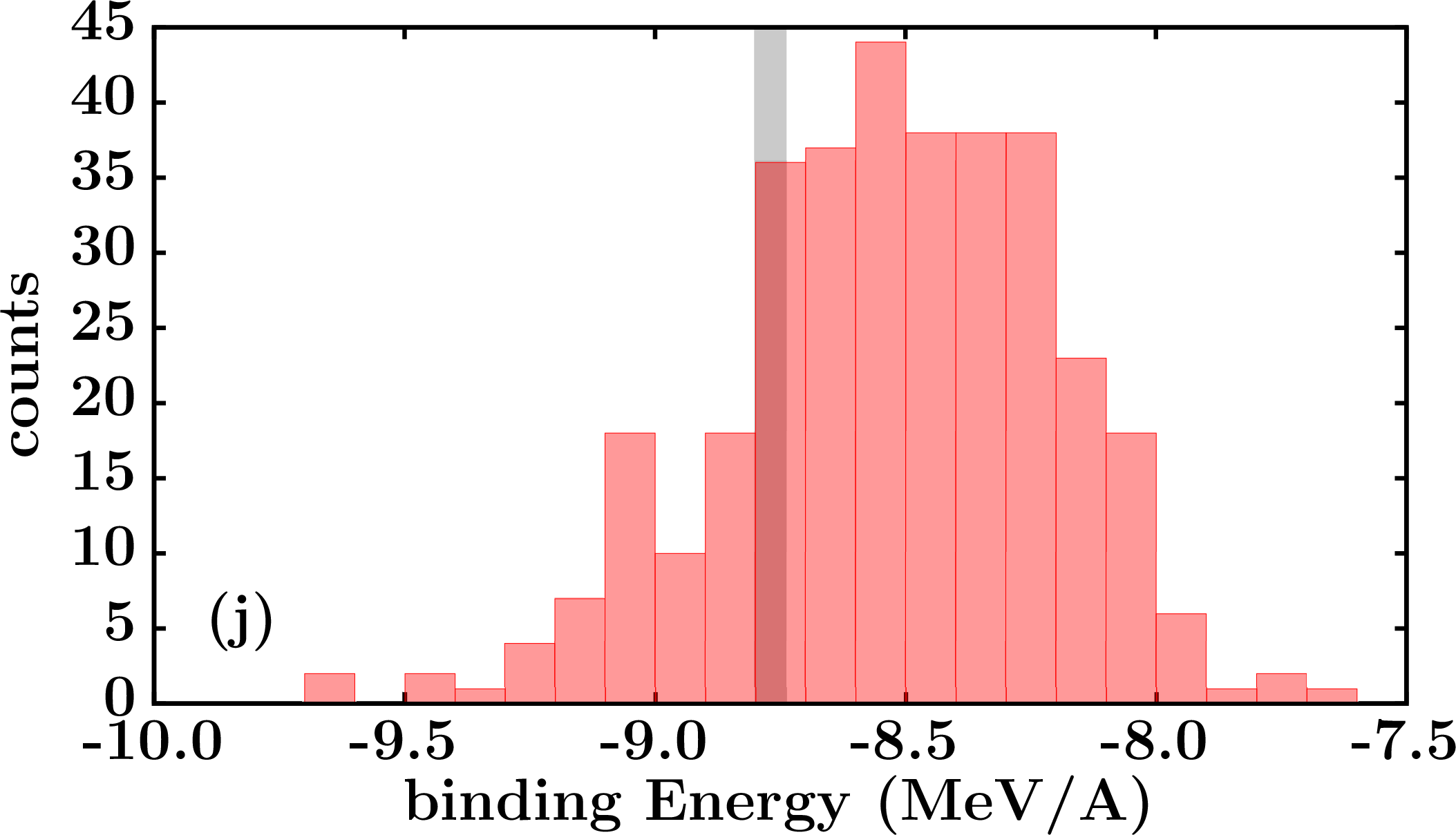}
        \label{DOM_ni64_BE}
    \end{minipage}
    \caption{\niFour: constraining experimental data and DOM fit. See introduction of
    Appendix C for description.}
    \label{DOM_ni64}
\end{figure*}

\begin{figure*}[!htb]
    \centering
    \begin{minipage}{0.4\linewidth}
        \centering
        \begin{minipage}[c]{0.5\linewidth}
            \centering
                \includegraphics[width=\linewidth]{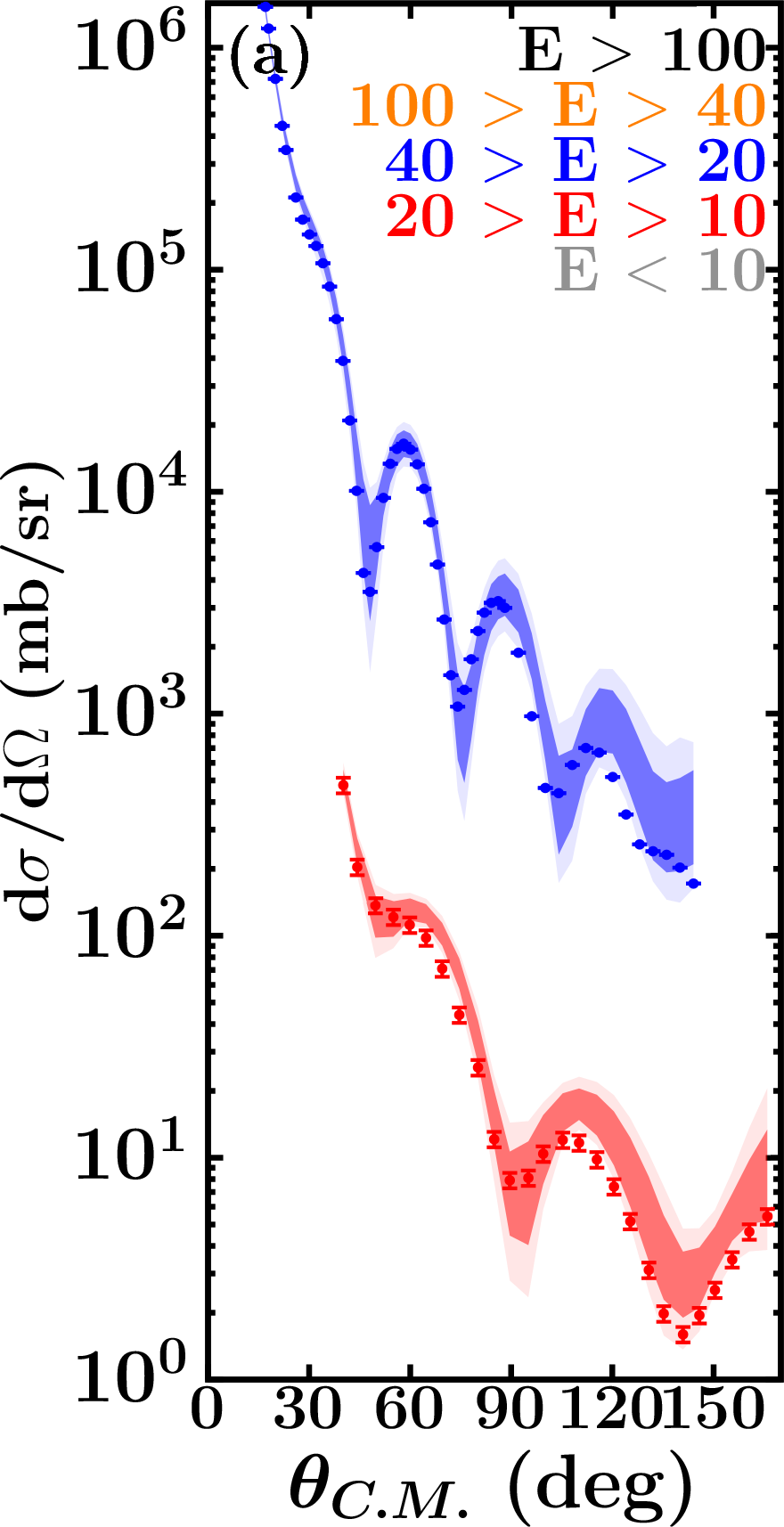}
        \end{minipage}
        \begin{minipage}[c]{0.45\linewidth}
            \centering
            No \snTwelve\ proton \\
            analyzing powers \\
            were available
        \end{minipage}
        \label{DOM_sn112_proton_elastic}
    \end{minipage}\hspace{6pt}
    \begin{minipage}{0.4\linewidth}
        \centering
        \begin{minipage}[c]{0.5\linewidth}
            \centering
                \includegraphics[width=\linewidth]{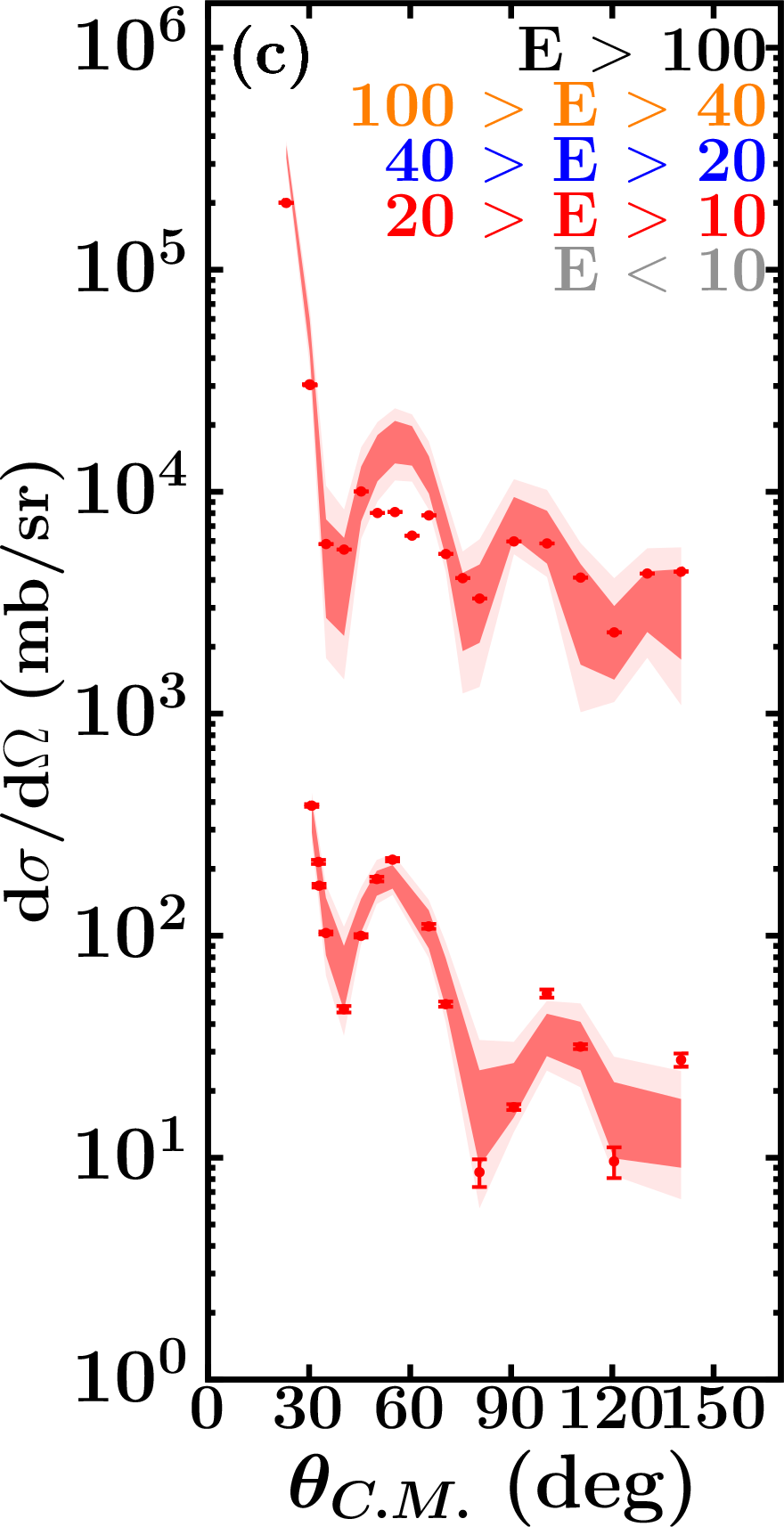}
        \end{minipage}
        \begin{minipage}[c]{0.45\linewidth}
            \centering
            No \snTwelve\ neutron \\
            analyzing powers \\
            were available
        \end{minipage}
        \label{DOM_sn112_neutron_elastic}
    \end{minipage}

    \vspace{10pt}

    \centering
    \begin{minipage}{0.4\linewidth}
        \centering
        \includegraphics[width=\linewidth]{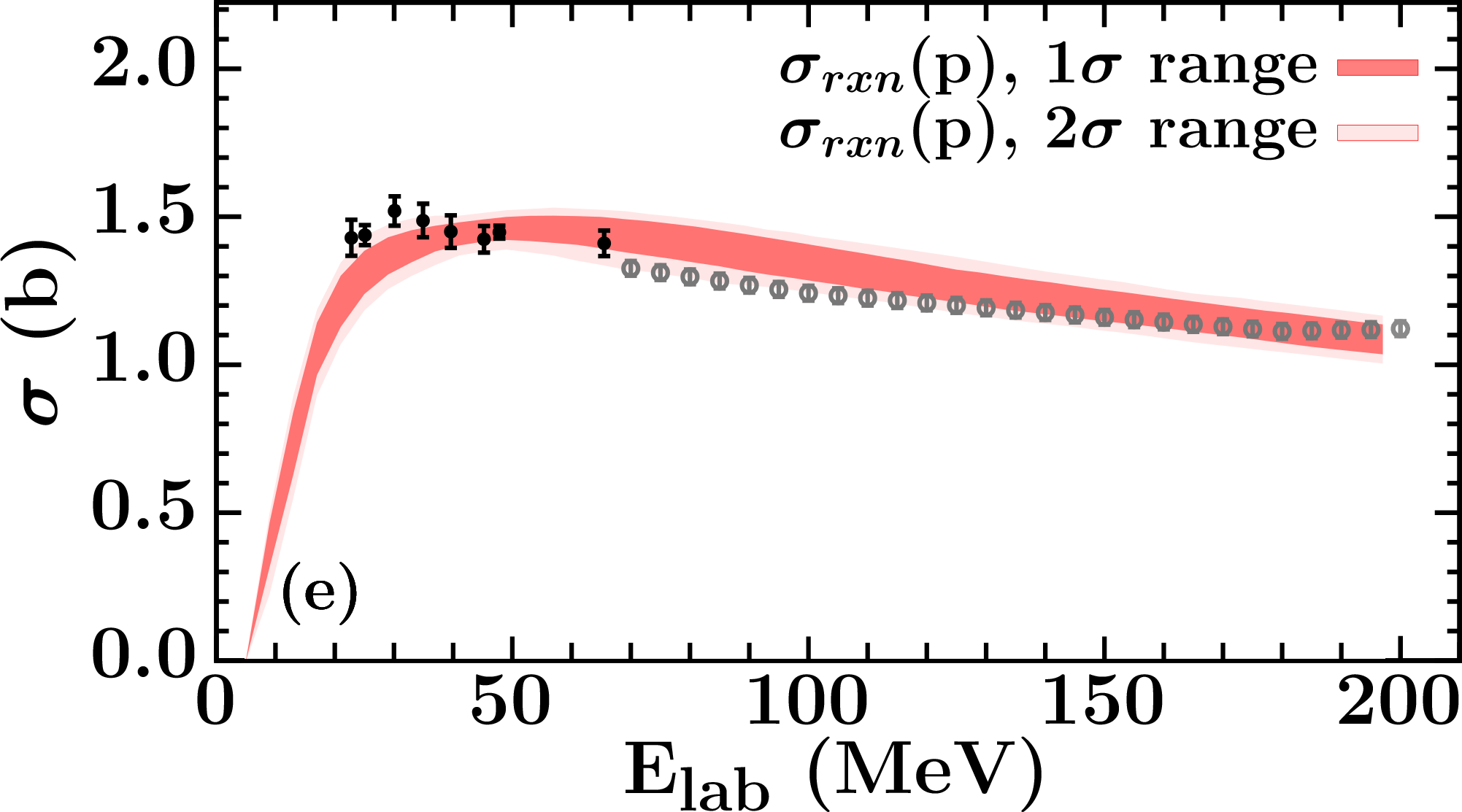}
        \label{DOM_sn112_proton_inelastic}
    \end{minipage}\hspace{6pt}
    \begin{minipage}{0.4\linewidth}
        \centering
        \includegraphics[width=\linewidth]{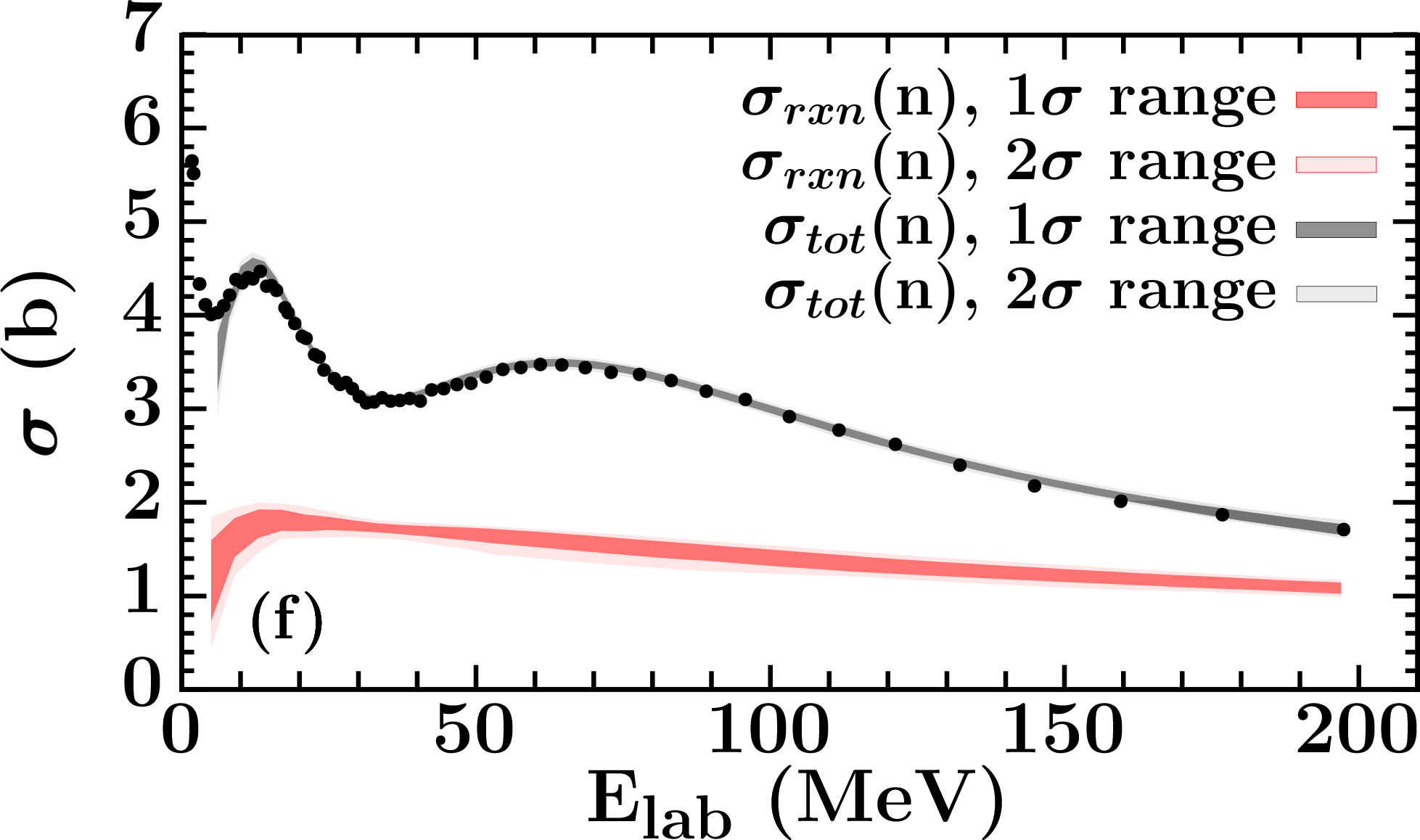}
        \label{DOM_sn112_neutron_inelastic}
    \end{minipage}
    \centering
    \begin{minipage}{0.4\linewidth}
        \centering
        \includegraphics[width=\linewidth]{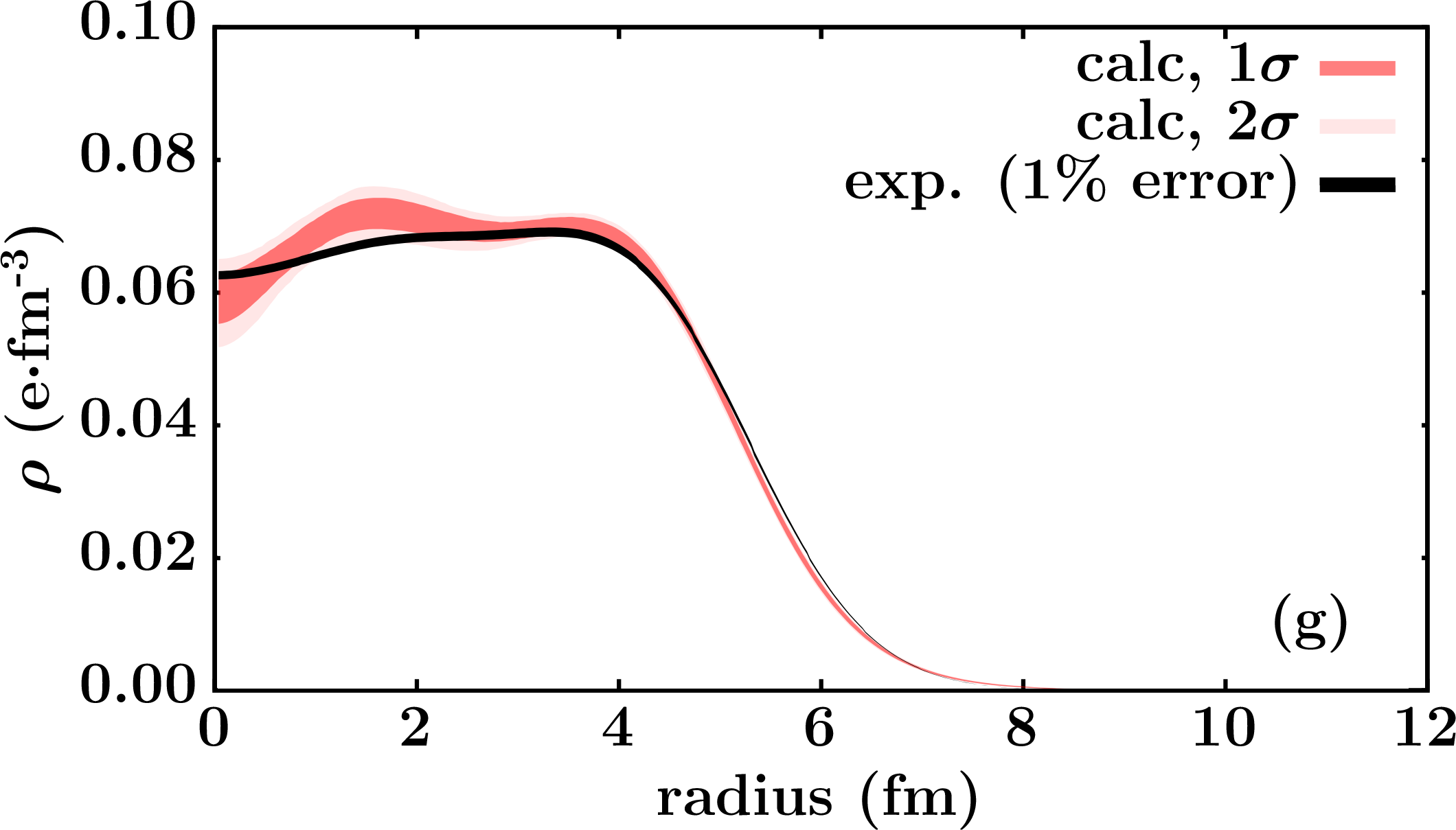}
        \label{DOM_sn112_chargeDensity}
    \end{minipage}\hspace{6pt}
    \begin{minipage}{0.4\linewidth}
        \centering
        \includegraphics[width=\linewidth]{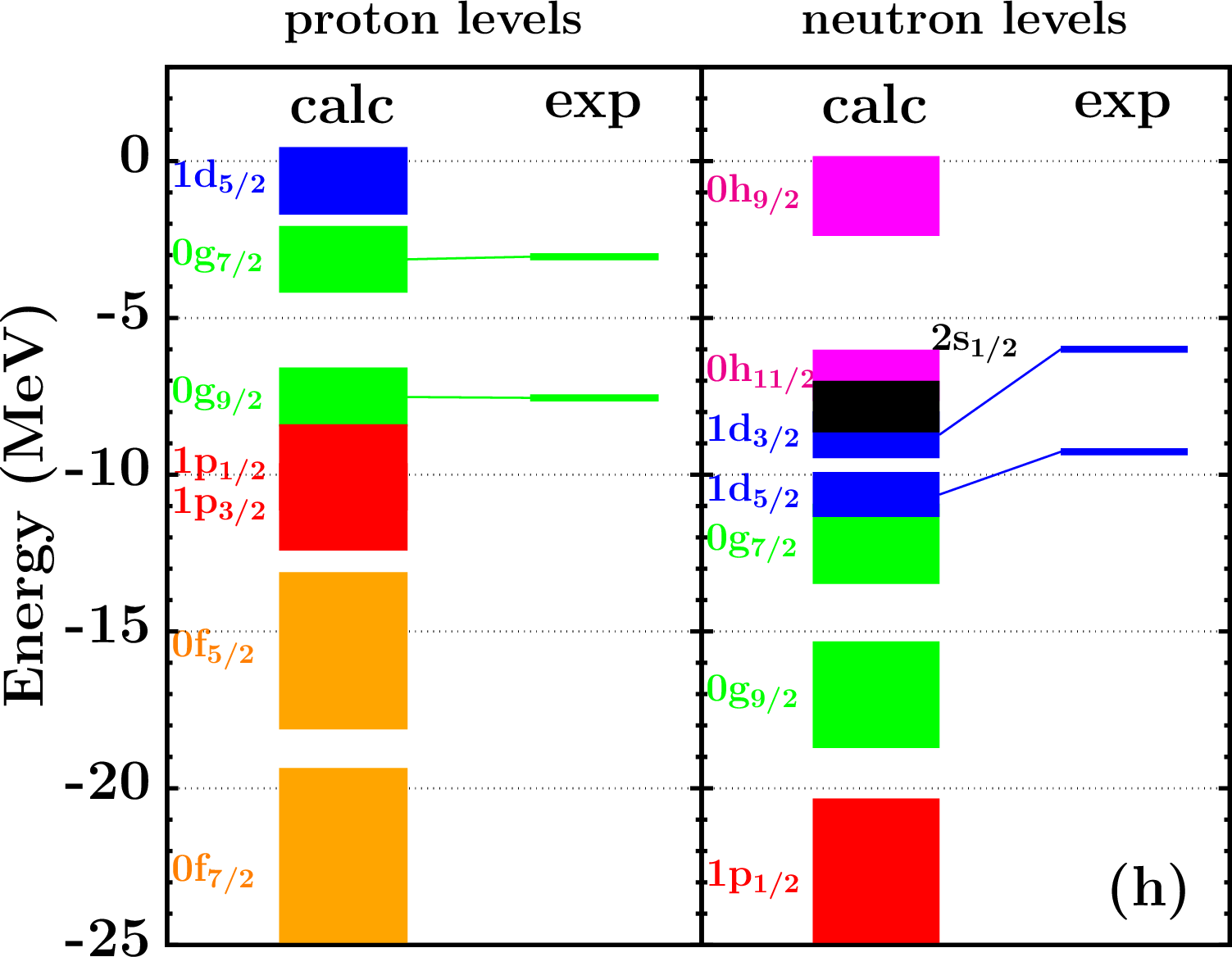}
        \label{DOM_sn112_SPLevels}
    \end{minipage}
    \begin{minipage}{0.4\linewidth}
        \centering
        \includegraphics[width=\linewidth]{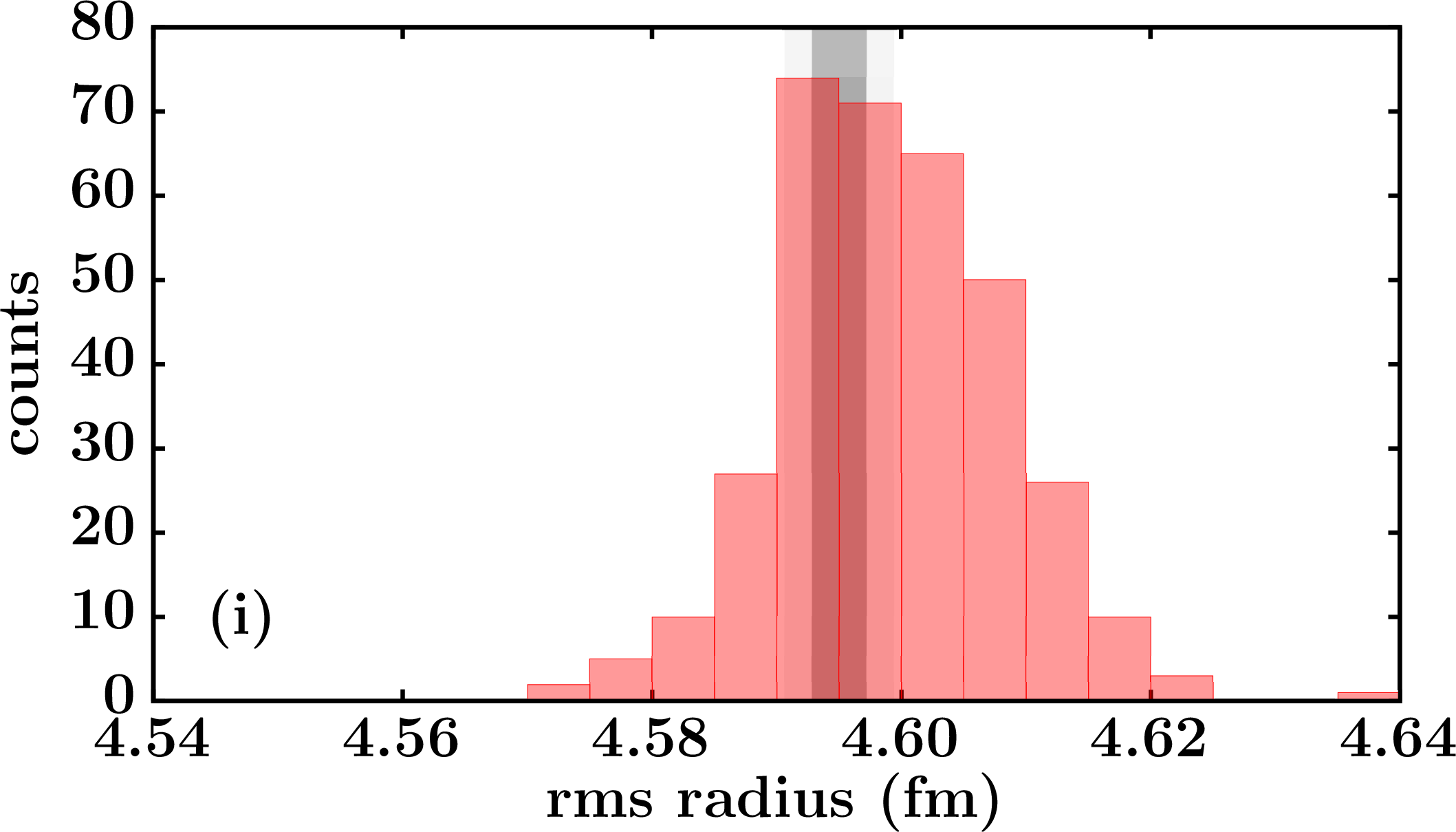}
        \label{DOM_sn112_RMSRadius}
    \end{minipage}\hspace{6pt}
    \begin{minipage}{0.4\linewidth}
        \centering
        \includegraphics[width=\linewidth]{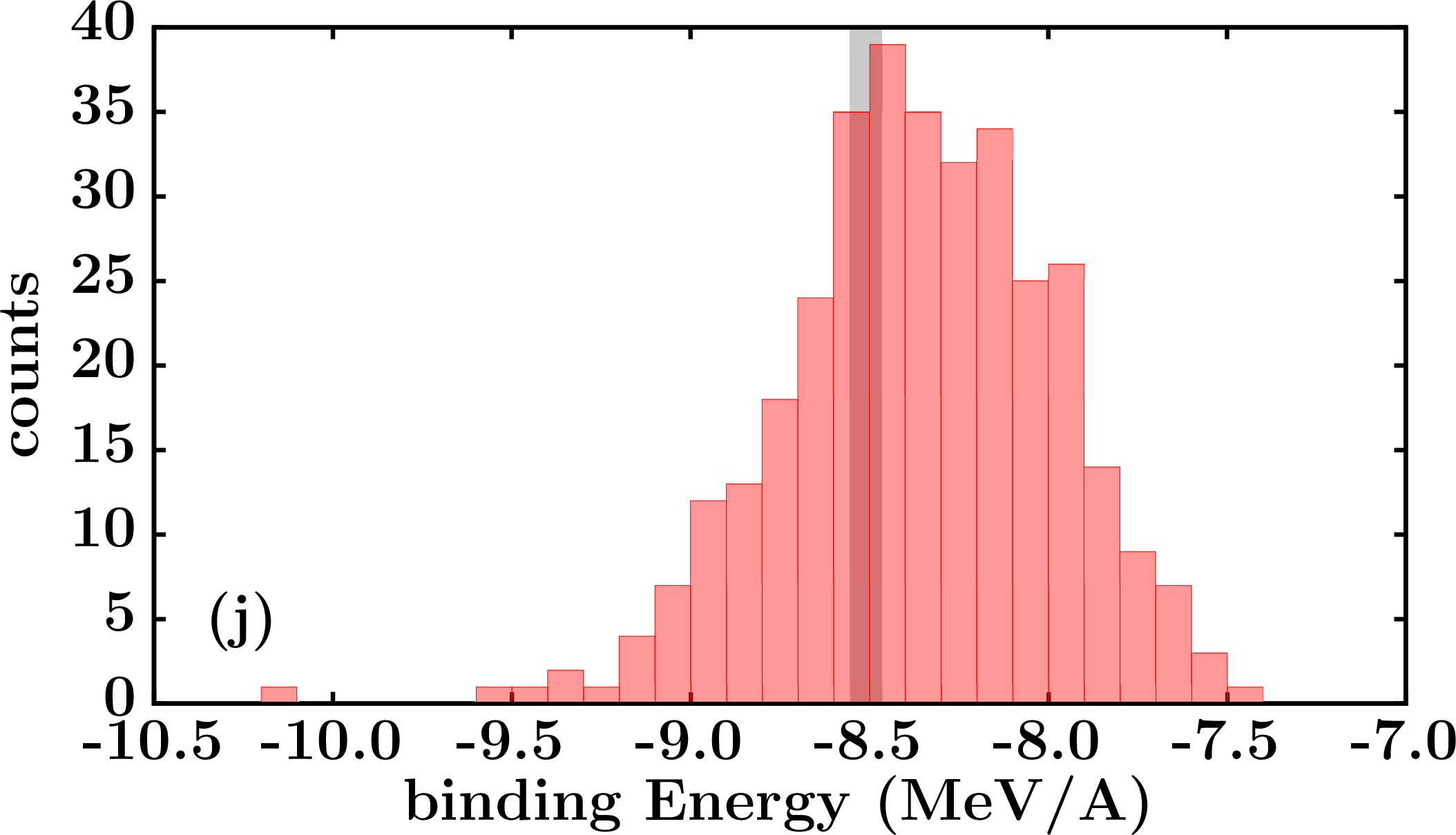}
        \label{DOM_sn112_BE}
    \end{minipage}
    \caption{\snTwelve: constraining experimental data and DOM fit. See introduction of
    Appendix C for description.}
    \label{DOM_sn112}
\end{figure*}

\begin{figure*}[!htb]
    \centering
    \begin{minipage}{0.4\linewidth}
        \centering
        \includegraphics[width=\linewidth]{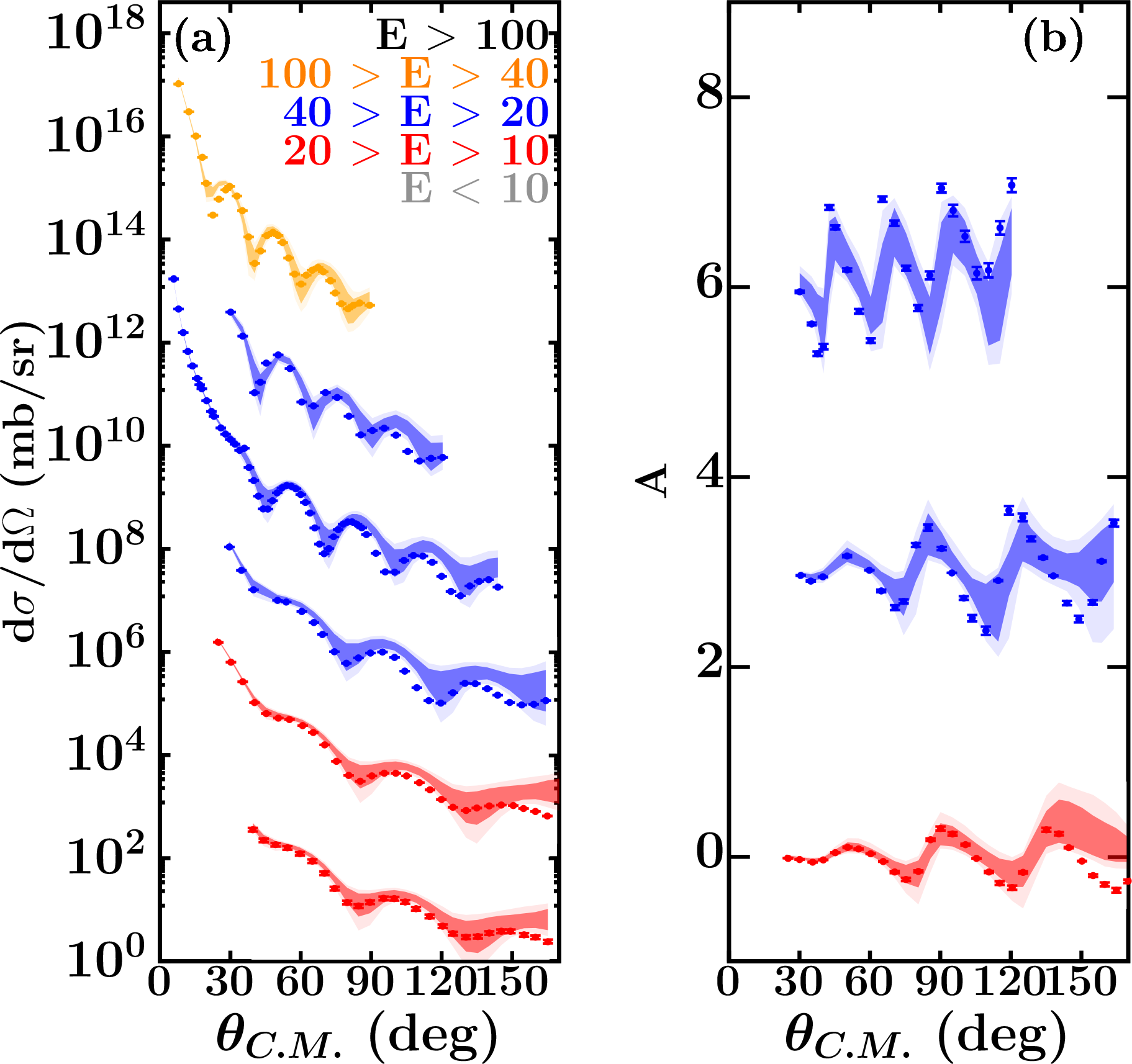}
        \label{DOM_sn124_proton_elastic}
    \end{minipage}\hspace{6pt}
    \begin{minipage}{0.4\linewidth}
        \vspace{-10pt}
        \begin{minipage}{0.5\linewidth}
            \includegraphics[width=\linewidth]{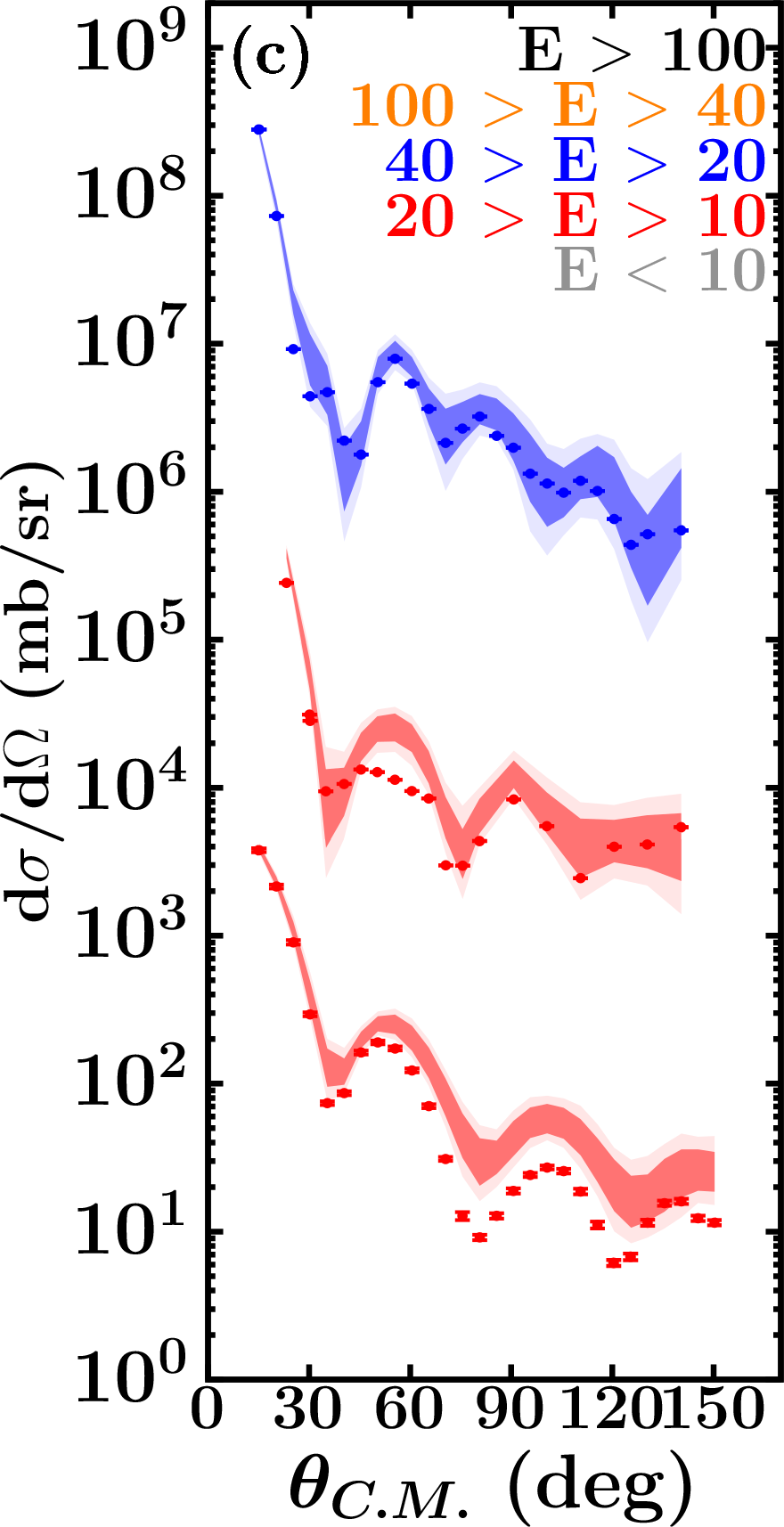}
        \end{minipage}
        \begin{minipage}{0.45\linewidth}
            \centering
            No \snFour\ neutron \\
            analyzing powers \\
            were available
        \end{minipage}
        \label{DOM_sn124_neutron_elastic}
    \end{minipage}
    \begin{minipage}{0.4\linewidth}
        \centering
        \includegraphics[width=\linewidth]{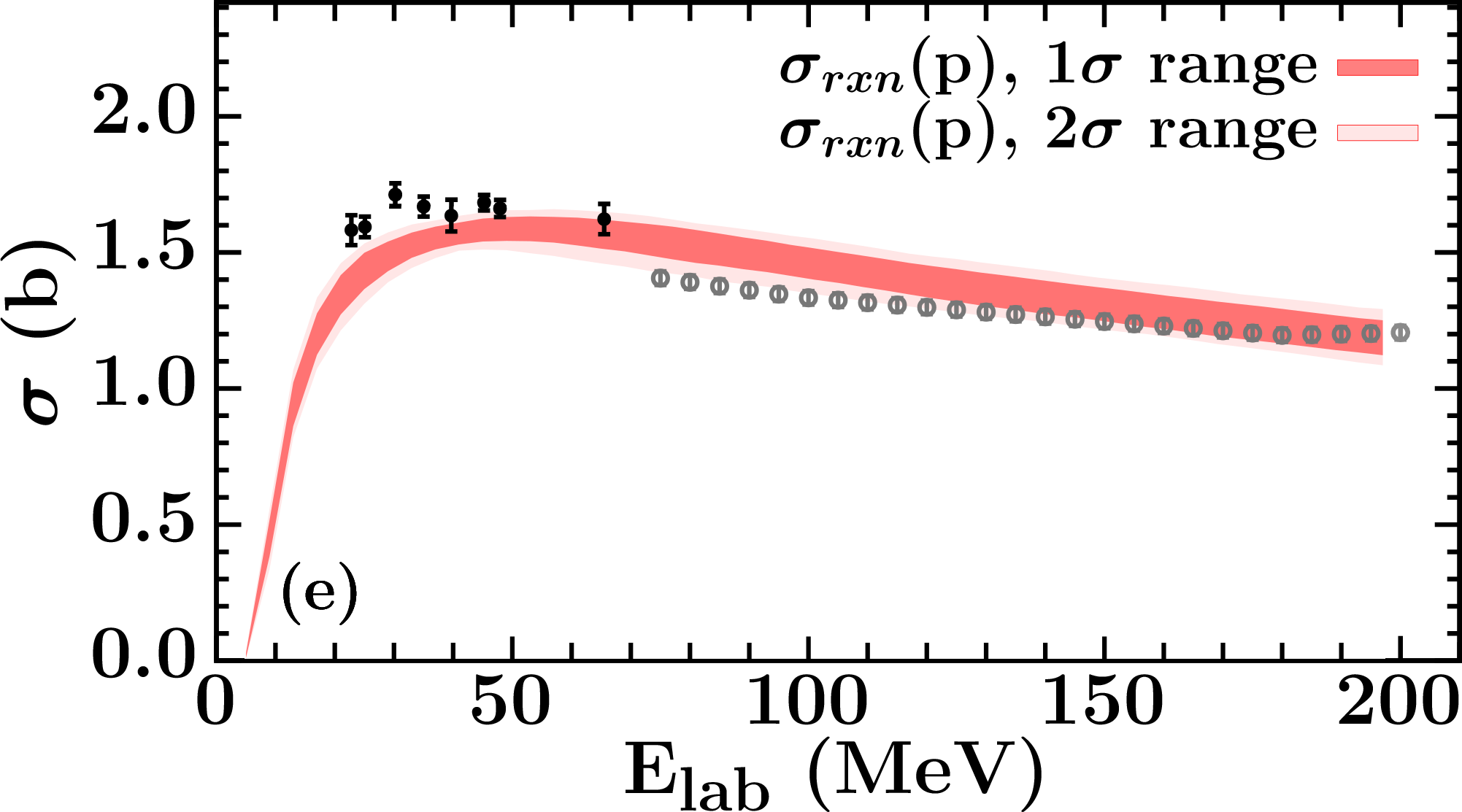}
        \label{DOM_sn124_proton_inelastic}
    \end{minipage}\hspace{6pt}
    \begin{minipage}{0.4\linewidth}
        \centering
        \includegraphics[width=\linewidth]{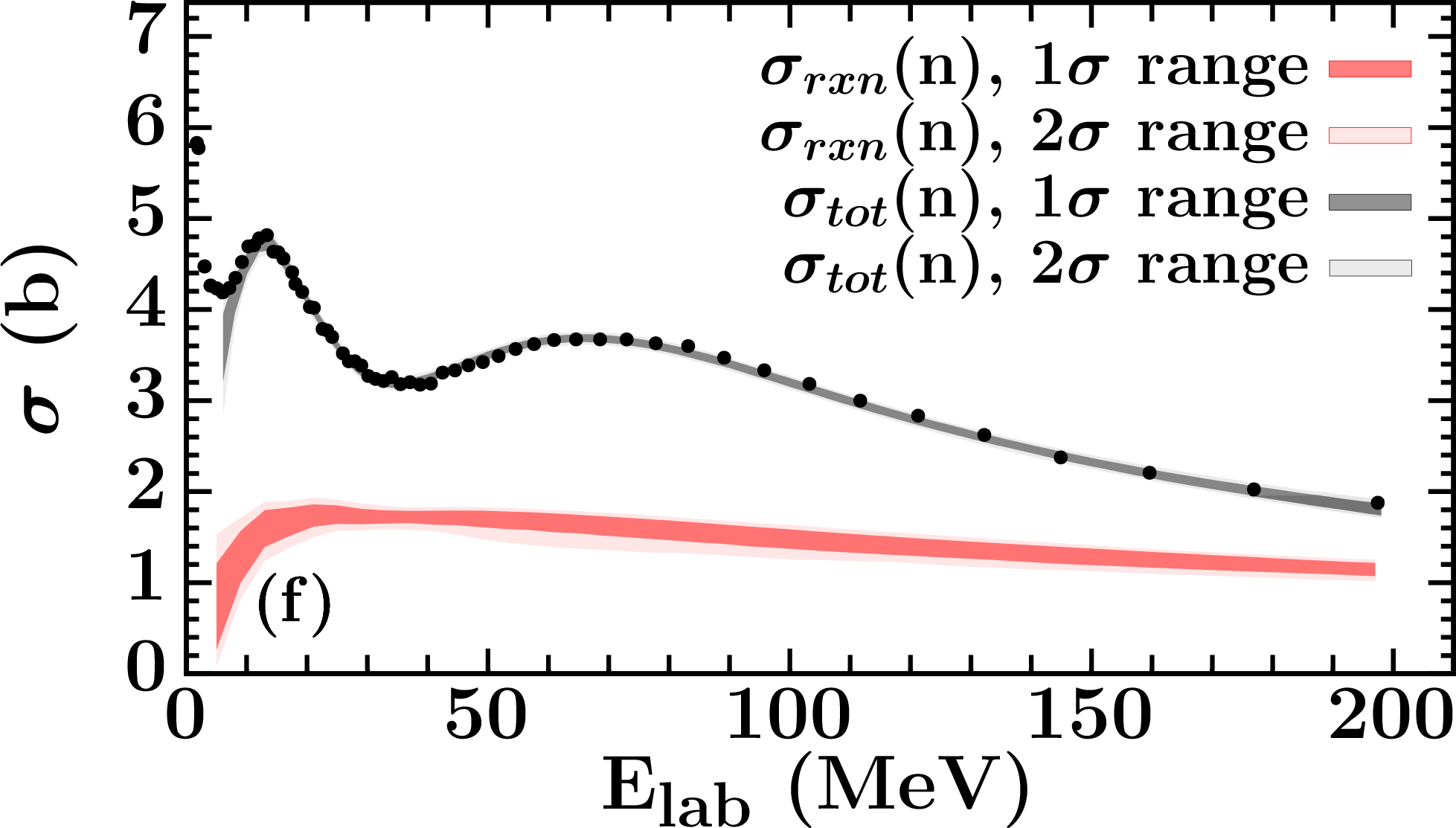}
        \label{DOM_sn124_neutron_inelastic}
    \end{minipage}
    \centering
    \begin{minipage}{0.4\linewidth}
        \centering
        \includegraphics[width=\linewidth]{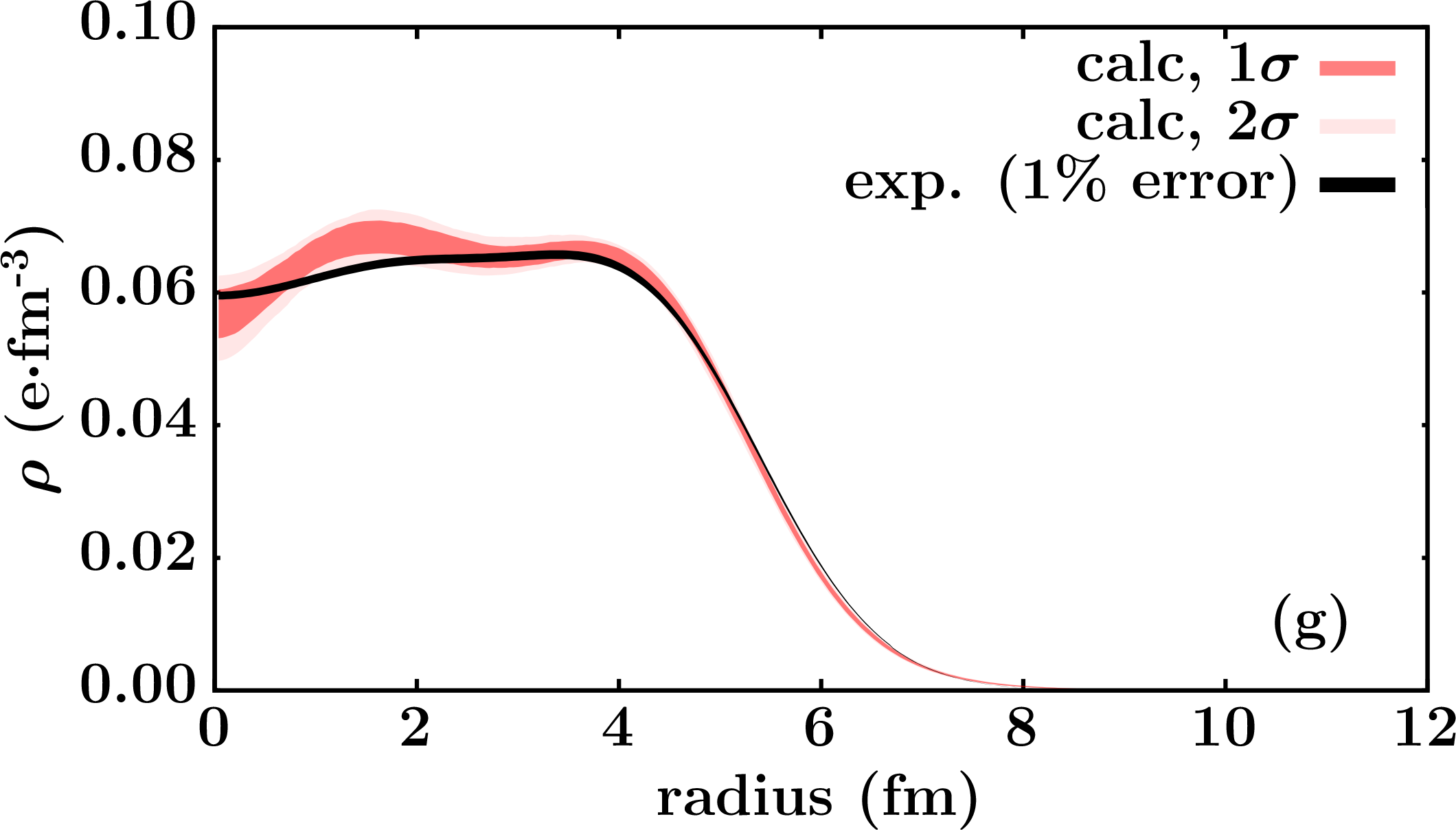}
        \label{DOM_sn124_chargeDensity}
    \end{minipage}\hspace{6pt}
    \begin{minipage}{0.4\linewidth}
        \centering
        \includegraphics[width=\linewidth]{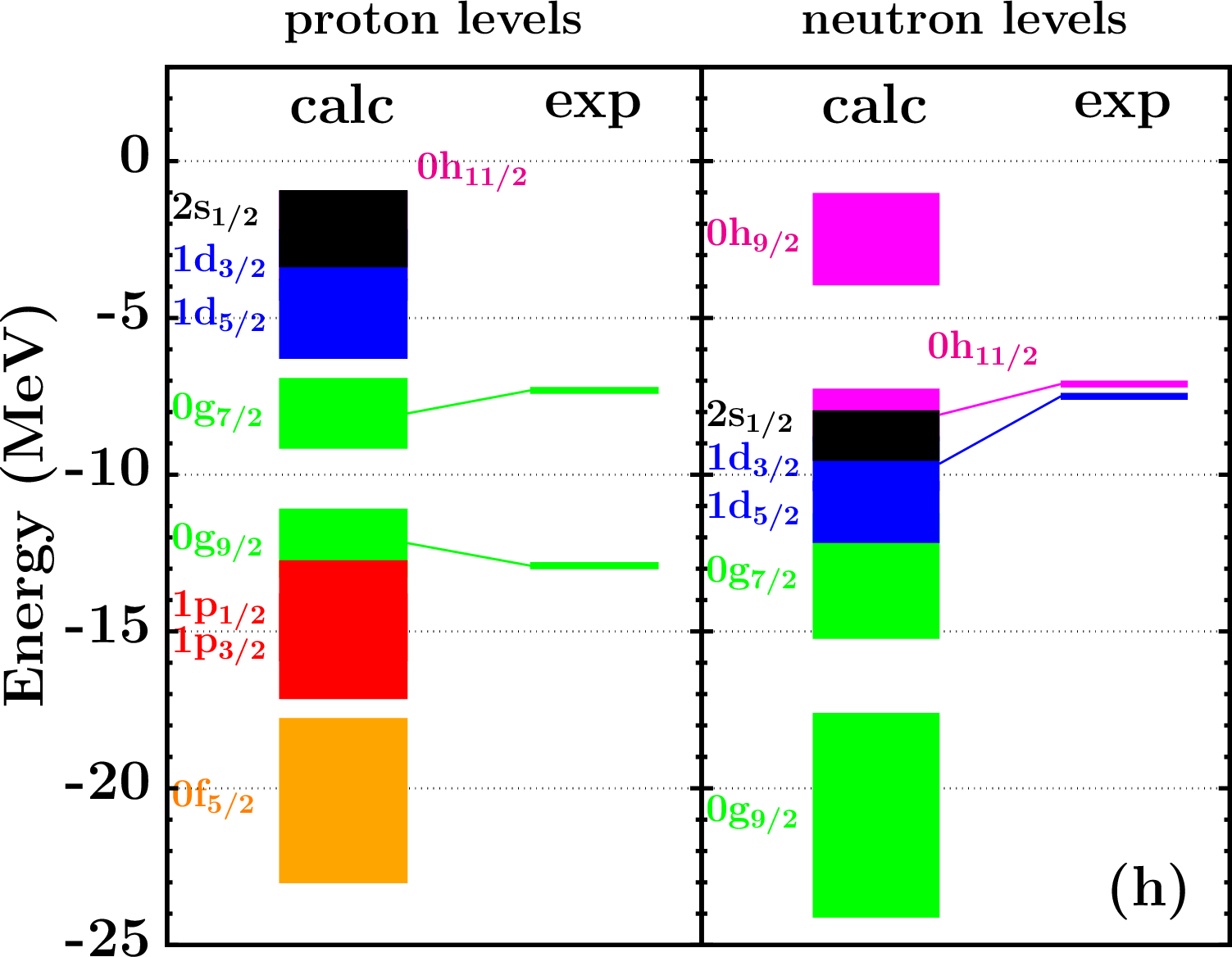}
        \label{DOM_sn124_SPLevels}
    \end{minipage}
    \begin{minipage}{0.4\linewidth}
        \centering
        \includegraphics[width=\linewidth]{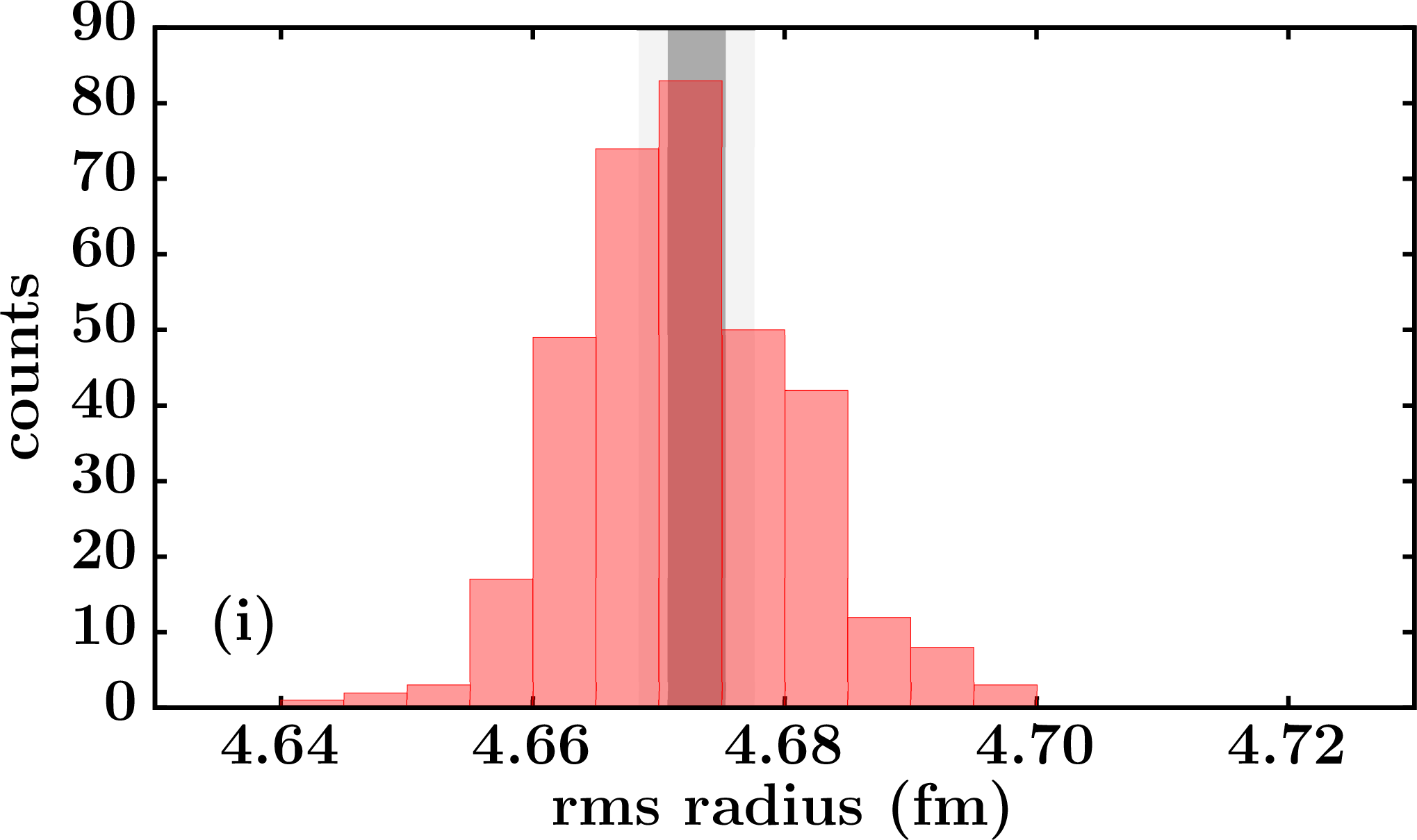}
        \label{DOM_sn124_RMSRadius}
    \end{minipage}\hspace{6pt}
    \begin{minipage}{0.4\linewidth}
        \centering
        \includegraphics[width=\linewidth]{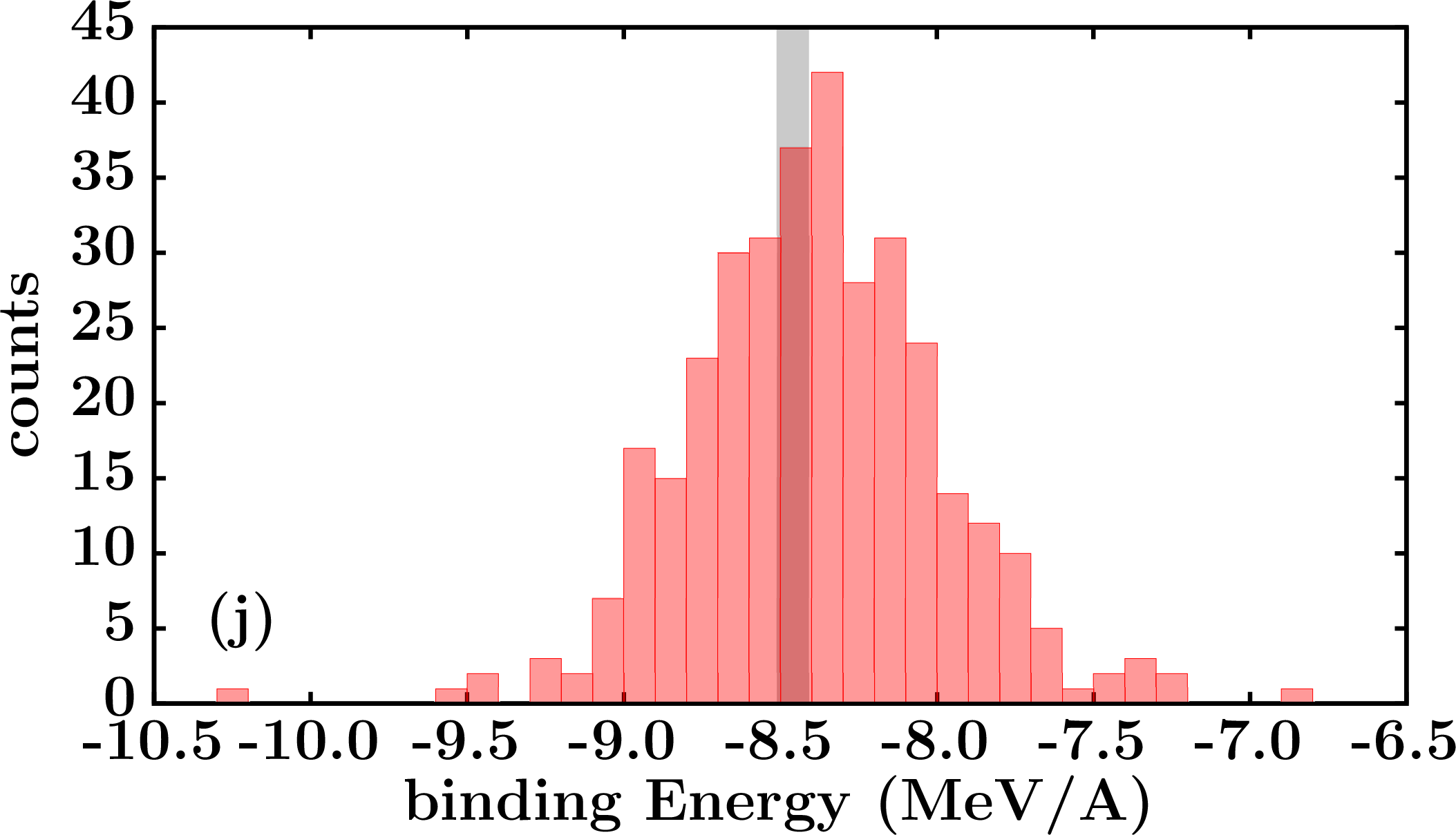}
        \label{DOM_sn124_BE}
    \end{minipage}
    \caption{\snFour: constraining experimental data and DOM fit. See introduction of
    Appendix C for description.}
    \label{DOM_sn124}
\end{figure*}

\begin{figure*}[!htb]
    \centering
    \begin{minipage}{0.4\linewidth}
        \centering
        \includegraphics[width=\linewidth]{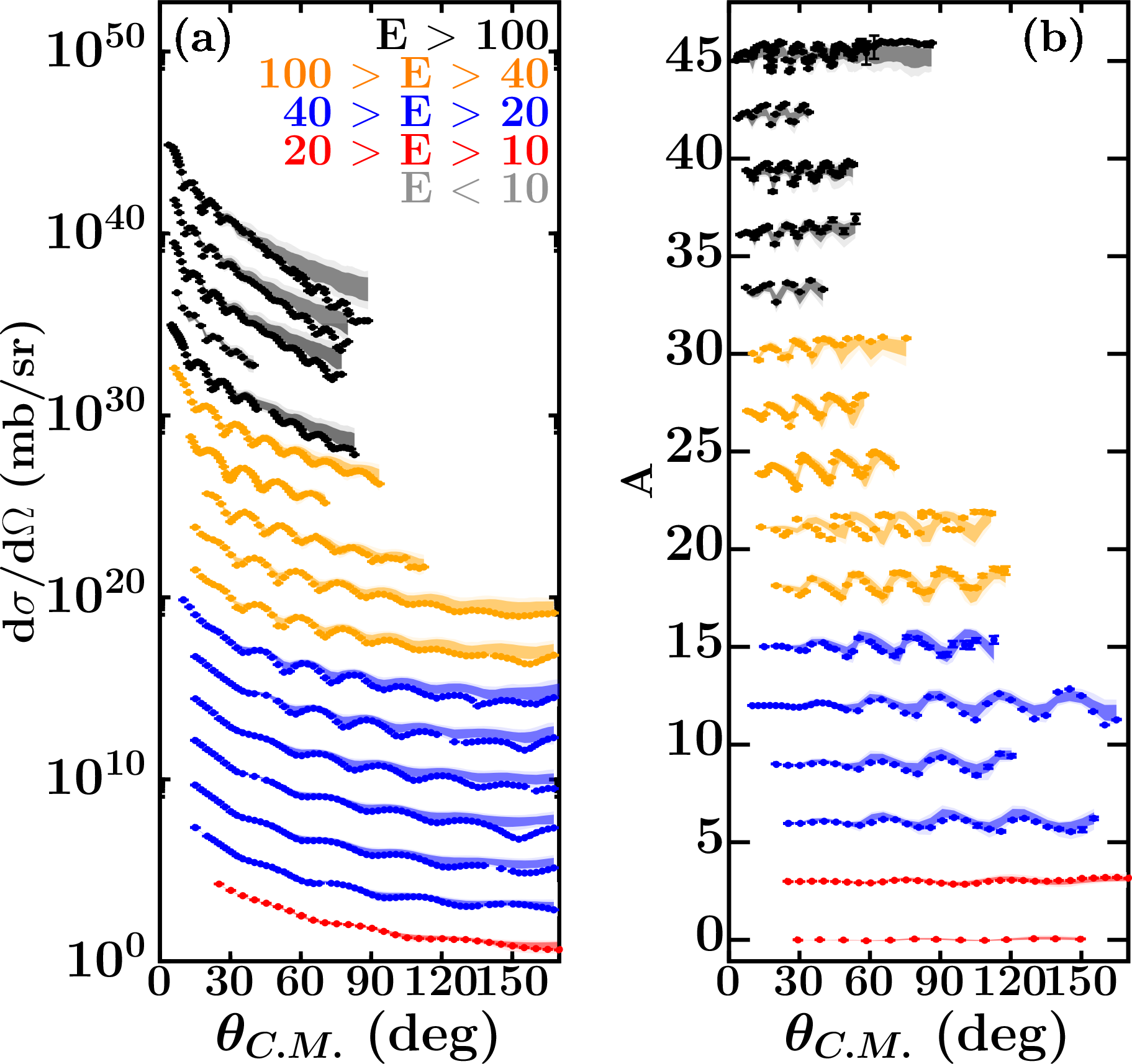}
        \label{DOM_pb208_proton_elastic}
    \end{minipage}\hspace{6pt}
    \begin{minipage}{0.4\linewidth}
        \centering
        \includegraphics[width=\linewidth]{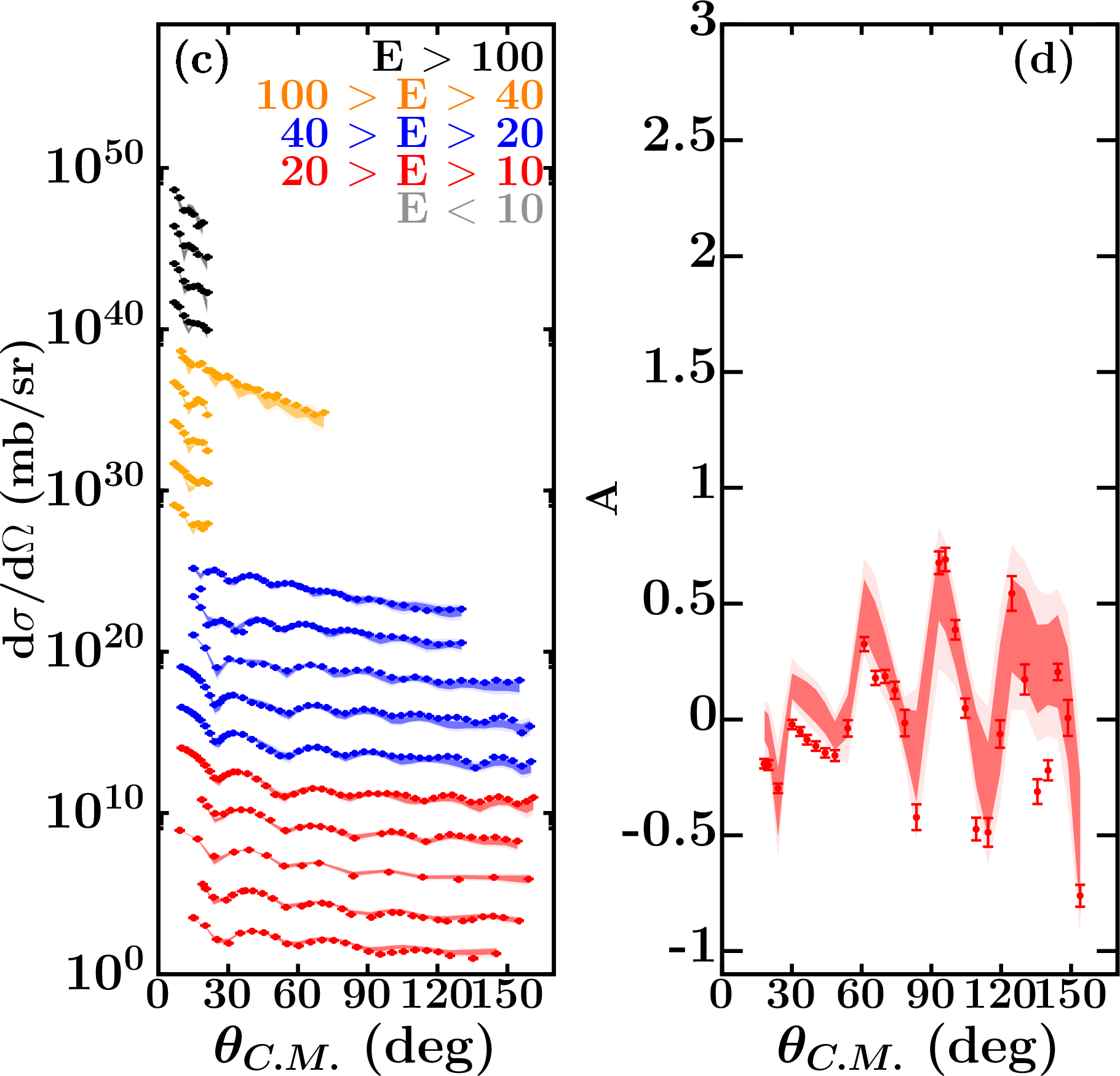}
        \label{DOM_pb208_neutron_elastic}
    \end{minipage}
    \centering
    \begin{minipage}{0.4\linewidth}
        \centering
        \includegraphics[width=\linewidth]{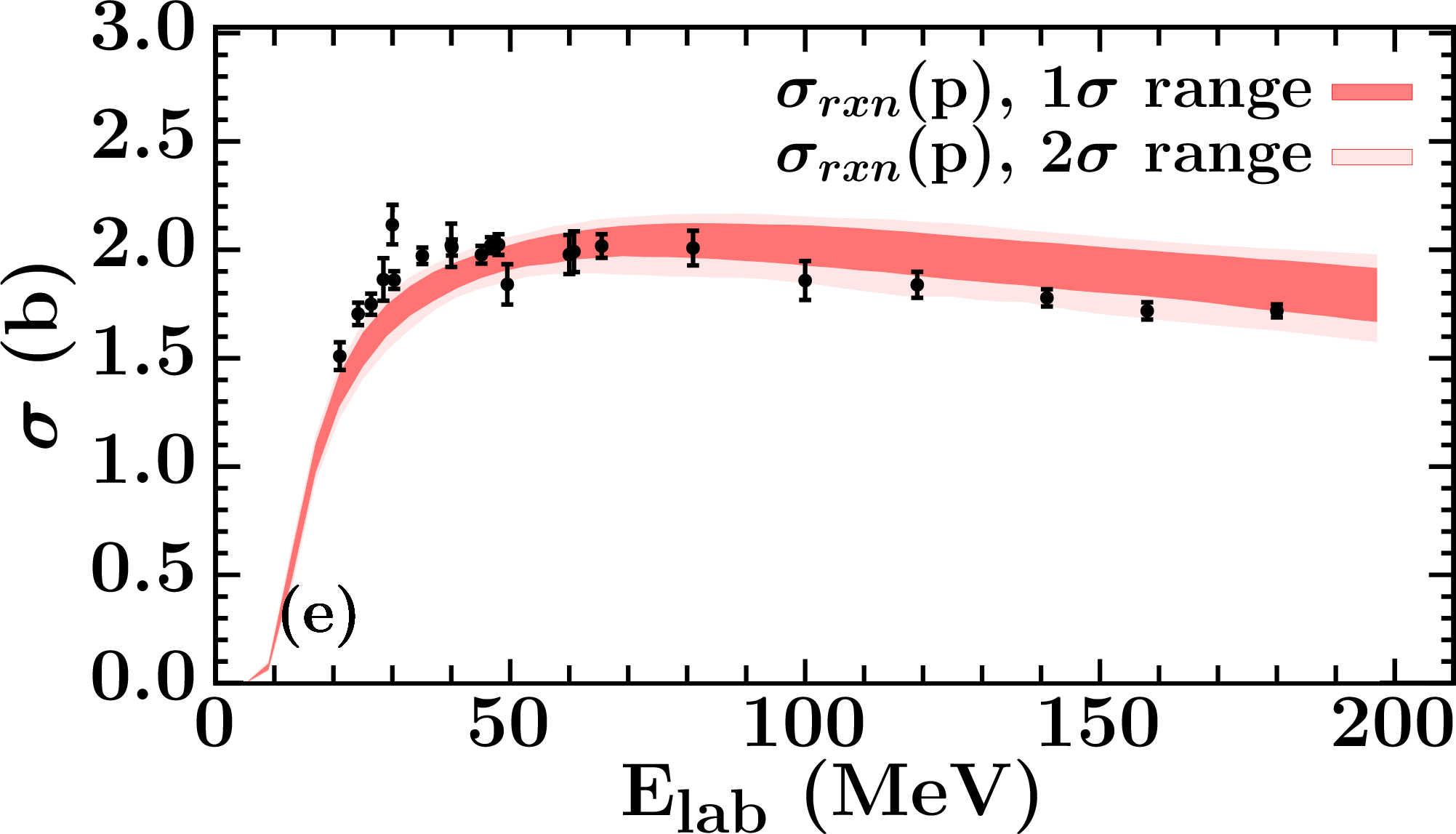}
        \label{DOM_pb208_proton_inelastic}
    \end{minipage}\hspace{6pt}
    \begin{minipage}{0.4\linewidth}
        \centering
        \includegraphics[width=\linewidth]{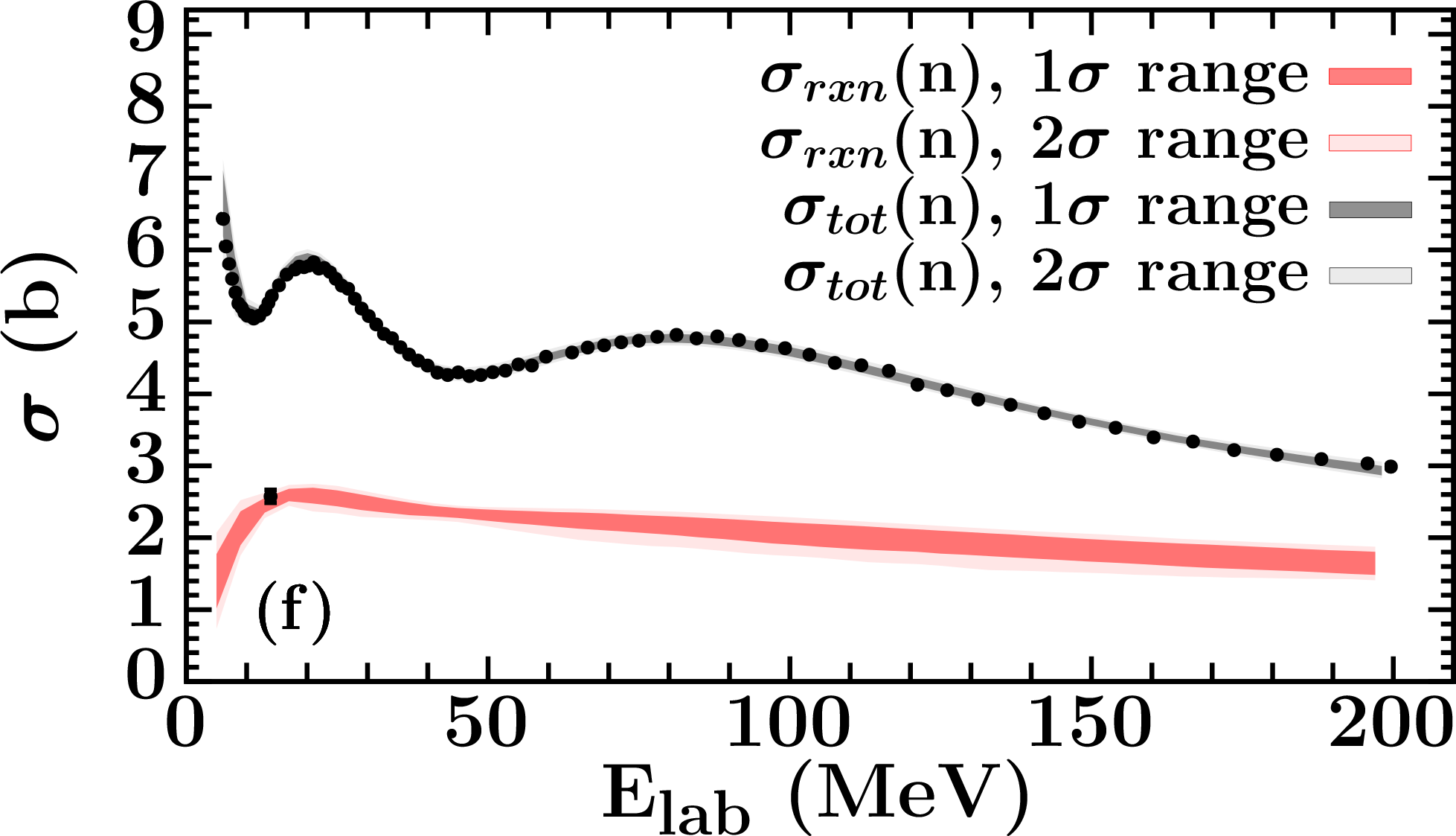}
        \label{DOM_pb208_neutron_inelastic}
    \end{minipage}
    \centering
    \begin{minipage}{0.4\linewidth}
        \centering
        \includegraphics[width=\linewidth]{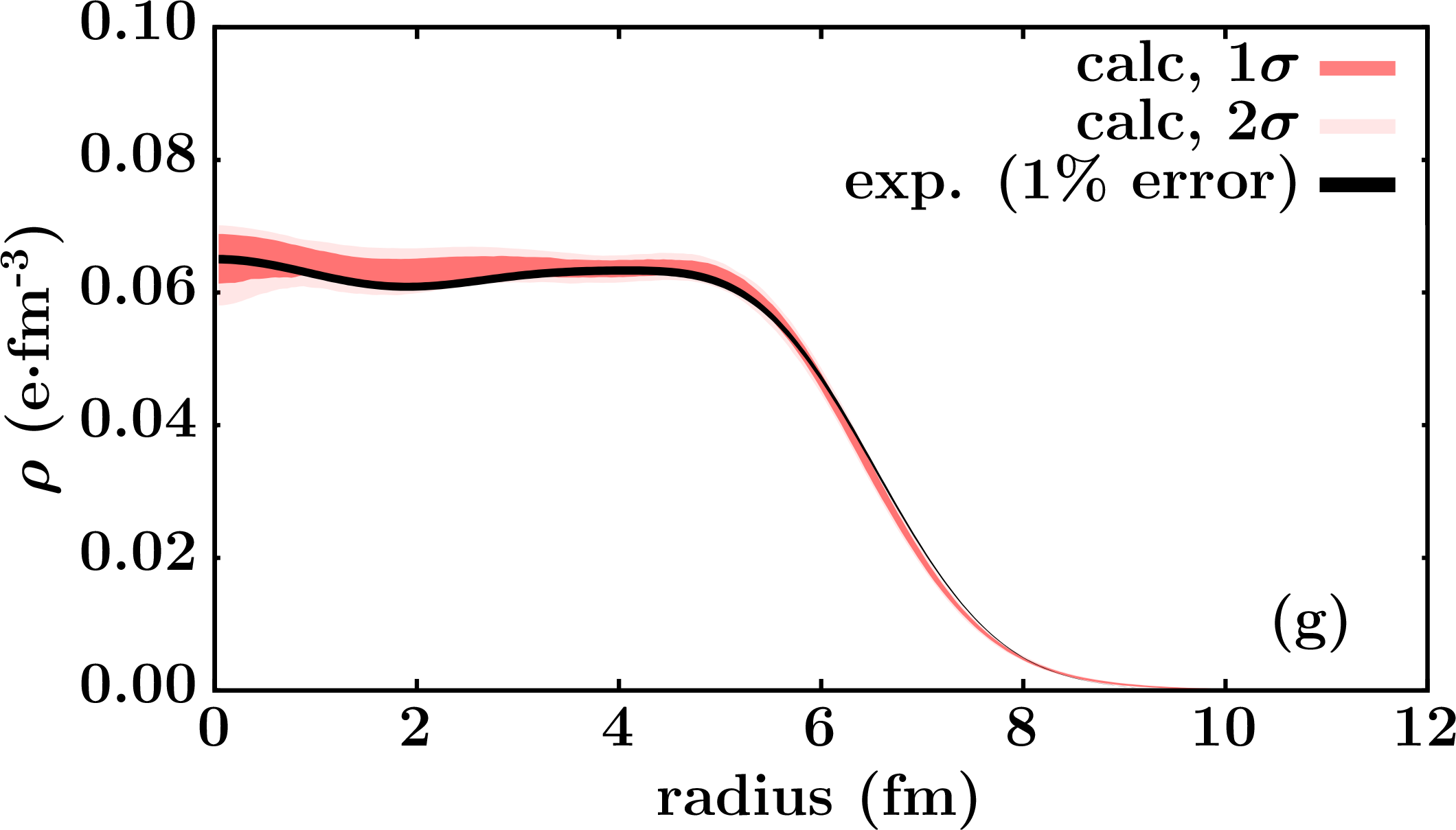}
        \label{DOM_pb208_chargeDensity}
    \end{minipage}\hspace{6pt}
    \begin{minipage}{0.4\linewidth}
        \centering
        \includegraphics[width=\linewidth]{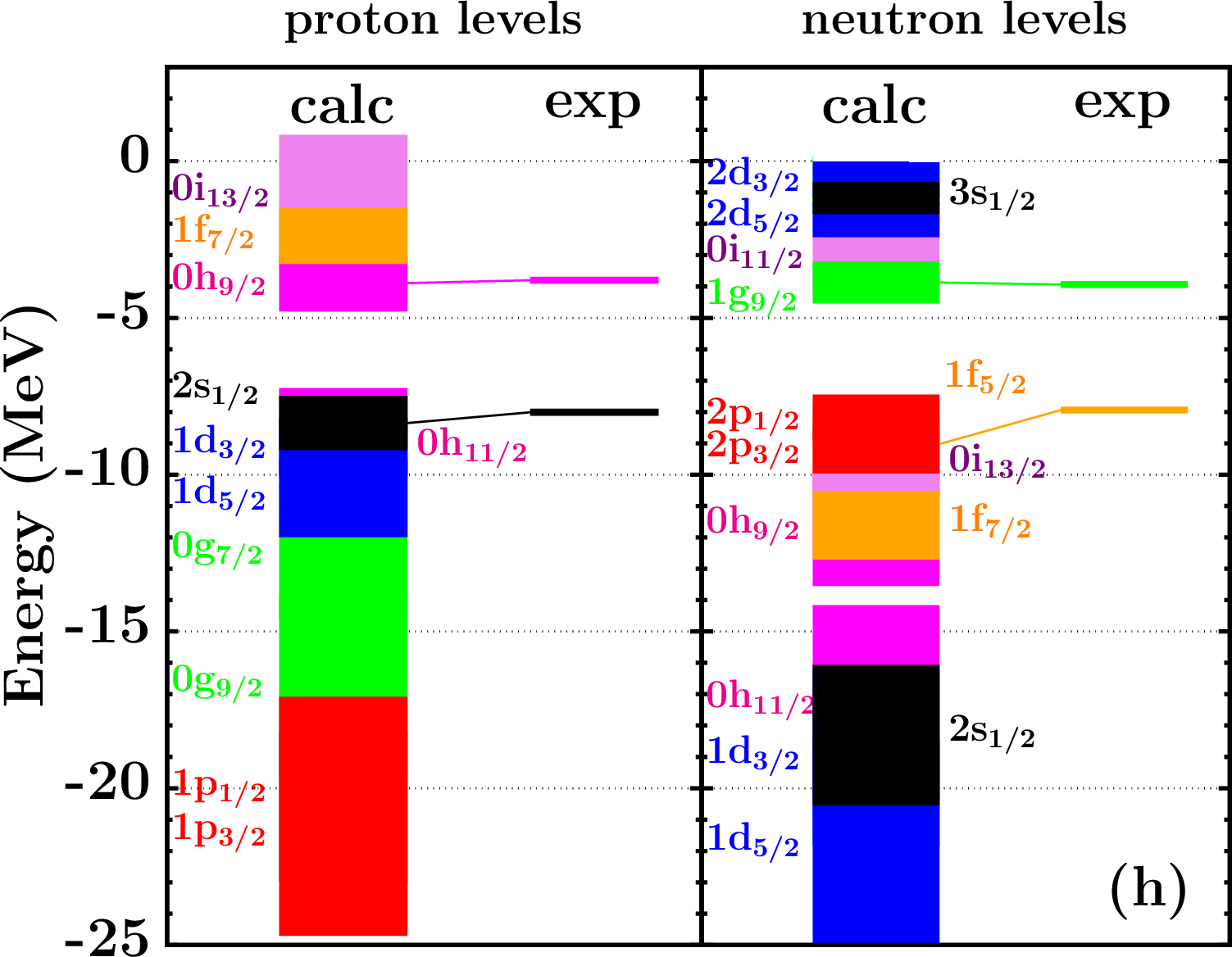}
        \label{DOM_pb208_SPLevels}
    \end{minipage}
    \begin{minipage}{0.4\linewidth}
        \centering
        \includegraphics[width=\linewidth]{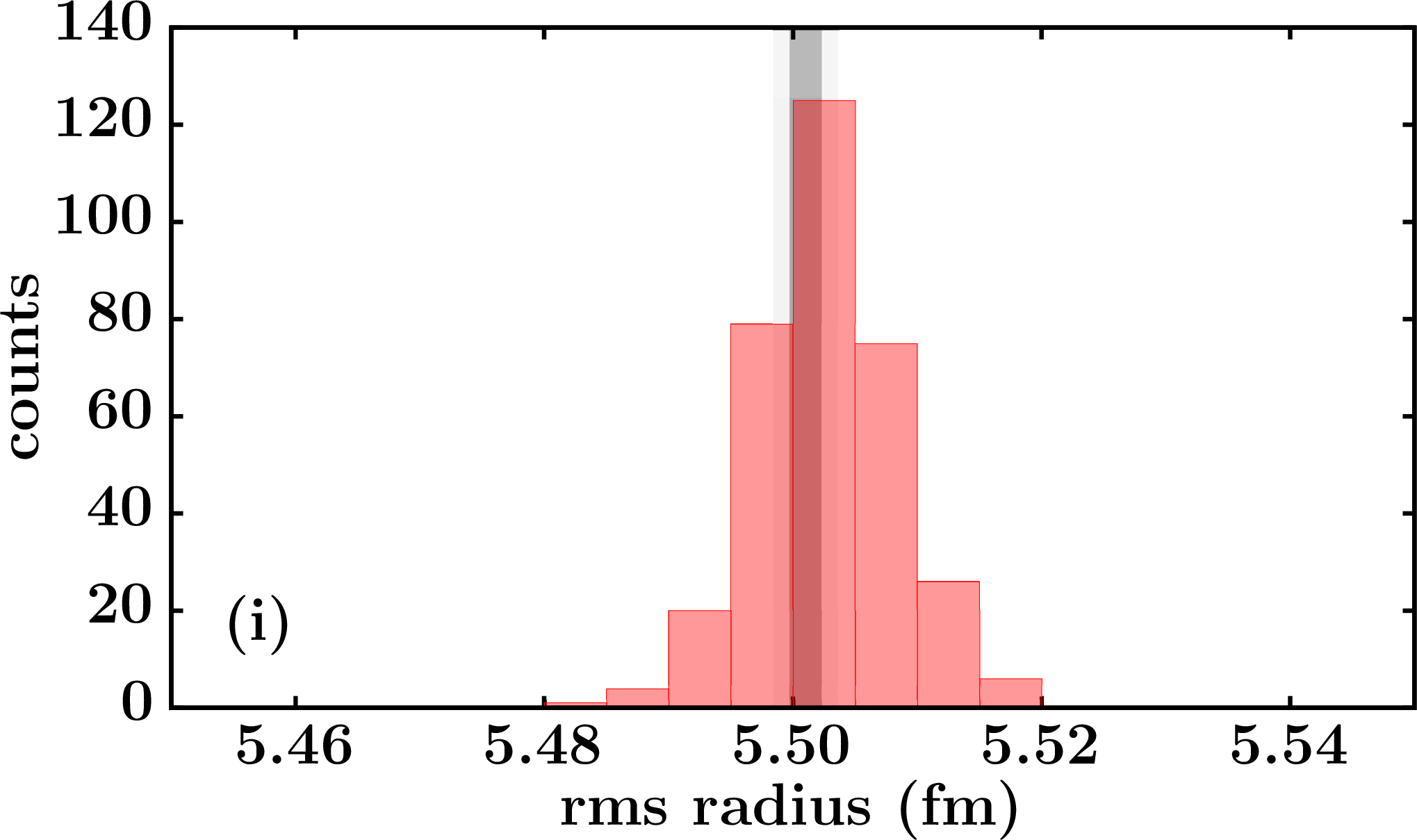}
        \label{DOM_pb208_RMSRadius}
    \end{minipage}\hspace{6pt}
    \begin{minipage}{0.4\linewidth}
        \centering
        \includegraphics[width=\linewidth]{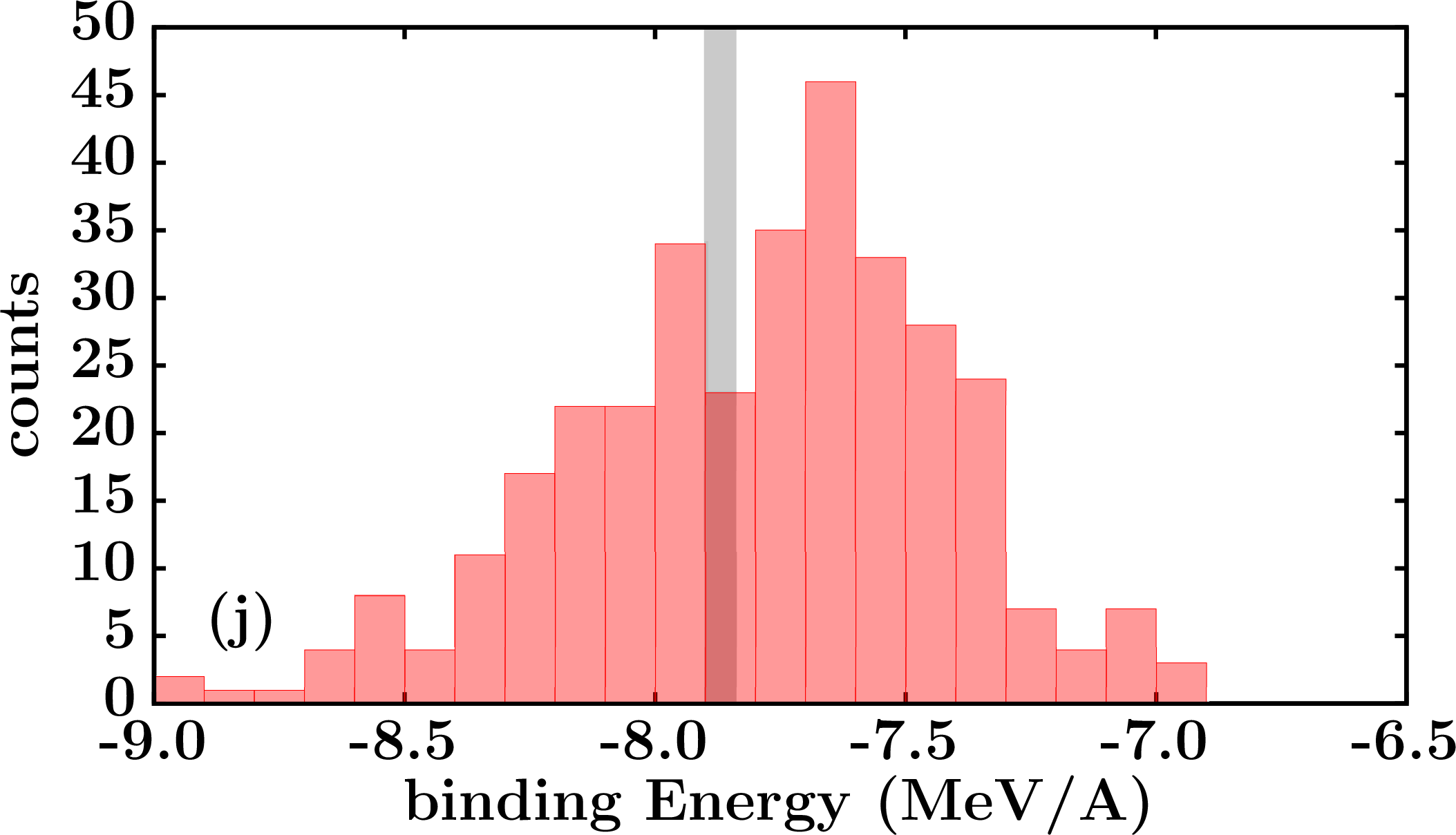}
        \label{DOM_pb208_BE}
    \end{minipage}
    \caption{\pbEight: constraining experimental data and DOM fit. See introduction of
    Appendix C for description.}
    \label{DOM_pb208}
\end{figure*}

\end{document}